# Severe plastic deformations, mechanochemistry, and microstructure evolution under high pressure: In Situ Experiments, Four-Scale Theory, New Phenomena, and Rules

Valery I. Levitas

*Iowa State University, Departments of Aerospace Engineering & Mechanical Engineering, Ames, Iowa 50011, USA*

**Abstract** *

This review focuses on recent breakthroughs in our understanding of complex and multifaceted interactions between high-pressure phase transformations (PTs) and chemical reactions (CRs), severe plastic deformations (SPD), and microstructure evolution. It is framed within the context of advanced multiscale mechanics and thermodynamics of materials subjected to stress and plastic strain tensors. Processes involving SPD and PTs and CRs under high pressures are prevalent for obtaining new nanostructured high-pressure phases and their bulk and surface processing, mechanochemical material synthesis, military applications, tribology, geology, and astrogeology. SPD under high pressure (a) significantly reduces the pressure for initiating and completing PTs and CRs, by one-two orders of magnitude; (b) leads to hidden metastable phases, which cannot be obtained by other methods, (c) reduces the PT pressure hysteresis, even down to zero, and (d) substitutes reversible PTs and CRs with irreversible ones, allowing engineering application of the obtained high-pressure phases. A novel concept of plastic strain-induced PTs and CRs under high pressure is explored in detail. These transformations initiate at defects (such as dislocation pileups, which have the strongest stress concentration and account for such drastic reduction in PT pressure), permanently generated during plastic flow. Four-scale theory and simulations (from atomistic to nano- and scale-free phase-field approach to macroscale), in situ experiments in traditional and rotational diamond anvil cells utilizing synchrotron-based X-ray diffraction, and their integration are discussed. The main rule is found that after SPD, the yield strength, minimum pressure for strain-induced PT, dislocation density, crystallite size, and microstrain reach /steady values, which are independent of plastic strain tensor and its path for broad classes of straining. However, these values differ for different but not yet defined, classes of straining representing the primary challenge. Various new phenomena are found and interpreted, and numerous misinterpretations and puzzles are resolved. Coupled analytical/computational/experimental approaches are developed for complete characterization of occurring processes and finding all heterogeneous scalar and tensorial fields. Various material classes (metals, ceramics, rocks, semiconductors, powders, etc.) are considered. Applications include suggested ways for economic defect-induced synthesis of high-pressure phases and nanostructures, high-pressure torsion, surface treatment (polishing, turning, cutting, etc.), high-pressure tribology, PTs and CRs in shear bands leading to severe transformation-induced and reaction-induced plasticity (TRIP/RIP) and self-blown-up SPD-PT-TRIP-heating processes, mechanisms of deep-focus earthquakes, and the appearance of microdiamonds in low-pressure-temperature Earth crust, and the mechanochemical origin of life beyond Earth. Unresolved issues and future directions are presented.

---

**Contents**

# 1. Introduction

## 1.1. Main processes and communities, and historical sketch

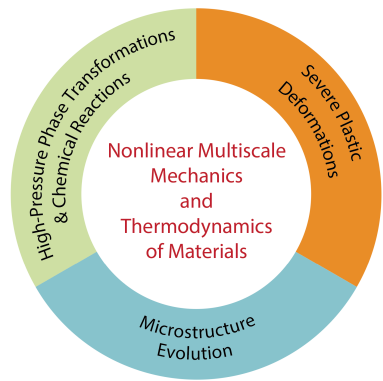

**Figure 1:** Chart showing the main processes discussed in this paper, their interactions, and the main approach at the center.

This review is devoted to understanding nontrivial interactions between high-pressure phase transformations (PTs), severe plastic deformations (SPD), and microstructure evolution from the point of view of advanced multiscale mechanics and thermodynamics of materials under general stress and plastic strain tensors (Fig. 1). Bridgman (1946 Nobel Prize in Physics for inventing high-pressure apparatuses and pioneering discoveries in high-pressure physics) invented principles of massive support (when loaded region is radially supported by surrounding material) and Bridgman anvils (Fig. 2a) [1–3] allowing to compress and study PTs, chemical reactions (CRs), deformation, and other processes in materials up to 10 GPa and even above, with WC anvils. It was found that under high pressure, even brittle materials can undergo large plastic deformations. To increase volume of compressed material, various types of anvil profiling were used with different shape of grooves (called cupped anvils [3], recessed anvils [4], etc.), see Fig. 2b, for relevant for this work grooves.

Next, Bridgman developed rotating Bridgman anvils, in which large plastic shear is superposed on axial compression [5–7]. He found that *plastic shear significantly promotes PTs and CRs in various classes of materials, i.e., reduces the PT/CRs pressure, accelerates CRs, and produces some PTs and CRs which were not observed under hydrostatic conditions.* Our main interest is in getting *comprehensive understanding of these phenomena by combining advanced in-situ experiment and multiscale theory.* Bridgman's studies were continued and extended in [8–17] for various organic and inorganic CRs. Bridgman also studied SPD under high pressure using various methods [2]. Grain refinement was also observed. However, generally, microstructure (grain size and dislocation density) evolution was mostly neglected by high-pressure community.

An alternative direction of application of SPD in rotating Bridgman anvils (called high-pressure torsion (HPT)) was in focusing on microstructure evolution, mostly producing nanograined materials [18, 19], see reviews [20–28]. Various PTs during HPT were also considered [23, 26–36], but still mostly from the point of view of microstructure evolution. All the above studies were based on ex-city (post-mortem) examination of the treated sample, like XRD, Raman and other types of spectroscopies, SEM and TEM, optical methods, etc.), with a few exceptions [37, 38]. Excellent review of Bridgman's works and following HPT works produced by SPD and nanostructured materials community was presented in [23]. In HPT, pressure was defined as applied force divided by anvil area (producing significant error due to strong pressure gradients) and plastic strain in terms of anvil rotation angle (i.e., neglecting material sliding with respect to anvil). FEM simulations of plastic flow (but not PTs) during HPP were performed to interpret some experimental results, see review [39]. High-pressure and SPD communities practically did not have interactions till recently.

The most advanced measurements and studies produced in rotational Drickamer apparatus (a development of Bridgman rotating anvils) were reported in geological literature [40–43]. Pressure and temperature reached 27.5 GPa and 2,150 K, respectively, for studying plastic flow in wadsleyite and ringwoodite [42]. *In situ* radial XRD was used to evaluate normal and shear stresses and texture, and X-ray radiography was utilized to determine normal and shear plastic strain. However, there is no free lunch: error bar for stresses and strains is very large (see Fig. 9a). Grain size was measured before and after experiment using SEM. In [41], polycrystalline behavior is modeled using viscoplastic self-consistent (VPSC) model and code [44, 45] to determine dominant slip system. Study in [43] was performed during wadsleyite to ringwoodite PT but without measuring kinetics. Kinetics of PT and grain growth was hypothetically described by time-dependent model without effect of plastic straining.

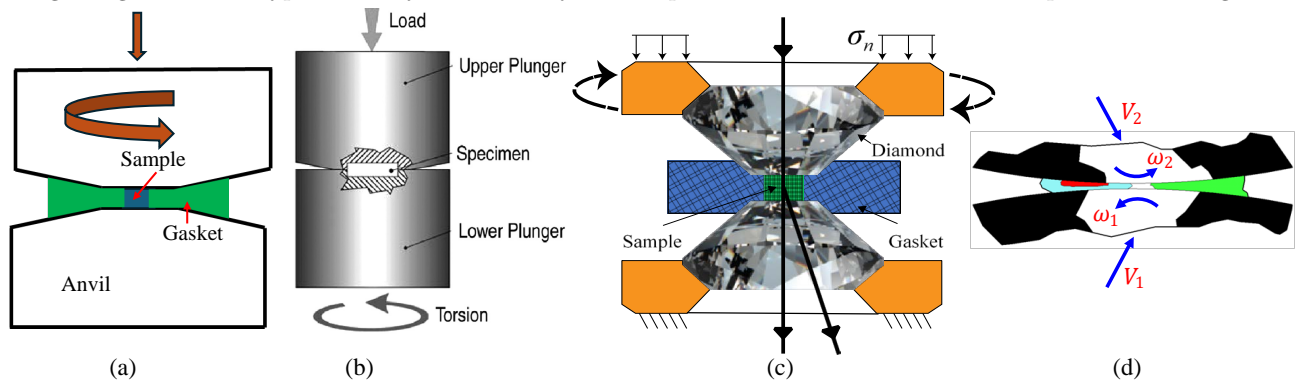

**Figure 2:** Schematics of Bridgman (without torque) and rotating Bridgman anvils (with torque) (a); rotating Bridgman anvils with grooves for quasi-constrained HPT [46] (b); DAC (without torque) and RDAC (with torque) [83] (c). (d) Scheme of deformation of material particles trapped between two colliding balls during ball-milling.

Independently, the author's group at the ISM worked on the development of the general continuum theory of large plastic deformations under high pressure and finding some general experimental rules [4, 47] and computational methods, namely, slip lines and FEM, for simulation of deformation of materials in Bridgman anvils, recessed anvils for diamond syntheses [4, 47–49], and diamond anvils [50, 51], and optimization of their design [52, 53], as well as modeling of industrial diamond synthesis [54, 55]. Microstructure evolution was not studied.

High-pressure research was revolutionized by invention of diamond anvil cells (DAC) [56], which not only allowed reaching pressures more than an order of magnitude higher (but in much smaller volume) but also to perform almost all possible studies in situ (XRD, X-ray absorption, Raman, pressure distribution, etc.), see reviews [57–60]. As next step, Blank added rotation to DAC [61–65], similar to rotating Bridgman anvils, and called this device shear DAC. The author's group at ISM was the second to apply rotational DAC (RDAC) for studying material behavior [66–68], but with theoretical insight from mechanics and materials. In 1999, the author brought RDAC to the USA and initiated RDAC-based research there. While Blank's group did not use in situ XRD in their studies [26, 64, 65, 69–73], the author's group intensively used in situ XRD [74–83], which gave significant advantages. Another difference is the author's group is continuously developing theory and computational approaches for interpretation, postprocessing, and guidance of experiments and connects PTs and microstructure evolution (see reviews [84, 85]), in contrast to Blank's group.

First in situ XRD study of interaction between strain-induced PT, namely, from hBN to superhard wBN, and defect structure (turbostratic stacking fault) was performed in [74, 86, 87]. Since 2017, the author participates in SPD and nanostructured material conferences and realized that there is also strong mutual interaction between strain-induced (and also pressure-induced, see [88]) PTs and grain size and dislocation density evolution. This resulted in the first in situ XRD measurements of distribution of these microstructural parameters along the sample diameter and finding the first rules of interaction between plastic flow, microstructure and PTs [81–83, 89, 90]. In situ Raman spectroscopy with RDAC was applied in [91, 91].

*1.2. Pressure-, stress-, and strain-induced PTs*

Independent of high-pressure efforts, the author's group has developed a strict continuum thermodynamic theory for PTs in elastoplastic materials [92–94] with corresponding analytical solutions and also FEM solutions at small [95] and large [96–99] strains, which was reviewed in [100, 101]. However, it was found in [100, 102] that this theory cannot describe even the order of magnitude of the reduction in PT pressure caused by plastic deformations in rotating Bridgman and diamond anvils. The first key message from [100, 102] was a *classification of high-pressure PTs and CRs*. Except for rare cases of homogeneous nucleation, heterogeneous nucleation at some defects (dislocations, disclinations, grain, subgrain, and twin boundaries) occur, which produce a concentration of tensor stress and provide some pre-existing surface energy. Temperature-induced PTs/CRs start predominantly at pre-existing defects without applied external stresses. Similarly, *pressure-induced PTs/CRs* occur by nucleation at the same pre-existing defects under applied hydrostatic pressure. *Stress-induced PTs/CRs* also initiate at the same defects when external nonhydrostatic stresses are applied but do not exceed the macroscopic yield strength. Since number of pre-existing defects is limited, one has to increase pressure/stresses to activate nucleation at defects with weaker stress concentrations.

If the PTs/CRs proceed during plastic flows and caused by plastic flow, they are classified as *plastic strain-induced* or *strain-induced* PTs/CRs. They occur by nucleation at new defects permanently generated during plastic flow. A similar classification (excluding pressure-induced PTs and any relation to high pressure) was used for martensitic PTs in steels [103] and was inspiring for the author. Strain-induced PTs in steels occur at shear-band intersection [103–106], but the magnitude of the thermodynamic driving force estimate due to this defect in [98] is not as high to explain drastic reduction in PT pressure.

The largest concentration of all components of the stress tensor $\sigma_{ij}$ can be generated at the tip of the dislocation pileups, where $\sigma_{ij} \sim N$ with $N$ for the number of dislocations in a pileup. Since $N$ can be 10 and even 100, dislocation pileup produces huge local stresses, increasing with increasing plastic deformation and $N$, which can drive the PTs and CRs at relatively low pressures. Thus, the PTs and CRs occurring during plastic flow in rotating Bridgman anvils or RDAC represent a separate class of *strain-induced PTs and CRs under high pressure*, which are very different from pressure- and stress-induced PTs and CRs and *require completely different thermodynamic and kinetic description as well as experimental characterization.* Due to prominent role of the defects, we suggest calling these PTs *defect-induced PTs under high pressure*, and corresponding material synthesis as a *defect-induced material synthesis*.

These works initiated study of what was called theoretical *high-pressure mechanochemistry* [100, 102, 107]. The analytical continuum theory in [100, 102] was developed at three scales: nanoscale (nucleation at the tip of dislocation pileup), microscale (derivation of the strain-controlled kinetic equation), and macroscale (deformational-transformational processes in DAC and RDAC).

During next two decades, atomistic simulations were added as the fourth scale; nanoscale and microscale (or scale-free) phase-field approaches (PFAs) were developed and applied for simulations of coupled phase and discrete dislocation microstructure evolution; and macroscale modeling and FEM simulations were performed for deformational-PT processes in DAC and RDAC, see review [84]. Atomistic and PFAs are needed for understanding the physical mechanisms and motivating microscale kinetic models. Macroscale FEM simulations are necessary for extracting quantitative information from the experiments. Since not all stress and plastic strain components can be in situ measured in DAC and RDAC, in situ XRD experiment and macroscale FEM simulations are combined to extract complete tensorial stress-strain and volume fraction of HPP fields [108, 109].

In addition, since *under compression in DAC without pressure-transmitting medium, large plastic deformations are generated, and PTs and CRs under such conditions should be treated as plastic strain-induced* theoretically [100, 102], experimentally [80–82] as well as computationally [110, 111]. This significantly expands the portion of high-pressure community involved in study of plastic strain-induced PTs and CRs (even if they still do not realize this), and requires *reconsideration of interpretation of numerous experiments in DAC.*

There is one more community producing mechanochemical synthesis of nanostructured materials via CRs and PTs by ball milling or mechanical attrition [112–119]. When material particles are hit multiple times by two balls (Fig. 2d), loading scheme is quite similar to that in RDAC, i.e., severe compression with shear, and observed effects are similar to those in rotating Bridgman anvils or RDAC mentioned at the beginning of Introduction. Similarity in material deformation and PTs under ball milling and HPT was analyzed in [120]. We will not go into details of how ball milling works, but wanted to mention that PTs and CRs produced by ball milling are plastic strain-induced, and many conclusions made here could be applicable for ball milling.

To summarize, there are 3 main processes: SPD, high-pressure PTs and CRs, and microstructure evolution, that strongly interact with each other; the main theoretical/computational tool, nonlinear multiscale mechanics and thermodynamics of materials (Fig. 1); and 3 communities, high-pressure (physics, chemistry, geophysics, and mechanics), SPD and nanostructured materials, and mechanics of materials, which weakly interact with each other. The goal of his review is to *present state of the art of relevant topics for each of 3 processes and their interaction from points of view of advance mechanics of materials.* There are excellent reviews on various aspects of SPD and nanostructured materials, including experimental studies [20–28, 121, 122], FEM simulations [39], and focused on PTs [36]; high-pressure research [58, 59] and RDAC research [64, 65, 123]. The author recently published reviews on theoretical aspects of pressure- and stress-induced PTs [101, 124] and strain-induced PTs under high pressure [84], on coupled experimental and theoretical aspects of strain-induced PTs under high pressure [85] and steady yield strength and microstructure during SPD [90]. Repetition with these papers will be minimized.

Processes occurring during SPD at high pressures are widespread in nature, high-pressure experiments (static experiments in Bridgman or rotating Bridgman anvils, DAC and RDAC or dynamic experiments), nature (geophysics and astrogeology), and modern high-pressure technologies. Examples are shown in Fig. 3; most of them will be discussed in proper places below.

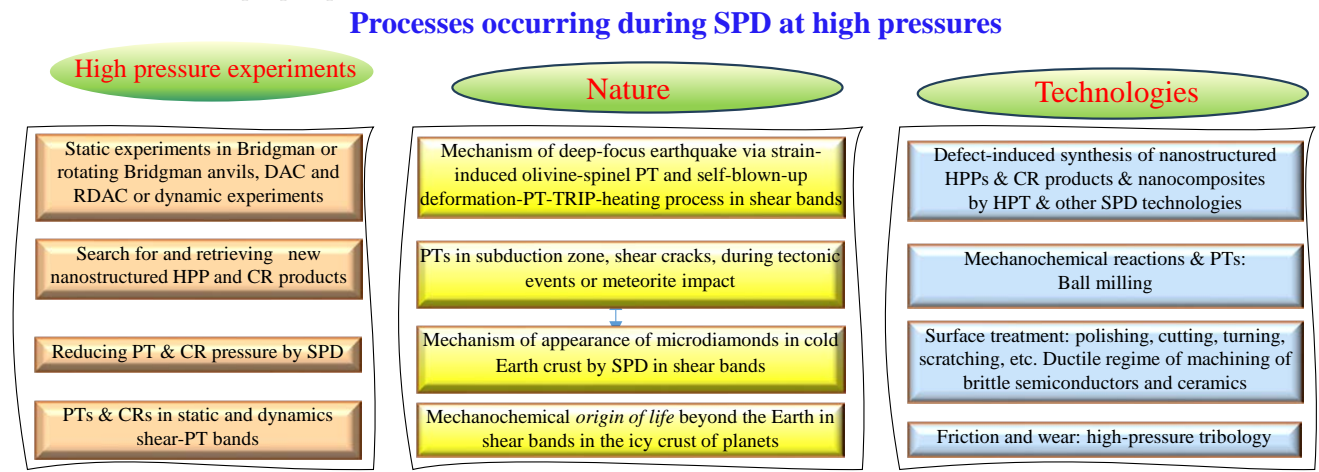

**Figure 3:** Processes occurring during SPD at high pressures.

*1.3. Pressure measurements*

The following main methods are used for pressure evaluation.

1. For HPT, pressure $p$ is reported as force $P$ divided by area $A$ of the flat portion of the anvil, $p_{av} = F/A$. Since large pressure gradients are present for unconstrained HPP (i.e., for anvils without grooves), pressure at the center can be more than 3 times larger than $p_{av}$ [125, 126]. For quasi-constrained HPT, pressure gradient is much lower; yet local pressure is not well defined. During HPT in [127], an averaged pressure was determined by detecting PT from Bi I to Bi II at 2.5 GPa and from Bi II to Bi III at 2.7 GPa. Also, FEM simulations were performed [39] for comparison with measurements.

2. Pressure distribution in DAC and RDAC is routinely measured using few-micron ruby particles spread over the sample surface and pre-calibrated pressure dependence of the shift of the $R_1$ ruby fluorescence line [62, 64, 66, 74, 128]. The most precise method is to locate pressure at points where the PT is just started, i.e., at the boundary between phases. In [62, 64, 66], the PT pressure is determined at steps, i.e., at small plateaus with almost constant pressure, which correspond to localized two-phase regions in which a PT occurs (see Section 3.2 and Fig. 19). In general, ruby-based methods disturb the stress-strain field of the sample material and shows pressure in the ruby particles rather than in the tested material, which may be essentially different. This is calibrated under hydrostatic conditions but used at nonhydrostatic conditions and plastic flow, which brings essential error.

3. The most currently popular method of determination of pressure distribution in *each phase* averaged over the sample thickness is based on in situ XRD measurement of X-ray patterns, calculating crystal cell volume, and utilizing known EOS. This method was routinely used for single-phase material even at megabar pressures [129–132]; it was applied for two-phase Ge, including gold particles [133, 134] and Zr [80] and even four-phase Si [81]. Pressures in phases may differ up to 5-7 GPa [81], which shows the importance of such measurements compared to ruby measurements. This method also has some problems. (i) Using the EOS obtained under hydrostatic conditions for strongly non-hydrostatic loading. (ii) For the axial XRD, axial strain/stress does not contribute to the measured X-ray patterns because crystallographic planes that are almost parallel to the beam contribute only.

4. Powder of some materials with well-defined EOS, like gold, is used to determine pressure. The drawbacks are mentioned above: gold disturbs local fields in material; pressure in gold is not equal to the pressure in the material phases, and with axial XRD, axial stress does not essentially affect measurements.

5. Raman spectra of the diamond were used for determination of the normal stresses at several points at the sample-diamond boundary for DAC [135–138] and RDAC [91]. The problem is that the effect of other components of the stress tensor on the Raman spectra is unknown, which lead to a scatter.

6. The 2D distributions of all components of the stress tensor were determined in a diamond cullet in contact with a sample using a nanoscale sensing platform integrating nitrogen-vacancy color centers [139]. This determines distribution of normal and shear contact stresses in the sample, which are continuous across the boundary. Some constants in the calibration methods have a scatter, which causes certain inaccuracy. FEM simulations and solution of the inverse problem were used for increasing accuracy and determining stress field in the entire anvil. Alternatively, axial stress and pressure within PTM were measured using immersed nano- or microdiamonds possessing nitrogen-vacancy color centers [140, 141].

7. The most promising combined CEA-FEM or experimental-FEM approaches [108, 109, 142, 143] are currently the only ways to determine all the tensorial stress-plastic strain fields in the sample and anvil, friction shear stress, and volume fraction of phases by fitting computational results to experiment and solving the inverse problem (see Sections 2.4 and 7.6). Method in [108, 109] does not use the EOS determined under hydrostatic loading to determine experimental pressure distribution. Fields of elastic radial and hoop strains are used as an experimental input.

We mostly focus on quasi-static material behavior at room temperature; however, some relevant references on high strain rate and temperature will be shortly analyzed as well. Presentation will be phenomena- and concept-centered with examples for different materials rather than specific material-focused. More complete material-focused information can be found in the cited papers. The author does not hesitate to show errors in their own works; everything evolves, and we correct our errors to make new ones. It is not a shame to be a fool but it is a shame to remain a fool. This also shows to other criticized authors that there is nothing personal here, just a desire to improve our understanding.

Direct tensor notations are used throughout this paper. Vectors and tensors $\boldsymbol{A} = \{A_{ij}\}$ with components $A_{ij}$ are denoted in boldface type; $\boldsymbol{A}\cdot\boldsymbol{B} = A_{ij}B_{jk} = A_{i1}B_{1k} + A_{i3}B_{3k}$ and $\boldsymbol{A}:\boldsymbol{B} = A_{ij}B_{ji} = A_{11}B_{11} + A_{12}B_{21} + A_{13}B_{31} + A_{21}B_{12} + A_{22}B_{22} + A_{23}B_{32} + A_{31}B_{13} + A_{32}B_{23} + A_{33}B_{33}$ are the contraction of tensors over one and two nearest indices; summation is assumed over the repeated indices. A superscript $-1$ denotes inverse operation, subscript $s$ designates symmetrization of the tensors, the indices 1 and 2 denote the values before and after the PT/CR; $\boldsymbol{I} = \{\delta_{ki}\}$ is the unit tensor with $\delta_{ij}$ for the Kronecker delta; $\delta_{ij} = 1$ for $i = j$ and $\delta_{ij} = 0$ for $i \neq j$. Any second-rank tensor can be decomposed $\boldsymbol{A} = A_0 \boldsymbol{I} + dev\boldsymbol{A}$ into spherical $A_0 \boldsymbol{I}$ ($A_0 = \frac{1}{3}(A_{11} + A_{22} + A_{33})$) and deviatoric $dev\boldsymbol{A} = \boldsymbol{A} - A_0\boldsymbol{I}$ parts. For true (Cauchy) stress $\boldsymbol{\sigma}$ and small strain $\boldsymbol{\varepsilon}$ tensors, these decompositions are

$$\boldsymbol{\sigma} = \begin{pmatrix} \sigma_{11} & \sigma_{12} & \sigma_{13} \\ \sigma_{21} & \sigma_{22} & \sigma_{23} \\ \sigma_{31} & \sigma_{32} & \sigma_{33} \end{pmatrix} = \begin{pmatrix} \sigma_0 & 0 & 0 \\ 0 & \sigma_0 & 0 \\ 0 & 0 & \sigma_0 \end{pmatrix} + \begin{pmatrix} s_{11} & s_{12} & s_{13} \\ s_{21} & s_{22} & s_{23} \\ s_{31} & s_{32} & s_{33} \end{pmatrix}; \quad \boldsymbol{\varepsilon} = \begin{pmatrix} \varepsilon_{11} & \varepsilon_{12} & \varepsilon_{13} \\ \varepsilon_{21} & \varepsilon_{22} & \varepsilon_{23} \\ \varepsilon_{31} & \varepsilon_{32} & \varepsilon_{33} \end{pmatrix} = \begin{pmatrix} \varepsilon_0 & 0 & 0 \\ 0 & \varepsilon_0 & 0 \\ 0 & 0 & \varepsilon_0 \end{pmatrix} + \begin{pmatrix} e_{11} & e_{12} & e_{13} \\ e_{21} & e_{22} & e_{23} \\ e_{31} & e_{32} & e_{33} \end{pmatrix}, \quad (1)$$

where $\sigma_0 = \frac{1}{3}(\sigma_{11} + \sigma_{22} + \sigma_{33}) = -p$ is the mean stress, $p$ is the pressure, $\varepsilon_0 = \frac{1}{3}(\varepsilon_{11} + \varepsilon_{22} + \varepsilon_{33}) = \frac{1}{3}\varepsilon_v$ is the mean strain, $\varepsilon_v$ is the volumetric strain, $s_{ij}$ and $e_{ij}$ are the components of the deviatoric stress $\boldsymbol{s}$ and strain $\boldsymbol{e}$ tensors. The same decomposition and meaning are preserved for small elastic, plastic, and transformational strains, and for deformation rate $\boldsymbol{d}$ and its contributions (even for finite strains). Symmetric second-order tensors have six independent components and often presented as 6D vectors, and called vectors.

## 2. Large plastic deformations under high pressure

*2.1. Perfectly plastic, isotropic, and strain-path-independent material behavior during SPD*

Various aspects of plastic flow under high pressure, like stress-strain curves and texture evolution, are broadly studied and discussed in literature [2, 20, 24, 26–28, 40–42, 132, 144–146]. *The main fundamental challenge in studying plasticity (and related microstructure evolution, see Section 2.7) is that they are affected not only by the achieved plastic strain tensor $\boldsymbol{\varepsilon}_p$ (which is already very complex) and pressure $p$, but also on the entire straining path $\boldsymbol{\varepsilon}_p^{path}$ and pressure path $p^{path}$* (e.g., various combinations of generally multiaxial non-proportional unequal compressions and shears and pressure). Pressure and its path are included here because, due to the plastic incompressibility constraint, only five components of $\boldsymbol{\varepsilon}_p$ are independent, and pressure is the reaction of the incompressibility constraint. Also, the effect of pressure on the microstructure is hysteretic, i.e., path-dependent (Section 4.3 and [88]). The vast number of independent loading parameters offers a faint prospect for identifying any overarching general rules. Here, we focus on proving the validity of the following hypothesis for material behavior during complex SPD: *after some critical level of plastic deformation, initially isotropic polycrystalline materials behave like perfectly plastic, isotropic, and plastic strain-path-independent with the corresponding limit surface of perfect plasticity.* This hypothesis was formulated, and to some extent justified, in [4, 47]. The detailed review on the topic, including normal pressure, is presented in [90]; here, the short version with some developments will be given.

Von Mises equivalent stress is defined as $\sigma_i = (3/2\boldsymbol{s}:\boldsymbol{s})^{0.5} = (3/2 s_{ij} s_{ji})^{0.5}$ and the equivalent plastic strain $q$, is defined by $\dot{q} = (2/3\boldsymbol{d}_p:\boldsymbol{d}_p)^{0.5} = (2/3 d_p^{ij} d_p^{ji})^{0.5}$, where $\boldsymbol{d}_p$ is the deviatoric parts of the the plastic strain rate. For 1D loading, for strain-hardening (or softening) behavior, the yield strength increases (or reduces) with increasing plastic strain. Perfectly plastic? behavior corresponds to the constant stress (Fig. 4a). For general 3D loadings, the evolving yield surface in the stress space, $f(\boldsymbol{\sigma}, \boldsymbol{\varepsilon}_p, \boldsymbol{\varepsilon}_p^{path}) = 0$, describes plastic behavior; inside this surface, deformation is elastic; for plastic straining, stress vector (tensor) should belong to the yield surface. Thus, material behavior depends on $\boldsymbol{\varepsilon}_p$ and the entire straining path $\boldsymbol{\varepsilon}_p^{path}$, and material acquires strain-induced anisotropy due to texture and internal stresses (back stresses). For hardening (softening) materials, the yield surface expands (shrinks) in the stressing direction. For perfectly plastic material, $f(\boldsymbol{\sigma}) = 0$ is fixed. According to the above hypothesis, the yield surface $f(\boldsymbol{\sigma}, \boldsymbol{\varepsilon}_p, \boldsymbol{\varepsilon}_p^{path}) = 0$ evolves during plastic straining but after some strain becomes independent of plastic strain and strain path, i.e., $f(\boldsymbol{\sigma}) = 0$. Function $f(\boldsymbol{\sigma})$ and surface $f(\boldsymbol{\sigma}) = 0$ can be the isotropic function and surface only, because for initially isotropic material, anisotropy could be strain-induced only. Indeed, since $\boldsymbol{\varepsilon}_p$ and $\boldsymbol{\varepsilon}_p^{path}$ are not arguments of $f(\boldsymbol{\sigma})$, there is no possible way to include the anisotropy. However, this is clearly contradictory because all materials after SPD possess strain-induced texture and back stresses, i.e., anisotropy. To resolve this contradiction, an additional fixed isotropic limit surface of perfect plasticity $\varphi(\boldsymbol{\sigma}) = 0$ was introduced in [4, 47] (Fig. 4b,c).

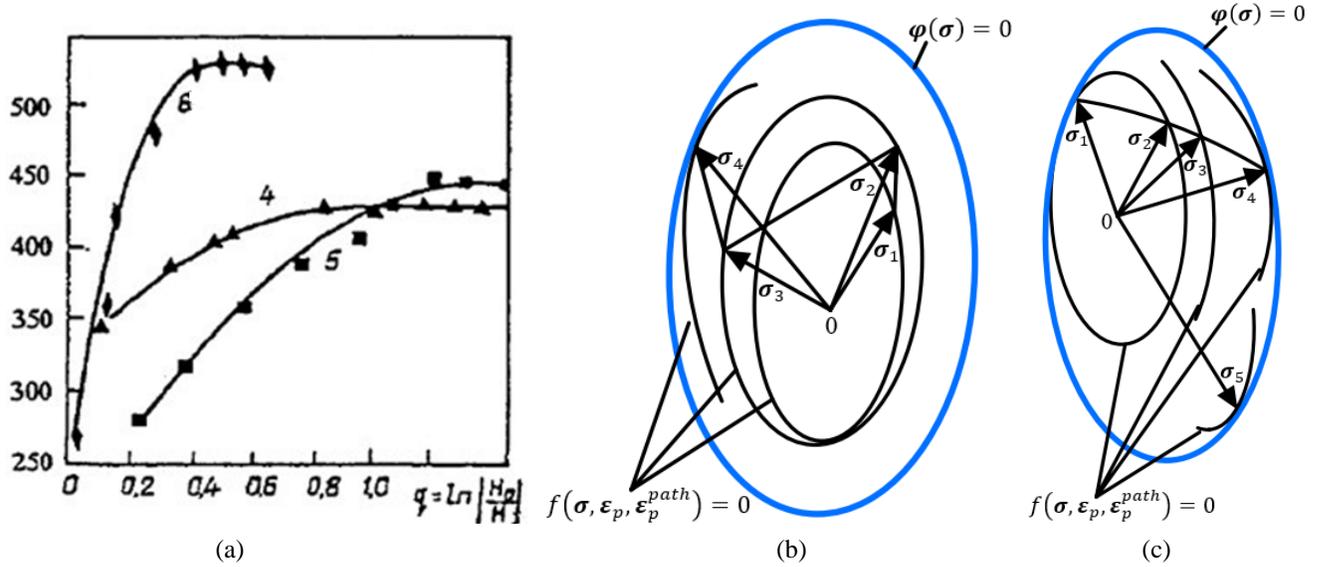

**Figure 4:** (a) Stress-strain curves for uniform compression of samples with cylindrical grooves on each face filled with three-layer lubricant for copper M2 (4), steel St 3 (5), and cast iron VCh-50 (6) [150]. Symbols are from the experiments, and curves are analytical approximations. Horizontal plateau corresponds to the transition to the perfectly plastic behavior. (b) Evolution of the yield surface $f(\boldsymbol{\sigma},\boldsymbol{\varepsilon}_p,\boldsymbol{\varepsilon}_p^{path}) = 0$ for non-monotonic straining until it reaches the isotropic limit surface of perfect plasticity $\varphi(\boldsymbol{\sigma}) = 0$, after which material behaves *like* perfectly plastic, isotropic, and plastic strain-path-independent. (c) After unloading and reloading in a different direction, material behaves according to its yield surface $f(\boldsymbol{\sigma},\boldsymbol{\varepsilon}_p,\boldsymbol{\varepsilon}_p^{path}) = 0$. However, after small strain increment, $q^* \simeq 0.1$, the stress vector $\boldsymbol{\sigma}$ and the yield surface $f(\boldsymbol{\sigma},\boldsymbol{\varepsilon}_p,\boldsymbol{\varepsilon}_p^{path}) = 0$ reach again $\varphi(\boldsymbol{\sigma}) = 0$, and the material again deforms *like* perfectly plastic, isotropic, and strain path-independent. Reproduced with modifications from [4].

Above some critical strain, for monotonic or nonmonotonic straining (i.e., loading paths without or with sharp changes in directions), the stress vector $\boldsymbol{\sigma}$ and the anisotropic yield surface $f(\boldsymbol{\sigma},\boldsymbol{\varepsilon}_p,\boldsymbol{\varepsilon}_p^{path}) = 0$ touch the fixed isotropic surface $\varphi(\boldsymbol{\sigma}) = 0$ and then move along it. This means that the initially isotropic materials deform *like* perfectly plastic and isotropic with a $\boldsymbol{\varepsilon}_p^{path}$-independent limit surface of the perfect plasticity $\varphi(\boldsymbol{\sigma}) = 0$ (Fig. 4b). However, actual evolving yield surface $f(\boldsymbol{\sigma},\boldsymbol{\varepsilon}_p,\boldsymbol{\varepsilon}_p^{path}) = 0$ can be found during unloading and loading in different directions (Fig. 4c). Nevertheless, after relatively small strain increment, $q^* \simeq 0.1$, the stress vector $\boldsymbol{\sigma}$ and the anisotropic yield surface $f(\boldsymbol{\sigma},\boldsymbol{\varepsilon}_p,\boldsymbol{\varepsilon}_p^{path}) = 0$ reach again the same surface $\varphi(\boldsymbol{\sigma}) = 0$, and the material again deforms *like* perfectly plastic, isotropic, and strain path-independent. Such a loading process that continues until $\varphi(\boldsymbol{\sigma}) = 0$, followed by unloading and then reloading until $\varphi(\boldsymbol{\sigma}) = 0$ is reached again, is called a quasi-monotonic loading in [4, 47]. The following *experimental confirmations* for such a material behavior were presented in [4, 47].

(a) *Monotonous loading.* A thin disc-shaped sample with the initial thickness $h_0$ was compressed between the Bridgman anvils down to the final thickness $h$ by force that increases to the same value $P$ for all $h_0$ [147–149]. Dependence $h(h_0)$ at $P = const$ was determined, which, after some increasing or decreasing branches, ended with plateau. These experiments were analyzed in [4, 47]. Taking into account complex and heterogenous stress-strain states and their evolution during compression and analyzing all possible types of $h(h_0)$ curves, it was concluded that the independence of $h$ on $h_0$ can be explained only if material behaves like perfectly plastic, isotropic, and independent of $\boldsymbol{\varepsilon}_p^{path}$. This was obtained for 43 materials studied in [147–149], including metals (In, Al, Pb, and Cu), oxides ($MnO_2$, $SnO_2$, $MoO_2$, $ZrO_2$, and $MgO$), chlorides ($NaCl$, $AgCl$, $KCl$), $KBr$, $KJ$, different types of pyrophyllite, including additions of NaCl and $Fe_2O_3$, $B$, $BN$, $B_4C$, two types of graphite, $Al(OH)_3$, $Ca(OH)_2$, $H_3BO_3$, marble,talc, kobushnendo clay, micarex, polystyrole, teflon, pressboard, and two types of cardboard. Plateau was not observed for fabric cloth-based laminate, which is initially anisotropic material; this shows a sufficient sensitivity of the method. Additionally, more general experiments in [4, 47] exhibited a plateau for blocked and compacted with the Bakelite varnish and polyvinyl alcohol limestones ($CaCO_3$). The magnitudes of the strain $m = ln(h_0/h)$ for transition to the perfectly plastic regime vary from 0.368 (compacted pyrophyllite powder) to 1.262 (armco-iron).

Using a special sample for homogeneous compression for large strains, stress-strain curves reaching perfectly plastic plateau were obtained for steels St 3, 12KhN3A, 45, and U8A, copper M2, and cast iron VCh-50 in [150]. Value $m$ was just 0.44 for cast iron, 0.82 for steel U8A, and in the range of 1.1 to 1.57 for the rest of the studied metals. The yield strength $\sigma_y$ increased from the initial to the steady value by a factor of 2 -2.5.

(b) *Quasi-monotonous straining.* The first experimental justification in [4, 47] of perfectly plastic, isotropic, and strain path-independent behavior for quasi-monotonic straining was based on reaching steady hardness after various deforming broaching regimes and compression and tension till fracture of cylindrical specimens [151], and by different surface strengthening techniques [152]. Deforming broaching consists of multiple (up to 40) broachings through bushing a broach with diameter larger than the hole diameter. By varying the friction conditions, geometric parameters, interference, and the number of runs, surface layer of bushing was subjected by several hundred percent of deformation with complex non-monotonic paths. The measured hardness $H$ of a thin layer on the internal surface of the bushings increased with a number of runs, reached the maximum, and did not change with further deformation for 6 ductile steels 10, 20, 30KhGSA, 40KhNMA, 38KhNMYuA, and U8, aluminum alloys D16T and AK6, and 3 low ductile cast iron V93 and V95, and magnesium alloy Ml.5pch (composition of each can be found in [90]). The steady hardness for these materials was independent of all parameters that were varied, i.e., of $\boldsymbol{\varepsilon}_p$ and $\boldsymbol{\varepsilon}_p^{path}$. What is more important, the same hardness for each of these 8 ductile metals was obtained in the broken neck of the samples for tensile, and for each of these 3 low-ductile metals in the broken surface of the compressed samples. Also, in [153], the same hardness as at fracture surfaces was obtained at facet formed when the apex of a conical sample is smashed along its axis. This adds completely different straining paths to the result of the independence of the steady hardness of the loading path. The known relationship for the yield strength $\sigma_y = kH$, where a factor $k$ is in the range from 1/3 to 0.386 [154, 155], implies that $\sigma_y$ is also getting steady and independent of $\boldsymbol{\varepsilon}_p$ and $\boldsymbol{\varepsilon}_p^{path}$, i.e., proves the above hypothesis of perfect plasticity under normal pressure. Loading after any SPD process followed by hardness measurement is nonmonotonic. Since for Vickers hardness measurement, strain increment is in the range 0.02 and 0.3, the value $q^*$ in the definition of the quasi-monotonic loading and necessary to reach the surface $\varphi(\boldsymbol{\sigma}) = 0$ after unloading and reloading in different direction (Fig. 4c) is of the same order of magnitude. Studying of the Bauschinger effect at tension-compression in [156] allowed to narrow this value to $q^* \simeq 0.1$ in [4, 47].

(c) *Steady torque and hardness during the HPT.* Pioneered by Bridgman [2] and followed by modern HPT efforts [20–23, 25–27], the initial increase in the torque during rotations of an anvil at constant applied force $P$ saturates and leads to a steady value. It was assumed in most of the early works that normal and shear contact stresses are distributed uniformly, pressure is equal to force divided by area, and the torque $M$ is simply expressed in terms of shear friction stress $\tau_f$, assumed to be equal to the yield strength in shear $\tau_y$. As will be discussed below, this is not the case. Still, qualitatively, these results are consistent with the transition to the perfectly plastic behavior, but the independence of the straining path and isotropic behavior was not discussed and could not be derived due to shear-dominated straining. However, there are serious doubts that steady torque implies steady $\tau_y$. Indeed, torque is proportional to $r^3$, i.e., it gets a major contribution from material outside of the anvil flat faces for Bridgman anvils and from the flash region for quasi-constrained HPT, where pressure is low, and $\tau_f < \tau_y$. Linear relation between the steady $M$ and $P$ in [157] was connected to the growth in Coulomb friction coefficient $\mu$ in the flash region.

FEM simulations of quasi-constrained HPT of Cu in [158] also implied that the relation between the torque and rotation angle and anvil $\varphi_a$ strongly depends on $\mu$ in the flash and anvil, which was found to be just 0.12. Therefore, shear stress-strain relations derived from $M - \varphi_a$ curves possess essential inaccuracy. Thus, the plastic strain for reaching the perfectly plastic behavior, $m$, from the HPT experiments is 5-35 for different metals [127, 157, 159] (compared to $m \leq 1.6$ for compression experiments reported above) because $M - \varphi_a$ curves weakly depend on $m$ and significant sliding near contact surfaces.

Since after HPT with SPD, a uniform hardness (and microstructure) is achieved in the external portion of the specimen, but the shear strain is proportional to the radius, the hardness (and microstructure) is independent of the plastic strain, but it is impossible to prove that it is independent of $\varepsilon_p^{path}$. In fact, Bridgman [2] claimed that saturation to zero strain hardening is reached for simple shear only. Note that steady hardness (and microstructure) was obtained for each of the methods of SPD, like equal-channel angular pressing, ball milling, twist extrusion, multi-directional forging, etc. [26, 27].

(d) *Torsion in RDAC*. The pressure distribution obtained with the ruby particles distributed over the sample surface did not visibly change at $P = const$ during twisting increment in RDAC for NaCl at 12 GPa in [62, 64] and after several compression-unloading-twisting paths ending with the same sample thickness for NaCl up to 12 GPa and for hardened stainless steel up to 35 GPa in [66]. This is consistent with perfectly plastic and isotropic behaviour in analytical [4, 47] and FEM [160] solutions. While for experiments in [62, 64] this does not prove the strain path-independence, experiments in [66] also prove the independence of $\varepsilon_p^{path}$ for the same paths. The change in pressure distribution due to change in $\sigma_y$ is well detectable experimentally [62, 64, 66].

*2.2. Determination of the pressure-dependent yield strength with HPT; concave yield surfaces*

Determination of the pressure-dependence of $\sigma_y$ is the major problem of the mechanics of plastic flow under extreme parameters. In particular, it determines (a) the maximum level of friction stresses, (b) the maximum pressure achievable in DAC and other high-pressure apparatuses, and (c) is the main material parameter in the modeling and simulation of stress-strain states of materials deformed in DAC, RDAC, various technological processes (forging, extrusion, HPT and other SPD technologies, high-pressure material synthesis, polishing, cutting, and ball milling) and in tectonic processes in planets including Earth. If material behaves like perfectly plastic, isotropic, and independent of $\varepsilon_p$ and $\varepsilon_p^{path}$, then its $\sigma_y$ and $\tau_y$ depend on pressure only; many experiments for single-phase materials are well-described by linear pressure dependence [4, 47, 89, 148, 161–164]. Then the pressure-dependent von Mises condition, the Drucker-Prager condition [165–167], is

$$\sigma_i = \sigma_y = \sigma_\infty + bp; \qquad \tau_y = \sigma_y/\sqrt{3} = (\sigma_\infty + bp)/\sqrt{3}. \tag{2}$$

Here, $\sigma_\infty$ means the yield strength at zero pressure and theoretically infinite plastic strains, i.e., after reaching perfectly plastic state. The key point is that *if the stress state of the sample compressed in DAC or Bridgman anvils is well described by this condition, then material behaves like perfectly plastic, isotropic, and independent of $\varepsilon_p$ and $\varepsilon_p^{path}$*. Note that the factor $\sqrt{3}$ comes from the application of von Mises or the DruckerPrager conditions. In many cases, the factor of 2 is used instead, which corresponds to the Tresca maximum shear stress criterion, which, however, is never used in FEM codes for 3D problems. Tresca, or its normal stress-dependent version, Coulomb-Mohr criterion, is used for 2D plane strain/stress and axisymmetric problems using slip line method [4, 47]. Under pressure, transition from one criterion to another involves $b$ [4, 47]. Consistency between the type of the yield condition and a factor connecting $\sigma_y$ and $\tau_y$ is important, because for FEM solution of the problem, $\sigma_y$ is used in the yield condition and $\tau_y$ is used for the description of the plastic contact friction. Interestingly, while an engineer, Tresca obtained his criterion experimentally, Tresca's prism in the 6D stress space is described by 6 different planes (equations), which is difficult to use for the solution of the boundary-value problems. That is why an applied mathematician, von Mises, substituted it with the cylinder that passes through edges of the prism and is described by a single equation. It was later found that for many ductile materials, the von Mises condition better describes experiments than the Tresca criterion [166, 167].

HPT at different pressures was routinely used to determine $\tau_y(p)$ and, consequently, $\sigma_y(p)$ for numerous materials using assumptions from Section 2.1(c) [2, 5, 23, 168–171]. The relationships $\tau_y(p)$ for different materials can be divided into four classes. Linear relationships are the most broadly observed ones, at least in some pressure range. For convex functions $\tau_y(p)$, the pressure hardening is less intense than for the linear one. In contrast, for concave upwards functions $\tau_y(p)$, the pressure hardening is more intense than for the linear relationship, e.g., for steels 45 and 17X18H9, W, pyrophyllite, paraffin, and graphite, among others [2, 5, 170, 171]. Bridgman [2] considered such a concavity to be a general property for the majority of materials above 10 GPa and at much lower pressures for noncrystalline organic substances. For the fourth class, $\tau_y(p)$ has some jumps or waviness caused by PTs (e.g., in Bi [5] or polymers [172]), CRs, or other structural changes, which also include concave parts. The main basic problem formulated in [173, 174] was that such concave yield surfaces contradict all main postulates of the plasticity theory [165–167], like von Mises, Drucker, Il'yushin postulates, and postulate of dissipation [4, 47], which require the yield surface to be nonconcave, and the plastic strain rate vector to be along the normal to the yield surface (associated flow rule). All theorems and principles of the plasticity theory, like existence and uniqueness of solution, and various extremum principles characterizing solutions of the boundary-value problems, are formulated for a nonconcave yield surface and associated flow rules. These theorems are not just mathematical exercises, but have important practical implications, as it was shown by semi-qualitative analysis of the slip lines for the solution of the compression of a sample in Bridgman anvils [4, 47]. Thus, for the concave function $\tau_y(p)$, slip-line field, which is found from the lateral boundary toward the sample center, cannot reach the center, i.e., plastic solution does not exist in the central region. In this case, material is rigid and center (within slip-line method), and thickness cannot be further reduced by rigid anvils. For elastoplastic model, deformation in the central region is elastic. This could explain (at least partially) experimentally-observed transition from plastic to elastic deformation at the center of the sample compressed to megabar pressures in DAC [131, 132] (Fig. 7c). However, more detailed FEM solutions for nonconcave $\tau_y(p)$ were not performed, and transition from plastic to elastic deformation at the center of the sample in [131, 132] was described in [142, 143, 175, 176] by models with linear $\tau_y(p)$ due to large elastic deformation of diamond anvils and cupping.

Theory suggested in [173, 174] and elaborated in [4, 47] for finite strains, is based on assumption that the dissipation rate $D(\boldsymbol{d}_p, \chi)$ and, consequently stress $\boldsymbol{\sigma}(\boldsymbol{d}_p, \chi) = \frac{\partial D}{\partial \boldsymbol{d}_p}$ depend on some structural parameter(s), describing PT, reactions, or change in dislocation structure, which depend on $\boldsymbol{\sigma}$, i.e., $\chi = \chi(\boldsymbol{\sigma})$. Then the non-associated flow rule and local and global extremum principles were derived using formulated dissipation postulate at the fixed structure $\chi(\boldsymbol{\sigma})$. Generally, this results in non-explicit constitutive equation for stress $\boldsymbol{\sigma} = \boldsymbol{\sigma}(\boldsymbol{d}_p, \chi(\boldsymbol{\sigma}))$. But if $\chi$ depends on pressure only, then for plastically incompressible material the constitutive equation $\boldsymbol{s} = \boldsymbol{s}(\boldsymbol{d}_p, \chi(p))$ is explicit and was used to describe the experimental concave yield surface for 2 steels and W.

The above concave functions $\tau_y(p)$ were not confirmed by more modern methods reported below. However, for each two-phase material with stronger HPP and/or with larger $b$ for HPP, e.g., for $\alpha$ and $\omega$ phases of Zr and Ti and $\alpha$ and $\varepsilon$ phases of Fe [177], the yield surface theoretically either has jump or smooth transition $\tau_y(p) = (1 - c(p))\tau_{y1} + c(p)\tau_{y2}$ with the volume fraction of the HPP $c$, i.e., it has concave parts, in which curve $\tau_y(p)$ is below a straight line connecting two points at the curve $\tau_y(p)$. In this case, all problems related to nonconcave yield surface and above models are in principle applicable. Such dependence is correct for pressure-induced PTs, for which $c = c(p)$. However, during plastic flow, strain-induced PTs occur [100, 102], for which $c$ is determined by integration over $q$ of the kinetic equation Eq. (37), which depends also on $p$, but one cannot say that $\sigma_y$ is concave function of $p$. That is why FEM simulations for Zr [108, 109] do not exhibit transition from plastic to elastic deformation during compression in DAC.

Note that associated flow rule for pressure-dependent yield surface leads to volumetric plastic dilatation (i.e., porosity), which is important, e.g., for soils and geomaterials under relatively low pressure. Usually, it overestimates dilation; that is why non-associated flow rules with lower dilation rate, which do not obey the above postulate of the plasticity theory, are used even for convex yield surfaces [167]. In models [173, 174], a non-associated flow rule is derived with explicit flow potential for concave yield surfaces using the postulate of dissipation. Under elevated pressure, the porosity is reduced or eliminated, e.g., by introducing additional cap that limits the yield surface in direction of the hydrostatic pressure [167, 178]. For high pressures of the interest here, porosity should be eliminated as soon as it appears. This can be achieved by non-smooth connection of the main (e.g., Prager-Drucker yield surface) and a cap, or just postulating plastic incompressibility conditions, as it is done in all FEM simulations under high pressure [39, 48, 49, 142, 143, 175, 179, 180].

Despite this, the conceptual idea introduced in [173, 174] to utilize structural parameters depending on stresses that are not thermodynamically conjugate to the plastic deformation rate was quite fruitful. A similar idea was used in [181, 182] to express dependence of the dislocation core structure (or Peierls barrier) for the chosen slip system on stresses other than the resolved shear stress for this system, which described the known non-Schmid effects and resulted in non-associated flow rules for single crystal plasticity.

*2.3. Simplified methods to determine the pressure-dependent yield strength in DAC*

The basis for a simplified analytical study of the radial pressure distribution within a thin disk-shape sample compressed by rigid anvils is the simplified equilibrium equation

$$\frac{\partial p}{\partial r} = -\frac{2\tau_{fr}}{h}. \qquad (3)$$

Here $r$ is the radial coordinate, $h$ is the thickness of the sample, and $\tau_{fr}$ is the radial component of the shear frictional stress $\boldsymbol{\tau}_f$ on the sample-anvils boundary. For a thin sample, the $|\boldsymbol{\tau}_f|$ usually assumed to reach its possible maximum value equal $\tau_y$ (which is not true for diamond anvils, see Section 2.6). For axisymmetric compression, there are only radial components of the friction. Let at the external radius of the anvil $r = R$, $p = \sigma_o + \sigma_y$ as the boundary condition, where $\sigma_o$ is the radial stress at $r = R$ due to the external support of the material at $r > R$. Elastic deformations and pressure inhomogeneity in the axial $z$-direction are neglected. Then Eq.(3) for $\sigma_y = const$ (i.e., $b = 0$) yields the linear pressure distribution with maximum at the center:

$$\frac{\partial p}{\partial r} = -\frac{2}{\sqrt{3}}\frac{\sigma_y}{h}; \qquad p = \sigma_o + \sigma_y\left(1 + \frac{2}{\sqrt{3}}\frac{R-r}{h}\right). \qquad (4)$$

For linear dependence (2), exponential pressure distribution is obtained [3]. It is interesting to note that usually plasticity relaxes stress concentrators, e.g., at notches, voids, or crack tips. The above problem gives an opposite example of how the contact friction and plastic flow produce strong stress (pressure) concentration, which is the basis of modern methods of producing very high pressures.

Eqs. (3)-(4) were derived in [183] and were broadly explored in the textbooks on metal forming [184, 185], and numerous papers [147, 186, 187], see the textbook [3]. They were routinely used at the ISM for design of Bridgman anvils and anvils for diamond synthesis with inclined surfaces [149, 188–190] and determination of $\sigma_y(p)$ for different materials in Bridgman anvils [148] and DAC [4, 47]. The most cited and inspiring for the community works on the determination of $\sigma_y(p)$ in DAC for $MgO$ based on Eq. (3) was [161] ; see also [162–164] for $NaCl$, mantle silicates, and rhenium. Note that the pressure gradient in this work was an averaged value over the radius (excluding central and peripheral regions), and pressure was taken at the sample center. In contrast, in [4, 47], the entire pressure distributions were fitted with the determined $\sigma_y(p)$.

In most works, except [4, 47, 89], care was not taken that the perfectly plastic behavior is reached; that is why the yield condition also depends on $\varepsilon_p$ and $\varepsilon_p^{path}$. Since this dependence was not taken into account, it was somehow attributed to the pressure dependence of $\sigma_y$, introducing significant error. Increasing pressure and, consequently, plastic strain reduced this error down to zero for $q > m$. This comment is related to most of the other methods described below. Recently [89], equilibrium Eq.(3) was generalized by considering normal stresses instead of pressure and their heterogeneity along $z$-axis:

$$\frac{\partial p}{\partial r} = -A\frac{\sigma_\infty + bp}{h}; \quad A = \frac{1(1 + 0.524b)}{\sqrt{3}(1 - 0.262b)} \rightarrow p = \left(p_0 + \frac{\sigma_\infty}{b}\right) exp\left(-Ab\frac{r - r_0}{h}\right) - \frac{\sigma_\infty}{b}, \qquad (5)$$

where $p_0$ is the pressure at point $r_0$. For Bridgman anvils, pressure distribution could not be determined. Still, a simple method was developed [4, 47] to extract $b$ and $\sigma_\infty$ separately by fitting to experimental $P(h)$ curves after they are getting independent of $h_0$ (i.e., reaching a perfectly plastic state). In addition to the simplified Eq. (5) with $A = 1/\sqrt{3}$, a more advanced numerical solution with the slip-line method was used, but results were quite close. Note that slip-line method is much faster than FEM but is limited to plane and axisymmetric problems and determination of the stress (but not strain) state only. Results were obtained for pyrophyllite ($\sigma_\infty = 0.04$ GPa and $b = 0.525$ GPa$^{-1}$), blocked limestone ($CaCO_3$, $\sigma_\infty = 2.44$ GPa and $b = 0.293$ GPa$^{-1}$), compacted with the Bakelite varnish limestone ($\sigma_\infty = 1.06$ GPa and $b = 0.332$ GPa$^{-1}$) and close numbers for the limestone compacted with the polyvinyl alcohols. The methods based on Eq. (5) with $A = 1/\sqrt{3}$ and the slip-line solutions were utilised in [4, 47] to fit distribution of pressure for a T301 stainless steel sample squeezed in DAC up to 170 GPa [191]; it was obtained $\sigma_\infty = 4.48$ GPa and $b = 10^{-4}$ GPa$^{-1}$.

The obtained data on $\sigma_y(p)$ were used for the sleep-line simulations of the stress state of the deformable gasket and optimization of the high-pressure apparatuses for diamond and BN synthesis and sintering with recessed (or capped) anvils and toroid type in [4, 47, 192–196, 218] and belt-type [195, 197] (some of them were included in the textbook [3]), and in DAC in [50, 52, 53]. Solutions were used as the boundary conditions to simulate with FEM stress-strain state, static and cyclic strength of the ceramic/metallic elements of these high-pressure apparatuses.

FEM solutions with obtained $\sigma_y(p)$ were found for elastoplastic flow in the high-pressure apparatuses with recessed anvils [48, 49, 54, 55, 198–200] and DAC [51]. The former (supplemented by models for electric and thermal conductivity and graphite-to-diamond PT) have been used for the modeling of the diamond synthesis process.

### 2.4. Coupled FEM-experimental methods for megabar pressures

The evolution of pressure distribution along the boundary between a diamond and sample was found using distributed ruby particles in [164, 191, 203] and with XRD and ruby particles [129, 130]. In [131], evolution of pressure distribution (averaged over the sample thickness, see Fig. 6a) and sample thickness profiles were measured for Re up to 300 GPa with XRD and X-ray absorption, respectively. In [132], similar but more precise measurements were performed for W up to 380 GPa. Large elastic deformation of diamond surfaces and samples, along with the resulting curved sample profiles and SPD, introduce additional challenges for simulations.

In [179, 180], FEM code NIKE2D [204] is utilized to simulate the plastic flow of a sample compressed by deformed diamond anvils. In [179], the effect of the $\sigma_y$ of a gasket on the maximum achievable pressure was studied, which increases with the increasing $\sigma_y$. With Supreme 63 tool metal with $\sigma_y = 2.5$ GPa for a gasket, achieving 460 GPa in experiment was claimed. Consistency with experimental results was obtained at 100 GPa. In [180], the effect of changing the cullet geometry, including double beveling diamonds, was investigated with FEM. In [175, 176], FEM simulation with NIKE2D code was extended by including pressure dependence of elastic moduli and varying pressure dependence of $\sigma_y$ of Re at $\sigma_\infty = 8$ GPa. Results were applied for analysis of experimental pressure and sample profile measurements in [131] up to 285 GPa, utilizing XRD and X-ray absorption, respectively. Transition from the plastic flow to elastic deformation of a sample at high pressures and shape the pressure (Fig. 6a,b), and sample profiles were qualitatively reproduced.

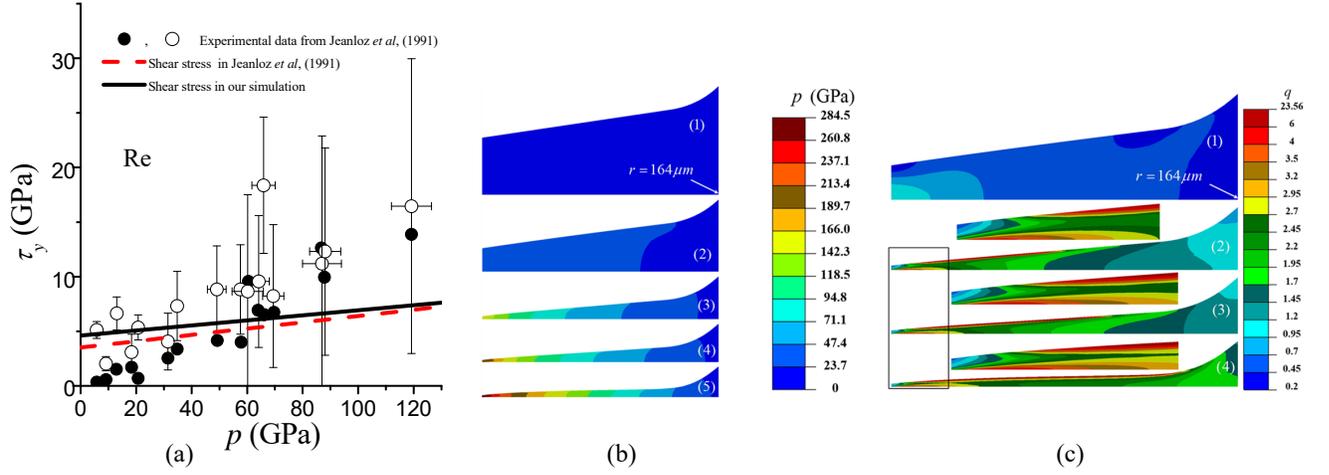

**Figure 5:** (a) Pressure dependence of the shear strength of Re in experiments [164] and approximations in simulations in [142] (solid line) and [164]. Solid symbols without error bar are obtained using Eq.(3), and open symbols with error bar are estimated from the offset between XRD and ruby measurements. The dashed line is an estimate $\mu/50$ in [164], where $\mu$ is the shear modulus. (b) and (c) Evolution of pressure and accumulated plastic strain distributions, respectively, in the quarter of a sample with an increasing applied force. In (c), the zoomed central part of a sample is shown above the sample. Reproduced from [142].

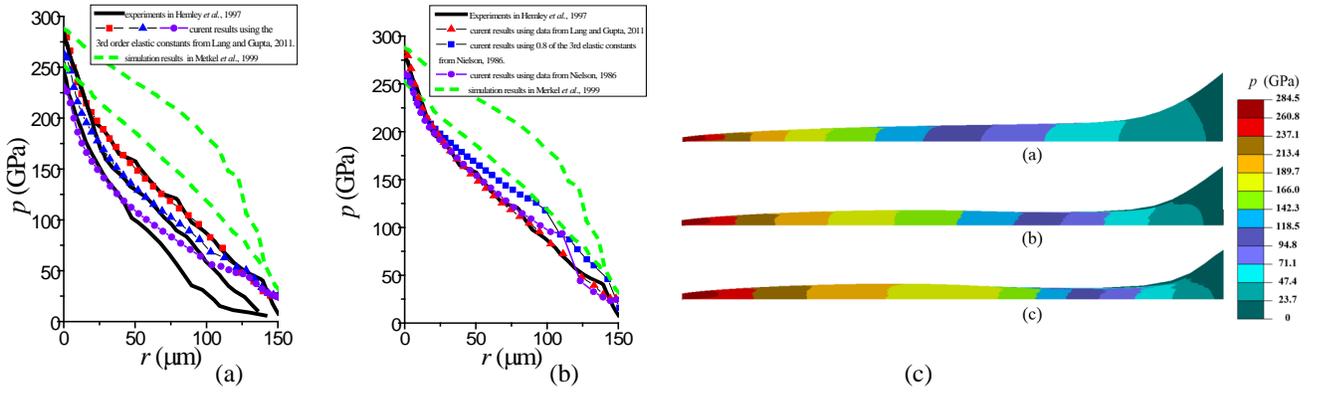

**Figure 6:** (a) and (b) Pressure distributions at the boundary between Re sample and diamond. Lines with symbols are the simulation results from [142] with different third-order elastic constants of diamond, adopted from [201] in (a) and from [201], [202], and 0.8 of the third-order constants from [202], respectively, in (b); black solid lines are experiments from [131]; the green dashed lines are the FEM results from [175]. (c) FEM-generated three different Re sample shapes and pressure distributions for the pressure in the center of 285 GPa. The third-order elastic constants are adopted from [201] in (a), from [202] in (b), and 0.8 of the third-order constants from [202] in (c). The sample thicknesses at the center for (a), (b), and (c) are 2.6, 3.2, and 3.6 microns, respectively. Reproduced from [142].

Combined XRD and FEM studies were performed on Ta (up to 100 GPa) and U (up to 190 GPa) [205]. Since material was not predeformed, $\sigma_y$ depends on $p$ and $q$. However, the FEM simulations were performed with $\sigma_y = const$ and assuming adhesion at the diamond-sample contact, utilizing NIKE2D code. Thickness without load was measured at the beginning and end of experiment and was linearly interpolated between. Strain was evaluated as $ln(h_0/h)$, $\sigma_y$ was evaluated using Eq. (4) and attributed to the maximum pressure for this $h$. For Ta, $\sigma_y$ unexpectedly reduces above 60 and 90 GPa for high and low strains, respectively, while in [206] and [207] it continues growing at least up to 100 GPa. In [208], $\sigma_y$ for Ta plateaus in the range of 50-60 GPa, and then grows again up to 276 GPa. Results have been obtained with radial diffraction up to 50 GPa and in the whole range with axial diffraction. Difference may be caused by different initial states of the sample and pressure-plastic strain paths, which shows the importance of either reaching a perfectly plastic state or controlling and characterizing pressure and plastic strain independently. The $\sigma_y$ for U in [205] grows until 100 GPa and then reduces with pressure/strain.

In [209], pressure distribution in Va was determined with XRD, and the sample thickness was measured with the X-ray absorption, like in [131]. Equilibrium Eq. (5) was applied for each radial point. The reduction in strength above 40 GPa was found and tentatively related to the second-order bcc-to-rhombohedral PT. However, this PT was not detected in XRD patterns in [209].

Note that in the FEM code NIKE2D code [204], used in [209], the system of equations for large elastoplastic deformations is oversimplified and not completely clear. Thus, flow theory equations for small strains with linear Hooke's law for elasticity and $\sigma_y(p)$, written in incremental form, were implemented. Strict finite strain kinematics based on the multiplicative decomposition of the deformation gradient into finite elastic and plastic contributions, nonlinear elasticity, and the finite particle rotations were not strictly taken into account. A strict derivation of the equations for large elastic and plastic deformations under high pressure was presented in [4, 47], with consecutive simplifications. For simulation in [48, 49, 51] equations were simplified for large plastic and elastic volumetric strains, but small elastic deviatoric strains. In the first FEM simulations of plastic flow in RDAC [160, 210], elastic strains were small.

In simulation in [142, 143], all equations are derived and implemented for the large elastic and plastic strains. The third-order isotropic Murnaghan potential was used for sample and third-order elastic potential for cubic diamond crystals, both in terms of Lagrangian strain. The model was implemented in the ABAQUS code [211]. Detailed computational algorithm was also given in [142], and mesh-independence of the solution was proven. Contact friction was described in [142] by the Coulomb friction law with a friction coefficient of 0.1 and adhesion if shear friction stress is smaller. In the megabar pressure range, in adhesion region, friction stress reached $\tau_y(p)$ and localized shear occurred, mimicking sliding within thin Re layer just below contact surface.

The solution [142] was well fitted to three experimental pressure curves from [131] up to 300 GPa (Fig. 6a,b), which also allowed determining $\sigma_\infty = 8.0$ GPa and $b = 0.04$ GPa$^{-1}$ (Fig. 5a), giving results surprisingly close to estimated $\mu/50$ in [164]. Note that the experimental estimates of $\sigma_y$ in Fig. 5a from [164] "can be considered reliable only to a factor of $\sim 2$ or better", mostly due to small number of ruby crystals. Fields of stress and elastic strain tensors in the sample and anvil were determined. The maximum thickness reduction was 30 times (Fig. 5b,c), with maximum $q = 23.56$ and transition from plastic to elastic compression at high pressure. With reduction of the third-order elastic constants of diamond from [202] by a factor of 0.8, experimentally observed cupping of the diamond is well reproduced (Fig. 6c). Difference between simulation of the same problem in [175, 176] and [142] and ability to quantitatively describe experimental results from [131] on of a Re in DAC are visible in Fig. 6a,b.

A more elaborated approach developed in [143] was applied for the description of the experimental results in [132] for W for four maximum pressures between 170 and 380 GPa (Fig. 7b,c). Fourth-order elastic potential for cubic crystals in terms of Lagrangian strain was used. Since higher-order elastic constants are not well-constrained by experiments and first-principle simulations, the third-order elastic constant of W and fourth-order elastic constants of diamond were slightly adjusted to obtain a better fit to experimental pressure distribution at three lowest pressure and sample shape at the highest pressure, respectively. Contact sliding starts when the friction stress reaches the Coulomb threshold or $\tau_y$, which is smaller.

Two material parameters in $\sigma_y = 1.8 + 0.1p$ GPa (Fig. 7a) and the Coulomb friction coefficient $\mu = 0.05 + 0.001\sigma_c$, where $\sigma_c$ (GPa) is the normal contact stress, were identified from the fitting of two pressure distributions for 170 and 240 GPa. Using these material parameters, excellent verification was demonstrated for two higher pressure distributions for 300 and 380 GPa and all four sophisticated sample thickness profiles (Fig. 7a), while none of the geometric parameters were used for the model calibration. After such advanced verification, the model was used to reproduce evolution of $\boldsymbol{\sigma}$ and $\boldsymbol{\varepsilon}_p$ fields in the sample and diamonds (Fig. 8). Obtained expression for $\sigma_y$ is much more justified than in [131, 132], which also possesses large scatter (Fig. 7). Several unexpected limitations have been found. (i) The expression for $\sigma_y$ is valid for $p \leq 225$ GPa only, because for higher pressures, material deforms elastically due to cupping of the anvil. (ii) The expression for $\mu$ is applicable for $\sigma_c \leq 37$ GPa only; for higher contact stresses, plastic friction or adhesion conditions are valid. (iii) The region with plastic friction is tiny for pressures 170-380 GPa, which makes practically inapplicable the method based on Eq. (5) for DAC because of large deformation of diamonds and adhesion zone. The only way to increase the above ranges is to apply torsion in RDAC, for which FEM simulations [212, 213] suggest that the adhesion zone is smaller and pressure at the center grows much faster (Fig. 17b,c).

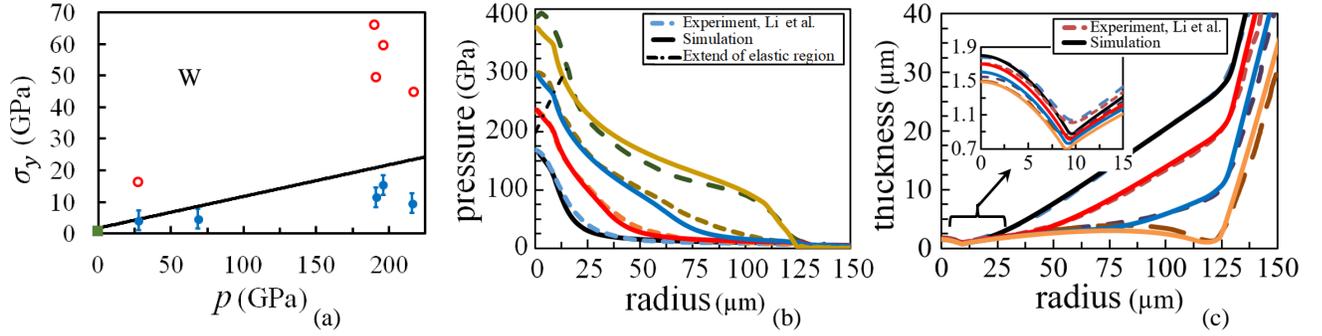

**Figure 7:** (a) Pressure dependence of the yield strength of W. Line is described equation $\sigma_y = 1.8 + 0.1p$ GPa obtained in [143] by fitting FEM simulations to experiments in [132]; symbols are from [131]. (b) and (c) Radial pressure distributions and corresponding sample/anvil thickness profiles, respectively, in experiment [132] (dash lines) and FEM simulations in [143] (solid lines). Dash-dot line in (b) designates the radius of the growing central region in which sample deforms elastically after plastic deformations at the initial stage of compression. Inset in (c) shows a zoomed central part of the sample thickness profile. Reproduced from [143].

Distribution of the friction stress is very complex (Fig. 8a) with several oscillations in a central cup region in the adhesion zone, which grows with increasing compression. Very heterogeneous fields in the sample are shown in Fig. 8b. Within the adhesion zone, radial velocity of W particle (equal to the diamond particles velocity) is directed toward the center; outside the adhesion region, sample particles away move from the center. Not only plastic strains, but also material particle rotations in Fig. 8b are large, reaching $46.8^o$. They lead to texture, and since isotropic flow theory describes experiments well, this is strong proof that material behaves like perfectly plastic and isotropic. The line $\dot{q} = 0$ shows the contour of the elastic region at the sample center.

Not only components $\sigma_{ij}$ and $p$, but also equivalent stress $\sigma_{eq}^{[110]}$ (normalized by stress-dependent theoretical strength) for compression in [110] direction near the diamond tip were calculated (Fig. 8c). This is a mandatory parameter to study strength of anvils under complex stress state; study of strength based on separate components of the stress tensor or non-justified strength criteria [176, 214–216] has a qualitative nature only. Fracture occurs at $\sigma_{eq}^{[110]} = 1$. Since for $p_{max} = 300$ GPa maximum $\sigma_{eq}^{[110]} = 0.32$, there is large safety factor for ideal diamond compressed in [110] direction. For complete strength analysis, similar distributions must be calculated for all other known fracture/slip planes with taking into account shear stresses, making optimization of the anvil's design and loading conditions for an ideal crystal possible. Fourth-order elasticity for diamond, along with crystal lattice instability criterion $det\boldsymbol{B} = 0$ (Section 5.3), was utilized in FEM simulations and analysis of the effect of diamond anvil geometry and friction condition on anvil performance in [217]. Since sample was not included, boundary conditions for anvil were oversimplified. However, using $det\boldsymbol{B} = 0$ as the strength criterion is a significant step forward. Introducing defects into FEM simulations will advance this procedure for real diamonds. Note that for maximum normal to the cleavage plane strain criterion, FEM simulations and optimization of geometry were performed in [50, 52, 53]. Since strength depends on the size of the anvil, a nonlocal size-dependent criterion was developed and applied for cemented carbide anvils [195–197, 218]. Parametric FEM study of stresses in and strength of sapphire anvils with connection experiments was performed in [219].

One of the surprising results in [143] is that despite the micron-size sample thickness and enormous stress (5 GPa/micron) and plastic strain gradients, experiments are well described with the scale-independent theory, implying the *scale-independence of elastoplastic properties.*

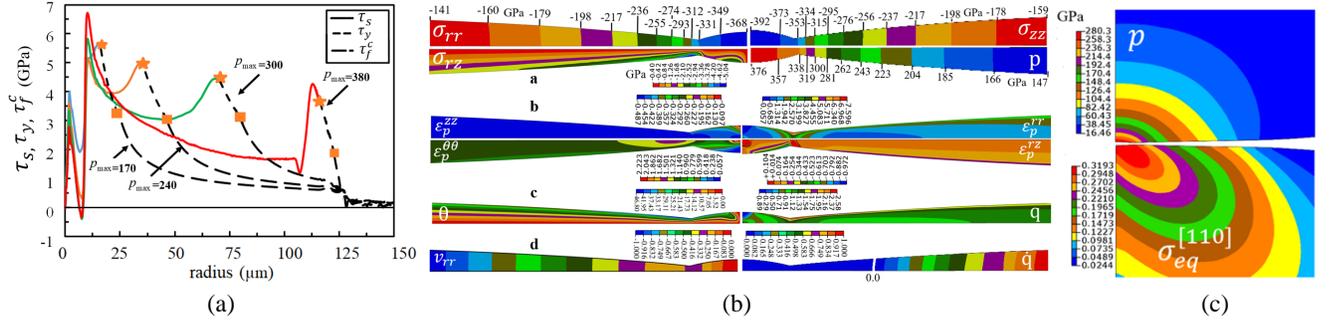

**Figure 8:** (a) Distribution of contact shear friction stress $\tau_f$ at the diamond-W surface for four maximum pressures shown near the curves. The dashed portion is for the Coulomb friction until $\tau_f$ reaches $\tau_y(p)$ (marked by squares). The dotted lines between stars and squares designate the plastic sliding zone with $\tau_f = \tau_y(p)$. The solid lines between stars and sample center designate the sticking (adhesion) zone. (b) Fields of components of the stress tensor $\boldsymbol{\sigma}$, pressure $p$, plastic strain tensor $\boldsymbol{\varepsilon}_p$, particles rotation angle $\theta$ (in degrees), accumulated plastic strain $q$, normalized radial velocity $v_{rr}$, and plastic strain rate $\dot{q}$ in the central part of a sample for $p_{max} = 300$ GPa. The thickness is magnified by a factor of 4. (c) Corresponding fields of pressure $p$ and equivalent stress $\sigma_{eq}^{[110]}$ for compression in [110] direction near the diamond tip. Reproduced from [143].

The much more precise and detailed correspondence in [142, 143] between advanced experiments [131, 132] and FEM simulations with a perfectly plastic and isotropic model, which is independent of $\boldsymbol{\varepsilon}_p^{path}$, provides much stronger support for such a material behaviour during SPD than in all previous works. Since plastic friction rule $|\boldsymbol{\tau}_f| = \tau_y(p)$ is completely determined by $\tau_y(p)$, plastic friction during SPD is also isotropic and independent of $\boldsymbol{\varepsilon}_p$ and $\boldsymbol{\varepsilon}_p^{path}$.

Even more advanced analytical-FEM-experimental methods have been developed in [108, 109] for Zr, but since they also involve $\alpha - \omega$ PT, they will be presented in Section 7.6. Note that radial pressure distribution in amorphous methanol-ethanol mixture in [220] was determined using Brillouin-oscillation frequencies, preliminary calibrated utilizing ruby particles. The pressure-dependence of $\sigma_y$ was found to scale with pressure-dependence of the shear and bulk moduli with factors of 0.040 and 0.012, respectively.

## 2.5. Determination of the yield strength with radial and axial diffraction

With XRD, $\sigma_y$ can be determined based on X-ray line shift (i.e., elastic macrostrain) [131, 146, 221–225] and line width (i.e., elastic microstrain) [208, 226–228] methods. It can be determined with radial diffraction (i.e., when beam is orthogonal to the load) [229–232] or axial diffraction (i.e., when beam is parallel to the load) [233]. Line shift method is more often used in the radial diffraction [144, 146, 224, 229–232, 234, 235] and line width method is mostly used with the axial diffraction [208, 226–228].

For X-ray research with radial diffraction [229–232], the beam passes through the entire sample diameter and gasket. To reduce the pressure gradient along the radius, small sample with relatively large height to diameter is used. X-ray transparent gaskets for radial diffraction are usually made of amorphous boron epoxy (which limits the maximum achievable pressure) or Be, which was used to measure strength of Os up to 280 GPa [234]. In contrast to the axial diffraction, radial diffraction allows measurements of how the *d*-spacings vary with the angle between the scattering and compression directions. In addition, pressure is determined using EOS of a witness metal. The procedure allows one to determine deviatoric stresses (difference between axial and radial stresses, which is interpreted as $\sigma_y$) and texture. The main challenge is to transition from elastic strains in a single crystal to stresses in single crystal (which requires reliable pressure-dependent elastic constants) and then to a polycrystalline sample. The theory was pioneered in [221, 222] using the Voigt (equal strain) or Reuss (equal stress) limits or their mixture with arbitrary proportion factor for transition from the single crystal to polycrystalline aggregate. The theory was advanced by including the plastic strain in [223] and utilizing viscoplastic [44, 45] and elasto-viscoplastic [236] self-consistent models. While the theory is applicable in the axial and radial diffraction geometry, it is more often used for the radial diffraction because more information is available. The self-consistent models were extensively used for ultrafine-grained metals [237, 238], Fe [239] (including PTs from bcc- to fcc- and hcp PTs up to 36 GPa and 1,000K), Ta [208], WC [240], and geomaterials [144–146, 224, 225, 235] (including determination of the stress-strain portioning in two-phase materials). The advantage of the radial diffraction with self-consistent modeling is that it also determines slip activities in different slip systems by fitting to the experimental texture. Still, indeterminacy in transition from single crystals to the polycrystalline aggregate remains. Since strain and pressure grow simultaneously, the strain hardening contributes to $\sigma_y$ but is attributed to the pressure. Also, stress heterogeneity along the radius, which increases with the reduction in sample thickness to radius ratio, limits plastic strains to several tens of percent. These problems could be resolved by strongly predeforming sample up to reaching steady hardness and microstructure; then small strain increment within DAC should be sufficient to reach the fixed surface of perfect plasticity and find corresponding $\sigma_y(p)$.

Note that the *SPD community* time to time also produces in situ studies with the X-ray beam orthogonal to the loading direction, e.g., for the edge of the sample during HPT [37, 38] and for the compression [241] and sliding [242] under high pressure. It was found in [38] that during unloading after HPT of Ni at 8 GPa, the dislocation density was decreased by 3 times and crystallite size was increased by more than twice. However, X-ray patterns and, consequently, stress fields, are very nonuniform along the diameter or beam (including undeformed material outside the die), and averaging gives a rough picture of the actual fields. Thus, pressure measured by XRD during SPD by sliding was by 2.5-4 GPa below than the load over the area [242].

XRD in perpendicular geometry is also used in *rotational Drickamer apparatus* [40–42, 243]. Ring-shaped sectored sample was utilized to reduce heterogeneity in stress-strain states. The total shear strain was evaluated from X-ray image analysis of a Pt or Mo foil marker located between sample sectors. To evaluate stress state, theory from [221], accounting for shear stress, was applied [243]. Unique stress-strain curves and textures up to 23 GPa and 2,150 K for various geomaterials were determined (Fig. 9a), which, however, exhibits a large error bar.

*RDAC* with nano-polycrystalline diamond anvils [244] or single crystal diamonds for pressure up to 135 GPa was developed in [245, 246]. To measure shear strain, an X-ray computed laminography technique [138] was elaborated for in situ tracking of the Pt strain marker placed along the sample diameter. To increase the contact friction, the radial and concentric circular grooves were produced on diamond culets [247], with radial grooves being more efficient. In [246], the stress-strain curves and texture evolution for MgO were presented for 45 GPa and 300 K and 65 GPa and 520 K; texture was also presented for 120 GPa and 515 K. To avoid XRD peaks from passing through Re gasket, the beam direction was inclined under $60^0$ to the load direction, which was taken into account in the theory of the type of [221, 243]. Inclined beam increases error in stress determination by including larger radial heterogeneity. Two stress-strain curves for each loading condition are presented, based on the shift of 200 and 220 peaks, which, despite a large error bar, exhibits clear transition to the perfectly plastic behavior. Stresses based on these peaks differ by a factor of 2-3. Thus, despite the uniqueness of the data, it is not clear which yield strength should be used in the constitutive equations.

With *axial XRD*, $\sigma_y$ can be estimated based on peak shifting [233] and broadening [80, 108, 208, 226–228]. Both methods contain various assumptions and can be fitted in a way that they produce comparable results [208, 227]. It is always desirable to check them with the independent method, like with pressure gradient method (provided that $\tau_f = \tau_y$), or based on the $\sigma_y$ or hardness $H$ ($\sigma_y = kH$, where a factor $k$ is in the range from 1/3 to 0.386 [154, 155]) at normal pressure [80, 227].

Axial diffraction allows generation of a 2D map of the pressure $p$ and differential stress $t$ (Fig. 9b,c) [233], using a submicron beam. Such a small beam probably does not contain a representative volume of a material (i.e., large enough to contain statistically representative number of crystals), which behavior does not change with increasing its size. That is why both $p$ and $t$ fields are very micro-heterogeneous and without clear patterns and correlation between $p$ and $t$. Also, $t$ varies from -5 to 5 GPa, which clearly shows that identification of $t$ with $\sigma_y$ may be very questionable. Most probably, material deforms elastically, like in Fig. 8b. Negative $t$ means that the axial stress is getting smaller than the radial stress, i.e., material is pushed to the center. As it often happens in various disciplines and experiments, *increased resolution produces too detailed information which prevents/complicates obtaining a bigger picture and rules.* Of course, increased resolution is required while focusing on the local submicron-scale phenomena in a single few grains.

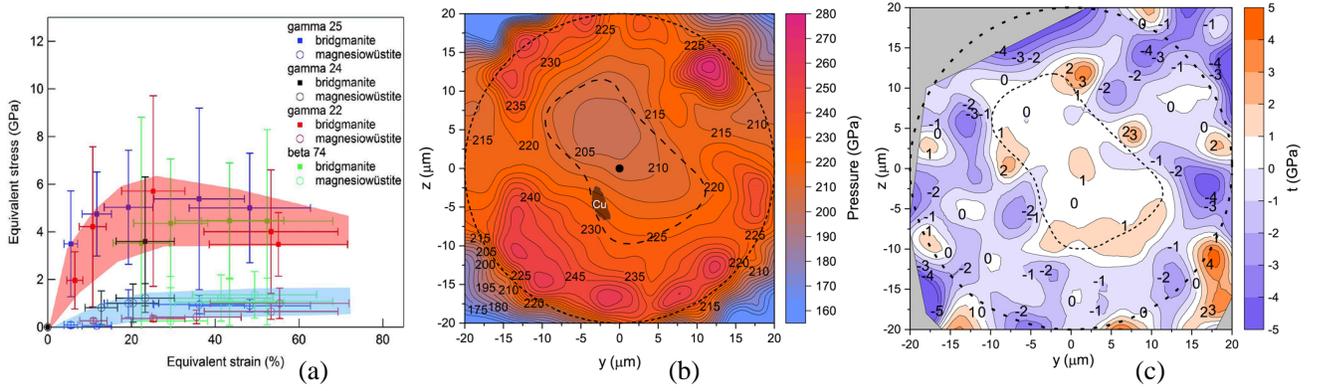

**Figure 9:** (a) Stress-strain curves in bridgmanite and magnesiowüstite obtained in rotational Drickamer apparatus at pressures 24-27.5 GPa and temperature 2,000-2,250 K [42]. Stress in bridgmanite was evaluated by averaging stresses obtained from diffraction peaks (110) and (112), and in magnesiowüstite utilizing peaks (200) and (220). (b) Contour pressure plot across the 40 microns culet (dashed circle) within Bi sample (irregular dashed line) and W gasket compressed in DAC, determined by an axial XRD 2D scan with a step of 2 microns [233]. The central black circle shows the size of the incident X-ray beam (0.85 microns full-width at half-maximum). (c) Corresponding contour plot of the uniaxial differential stress $t$ calculated utilizing peak shifting [233].

### 2.6. Increasing contact friction: toroid-type apparatuses

*Frictions stress in many cases is smaller than* $\tau_y$. Determination of $\tau_y$ using equilibrium Eq. (3) is based on the assumption that $\tau_f = \tau_y$. This was doubted in [248], where the results of FEM simulations for the Ta ring sample compressed in DAC were fitted to the displacements of its internal and external radii with $\tau_f \leq 0.43\tau_y$ for relative compression of 35%. External and internal applied pressure was up to 6 GPa. However, these results may not be representative. First, compression is relatively small and is not sufficient for reaching maximum possible friction [160], especially since $\tau_y$ grows with strain. The initial and final thicknesses are 40 and 30 $\mu m$, respectively. Initial internal and external radii are 40 and 125 $\mu m$, respectively, and length between them is 85 $\mu m$. In the deformed state, from the plastic incompressibility condition, this length roughly increases by $\sqrt{1.35}$ to 99 $\mu m$. Since external radius increases and internal decreases, the neutral radius and adjacent sticking zone is between them. That is why roughly half of 99 $\mu m$ (i.e., 50 $\mu m$) flows in each direction, and its ratio to the hight is just 1.7. This is far less than at lest 3-4 required for a "thin" sample, the effect of cylindrical surfaces and sticking zone is strong, and region with maximum possible friction is very small or does not exist at all. This ring-shaped sample and small compression are not a good choice for checking whether the condition $\tau_f = \tau_y$ is reached.

FEM simulations in [249] prove that all neglected terms in transition from the exact equilibrium equation to Eq. (3) are indeed small at the contact surface (which is not true for the beveled anvils, see Fig. 17a from [212]), but if boundary condition for friction stress $\tau_f < \tau_y$, than there is significant error in determination of $\tau_y$. This is, however, clear without simulations because Eq. (3) determines $\tau_f$.

In [80], $\sigma_y$ for strongly predeformed $\alpha$ and $\omega$ Zr determined using the pressure gradient method was significantly lower than the yield strength determined by hardness at the normal pressure. Also, the pressure distribution in [80] is linear in each phase and mixture, i.e., corresponds to pressure-independent $\sigma_y$ the same in both phases, while $\sigma_{y\omega} \gg \sigma_{y\omega}$. At the same time, the calculated $\sigma_y$ at increasing load grows by a factor of 2 during the PT. Should both phases have the same $\sigma_y$, it should be independent of plastic strain for strongly predeformed material, and pressure change is minor. Both contradictions implied that $\tau_f < \tau_y$. The absence of signatures of the PT at the pressure distributions was clarified by the approximately equal friction coefficients of Zr phases. Advanced CAE-FEM methods developed in [108] determined refined $\tau_f(r)$ distributions, which were found to be at least two times lower than $\tau_y(p)$ for each Zr phase and mixture. For each phase and mixture, the friction stress rule was found in the form $\tau_f/\tau_y = A + Bp$ by approximating data for all radial points and loads; however, there was a large scatter in the data.

Low friction at the contact with diamond was also confirmed by large relative sliding between sample and diamond. Material displacements in DAC/RDAC were determined by mapping displacements of ruby particles spread over one [63] and two [250] anvil-sample surfaces, qualitatively in [63] and quantitatively in [250]. In [250], the rotation angle of Zr was found to be up to 5 times lower than that of an anvil. This illustrates the error of the evaluation of shear strain using the rotation angle of an anvil. Note that prescribing material displacements as boundary conditions instead of adhesion or $\tau_f$ in FEM simulations of the deformation in DAC and RDAC [110, 125, 126, 160, 251–253] could offer increasing accuracy, but still has not been implemented.

Contact friction during compression and torsion of materials is a complex and multistage process described in the metal forming textbooks [184, 185]. Asperities of the deforming tool penetrate into material during compression, which is similar to the hardness measurement and requires applied stress $\sim 3\sigma_y$. For large asperities, after their complete penetration into material surface, sticking (adhesion) of the material to the anvil occurs, which is formalized in the condition that the relative displacement of material with respect to anvils is absent. In this case, a thin shear band appears within material near the contact surface, which is equivalent to the internal sliding, resulting in $|\boldsymbol{\tau}_f| = \tau_y$. However, even with full adhesion conditions, FEM simulations [160] show that *pressure at the center should reach $\sim 9\sigma_y$ until condition $\tau_f = \tau_y$ is reached in the entire sample,* excluding small central part $r/R < 0.2$. This pressure may reduce significantly if a higher $h_0/R$ is used.

However, for a traditionally polished diamond culet surface, the asperities are as small as $\pm 10$ nm (Fig. 11a); complete penetration may not occur, and friction coefficient is low. If Coulomb friction is allowed for $\tau_f = \mu p < \tau_y$, external ring with Coulomb friction exists up to very high pressure at the center, like in Fig. 8a. Thus, for W, $\mu = 0.005 + 0.001\sigma_c$, and $\tau_y$ is reached at 37 GPa [143].

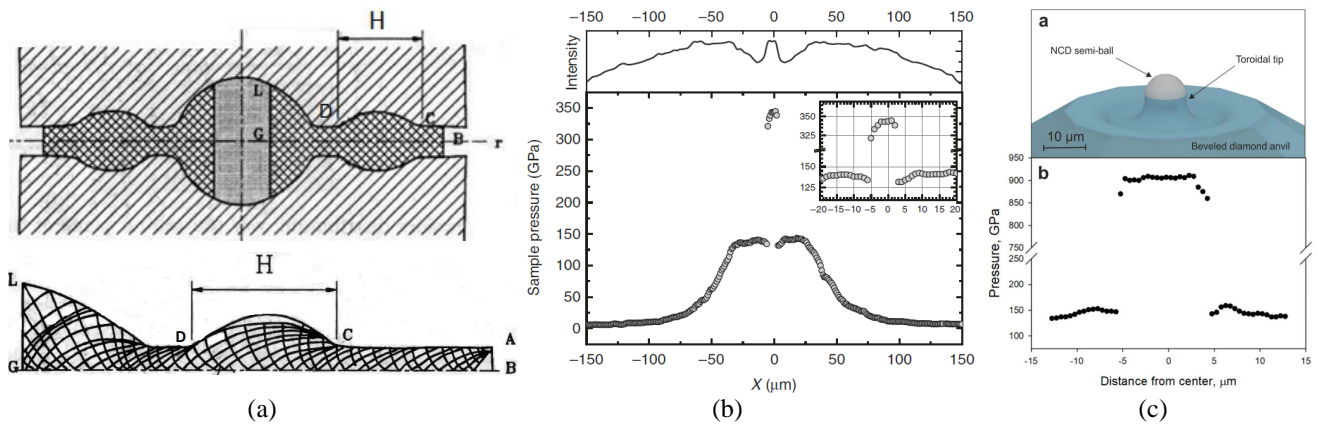

(a) (b) (c)

**Figure 10:** (a) Schematic of toroidal high-pressure apparatus with the sample at the center surrounded by a gasket (top) and slip-line field in the gasket [4]. (b) Top: X-ray transmission intensity plot along the culet diameter for toroidal DAC characterizing sample thickness profile with toroidal portion. Bottom: Pressure profile showing large jump at the toroidal tip with largest bevel angle [257]. (c) Schematic of lower part of the two-stage toroidal DAC (top) and pressure profile with a large jump at the toroidal tip with largest bevel angle [259]. Increased pressure in toroidal DAC is due to increased friction due to sliding inside of the sample and large bevel at the toroidal tip causing large pressure gradient (see Fig. 17).

*Toroid type apparatuses.* An important step in the development of high-pressure apparatuses with recessed

(capped) anvils was the introduction of toroid-type apparatuses [3, 254, 255], in which toroidal and multiple toroidal grooves were made around the central recess (Fig. 10a). According to [255], the pressure generated in the tore region strongly reduces the flow of the central part of a gasket and the shear stresses in the anvils. The latter is doubtful, because we are not aware of any FEM solution supporting this, and any groove produces stress concentrator. The slip-line solution for the stress state of the gasket in the toroidal apparatus was given in [194] and reproduced in books [3, 4, 47]. Based on the slip line field in the gasket in Fig. 10a, independent of the friction level between anvil and gasket, sliding within the groove occurs along the upper slip line CD, and the shear stress along each slip line is equal to $\tau_y(p)$. This effectively increases friction and pressure gradient and reduces radial flow from the center, especially at the initial compression stage when friction stress in other places is low. Such an increase of friction is also important during unloading, because pressure in the thicker sample and central part of the gasket reduces more slowly than in the thin part of the gasket, which may lead to uncontrolled dynamic pressure release; gasket material in the tear stabilizes the unloading process. Increased thickness of the gasket in the tear also leads to slower stress release during unloading and provides additional support to central part of an anvil, where stresses are still high. The key point for optimization of the geometric parameters of a groove is that some gasket material must be between anvil and the upper slip line; otherwise, friction will be determined by the friction coefficient between anvil and gasket. Results of numerical optimization were confirmed experimentally and patented [256]. At the same time, the larger the groove, the larger the stress concentrator in the anvil. That is why multiple smaller grooves were suggested in [4, 47, 194]. This was realized in [247] for DAC by making multiple concentric circular grooves on the diamond culets. For torsion in RDAC, radial grooves have been produced in [247] to reduce relative sliding. Still, friction in the smooth part of the anvils is traditionally low.

Toroidal DAC was recently developed (Fig. 10b,c) and applied in [208, 233, 257, 258] and (including double stage compression with nanocrystalline diamond hemisphere for pressures up to 900 GPa) [259] for increasing achievable pressure. As follows from FEM simulations [212, 213], in addition to increased friction, a large bevel angle of anvil produces the main effect (Fig. 17a).

*2.7. DAC/RDAC with rough diamonds: steady pressure-dependent yield strength and microstructure*

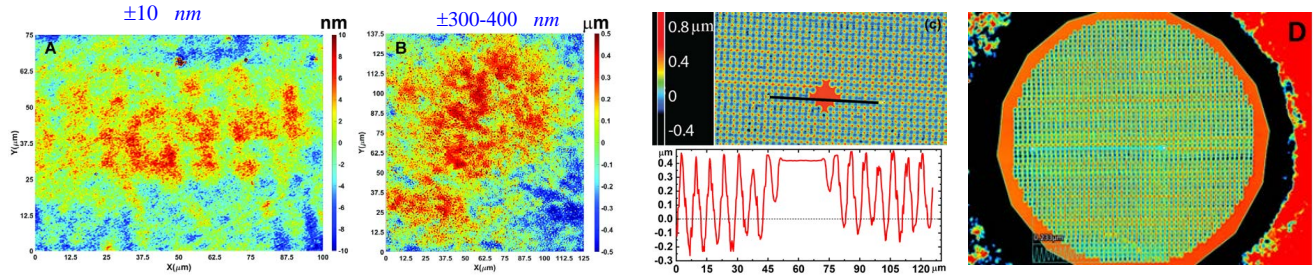

**Figure 11:** 2D map of asperities profiles for standard smooth (a) and rough polished (b) diamond culets [89]. 2D periodic asperities profile at the central part (c) and the whole (d) diamond culet produced by laser ablation [83]. The 1D curve in (c) was measured along the black line.

Suggestion from [4, 47, 194] to produce multiple small grooves to increase friction to $\tau_y$ were realised in [81, 89] by rough polishing (Fig. 11b) or lapping and in [83] by lased ablation (Fig. 11c) of the diamond culets for DAC/RDAC. Instead of traditional $\pm 10$, nm asperities (Fig. 11a), $\pm 300 - 400$ nm asperities are produced. In situ X-ray diffraction and absorption were utilized in this work to obtain radial distribution of the pressure, volume fraction of phases, microstrain, crystallite size, and dislocation density, using Rietveld refinement with MAUD software [260] and Williamson method for dislocation density [261]. Fig. 12a shows pressure distributions in strongly predeformed $\omega$-Zr after transformation from $\alpha$-Zr during compression with rough anvils. All 4 curves are well described by Eq. (5), giving $\sigma_y = 1.24 + (0.0965 \pm 0.0016)p$ GPa. Value 1.24 GPa is consistent with HV/3, where hardness HV=3.72 GPa. Taking into account variety and complexity of plastic strain $\varepsilon_p$, straining path $\varepsilon_p^{path}$, and pressures in different points at different loads, and independence of the obtained equation of $\varepsilon_p$, $\varepsilon_p^{path}$, and corresponding parameters of the strain-induced anisotropy, results in Fig. 12a presents the most complete confirmation of the isotropic perfectly plastic behavior of Zr, which is independent of $\varepsilon_p$ and $\varepsilon_p^{path}$.

The yield strength can be presented as a sum of the Taylor term due to dislocation density $\rho_d$ and Hall-Petch term due to grain/crystallite size $d$ [89, 262, 263]:

$$\sigma_y = \tilde{\sigma}_y(p) + \alpha \rho_d^{0.5} + \beta d^{-0.5} \tag{6}$$

with $\tilde{\sigma}_y(p)$ for a pressure-dependent contribution. That is why steady $\sigma_y$ is expected to be connected to the steady grain/crystallite size and dislocation density. Indeed, Fig. 12b,c shows for strongly predeformed Zr, after quite heterogeneous and different at different pressures crystallite size and dislocation density, they are getting practically independent of radius and applied pressure after completion the PT to $\omega$-Zr and further compression in DAC with rough anvils up to 14 GPa [89]. Since $\varepsilon_p$, $\varepsilon_p^{path}$, and pressure essentially change along the radius and increasing compression, this implies that *steady crystallite size and dislocation density are reached, which are independent of $\varepsilon_p$, $\varepsilon_p^{path}$, and pressure.* Along with steady $\sigma_y(p)$, the above results represent the *first in situ obtained rules for strength and microstructure evolution during SPD.* Note that since dislocation density in Rietveld refinement is determined in terms of crystallite size and microstrain, this rule also implies that the *same rule is also valid for the microstrain.*

Similar rules were extended for minerals after SPD in RDAC with rough diamonds for olivine and ringwoodite: the crystallite size and microstrain in olivine and crystallite size in ringwoodite (in the course of incomplete olivine-ringwoodite PT) became steady and independent of $\varepsilon_p$, $\varepsilon_p^{path}$, $p$, and even PT progress [83]. Note that compression with torsion provides a much richer class of straining programs than compression alone. The crystallite size of the ringwoodite was 5-10 nm, consistent with the grain size for ringwoodite of 12-20 nm in [43] after small plastic strains at 1,173 K.

Steady grain size, dislocation density, and consequently or hardness after SPD are well-known for metals and ceramics from ex situ experiments [4, 20–22, 24–27, 47, 264–266]. The main physical mechanisms leading to steady microstructure are the dynamic recovery and recrystallization [23] and grain boundary migration[22]. However, in most of these papers, there are also examples in which steady states were not reached. Pressure independence of the grain size and/or dislocation density and/or hardness after HPT was observed for single-phase low- and high-pressure states in Zr, Ti, Fe, Ni, V, Hf, Pt, Al, Au, and Cu-30%Zn [29, 30, 265]. However, these measurements were performed at normal pressure as function of pressure during preceding HPT, which was defined as force divided by area, despite strongly heterogeneous pressure distribution. In contrast, results in [89] were obtained in situ under pressure as function of local pressure at each point, also generating big data. For instance, independence of hardness, and, consequently, $\sigma_y = HV/3$ of pressure after SPD does not imply that $\sigma_y$ is independent of pressure. Similarly, the independence of $d$ and $\rho_d$ at normal pressure of the pressure during SPD does not mean that they are pressure-independent. That is why in situ results in [89] are much more informative and precise. Also, new rules were formulated in the language of theory of plasticity in terms of yield surface, $\varepsilon_p$, and $\varepsilon_p^{path}$ rather than technological language, which is important for enriching basic plasticity and the development and application of plastic models to numerous SPD processes. In fact, in situ measurements in [38] with radial XRD, while with error due to passing through and averaging over the zones with very different stress-strain states, demonstrated that during unloading from 8 GPa after HPT, $d$ increases and $\rho_d$ decreases by a factor of 2, which underscores the importance of in situ measurements. In [88], a strong change in microstructure under hydrostatic loading/unloading of Zr was found.

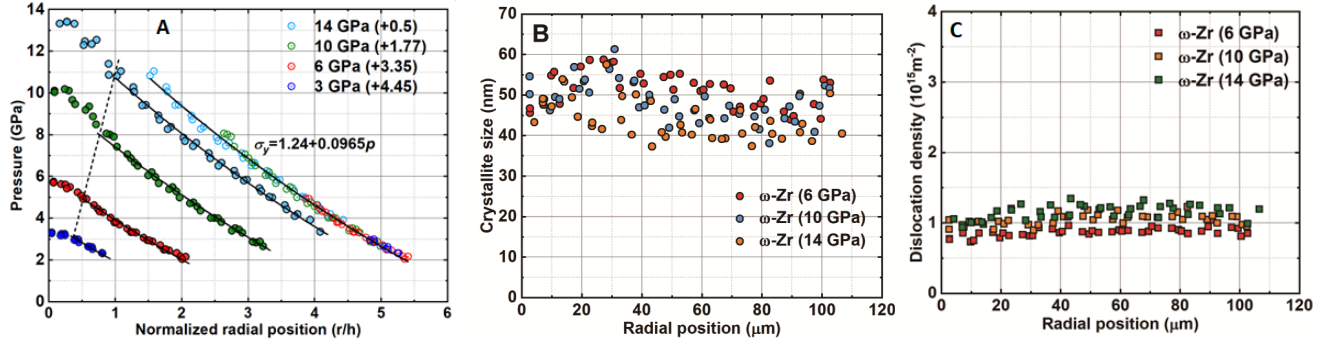

**Figure 12:** (a) Pressure distributions in $\omega$-Zr for four different loads (symbols, marked by maximum pressure at the center) are well described by Eq. (5) (lines) with $\sigma_y(p)$ shown in the picture Dashed line presents the boundary of the central region is not used for fitting, where friction stress reduces to zero at the symmetry axis The upper curve contains data for all four loadings obtained by shifting symbols from each individual curve (which is not prohibited by differential Eq. (5)) along the horizontal axis by value presented in brackets. (b) and (c) Distribution of the crystallite size and dislocation density, respectively, in single-phase $\omega$-Zr along the radius for three loadings, showing reaching steady microstructures, independent of $\boldsymbol{\varepsilon}_p$, $\boldsymbol{\varepsilon}_p^{path}$, and pressure. Reproduced from [89].

### 2.8. Multiple steady microstructures and yield strengths

The major problem related to the steady state rules was formulated in [89]. Different steady states were found for $\alpha$-Zr. Thus, after multiple rollings, compression in DAC with smooth and rough anvils, crystallite size was 75, 65, and 48 nm, respectively, and dislocation density was 1.00, 1.26. and 1.83 $\times 10^{15} m^{-2}$, respectively. Based on experiments, each of these numbers was claimed to be independent of $\boldsymbol{\varepsilon}_p$ and $\boldsymbol{\varepsilon}_p^{path}$, which is clearly contradictory; if it were independent, then we should have a single value for the crystallite size and for dislocation density. It was very unexpected that changing just in friction can cause such changes in the steady microstructure. However, from the technological perspectives, it was well known that different SPD technologies lead to different steady dislocation densities and grain sizes (Fig. 13b,c), and corresponding hardness, and consequently, $\sigma_y$ and different isotropic surfaces of perfect plasticity $\varphi^i(\boldsymbol{s}) = \sigma_y^i(p)$ (Fig. 13a). However, translation of these technological results into the language of theory of plasticity and microstructure evolution revealed disappointing implication that *there are no currently consistent theories for continuum plasticity, dislocations, and grain size for SPD*. Revealing the existence of the isotropic limit surface of perfect plasticity independent of $\boldsymbol{\varepsilon}_p$ and $\boldsymbol{\varepsilon}_p^{path}$ transformed the most complex theory of large-strain plasticity into the simplest one, which can be relatively easily calibrated. Now, this euphoria is faded by simple question: which exactly $\sigma_y$ (and steady $\rho_d$ and $d$) should be used? However, the situation is not too bad. Specific values of $\sigma_y^i(p)$, $d^i$, and $\rho_d^i$ for each steady states are well defined and independent of $\boldsymbol{\varepsilon}_p$ and $\boldsymbol{\varepsilon}_p^{path}$ for some paths, which were used to prove the path-independence of each of the steady states; however, they are different for other straining pathes.

This leads to formulation of new key problem in the theories of plasticity and microstructure evolution under SPD: for which classes of plastic strain $\boldsymbol{\varepsilon}_p$, $\boldsymbol{\varepsilon}_p^{path}$, $p$, and pressure path $p^{path}$ material deforms according to each of the steady microstructure and surfaces of perfect plasticity $\varphi^i(\boldsymbol{\sigma}) = 0$ and for which of $\boldsymbol{\varepsilon}_p$, $\boldsymbol{\varepsilon}_p^{path}$, $p$, and $p^{path}$ classes there are jumps from one steady microstructure and $\varphi^i(\boldsymbol{\sigma}) = 0$ surface to another? This is an unsolved problem, and some steps toward its solution were suggested in [90]. Examples of loading paths are shown in Fig. 50c.

Despite this problem, the concept of perfectly plastic isotropic behavior, independent of $\boldsymbol{\varepsilon}_p$ and $\boldsymbol{\varepsilon}_p^{path}$, and the corresponding steady microstructures retains not only fundamental but also applied importance. Indeed, for each of the SPD techniques in Fig. 13b,c, and physical experiments leading to the deduction of their independence of $\boldsymbol{\varepsilon}_p$ and $\boldsymbol{\varepsilon}_p^{path}$, sufficiently broad classes of $\boldsymbol{\varepsilon}_p^{path}$ were used but not fully classified. Therefore, for these SPD processes and experiments, theories with the single surface $\varphi(\boldsymbol{\sigma}) = 0$ and corresponding value of steady $\sigma_y(p)$, $\rho_d$, and $d$ determined for this process are applicable. Indeed, determined with such a procedure the simplest isotropic surface of perfect plasticity $\sigma_i = \sigma_y(p)$ reproduces very sophisticated experimental evolution of distributions of the various parameters like pressure, radial and hoop elastic strains, and complex shape of the sample [89, 108, 109, 142, 143] (Figs. 6a,b, 7b,c, 12a and 49). Note that very small strain hardening (0.42 MPa for Fe and 0.22 MPa for Ni) for *ultra-SPD* ($q > 1,000$) was revealed and analyzed in [268, 269] and [90].

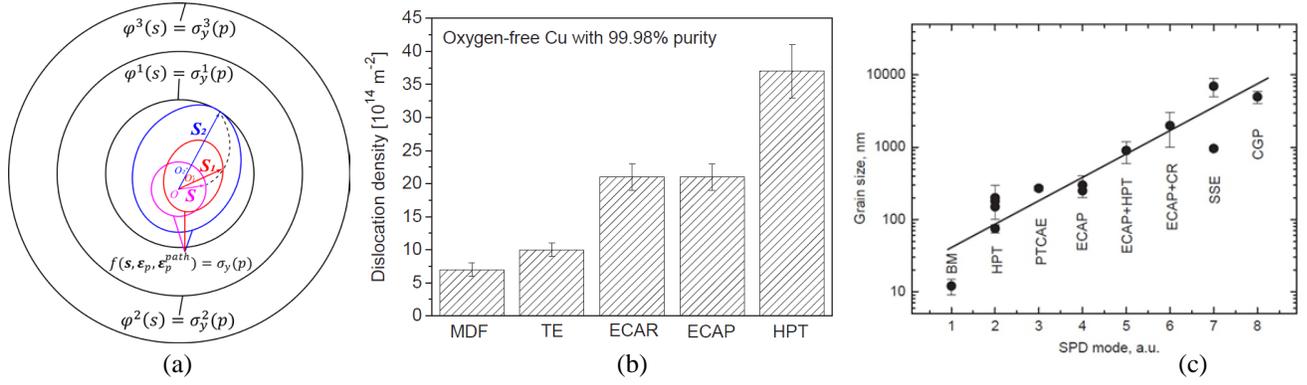

**Figure 13:** (a) Evolution of the deviatoric stress vector $\boldsymbol{s}$ and the yield surface $f(\boldsymbol{s},\boldsymbol{\varepsilon}_p,\boldsymbol{\varepsilon}_p^{path}) = \sigma_y(p)$ in 5D deviatoric stresses space at constant $p$ until they reach the fixed isotropic surface of perfect plasticity $\varphi(\boldsymbol{\sigma}) = \sigma_y^1(p)$, after which the material behaves like perfectly plastic, isotropic and independent of $\boldsymbol{\varepsilon}_p$, and $\boldsymbol{\varepsilon}_p^{path}$ [89]. There are several other fixed surfaces of perfect plasticity $\varphi^i(\boldsymbol{s}) = \sigma_y^i(p)$ with different yield strengths corresponding to different steady microstructures. They can be reached for some yet unknown classes of $\boldsymbol{\varepsilon}_p$, and $\boldsymbol{\varepsilon}_p^{path}$. (b) and (c) Steady-state dislocation density [267] and grain size [36] in Cu after various SPD modes. Designations: multi-directional forging (MDF), twist extrusion (TE), equal channel angular rolling (ECAR), equal-channel angular pressing (ECAP), ball milling (BM), planar twist channel angular extrusion (PTCAE), ECAP with following HPT (ECAP+HPT), ECAP with following cold rolling (ECAP+CR), simple shear extrusion (SSE), and constrained groove pressing (CGP).

There are debates at different meetings and in the literature about how to distinguish large plastic deformation from SPD. Giving any number, like $q > 10$ or 100, looks arbitrary. We suggest using $q > m$ as a definition of SPD, i.e., the strain at which the steady state in microstructure or hardness or $\sigma_y$ is reached for the given SPD technique or plastic strain paths. After this, not a lot of changes happen. The ultra-SPD can be defined by $q$ for which a meaningful (e.g., 5 or 10%) deviation from steady state is observed. Each material and SPD technique (straining path) have own SPD ultra-SPD limits. Note that independence of $\boldsymbol{\varepsilon}_p$ and $\boldsymbol{\varepsilon}_p^{path}$ and multiple steady states are also applicable to the minimum pressure $p_\varepsilon^d$ and kinetics for the strain-induced PTs [80, 108, 109] (Section 7.5).

## 2.9. Turbulent-like plastic flow in solids

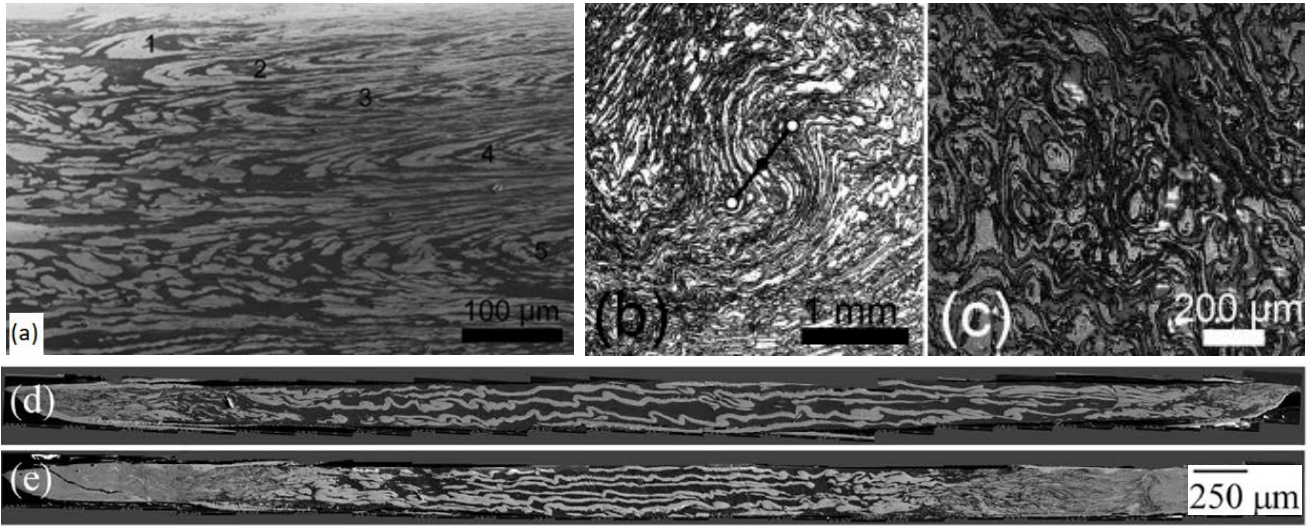

**Figure 14:** Images of the turbulent-like plastic flow during HPT. (a) TEM image of the shear patterns in the austenite/ferrite duplex stainless steel in the cross-section passing through rotation axis after 16 anvil rotations. (b) and (c) Optical images on the top sample surface of the same steel after 6 revolutions [271]. (b) A double-swirl pattern. The black dot is the centre of the sample. (c) The vortices are slightly away from the sample center. (d) and (e) SEM images of layered Al/Ni sample after 2 (d) and 5 (e) revolutions of the anvil, respectively [274].

During SPD with HPT, traditional laminar flow loses its stability and transition to the hierarchical turbulent-like plastic flow at different scales with different patterns occurs on the surface of austenite/ferrite stainless steel, high-purity Al, and Zn-22%Al eutectoid alloy [270, 271] and in the cross section passing through the rotation axis of the duplex stainless steel, Al-Mg, Al/Cu, Al/Ni, and Al/Co laminates [271–274] (Fig. 14). Similar patterns are observed in reacting shear bands within solids [275] and at the friction surfaces [276, 277]. They lead to fragmentation of stronger and more brittle materials, large local rotations, and intensified plastic flow and mixing. Due to qualitative change in the character of the plastic flow, the vortices may contribute to the jump from one steady microstructure to another (Section 2.8). They also may activate the 'ROLLER' mechanism of the intensification of the CRs under severe shear [278]. Attempts of the simulations of the main observed experimental features are presented in [273, 274] with FEM and [279] with isotropic moveable automata technique. In these papers, fragmentation of the layer with larger viscosity [273] or strength [274, 279] was considered, followed by formation of vortices. Understanding of the turbulent-like flow in solids is still in its infancy, and its incorporation in the constitutive models at different scales promises a qualitative leap.

*2.10. Analytical and FEM modeling of torsion under high pressure*

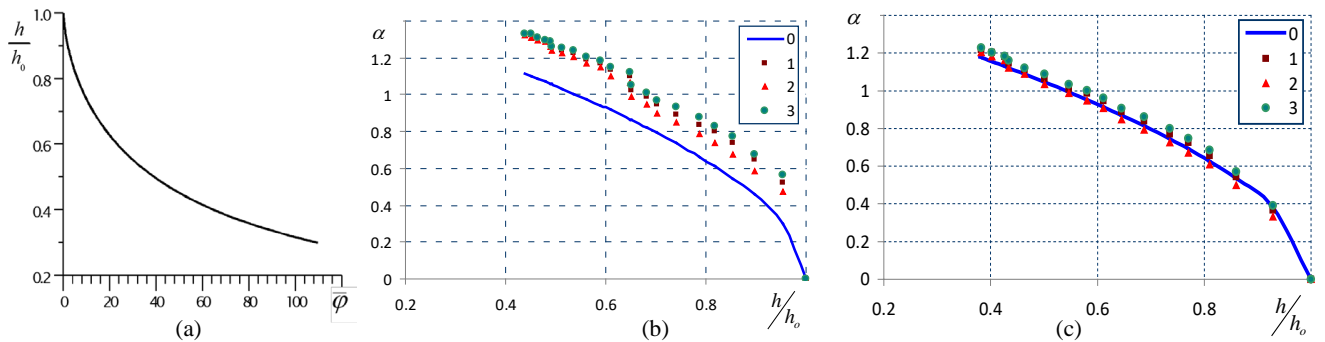

**Figure 15:** (a) Relative thickness of specimen $h/h_0$ as a function of the angle of relative sliding between an anvil sample material according to Eq. (10) [102]. (b) and (c) Angle $\alpha$ between the friction stress vector $\boldsymbol{\tau}_f$ and radial direction versus the current to initial thickness ratio of a sample under constant averaged axial stress $F$ over the culet radius $R$ normalized by $\sigma_y$ (b) $F = 3.68$ and (c) $F = 4.29$ for $r/R = 0.6$ (1), $r/R = 0.4$ (2), and $r/R = 0.8$ (3) [160]. Symbols are the FEM solution, and line corresponds to Eq. $(8)_3$.

*Flat culet.* The first analytical solution for the unconstrained cylindrical sample twisted between rigid rough anvils at $P = const$ and lateral pressure $\sigma_0$ was presented in [92] and, with improved consideration of surface sliding and more elaborated, in [100, 102]. The radii of a sample and anvil are equal; $|\boldsymbol{\tau}_f| = \tau_y = \sigma_y/\sqrt{3}$; $\boldsymbol{\tau}_f$ is *collinear to the relative sliding velocity* of material with respect to anvil. That is why

$$\tau_{rf} = \tau_y \cos\alpha, \tag{7}$$

where $\alpha$ is the angle between $\boldsymbol{\tau}_f$ and the radius $r$, i.e., rotation of an anvil reduces friction in the radial direction. Integration of Eq.(3) for this case with allowing for $P = const$ results in

$$p = \sigma_o + \sigma_y\left(1 + \frac{2}{\sqrt{3}}\frac{(R-r)}{h_0}\right); \qquad P = \pi R^2\left(\sigma_o + \sigma_y\left(1 + \frac{2}{\sqrt{3}}\frac{R}{3h_0}\right)\right); \qquad \cos(\alpha) = \frac{h}{h_0}, \tag{8}$$

with $h_0$ for the initial thickness before rotation. Eq.(8) coincides with Eq.(4) and means that the pressure distribution is independent of the anvil rotation. This is in agreement with the experiments for NaCl [62, 66] and X18H10T hardened steel [66]. The following differential equation was derived

$$d\bar{\varphi} = \frac{0.204\tilde{m}}{1 + 0.204\tilde{m}}d\varphi_a = -\frac{dh}{h}\sqrt{\left(\frac{h_o}{h}\right)^2 - 1}, \tag{9}$$

where $\varphi_a$ and $\bar{\varphi}$ are the angle of rotation of an anvil and the relative rotation angle of an anvil with respect to the sample, respectively, and $\tilde{m} = R/h_0$. After integration, we obtain

$$\bar{\varphi} = \frac{0.204\tilde{m}}{1 + 0.204\tilde{m}}\varphi_a = \sqrt{\frac{h_o^2}{h^2} - 1} - Arccos\frac{h}{h_o}. \tag{10}$$

The plot of $h/h_0$ vs. $\bar{\varphi}$ is shown in Fig. 15. The plot of $h/h_0$ vs. $\varphi_a$ looks similar, but with rescaled $\varphi$ axis and qualitatively corresponds to experimental data presented in [23, 66, 280] for various materials. Thus, the *rotation of an anvil reduces $\tau_{rf}$, which leads to a decrease in the sample thickness.* This, in turn, promotes the strain-induced PT, which will be discussed in Section 8.2. The thickness reduction is large at small $\varphi_a$ (for $\bar{\varphi} \to 0$, $\frac{dh}{d\bar{\varphi}} \to \infty$) and decelerates sharply with increasing $\bar{\varphi}$. For the same $\varphi_a$, sliding and the thickness reduction grow with increasing $\tilde{m}$; for $\tilde{m} = 5 - 10$, $\bar{\varphi} = (0.505 - 0.671)\varphi_a$. In the limit $\tilde{m} \to \infty$, $\bar{\varphi} \to \varphi_a$, only sliding without the torsion of the material occurs; still, the thickness reduction occurs and reaches its maximum. Analytical solution to the similar problem for $|\boldsymbol{\tau}_f| \leq \tau_y$ followed by some FEM testing was recently suggested in [281].

The first FEM solutions for torsion in RDAC were presented in [160, 210]. Small elastic strains and $\sigma_y(p)$ were assumed. While comparing the above analytical solution with the FEM results, in particular, in Fig. 15b,c, the following differences should be kept in mind. In FEM gasket is in contact with the conical diamond surface (which is no the case for analytical solution), which takes up to 11% of the total normalized force $F$ for $F = 4.29$ and up to 24% for $F = 3.68$ in Fig. 15b,c from [160]. Complete adhesion was imposed along the entire contacted surface, in contrast to plastic sliding in the analytical solution. Still, large plastic shears localized near the contact surface mimic contact sliding. Note that the slip was observed in HPT with metallic/ceramic anvils and large friction by vanishing the line drawn on the sample surface [23, 282]. For compression with $F = 4.29$ and following torsion, $|\boldsymbol{\tau}_f| = \tau_y$ along the major part of the culet, excluding central part with $r/R < 0.2$ for compression, which reduces almost to zero with increasing torsion. For this case, *linear pressure distribution, which does not change during the anvil rotation*, predicted analytically, is confirmed with FEM. Also, FEM solution confirms relation (10) between $\varphi_a$ and $h/h_0$ obtained from the curve in Fig. 15a and Eq. $(8)_3$ between angle $\alpha$ and $h/h_0$ (which is independent of $r$) in Fig. 15c. Confirmation is remarkable, taking into account the difference in problem formulations.

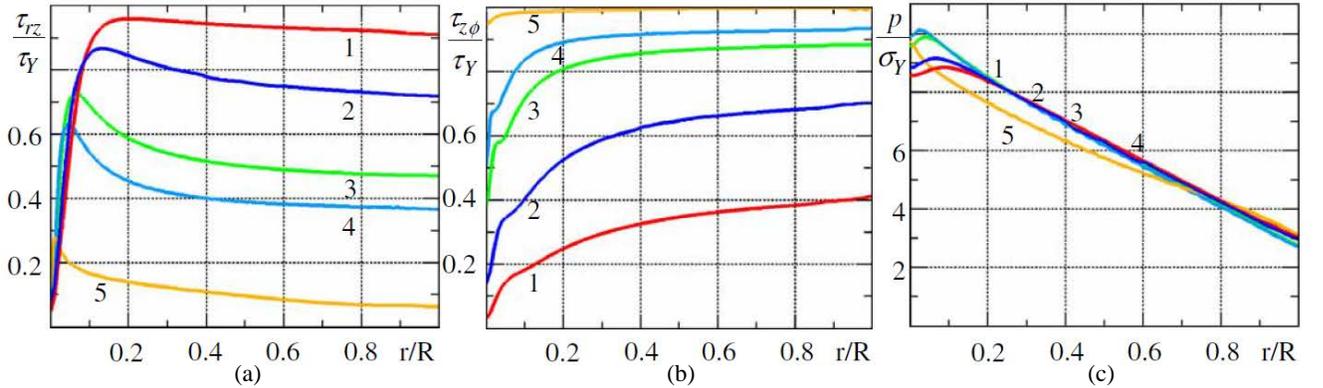

**Figure 16:** Radial distribution of the radial $\tau_{rz}$ (a) and circumferential $\tau_{\varphi z}$ (b) shear stresses and pressure (c) at the contact surface for rotation angles $\varphi_a = 0.03$ (1); 0.21 (2); 0.958 (3); 1.45 (4), and 2.034 (5). The magnitude $|\boldsymbol{\tau}_f| = (\tau_{rz}^2 + \tau_{\varphi z}^2)^{0.5} = \tau_y$ [160]. Decrease in pressure in (c) for $\varphi_a = 2.034$ is due to an increase in the part of the axial force for $r > R$.

For compression with $F = 3.68$, $|\boldsymbol{\tau}_f| = \tau_y$ for $r/R > 0.8$ only. During torsion, $|\boldsymbol{\tau}_f|$ increases until reaches $\tau_y$, but due to significant increase in the portion of force acting on the conical part of the anvil, pressure gradient slightly reduces during the torsion. In another simulation with $|\boldsymbol{\tau}_f| = \tau_y$ for $r/R > 0.5$ for compression in [210], during torsion, pressure significantly grows in the central region during the torsion due to the increase in $\tau_{fr}$. This is the *pressure self-multiplication effect*, which is the increase in pressure at $P = const$. It was earlier observed experimentally during the PT [62, 64, 66] and explained in [92, 100, 102] by PT to a stronger HPP. Also, FEM

relationship between angle $\alpha$ and $h/h_0$ has the same trend but shifted up in comparison with analytical Eq. (8)$_3$ in Fig. 15b, which is not surprisingly due to variable magnitude of friction and force on the culet.

For constrained HPT of a rigid-plastic disc with a power law equivalent stress-strain curve, an elegant self-similar analytical solution was found [283], provided that the friction on the cylindrical wall is not very large. In a self-similar regime, stresses, strains, and torque are power-law functions of $\varphi_a$ with different exponents. FEM simulations and experiments confirmed the existence of a portion of a solution with the self-similar regime, outside of which only FEM solutions are applicable. More advanced analytical solution for the constrained HPT with complete adhesion was obtained and analyzed in [284, 285] based on a strict theory in [4] with finite elastic and plastic strains, nonlinear Mooney-Rivlin elasticity, and perfect plasticity with $\sigma_y(p)$. FEM results for constrained HPT are also presented in [286, 287]. In [287], they were used to determine stress-strain curves for Al for $q \leq 11$. Significant correction on friction was included to get results compatible with the compression test for $q \leq 1$. Despite large $q$, a perfectly plastic plateau was not reached, probably due to errors involved in stress determination. FEM solution for porous material under constrained HPT based on the Beygelzimer model was presented in [288].

Starting with the first study in [289], there are numerous FEM solutions for HPT with metallic anvils. They are well reviewed in [39], and that is why they will be mentioned very shortly here. The most cited papers are for quasi-constrained HPT with focus on plastic flow [290] and corresponding temperature increase [291–293]; see also [294] with study of the effect of the geometric parameters and [295, 296]. Unconstrained HPT was treated for copper in [297] and polymer in [298]. The effect of the double-sided HPT was investigated experimentally and with FEM in [299]. Microhardness distribution was simulated in [127, 158, 300] in comparison with experiments.

An advanced dislocation-based model from [301] was applied for unconstrained [302] and in much more detailed paper [303] to quasi-constrained HPT; see also [304]. A combination of the pressure measurement using known PT pressure in Bi with FEM simulation in [127] allowed us to analyze the actual pressure distribution in copper during quasi-constrained HPT. A combined Coulomb and plastic friction model and stress-strain curve with transition to perfect plasticity in [158] resulted in even better description and analysis of the same experiments in [127], in addition to results described in Section 2.1(c). Plastic strain distribution is not well described by a simplified torsion equation. Pressure after compression in [158] has large gradient with maximum of 2.6 GPa at the center, but reduces to 2.3 GPa with small pressure gradient for quasi-steady state after torsion due to increase in the flash area. Thus, processes like PTs and microstructure evolution cannot be characterized by measurement in the steady state only.

*Beveled culet: pressure self-focusing effect.* Beveled culet is often used for megabar pressures to compensate cupping due to large elastic deformation of anvils. A generalization of an equilibrium Eq. (3) for this case is [3, 212, 213]

$$\frac{d\sigma_{rr}}{dr} = -\frac{2\tau_r}{h} + \frac{2\tan\alpha(\sigma_c - \sigma_{rr})}{h} + \frac{\sigma_{\theta\theta} - \sigma_{rr}}{h} \quad \rightarrow \quad \frac{d\sigma_{rr}}{dr} \simeq \frac{dp}{dr} \simeq -\frac{2\tau_r}{h} = -\frac{2\tau_r}{h_0 + 2r\tan\alpha}. \tag{11}$$

Here, $\alpha$ is a bevel angle, $\sigma_{\theta\theta}$ and $\sigma_c$ are the hoop and normal to contact stresses, and $h_0$ is the sample thickness at $r = 0$. All stresses are at the diamond-sample contact, where $|\boldsymbol{\tau}_f| = \tau_y$ is assumed at megabar pressure. Therefore, it follows from the von Mises yield condition that $\sigma_{\theta\theta} \simeq \sigma_{zz} \simeq \sigma_c \simeq \sigma_{rr} \simeq p$, which enables the transition from Eq. (11)$_1$ to (11)$_2$. Eq. (11) was analyzed in [212, 213] with the goal to maximize pressure gradient, and, consequently, $p$ near the sample center. It is evident from Eq. (11) that a very large pressure gradient could be achieved at $r = 0$ for a small $h_0$. The larger $\alpha$ is, the larger the pressure gradient growth toward the sample center. However, the maximum $\alpha$ is constrained by diamond fracture at relatively low pressures and elastic deformations, and does not usually exceed $8.5^o$. Along with analytical treatment, FEM solutions for DAC and RDAC have been found for Re in [212, 213]. Fully nonlinear formulation with large elastic and plastic strains and material rotations and nonlinear elasticity for Re and diamonds is implemented (similar to [142] but without axial symmetry). Coulomb friction with $\mu = 0.1$ was used until $|\boldsymbol{\tau}_f| = \tau_y$ is reached; $\boldsymbol{\tau}_f$ is *collinear to the velocity of relative sliding*.

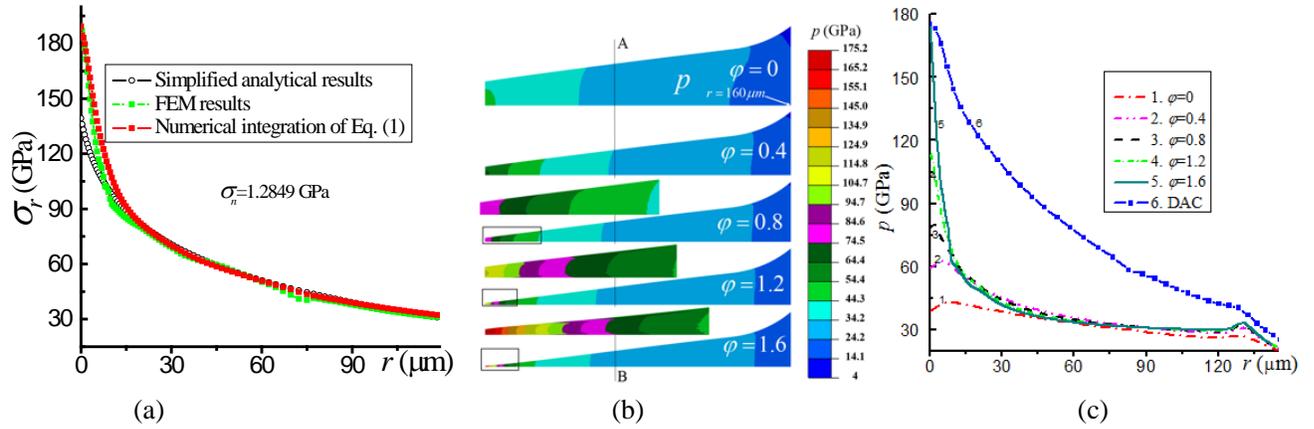

**Figure 17:** Pressure self-focusing effect in DAC and RDAC. (a) Radial distribution of the radial stress $\sigma_r$ obtained with FEM, simplified Eq. (11)$_2$, and integration of the equilibrium Eq. (11)$_1$ with FEM input. (b) Evolution of the quarter of the sample profile and 2D pressure field during anvil rotation in RDAC. (c) Corresponding radial pressure profile for torsion in RDAC and compression in DAC to the same maximum pressure [212].

Strong growth in the pressure gradient toward the center due to gradient in the thickness is evident in Fig. 17. Such a growth was called in [212, 213] the *pressure self-focusing effect*, to be distinguished from the pressure self-multiplication effect due to PT to a stronger phase revealed in [62, 64] and explained in [92, 100, 102]. FEM solution gives significantly higher pressure gradient and $p_{max}$ than Eq. (11)$_2$. However, a more complete Eq. (11)$_1$, integrated numerically with the stress input from FEM, reproduces FEM solution quite well (Fig. 17a). A non-trivial effect of the pressure hardening coefficient $b$ in $\sigma_y = \sigma_\infty + bp$ was found in [212, 213]. While comparing the results for $b = 0$ and $b = 0.04$ at the same sample thickness, essentially larger pressure gradient was obtained for $b = 0.04$, as expected from Eq. (11). However, while comparing for the same $P$, the pressure gradient and $p_{max}$ are essentially greater for $b = 0$, because of the thickness reduction at the center by a factor of 4.

An increase in the center pressure gradient is found in DAC experiments on Re and W [131, 132], which are well described with FEM in [142, 143]. This is considered a conceptual proof of the pressure self-focusing effect. Another, much stronger proof comes from the huge pressure gradient due to large bevel at the toroidal tip of toroidal DAC (Fig. 10b,c). Such an explanation of the increased pressure in the toroidal DAC was lacking in literature.

Evolution of the contact normal and friction stresses and stress and plastic strain fields in the Re sample in RDAC and in diamond anvil have been analyzed in [213] for different loading and friction conditions, material, and geometric parameters. *Torsion in RDAC is much more effective way to reduce sample thickness and consequently to increase pressure gradient and pressure* at the center (Fig. 17b,c). In contrast to the flat culet, for which pressure distribution does not change during anvil rotation at $P =$const (Eq.(8) and Fig. 16c), for beveled culet $p_{max}$ drastically grows due to drastic reduction in thickness. The total force for RDAC to produce the same $p_{max}$=175 GPa like in DAC is much lower and causes smaller anvil bending and tendency to cupping, and reduces chances to encounter defects in the anvils and breaking them. Thus, *RDAC offers a promising way to increase $p_{max}$*, although in a tiny volume. *Toroidal RDAC is expected to be even more effective*.

*2.11. Summary and perspectives*

The main experimental challenges in studying materials' behavior in DAC and RDAC are strong heterogeneity of stress and plastic strain tensor fields in radial and (to lesser extent) axial directions, and that very limited number of parameters (like pressure averaged over the sample thickness and sample thickness profile) can be measured only. Recollecting Robert Penn Warren saying "You have to make the good out of the bad because that is all you have got to make it out of," this challenge is transformed into an opportunity to generate big data from a single experiment to develop coupled experimental-FEM approaches.

As a result, the following main rule was proved for many materials, in some cases based on in-situ experiments: *after some critical level of plastic deformation, the yield strength, dislocation density, crystallite size, and microstrain reach steady values (not necessarily simultaneously), which are independent of $\boldsymbol{\varepsilon}_p$, including its mode and $\boldsymbol{\varepsilon}_p^{path}$; dislocation density, crystallite size, and microstrain are also independent of pressure and $p^{path}$; for monotonous and quasi-monotonous loading, materials also deform like isotropic and perfectly plastic with fixed surface of perfect plasticity $\varphi(\boldsymbol{\sigma}) = 0$.*

However, different steady states for all the above parameters were found after different SPD processing, which led to formulation of new outstanding problem: *for which classes of $\boldsymbol{\varepsilon}_p$, $\boldsymbol{\varepsilon}_p^{path}$, $p$, and pressure path $p^{path}$ material possesses one steady values of dislocation density, crystallite size, microstrain, and yield strength ($\varphi(\boldsymbol{\sigma}) = 0$ yield surface), and for which of $\boldsymbol{\varepsilon}_p$, $\boldsymbol{\varepsilon}_p^{path}$, $p$, and $p^{path}$ classes there are leaps from one steady microstructure and yield strength to another?* Some steps toward its solution were suggested in [90]. Note that very small strain hardening for *ultra-SPD* ($q > 1,000$) was revealed and analyzed, which does not compromise the above rule for smaller strains (see also [90]).

Advanced models and FEM approaches have been developed to study the mechanical behaviour of a sample. They are coupled to experiments to extract all material properties, including $\sigma_y(p)$, higher order elastic constants (at megabar pressures), contact friction, and then calculate stress and plastic strain fields. This is done along the statement in [124] that the ideal characterization should contain complete information which is required for simulating the same experiments. At megabar pressures, contact friction stresses $\tau_f$ are very nontrivial, with small plastic friction zone (which makes it impossible to determine $\tau_y(p)$), and growing adhesion zone (where stresses may oscillate) in expense of mostly Coulomb friction. DAC and RDAC with rough diamonds have been developed to maximize the contact friction and ensure that after some straining $\tau_f = \tau_y$ in the major part of the sliding zone. DAC/RDAC with rough diamonds intensify plastic flow, nanostructure evolution, and (see [82, 89] and Section 7.5.1) strain-induced PTs. They are similar to toroidal DAC but with numerous stochastic or arranged grooves.

Among various methods to determine $\sigma_y(p)$ discussed here, the most practically attractive is to measure its distribution in each phase along the diameter or entire area of the sample using axial XRD and line width (i.e., elastic microstrain) method. However, much better theoretical equations for extracting $\sigma_y$ and their verification are required. Understanding of the observed turbulent-like flow in solids during SPD is one more outstanding problem, and its incorporation in the constitutive models at different scales promises a qualitative progress in simulation and understanding of microstructure and mechanical and mechanochemical behavior. For DAC and RDAC with a beveled culet, pressure self-focusing effect was predicted theoretically, which is consistent with some existing experiments. It suggests a promising way to increase pressure in the central region, especially in RDAC.

## 3. Experimental Mechanochemical Phenomena

PTs in solids are mostly characterized by equilibrium pressure-temperature phase diagrams [305], each line of which is obtained from the equality of the Gibbs energy of the corresponding phases or zero driving force for the PT. The phase equilibrium conditions in elastic materials have been extended for the general tensor stress in geological [306–309], material science [310–313], continuum mechanics [314] literature, and summarized in a review [315] and book [316]. However, in experiments, especially at room and low temperatures, PTs occur under significant deviation from equilibrium pressure $p_e$ [77, 317–319], i.e., with large hysteresis. For example, at room temperature, the graphite-diamond equilibrium pressure is 2.45 GPa; however, due to hysteresis, the PT is observed only at 70 GPa [318] for crystalline precursor and at 80 GPa [320] or 100 GPa and 400 $^oC$ [321] for a glassy carbon. While the superhard cBN is stable at ambient conditions, [319], disordered low-pressure hBN does not transform even at 52.8 GPa [77]. At continuum level, this deviation is formalized through introducing an athermal threshold $K$ in Eq. (17) for the thermodynamic driving force $X$. Some stable phases, e.g., ringwoodite [83] or Si II in 30 nm particles [81], do not appear under hydrostatic loading, because hysteresis is so high that other phases appear first.

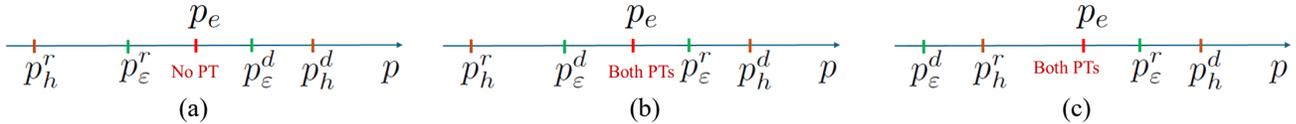

**Figure 18:** Characteristic pressures for three types of transformational behavior. (a) Relatively slight promotion of direct and reverse PTs by plastic straining, $p_\varepsilon^d > p_\varepsilon^r$. (b) Strong promotion of direct and reverse PTs by plastic straining, $p_\varepsilon^d < p_\varepsilon^r$. (c) Very strong promotion of direct PT by plastic straining, $p_\varepsilon^d < p_h^r$.

Let $p_h^d$ and $p_h^r$ be the minimum pressures for direct and reverse PTs under hydrostatic conditions, which are located on both sides of $p_e$ (Fig. 18), and $p_\varepsilon^d$ and $p_\varepsilon^r$ are the minimum pressure for direct and maximum pressure for the reverse strain-induced PTs. The pressure hysteresis under hydrostatic loading is defined as $H_p = p_h^d - p_h^r$. Due to hysteresis, it is practically impossible to determine $p_e$ experimentally. It is conditionally defined as $p_e = 0.5(p_h^d + p_h^r)$ or by extrapolation from the high temperature $p_e$, since $H_p$ reduces down to zero at high temperatures, when thermal fluctuations can overcome all barriers. Since plastic straining promotes both direct and reverse PTs, i.e., reduces the PT pressure for direct PT and increases the PT pressure for the reverse PT (Fig. 18a), pressure hysteresis $H_\varepsilon = p_\varepsilon^d - p_\varepsilon^r$ reduces. This led to hope that plastic straining can localize the phase equilibrium pressure [62, 64, 322]. However, for strong effect of the plastic straining (Fig. 18b), one may observe $p_\varepsilon^d < p_e < p_\varepsilon^r$, i.e., direct strain-induced PT may occur below $p_e$ and reverse strain-induced PT may occur above $p_e$. For very strong promotion of direct PT by plastic straining, one may observe even $p_\varepsilon^d < p_h^r$. This seems contradictory, because strain-induced HPP should immediately transform back below $p_h^r$ via pressure-induced PT. However, all four characteristic pressures depend on evolving microstructure (grain size and dislocation density) and volume fraction of HPP $c$, and strain-induced PT may reduce $p_h^r$ and avoid or strongly retard the inverse pressure-induced PT.

Note that reports on *hydrostatic* loading under pressure above solidification pressure of the PTM should be considered with sufficient irony. At room temperature, the highest solidification pressure of 12.1 GPa at room temperature has He [323]. Such experiments are called "quasi-hydrostatic", which does not say how high non-hydrostatic stresses are applied. In particular, by changing PTM from more to less hydrostatic, the start pressure for both direct-reverse $\alpha \leftrightarrow \varepsilon$ PTs in Fe varies between 10 and 16 GPa [324]. The pressure for completing the PT varies between 15.3 and 25.4 GPa for the $\alpha \to \varepsilon$ PT and between 3.7 and 8 GPa for the reverse $\varepsilon \to \alpha$ PT. The effect of the PTM partially explains the large experimental scatter in PT pressures in Fe (summarized in [317, 324]), with $p_e = 6 - 14$ GPa, start and finish pressures for $\alpha \to \varepsilon$ PT from 8.6 to 16 and from 14 to 25.4 GPa, respectively, and the $\varepsilon \to \alpha$ PT beginning and end pressures between 7 and 16.2 GPa and 1 and 8.5 GPa, respectively. Similar studies were performed for $\alpha - \omega$ PT in Ti in [325] and Zr in [326]. Surprisingly, if for coarse-grained Ti, the PT in [325] was reduced from 10.5 (in argon) and 10.2 GPa (in 4:1 methanol-ethanol PTM) to 4.9 GPa without PTM, in [327] strongly pre-deformed nanocrystalline Ti starts to transform at 10.0 GPa in 4:1 methanol-ethanol PTM and 9.9 GPa without PTM. However, in our experiments on strongly pre-deformed by multiple rolling till steady microstructure nanocrystalline Ti (Lin, Yesudhas, and Levitas, unpublished), $\alpha - \omega$ PT in Ti under plastic compression starts at 2 GPa and completes at 6 GPa.

One can range the degree of non-hydrostaticity based on the yield strength of the PTM. For example, it was found in [233] that Bi at 200 GPa has low yield strength below 1 GPa (Fig. 9) and is the most hydrostatic PTM. This still does not say how to characterize heterogenous tensorial stress state in the sample due to its irregular shape and unclear friction condition. Even worse, sample starts deforming plastically even before both anvils start touching it directly. When thickness of PTM layer between anvil and sample is getting small, its stress state is close to that for a sample between two anvils (Sections 2.2 and 2.4), i.e., it has large pressure gradient and reaches pressure many times higher than the yield strength of PTM. That is why one may have both plastic strain and stress-induced PTs occurring simultaneously, which are more difficult to analyze quantitatively, and this has never been done. Below, we analyze some experimental phenomena, which occur during plastic flow under high pressure.

*3.1. Plastic straining drastically reduces the PT and CR pressures*

Numerous examples of the significant reduction in PT and CR pressure due to HPT with rotating Bridgman anvils, starting from Bridgman's work [6, 7], are reviewed in [23, 64, 65, 100]; for organic reactions in [11–14, 16], oxides decomposition in [17], for RDAC in [64, 65, 84, 85, 100]. We will not repeat them. Methods based on detecting PTs/CRs, e.g., by jumps in the torque or electric conductivity, are not generally reliable. We will focus on in situ XRD results and ex situ results for irreversible transformations for which retrieved HPP can be characterized by XRD and/or TEM. Also, one needs to distinguish different methods of determination of the PT pressure, see Section 1.3.

Table 1 contains the most prominent to date cases of the reduction in PT pressures for various classes of materials, many of them will be considered in more detail in later Sections. The largest reduction is for the irreversible graphite-cubic diamond ($\times$ 100) and reversible graphite-hexagonal diamond ($\times$ 50) PTs and for Si-I$\rightarrow$Si-II PT for 100 nm particles. However, some PTs (marked with $\times\infty$, i.e., infinite reduction in PT pressure) were not observed at hydrostatic loading at all (e.g., PTs olivine$\rightarrow$ringwoodite, Si-I$\rightarrow$Si-III PT for 30 nm particles) but transform to other phases. These results can also be classified as the changed in PT path considered in Section 3.3.

An important point for comparison pressure- and strain-induced PTs is that they should be done for exactly the same material with the same initial microstructure, as it is illustrated for $\alpha \rightarrow \omega$ PT in Zr. Another point is that if pressure is not determined locally in situ but as force/area (typical for HPT with metallic or ceramic anvils), then, according to [125, 126], it should be multiplied by a factor of 3 or higher.

It is evident from Table 1 that PTs in graphite, Zr, and Si occur well below $p_e$. The PT pressure for Si-I$\rightarrow$Si-II PT under plastic straining is also lower than the reverse PT pressure under hydrostatic loading (while reverse PT Si-II$\rightarrow$Si-I does not occur, but Si-II transforms to other phases, Si-XII and III). The same was claimed (but without XRD proof) for semiconductor$\rightarrow$metal PT in InTe, InSb, Ge-I$\rightarrow$Ge-II, and Si-I$\rightarrow$Si-II [330–332]. Thus, the sequence in PT pressures for these PTs corresponds to Fig. 18b and c. Significant reduction in PT pressure in RDAC was obtained for synthesis of highly energetic polymeric phases of nitrogen and sodium azide [333, 334].

Drastic reduction in PT and CR pressures cannot be explained by the thermodynamic effect of the applied deviatoric (shear) stresses (Section 4.4). It is explained in Section 6 by nucleation at the tip of the dislocation pileup, producing very strong concentration of all tensor stress components. In early works, some qualitative mechanisms were suggested based on the effect of elastic (but not plastic) shear [335–338], "ROLLER" [278] and "CONMAH" [339] models. Intensification of CRs is connected to fracture and appearance of 'fresh' (i.e., not passivated by oxides and other protective films) surfaces and intensified mixing [113]. It was suggested in [14] that stressed chemical bonds are chemically activated and react with other molecules under flow-induced collision. Alternatively, high elastic bond energy transfers to an oscillatory energy of molecules, increasing their reactivity. Two potential mechanisms of reaction acceleration were suggested [8]: one is accumulation of elastic energy (which was rejected for experiments with quartz and calcite) and bond breakage that produces nucleation sites.

Previously, it was considered that plastic shear reduces the PT pressure not only in comparison with hydrostatic, but also compared to nonhydrostatic compression [5–7, 23, 31, 32, 64]. Plastic compression without the rotation of an anvil also significantly reduces PT pressure compared to hydrostatic loading, e.g., for irreversible PT in $\gamma-Al_2O_3$ from over 50 to 35 GPa [340] and for $\alpha \rightarrow \omega$ PT in Zr [32, 329]. As it is discussed in Section 7.5

Table 1: Examples of reduction in PT pressure due to plastic straining

| Material | Hydrostatic, GPa | Strain-induced GPA | Loading | Reduction | Source | Method |
|---|---|---|---|---|---|---|
| Highly disord. hBN →wBN[1] | NT<52.8 GPa $p_e$<0 | 6.7 GPa, 300° | RDAC | >×8, ∞ | Ji, Levitas., Zhu … PNAS, 2012 | In situ XRD |
| highly ordered rBN to cBN[2] | 55 GPa $p_e$<0 | 5.6 GPa | DAC | ×10 | Levitas, Shvedov PRB, 2002 | TEM |
| α→ω commercially pure Zr[3] | 6.0 GPa $p_e$=3.4 | 0.67 GPa 1.36 GPa | Rough-DA Smooth-DAC | ×9 ×4.4 | Lin, Levitas, Pandey.. MRL2023 | In situ XRD |
| α→ω PT ultra-pure Zr[4] | 5.4 GPa $p_e$=3.4 | 1.2 GPa | DAC RDAC | ×4.5 | Pandey, Levitas Acta Mat 2020, 196, 338-346 | In situ XRD |
| α→ω annealed ultra-pure Zr | 5.4 GPa $p_e$=3.4 | 2.3 center 1.2 edge | DAC RDAC | ×2.3 ×4.5 | | In situ XRD |
| α→ω PT in commercially pure Zr[5] | 6.0 GPa $p_e$=3.4 GPa | 1 GPa (>3GPa) | Compression | ×2 | Srinivasarao, Zhilyaev Scr Mat 2011, | Ex situ XRD |
| | | 0.25 (>0.75 GPa) | HPT 5 turns | ×8 | Zhilyaev… Mat. Sci. Eng. A 2011 | |
| Graphite-cubic diamond[6] | 70 GPa $p_e$=2.45 GPa | 0.7 GPa | RDAC | ×**100** | Gao, Ma, An, Levitas, Zhang, Feng, Chaudhuri, Goddard, Carbon 2019, 146, 364-368 | In situ XRD |
| graphite-hexag. diamond | 20 GPa $p_e$=2.45 | 0.4 GPa | RDAC | ×**50** | | In situ XRD |
| Si I-Si II 30 nm 100 nm particles 1 micron[7] | NT 16.2 GPa 13.5 GPa | 4.3 GPa 0.3 GPa 2.6 GPa | DAC DAC, RDAC DAC | × ∞ ×**54** ×5.2 | Yesudhas S., Levitas, V.I., Lin F., Pandey K.K., Smith J., Nature Communications 2024, 15, 7054 | In situ XRD |
| Si I-Si III 30 nm 100 nm particles 1 micron | NT NT NT; $p_e$=10 | 6.7 GPa 0.6 GPa NT | DAC RDAC DAC | × ∞ × ∞ | | In situ XRD |
| olivine → ringwoodite[8] | NT $p_e$=15 GPa | 15 GPa | RDAC Rough-DA | × ∞ | Lin, Levitas, Yesudhas … GRL 2025 | In situ XRD |

Designations: NT means not transformed because material transformed to higher-pressure phases; NT ≤ 52.8 GPa means did not transform up to ≤ 52.8 GPa. [1]Highly 2D-disordered (concentration of turbostratic faults $s \simeq 1$) hexagonal hBN to superhard wurtzitic wBN PT [77]; [2]PT from highly ordered and textured rhombohedral rBN to cubic cBN after rotational plastic instability in DAC; pressure could essentially grow during PT [328]; [3]α → ω PT in strongly predeformed commercially pure Zr; $p_\varepsilon^d$ is the same along the radius for both smooth-DA and rough-DA, i.e., it is independent of $\varepsilon_p$ and $\varepsilon_p^{path}$ [89]; [4]α → ω PT in strongly predeformed ultra-pure Zr; $p_\varepsilon^d$ is the same along the radius for both smooth-DA and rough-DA, i.e., it is independent of $\varepsilon_p$ and $\varepsilon_p^{path}$ [80]; [5]α → ω PT in strongly predeformed commercially pure Zr [32, 329]; pressure is determined as force/area and, according to [125, 126], should be multiplied by > 3; [6] Irreversible graphite-cubic diamond and reversible graphite-hexagonal diamond PTs [79]; [7] Si-I→Si II and Si-I→Si III PT for 30 nm, 100 nm, and micron particles [81]; [8]Reversible olivine→ringwoodite PT [83].

for strongly plastically predeformed Zr, $p_\varepsilon^d$ for the strain-induced $\alpha - \omega$ PT is independent of the plastic tensor strain, its mode, and path. It was also obtained in Section 7.6 that the entire strain-controlled kinetic equation is independent of plastic tensor strain, its mode, path, and deviatoric tensor stress, as well as pressure $p$-plastic strain $q$ paths.

Retainable at normal pressure $\beta$-Zr (along with $\omega$-Zr) was obtained by HPT at 3-6 GPa in [31], at 1 GPa (just with compression) in [32], and even at 0.5 GPa (after 5 anvil turns) in [329], instead of reversible PT at 30-35 GPa under hydrostatic loading. Published X-ray patterns on retrieved sample show clear $\beta$-Zr peaks. While the PT pressures for $\beta$-Zr should be multiplied by a factor of 3-5 due to strong pressure gradient [125, 126], the reduction in PT pressure and, especially, retrieving $\beta$-Zr are remarkable. Unfortunately, no other groups could repeat these results. In situ study of plastic straining for ultra-pure Zr in [80] up to 13 GPa and $200^o$ rotation and of exactly the same commercially pure Zr (kindly sent to the author by M.T. Perez-Prado and late A.P. Zhilyaev) up to 16 GPa [82, 89] did not show traces of $\beta$-Zr. The DFT simulations also demonstrate phonon instability in $\beta$-Zr at 0 K below 26 GPa [341]. Thus, results in [31, 32, 329] remain a mystery.

A similar mystery is obtaining HPT at averaged pressure of 5 GPa (local pressure could be more than 3 times larger) and 0.25 to 10 revolutions of $\beta$-Ti [342] and retaining it at normal conditions for the first time. This work reports a small amount of $\beta$-Ti with the grain size of 150 nm within mixture of $\alpha$ and $\omega$ Ti determined from the transmission Kikuchi diffraction micrographs. Drastic reduction in PT pressure was interpreted with the help our mechanism from [100, 102]. The $\beta$-Ti could be an intermediate phase on the pathway from $\alpha$ to $\omega$, and could be stabilized by small grain size and low-energy phase interfaces between these phases pinned by defects.

There are also various examples that plastic straining practically does not reduce the PT pressure [6, 7, 23, 74, 86, 327]. For PT hBN$\rightarrow$wBN, this was explained in [74, 86] by a simultaneous occurrence of two opposite processes: traditional promotion of PT due to strain-induced defects with strong stress concentrators is compensated by increase in the PT pressure due to increase in concentration of the degree of 2D disordering (see Section 7.4).

In early geological literature, PT from the orthoenstatite to clinoenstatite in $MgSiO_3$ is promoted by twisting of an anvil [10] and occurs in the region of stability of the orthoenstatite. In particular, clinoenstatite was obtained at 1 GPa and 920 $^0$C, i.e., 9.7 GPa below the $p_e$ or 254 $^0$C above the phase equilibrium temperature. This result also shows that *shear affects some PTs, even at relatively high temperature (at $\simeq 0.63 T_m$)*, for which the yield strength is 2 times lower than at room temperature.

Various new HPPs with unique properties have been discovered experimentally: new superhard phases of carbon [343, 344], high-temperature superconductors [345–349], $BC_2N$ [350, 351], $BC_5$[352], $B - BN$ [353], polymeric nitrogen [354] and $CO_2$ [355] (which are expected to be highly energetic materials), and ionic boron [356]. PT pressure for most of these phases is too high for technological applications. Based on the above results on HPT and in RDAC, SPD can significantly reduce the required pressure.

*3.2. Plastic straining reduces the pressure hysteresis for the PTs and CRs; steps at pressure distribution*

As it was mentioned around Fig. 18, promotion of each or both direct and reverse PTs implies reduction in pressure hysteresis $H_\varepsilon = p_\varepsilon^d - p_\varepsilon^r$. This led to hope that plastic straining can localize the phase equilibrium pressure [8, 9, 62, 64, 322, 357]. In geological literature, utilizing grinding in mortars and mechanical mixers [357] or more qualitative experiments with anvils rotating oscillatory by $\pm 2^o$ [8, 9], the rutile-orthorhombic PT $PbO_2I \leftrightarrow PbO_2II$ and $MnF_2I \leftrightarrow MnF_2II$, quartz$\leftrightarrow$coesite PT in $SiO_2$, litharge $\leftrightarrow$ massicot PT in $PbO$, calcite $\leftrightarrow$ aragonite PT in $CaCO_3$, as well as reaction 2 jadeite$\leftrightarrow$nepheline+albite have been studied. The main conclusion was that superposed shear does not change the PT pressure-temperature equilibrium line, but the transformation starts at a temperature lower by $100 - 300^0 C$ (which is the same as reduction in PT pressure for the same $T$) than under hydrostatic loading and thermally activated kinetics is significantly accelerated by a factor of 3 to 30 for PT in $PbO_2$ and 50-200 for the reaction in jadeite. Since deviation of PT start pressure from $p_e$ reduces both for direct and reverse transformations, especially for low temperature, pressure hysteresis is lowered to almost zero. Thus, this case corresponds to Figure 18a, but maybe much larger plastic straining will lead to case in Figure 18b or c. Such a behavior can be described by a thermally activated nucleation at dislocation pileup (Section 6.5).

For PT semiconductor↔metal in $InSb$, the $H_\varepsilon = 1.0$ GPa with plastic straining vs. 1.75 GPa under hydrostatic loading [331]. Pressures for direct and reverse strain-induced PTs for commercially pure Zr and Ti in rotating Bridgman anvils were practically the same in the range of 2-2.5 GPa, implying zero hysteresis [64, 322]. In DAC and RDAC, narrow *plateaus or steps,* i.e., regions with almost constant pressure, were observed at the heterogeneous pressure distribution (Fig. 19) [62, 64, 66, 358, 359]. They correspond to a localized two-phase region (diffuse interface) in which PT occurs; therefore, the pressure value at these steps corresponds to $p_\varepsilon^d$ for the direct PT and $p_\varepsilon^r$ for the reverse PT. FEM simulations on behavior of the sample in DAC and RDAC in Section 8.3 show that for equal and stronger HPP and fast enough kinetics, a plateau at the step indeed corresponds to narrow two-phase region for which $p = p_\varepsilon^d$.

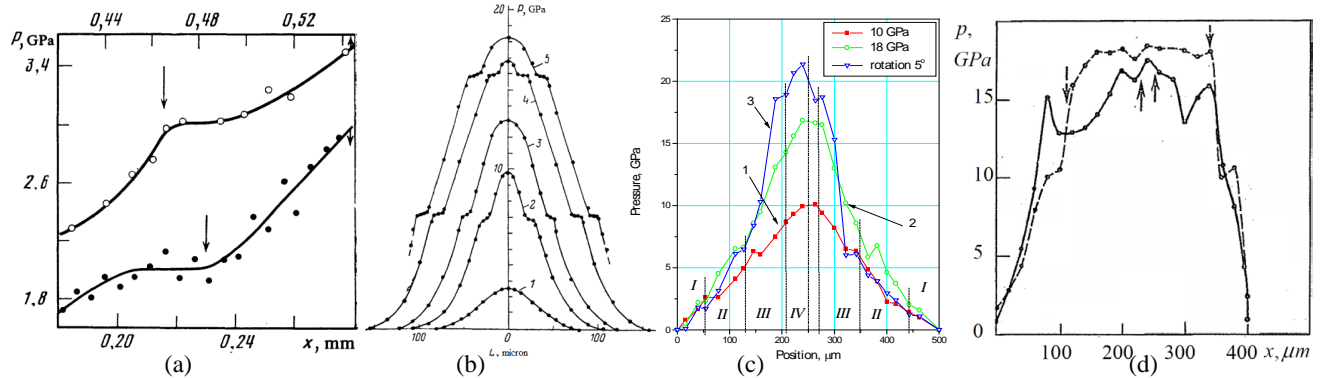

**Figure 19:** Plateaus (steps) at pressure distributions measured by ruby fluorescence. (a) At $B1 \to B2$ phase boundary in KCl; hollow circles are after compression and solid circles are after torsion by $15^o$ [62]. The arrows here and in (d) indicate the location of a visible phase boundary. (b) At I→II and II→III boundaries in PbTe for different degrees of compression [358]. (c) At phase boundaries between 4 different phases of fullerene $C_{60}$ after compression and torsion [66]. (d) Oscillating pressure distributions for PT in ZnSe; solid and dashed lines are before and after torsion by $11^o$, respectively [63]. In (c) and (d), pressure growth in the transforming region during torsion at constant axial force, exhibiting pressure self-multiplication effect.

Remarkably, for I→II and II→III PTs in PbTe and II→III PT in $C_{60}$, $p_\varepsilon^d$ does not change during the loading with changing position of the phase boundary (Fig. 19b and c). Based on discussion in Section 7.5.1, since plastic tensor strain, its mode, and path vary significantly with radius (see FEM simulations [110, 125, 126, 251, 360]), we arrive at the following rule: *the minimum pressure for the strain-induced PTs, $p_\varepsilon^d$, is independent of the plastic tensor strain, its mode, and path.* However, such a conclusion was not made in [66, 358] but was made first in [80].

For $B1 \to B2$ PT in KCl after the twisting an anvil by $15°$, the pressure at the plateau was $1.8 \pm 0.1$ GPa for both direct and reverse PTs, i.e., $H_\varepsilon = 0$ [62]. Without twisting, the pressure at the step was 2.4-3.0 GPa for the direct and 1.2-1.4 GPa for the reverse PT, i.e., with significant hysteresis [62, 64]. It was claimed in [62, 64] that the pressure at the step under torsion in RDAC can be considered as a phase equilibrium pressure. This is, however, generally not true due to several reasons.

1. While with decreasing hysteresis in Figure 18a, gap between $p_\varepsilon^d$ and $p_\varepsilon^r$ reduces, nothing implies that it should close exactly at $p_e$. For most results in Table 1, $p_\varepsilon^d < p_e$, which corresponds to Figure 18b or c.

2. As it will be shown in Fig. 45a and following discussion analyzing kinetic Eqs. (37)-(38) for strain-induced PTs, for any pressure in the range $p_\varepsilon^d < p < p_\varepsilon^r$, there is a corresponding steady volume fraction of the HPP. Starting from it, any infinitesimal pressure increase (or decrease) and plastic deformation results in LPP→HPP (or HPP→ LPP) PT and zero pressure hysteresis. This implies that zero hysteresis is not connected to $p_e$. Since $p_e$ is not present in Eqs. (37)-(38), it cannot be determined from the macroscopic strain-induced experiment.

3. In experiments on PTs in ZnSe [63] (Fig. 19d) and CuI [359], oscillatory pressure distribution is observed,

with several steps, which are not related to the phase boundaries. FEM simulations on behavior of the sample in DAC and RDAC in Section 8.3 show that for essentially weaker HPP than the LPP, leading to material instabilities, oscillatory behavior like in experiments in Fig. 19d is found with pressures at the steps not equal to $p_\varepsilon^d$ and do not have any meaning, and are determined by the mechanics of plastic flow and PT.

*3.3. Plastic straining leads to new PTs/CRs, changes in transformation paths, and retaining metastable HPPs*

Energy landscape in atomistic simulations practically for each material has many local minima corresponding to new metastable phases (reaction products), which may potentially have useful properties. Many new phases have been predicted computationally [361–365] but could not be synthesized under quasi-hydrostatic conditions, because of high kinetic barriers or because the low-barrier transformation path could not be realized. Plastic straining is one of the technologies that may allow reaching some of them. Many examples of new PTs and CRs produced by HPT are presented in [5–7, 14–16, 23]; many are produced by ball milling [113–119]. Synthesis of new ceramics by HPT is reviewed in [28]. Examples of new phases obtained in RDAC include the rhombohedral phase of GeTe [366], high-density amorphous phase of SiC [78], CuI-VIII$'$ phase [359], phases IV [66, 71] and V [71] of fullerene $C_{60}$, orthorhombic diamond phase [79], and superhard carbon onion phase [65, 73]. Some of them were later produced without plastic shears at high pressures and temperatures, like phases V of fullerene [70] and carbon onion [65], both claimed to be harder than diamond. An unidentified superhard phase of BN was found in RDAC after PT hBN→wBN, which scratches diamond anvils; however, wBN cannot scratch diamond [74]. A new superhard phase of single wall carbon nanotube was obtained in RDAC in [367].

In many cases, *plastic straining changes the transformation path*. The best illustration can be shown for Si and Ge, which have more than a dozen of phases under pressure. Thus, instead of semiconductor-metal PTs Si-I→Si-II and Ge-I→Ge-II, transformations Si-I→Si-III→Si-II and Ge-I→Ge-III→Ge-II were obtained in RDAC [332]. It was refined in [368] with TEM that first Si-I→Si-IV→Si-III transformations occur at 2-4 GPa, but Si-I→Si-III→Si-II takes place at higher pressures. At the same time, in situ XRD study in [81], Si-II always precedes Si-III and Si-IV was not detected, while in 30 nm, Si particles the PT Si-I→Si-XI occurs under hydrostatic conditions, with plastic straining PT Si-I→Si-II→Si-II+Si-XI were found in [81]. Si-I single-crystal nanopillar compressed in $<111>$ direction undergoes formation of a shear band, leading to Si-IV (diamond-hexagonal) strip within this band shear band at 18 GPa, accumulation of dislocations in it, following by amorphization [369]. PT to Si-IV and amorphous Si serve as additional mechanisms of plastic straining. Usually, Si-IV and amorphous Si do not appear during hydrostatic loading or in RDAC [81], but Si-IV was found with TEM after unloading in RDAC [368].

Under compression in DAC [370], Si-I→Si-XI PT was obtained at 12.3 GPa in Al-1%Si alloy, i.e., missing Si-II. In Al-4%Si alloy, Si-XI+Si-V appear at 13.4 GPa, while usually Si-V requires higher pressures than Si-XI [81]. Apparently, Al matrix and local interfacial defects and stresses at Al - Si-I interfaces change local strain-stress paths leading to a different PT sequence.

As it was mentioned in Table 1, some PTs, such as olivine→ringwoodite and Si-I→Si-III PT (marked with $\times\infty$) were not observed at hydrostatic loadings because of transformation to other phases. Under plastic straining both these PTs were obtained in [81, 83]. Similarly, Si-II does not transform back to Si-I under hydrostatic loading-unloading (instead it transforms to Si-XII+Si-III);but it transforms for 100 nm particles after plastic straining in [81]. This PT occurs because direct Si-I→Si-II PT was obtained at pressure as low as 3.5 GPa (instead of >10 GPa), when alternative phases are unstable. In [330], PT I→ II in InTe was obtained in RDAC rather than I→II$'$ →II under hydrostatic loading.

One important manifestation of the change in PT path is *separation of the chosen phase* or phases which usually appear with some other undesirable phases. For example, under hydrostatic loading of micron and 100 nm Si particles, Si-II and Si-XI appear together; with plastic straining, Si-II appears at 4-5 GPa lower pressure than Si-XI. Another important problem is synthesis of single-phase Si-III due its unique properties [371, 372]. Usually, it appears during pressure release together with Si-XII. Special technologies were developed: quenching from 14 GPa and 900 K during 3 days or in Na-Si system at 9.5GPa and 1000 K [371] or by 2 compression/slow (within 4 hours) pressure release cycles to 13 GPa at normal temperature [372]. In contrast, in RDAC [81], it was obtained in a few minutes by unloading from 11.6 GPa; after higher pressures, relatively large Si-III region was obtained. Nanocomposite of semiconducting phases Si-I+Si-III, obtained by unloading after 4.4 GPa and torsion by $24.6^o$, may have interesting electronic properties.

In some cases, *plastic straining substitutes a reversible PT with an irreversible PT.* For example, the reversible PT from graphite to hexagonal diamond has been found at 16-30 GPa; plastic shearing in RDAC at 17 and 19 GPa made this PT irreversible [373]. The irreversible PT from rBN to cBN under hydrostatic loading was reported at 55 GPa [374], but with plastic straining, it was started at 5.6 GPa [328, 375]. Si-II was obtained and retained after plastic straining in RDAC at 4.4 GPa [81]. Stabilization of metastable high-pressure ceramic phases by HPT was recently reviewed in [28]. Retaining of metastable phases at normal pressure is the main condition for their *use in engineering applications.* This is especially important for high-temperature superconductors [345–348], polymeric nitrogen and $CO_2$ [354, 355], which so far cannot be retained at normal pressure.

Stabilization of HPPs at normal pressure means increased pressure hysteresis, which contradicts the results in Section 3.2. This contradiction was resolved in [100, 102] by differentiating between strain-induced and pressure-induced PTs. Reduction in hysteresis is observed for plastic strain-induced PTs and means reduction in $p_\varepsilon^d - p_\varepsilon^r$. Stabilization of HPP during unloading is the prevention of the reverse pressure-induced PT, meaning reduction in $p_h^r$ below zero pressure. These two are completely independent material parameters, which resolves the contradiction.

*3.4. Kinetics of plastic strain-induced PTs and CRs*

*Time-independent kinetics.* The general wisdom is that during plastic strain-induced PTs [5, 6, 23, 26, 64, 100, 102, 330, 376] and CRs [7, 14, 16, 377], the *volume fraction of the HPP is an increasing function of the plastic strain; when the shearing (torsion) stops, PTs/CRs stop as well.* This means that time is not an essential parameter, and plastic strain plays the part of a time-like parameter. This assumption led to the derivation of the first general kinetic equation (37) for strain-induced PTs and CRs in [100, 102, 378] utilizing the accumulative plastic strain $q$ as a time-like parameters. This kinetic equation was experimentally confirmed [80, 82, 108] and further generalized [109] for $\alpha - \omega$ PT in Zr.

At the same time, there are statements about strong acceleration of PTs and CRs by plastic deformation [8, 9, 14, 16, 377], which may mean that time is an important parameter. Since large shear strains also strongly accelerate mass transport and mixing, the effective diffusion coefficients increase by 8-10 orders of magnitude compared to the same conditions but without shear, and by 3-5 orders of magnitude compared to the liquid at the same temperature [377]. Thus, *mass transport is not a rate-limiting process, which allows excluding diffusion and mixing from the kinetic equations for the CR rate.* The reaction rate under shear is by 2-6 (for some polymerization reactions in acrylamide, styrene, and butadiene [14, 16]) or even 3-8 [377] orders of magnitude larger than in the liquid. Activation energy in the expression for CR rate is zero (i.e., barrierless nucleation and growth occur), leading to independence of the reaction rates of temperature in the range [-196; 200 $^o$C] [377]. In [14], the CR rate slightly decreases with a decrease in temperature. The yield of polymer increases with increasing plastic shear and is independent of the shear rate. Note that the estimated temperature increase in rotating Bridgman anvils [14, 16, 291] shows that even at 10 turns with strain rate of 3 RPM for iron, temperature increase is below 100 K, which cannot produce such high CR rates. A hypothesis about the existence of hot spots with molten material could not explain high rates of CR either [14, 16]. It was also reported in [14] that the yield of polymer at fixed pressure and temperature is determined by the magnitude of the shear strain and is independent of the shear rate.

It looks like a contradiction: kinetics is time independent but there is acceleration of PTs and PTs. Resolution is quite simple. If time is not a parameter, then $dc \sim dq$, and dividing by time increment $dt$, one obtains $dc/dt \sim dq/dt$, i.e., transformation rate is proportional to the strain rate. Since comparison is done for room temperature, for pressure-induced CRs, reaction completion time could be very long. However, for torsion with reasonably large rotation rate, reaction can be completed within minutes or even seconds, which gives a drastic increase in averaged reaction rate despite the time-independent kinetics.

At the same time, it is written in [8, 9] that the thermally activated kinetics was accelerated by a factor of 3 to 30 for PT in $PbO_2$ and 50-200 for the reaction in jadeite by applying oscillatory torsion (Fig. 20a,b). Kinetic curves $c$ versus $t$ were obtained for $PbO_2I \leftrightarrow PbO_2II$ PT at 4 GPa and 25, 150, and 400 $^oC$ with and without torsion. PT is accelerated with increasing temperature (despite the $p_e$ growth), and at each temperature, with application of torsion. During torsion, time can be substituted with accumulative shear (if torsion rate would be given); but since PT occurs without shear as well (at least for >150 $^oC$), time is an essential parameter, along with plastic strain. The difference with other materials is that $PbO_2$ has low melting temperature (901 K at 4 GPa and 773 K at 8 GPa [379]), and the homologous temperature $T_h = 0.47$ at 150 $^oC$ (probably, above the recrystallization temperature $T_r$) and at 400 $^oC$, $T_h = 0.75$ is above $T_r$. Based on the dislocation pileup-based mechanism [100, 102] (Sections 6.2 and 6.5), just few dislocations may pileup and produce stress concentration and nucleation before they annihilate. In this case, nucleation is thermally activated (in contrast to barrierless nucleation for larger number of dislocations) and thermally activated kinetics is not surprising. At 25 $^oC$, when $T_r = 0.33$, many more dislocations can be pileup, causing barrierless nucleation. Since very low $c$ was obtained at 25 $^oC$ within 43 hours, time may not be considered an important parameter.

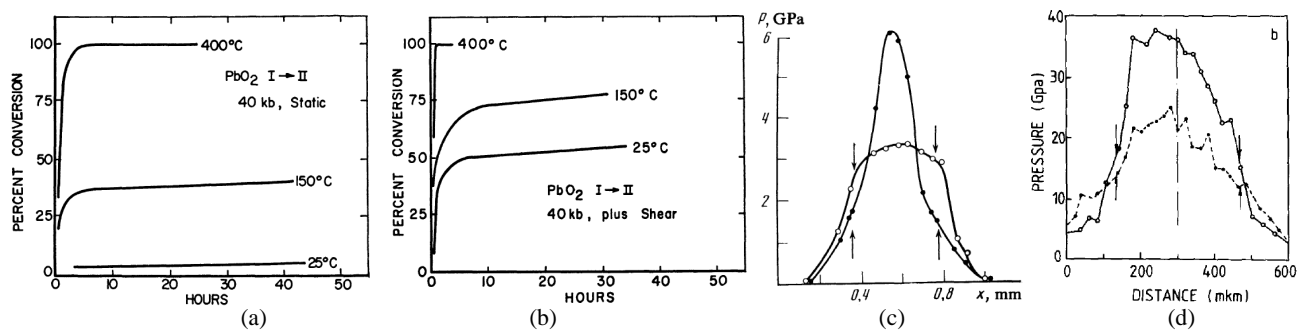

**Figure 20:** Kinetics of $PbO_2I \leftrightarrow PbO_2II$ PT at 4 GPa and 3 temperatures without (a) and with (b) oscillatory torsion [9]. (c) Pressure distributions after compression (hollow circles) and torsion by $15^o$ (solid circles) during $B1 \rightarrow B2$ PT in KCl [62]. The arrows here and in (d) indicate the location of a visible phase boundary. (d) Pressure distributions after compression (dashed line) and torsion (solid line) during $IV \rightarrow V$ PT in fullerene [71]. In (c) and (d), pressure growth in the transforming region during torsion at constant axial force, exhibiting pressure self-multiplication effect.

However, very recently, *time dependence of strain-induced PT* at $q = const$ was revealed in situ for $\alpha - \omega$ PT in Zr [82, 89] (see Section 7.5.3) at room temperature, i.e., >1,800 K below the melting temperature. To explain the surprise, a PFA solution from [380, 381] (Figs. 34 and 36), was used. After nucleation, growth of HPP occurs for some time accompanied by various processes. Aside from the tip, this growth represents a stress-induced PT. It was routinely hypothesised that growth time until steady state is much shorter than the measurement time and only steady configurations were detected. However, due to in situ measurements, it was detected in [82, 89], and steady state was reached in ~1 hour. This observation implied necessity of the developing advanced *combined strain- and time-dependent kinetics*, i.e., *theory for combined strain- and stress-induced PTs*.

*Effect of nonreacting matrix.* It was found [15, 382] that the rate constant for reactions in various organic compounds within nonreacting matrix increases linearly with the yield stress of the matrix. Thus, matrix with a high $\sigma_y$ increased the reactivity, reduced the CR initiation pressure and produced CRs in those compounds which did not react otherwise. For example, adamantane matrix lowered the minimum pressure for polymerization of succinic acid dinitrile from 10 to 3 GPa and naphtaline from 4 to 2 GPa; decomposition pressure for $Mo(CO)_6$ was reduced from 7 to 3.5 GPa and for $W(CO)_6$ from 5 to 2.5 GPa [15]. Accordingly, a matrix with the lower $\sigma_y$ than that for initial reagents decelerates the CR.

In the private discussion in 1989 in Kyiv (Ukraine), A.A. Zharov told the author that mechanochemical community cannot explain the effect of non-reacting matrix on CR; the problem was qualitatively resolved in [100, 102, 378]. Namely, stronger nontransforming phase increases rate of PT and CR by localizing and increasing plastic strain in the LPP, however adding weaker phase reduces the PT and CR rate by localizing plastic strain in it and reducing it in LPP. Larger plastic straining within a stronger matrix also reduces $p_\varepsilon^d$ by reducing grain size and increasing defect densities, as it was observed for Zr [80, 82, 89], Si [81], and olivine [83]. Similar, if HPP is stronger (or weaker) than the LPP, this localizes plastic strain in LPP (or HPP) promoting (suppressing) transformations. Ruby particles used for pressure measurement play a similar role. However, the linear correlation of the rate constant with the yield strength remains to be explained. In addition, significant part of the promoting effect of the nontransforming matrix with the higher $\sigma_y$ or suppressing effect of the matrix with the lower $\sigma_y$ could be related to the pressure growth or decrease at the center of the sample during rotation at the fixed averaged force (Sections 3.5 and 8), since only averaged pressure was reported in [15, 382].

An interesting study on PT, grain refinement, and defect evolution in eutectic Al-Si alloys under compression in DAC was performed in [370] following the idea from [100, 102] that strain-induced PTs can be studied in DAC. It appears that PT in Si particles embedded in softer Al matrix behave differently than pure Si. A priori, weak matrix should suppress stress-induced PT [15, 100, 102, 382]. Also, strong stainless steel gasket provides relatively high $p$-low $q$ loading path [81]; thus, the effect of plastic straining is expected to be weaker than for pure Si. Surprisingly, instead of Si-I→Si-II→Si-XI PT, Si-I→Si-XI PT was obtained at 12.3 GPa for Al-1Si alloy. For Al-4Si alloy, in addition to absence of Si-II, Si-XI appears simultaneously with Si-V at 13.4 GPa, while in all previous (and later [81]) experiments, Si-V appears at higher pressures than Si-XI. Also, pressure in Si-V and Si-XI was 4 GPa lower then in Al. Complex Al-Si interfaces, which have undergone significant structural changes and grain refinement, presumably, produce nucleating defects promoting Si-XI and Si-V instead of traditional Si-II.

Pressure-temperature *transformation diagram* for Si and Ge was supplemented by pressure-shear strain plane at room temperature in [64, 332], introducing shear-induced Si-III and reduction in PT pressure for Si-I→Si-II and Si-II→Si-III PTs with shear. Based on in situ studies in [81], this diagram in $p_\varepsilon^d - q$ plane should have a different sequence of Si-II and Si-III PT pressures and should include Si-XI and Si-V. Also, $q$ or shear is not a physical parameter because material is usually predeformed at normal pressure during producing or intentionally, and zero point of $q$ is not defined. Also, $p_\varepsilon^d$ strongly depends on the microstructure, and plot $p_\varepsilon^d$ versus $d$ or $\rho_d$ are much more physical (like in Section 6.3 and Fig. 31a [81].) Even the entire kinetic equation for strain-induced PTs/CRs could have more meaningful form if some relevant microstructural parameter ($d$, $\rho_d$, or concentration of TSF for layered materials) serves as a time-like parameter instead of $q$, until these parameters reach a steady state.

*3.5. Pressure self-multiplication and self-focusing effects*

During torsion at a constant compressive force, for some PTs, the *pressure increases in the transforming region at the center of a sample despite the volume drop due to PT* (Figs. 19c and d and 20c and d) [62–64, 66, 71, 359, 383]. This phenomenon was coined "the pressure self-multiplication effect." Similarly, pressure decreases for the same materials in the transforming region in the course of the torsion-induced reverse PT at constant force, despite the volume increase. Examples include: pressure growth (drop) on $50 - 70\%$ due to direct (reverse) $B1 - B2$ PT in KCl (Fig. 20c) [62], also reproduced in [66]; for PT in PbTe from NaCl to GeS structure from 4.6 to 6.6 GPa and from GeS to CsCl structure from 9.6 to 12 GPa [383]; for PT in fullerene IV to V from 21 to 32 GPa and from 23 to 38 GPa [71] (Fig. 20d), and from phase III to IV from 17 to 21.5 GPa [66] (Fig. 19c); and for PT in ZnSe [63] (Fig. 19d). Inversely, under torsion, pressure drops from 34 to 24 GPa during PT to new CuI-VIII' phase [359], from 6.5 to 5.7 GPa for semiconductor-metal PT in GaSb [64], and from 9 to 7 GPa for PT from Ge-III to Ge-II PT [332].

The pressure self-multiplication effect at constant force was also obtained even without the torsion. During PT $B1 \to B2$ in *KCl* induced by heating from 300 to 600 K pressure increased from 6 GPa by 30 % [384]. During PT from rBN to cBN caused by new phenomenon, rotational plastic instability, pressure increased from 5.6 GPa to an estimated 60-76 GPa [328]. All these results represent a seeming formal violation of the Le Shatelie principle because reduction (increase) in volume has to lead to a pressure drop (growth).

A theoretical interpretation of this effect was first obtained analytically in [92, 100, 102] (Sections 8.1 and 8.2) and with FEM simulations in [251, 385] (Section 8.3). According to Eq. (3), for equal yield strength of phases, pressure distribution does not change during the PT (Fig. 52). For a stronger HPP at the center, pressure gradient and consequently pressure in HPP grow at the center, despite the transformational volume decrease, which is compensated by the thickness reduction during the torsion. Pressure in the LPP slightly reduces to keep the same axial force. Similar, for weaker HPP, pressure gradient and pressure decrease in the transforming region. For compression without torsion, one needs some way to produce axial displacement at constant force to compensate transformation volume reduction; then the same mechanism will cause pressure self-multiplication/demultiplication effects [100, 102]. This mechanism is usually related to the reduction in the yield strength, e.g., by heating of KCl [384] or inducing rotational plastic instability for BN [328]. Thus, there is no any violation of the Le Shatelie principle; pressure growth is explained by a specific character of plastic flow, which is neglected in classical thermodynamics.

Originally [62], the increase in Young's modulus after PT was claimed to be responsible for the pressure self-multiplication effect. However, plasticity was not included and the equilibrium equation was not met, which led to such a wrong conclusion. Even at infinite moduli (like in model [92, 100, 102]), the pressure is limited by the solution of the plastic equilibrium problem. For $\sigma_{y1} > \sigma_{y2}$ or material flowing to the center, the pressure in the HPP reduces irrespective of the increase in elastic moduli. Still, Blank's group continued using interpretation from [62] neglecting more recent and advanced theories (see [65]).

When sample is placed in a gasket, a homogeneous pressure self-multiplication effect (small, almost homogeneous pressure increase) during torsion-induced hBN→wBN PT was obtained experimentally in [74] based on a simplified analytical design optimization. It was also reproduced in FEM simulations in [360, 386], see Section 8.3.

Strong pressure growth in the central region of the sample due to beveled culet, even without PT, during compression and especially torsion was predicted analytically and confirmed with FEM in [212, 213] (Section 2.10 and Fig. 17). Based on analytical solution to Eq. (11), it was concluded that strong reduction in sample thickness toward the center leads to very large pressure gradient and pressure at the center, which was called *pressure self-focusing effect*. FEM solution gives slightly larger pressure gradient. Torsion leads to drastic increase in pressure gradient for the same maximum pressure, and, consequently, essentially lower axial force or much larger maximum pressure for the same load. Conceptual experimental proof of the pressure self-focusing effect in DAC was claimed in [212, 213] for Re and W [131, 132], which are well described with FEM in [142, 143]. We even see much stronger proof from the huge pressure gradient due to large bevel at the toroidal tip of toroidal DAC (Fig. 10b,c).

3.6. *Phase transformation rBN→cBN induced by rotational plastic instability*

A sudden irreversible martensitic rBN→cBN PT was obtained during compression in DAC without a gasket at 5.6 GPa before the PT [328, 375], while under hydrostatic conditions, such a PT was recorded at 55 GPa only [374]. Analysis showed that a new phenomenon, *PT induced by rotational plastic instability*, is responsible for this drastic PT pressure reduction. For a very highly-textured rBN, C-axes of crystallites was aligned along [0001] rBN texture axis and compression direction with deviation within 3°. Sample was compressed elastically up to about 3.5 GPa, without a residual anvil indent. At recompression in a range of 4.2 - 5.6 GPa, small strips, presumably the boundaries between misoriented crystal regions became visible in transmitted light. At fixed load at pressure 5.6 GPa at the center (measured with ruby particle), an amount of stochastically located strips rapidly increased, and in $5-7$ s a sudden irreversible martensitic PT from rBN to cBN was recorded in the central region of the sample with dynamic reduction of the sample thickness by a factor of 10. The electron-diffraction patterns of spalled particles confirmed significant amount of cBN with the microhardness of 49.6 GPa, typical of cBN.

The following deformation and PT scenario was considered here. Below 4.2 GPa, rBN is compressed elastically because shear stresses are close to zero at the weak basal slip and twinning planes and much stronger prismatic slip and twinning systems are not activated; stresses are quasi-homogeneous along the radius. During the rotation of crystals with respect to the uniaxial stress, the yield stress in axial direction reduces from the maximum to the minimum values. In a range of 4.2 - 5.6 GPa, reorientation of crystals starts, leading to an increase in shear stresses along the basal plane and reduction in the yield strength in axial direction along with its increase due to strain hardening. At 5.6 GPa and constant force, before the PT, rotational softening is getting greater than strain hardening and sample cannot equilibrate such a load, so unstable dynamic compression accompanied by further crystal rotation and softening takes place. Due to reduction of $h/R$, pressure gradient is developed and stresses at the center increase according to Eq. (4) and decrease at the external radii to keep the same force. Large plastic straining, along with increase in pressure, lead to the initiation of the plastic strain-induced PT to cBN. Since the yield strength of the cBN is much higher than that for hBN, strong pressure self-multiplication occurs, pressure and plastic strain grow, further promoting the PT (almost) to completion in the central part. Process occurs until the thickness for which the heterogeneous stresses in the sample equilibrate the applied force.

Note that while stress redistribution was properly described analytically in [328], the PT was treated as stress-induced, further developing and specifying the large-strain PT criterion from [93]; the theory for strain-induced PT was not yet developed. While theoretical estimates in [328] gave similar values of the estimated athermal threshold for all 4 experiments, treatment of this PT as the plastic strain-induced could give much lower PT pressure; however, experiments in [328] are not sufficient for such a treatment, because it is not known at which thickness/pressure the PT starts and finishes. Performing similar experiments with the gasket, i.e. preventing system instability, along with in situ measurements, could help to get much more complete view on this phenomenon and PT. Similar phenomenon is expected for other layered systems, like graphite and hexagonal or rhombohedral BCN.

*3.7. PTs and CRs in shear bands: TRIP, RIP, cascading structural changes, and self-blown-up deformation-PT-heating phenomena*

Shear banding is broadly spread material instability under normal, high static and dynamic pressures. Due to accumulation of SPD, it leads to various strain-induced PTs and CRs, see examples in Fig. 21. Thus, exothermic CRs in Nb-Si and Ti-Si powder mixtures within dynamic shear bands have been investigated in [387, 388]. Shear-PT bands in fcc fullerene $C_{60}$ were discovered after torsion in RDAC [72], consisting of nanocrystals of the triclinic, monoclinic, and hcp $C_{60}$, linearly polymerized fullerene and polytypes, and fragments of amorphous structures. Five HPPs have been retained after unloading. Since HPP of $C_{60}$ are stronger than the parent LPP, localized shear deformation contradicts the main condition for shear banding, namely softening (or lack of hardening). PT from olivine to spinel caused by shear banding (or which causes shear banding) [389] is considered to be one the main mechanisms of the deep-focus earthquake [83, 389–393]. Under uniaxial compression up to 18 GPa, single crystal $<111>$-oriented Si I nanopillar exhibited shear banding, a PT to Si-IV band within the shear band, and nucleation and accumulation of dislocations leading to amorphization of Si-IV [394]. Deformation-amorphization bands in Si-I are also observed during shock loading [395–397], scratching and machining [398–401]. In addition to the thermodynamic reasons caused by the elimination of the energy of dislocations and other defects, amorphization is also considered an additional carrier for plastic deformation [402–404]. Shear banding leading to amorphization in various classes of materials is reviewed in [395, 402].

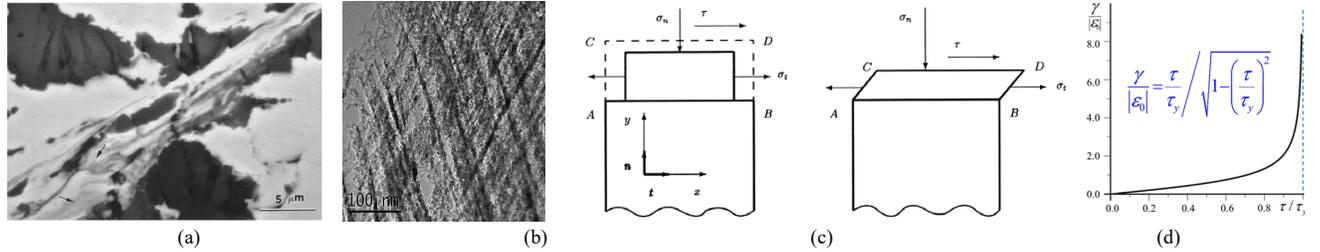

**Figure 21:** (a) Initiation of CR within a shear band in Ti-Si powder mixture [388]. (b) Shear-PT-TRIP bands in the fcc fullerene $C_{60}$ consisting of 5 HPPs after torsion in RDAC, despite the stronger HPPs [72]. (c) Schematic of PT/CR in a shear band ABCD within undeformed half space (below the line AB); left: initial state (dashed line) and state after volumetric transformation strain (solid line) leading to displacement discontinuity; right: plastic deformation restores displacement continuity in the actual state, while shear stress produces TRIP/RIP shear [94]. (d) Analytical relationship between TRIP/RIP-induced shear strain $\gamma$ normalised by transformation volumetric strain $\varepsilon_0$ and shear stress $\tau$ normalized by $\tau_y$ within transformation-shear band [92, 93, 388]. Different multipliers close to 1 may appear in the right side of equation for different formulations.

A formal problem on the PT in a thin layer, in particular, in shear band, was solved analytically in [92–94] at small strain (including 3D solution in [93]) and in [405] at large strains. Such a problem also models PT at a moving phase interface. In addition to thermodynamic aspects, it was obtained that the transformation volume reduction tends to cause a displacement discontinuity at the phase interface. During the restoration of displacement continuity by plastic deformation, large plastic shear $\gamma$ occurs under action of the applied shear stress $\tau$ described by the expression in Fig. 21d. Such a plastic deformation is nothing but TRIP phenomenon. TRIP is produced during PT by internal stresses due to transformation strain, in particular, its volumetric part, combined with external stresses. Existence of the RIP phenomenon (which represents a new mechanism of plastic flow in solids), in a shear band caused by transformation volume reduction during CRs, based on a similar finite-strain solution, was predicted in [388] and utilized for the explanation of acceleration of CRs by plastic straining. It is visible in Fig. 21d that plastic flow occurs at arbitrary (even at infinitesimal) shear stress below $\tau_y$. Plastic straining takes place due to the variation of volumetric transformation strain $\varepsilon_v$ (designated in [92] as $\varepsilon_0$) and is proportional to $\varepsilon_v$ (independent of the source of $\varepsilon_v$, i.e., for any PT or CR). This result was used in [72] as the only reason for the possibility of obtaining HPPs of $C_{60}$ in a shear band that are stronger than the parent LPP. When $\tau \to \tau_y$ (which is the case in a shear band), plastic shear $\gamma \to \infty$, so very high plastic TRIP/RIP shear can be accumulated. Such a large RIP could lead to very intense heating, more than 1,500 K in adiabatic approximation, which is another contribution for acceleration of CRs [388]. The concept of the *effective temperature*, which takes into account temperature increase during the nucleation event, was introduced in [94, 388]. At the same time, such heating due to severe TRIP suppresses most martensitic PTs and, consequently, heating and TRIP, i.e., it decelerates PT and the accompanied processes.

Experimentally, TRIP in RDAC was revealed in [74, 87] during hBN→wBN PT, which possesses a large $\varepsilon_v = 0.34$. It was found that the strain-induced 2D disordering, which is a measure of plastic straining similar to dislocation density for metals, grows much faster due to change in volume fraction $c$ of HPP than due to applied shear. Since TRIP is proportional to $c$, it was concluded that TRIP is 20 times larger than the traditional plasticity.

The cascading mechanism of structural changes during plastic flow and hBN→wBN PT in RDAC was also suggested/revealed in [74, 87]. Plastic straining generates both TSFs that suppress the PT, and nucleating defects like dislocation pileups that promote the PT. A PT produces severe TRIP under shear stresses; TRIP, similar to traditional plasticity, generates the TSF and new dislocation pileups; these dislocation pileups again run the PT, which produces TRIP, and so on. This mechanism represents a positive mechanochemical feedback between PT, change in volumes, and plastic straining/TRIP. It was mentioned in [74, 87] that similar processed in shear bands can be caused by CRs and RIP, and they should be important for explanation of the earthquakes in [390–392].

Further development was performed in [393]. More advanced analytical 3D solution for coupled deformation-PT-heating problem in a shear band was obtained with focus on PT olivine-spinel, which was treated like plastic strain-induced (in contrast to stress-induced PT in the previous papers). This solution predicted conditions for severe TRIP and self-blown-up deformation-PT-heating process due to positive thermomechanochemical feedback between TRIP and strain-induced PT. This solution allowed quantitatively analyzing the suggested mechanism of deep-focus earthquake (Section 10.1).

Advanced theoretical finite-strain study of shear banding in thermoviscoplastic material with stress-induced PTs with application to Fe and steel was performed in [406]. Strain hardening and thermal softening are included, and the effect of PT and variation of material parameters was analysed. Atomistic simulations in [407, 408] revealed mechanism of formation of amorphous shear band in boron carbide. In [409], shear-induced formation of an amorphous band via virtual melting mechanism [410–413] was revealed by MD simulations. Corresponding thermodynamic criterion was derived and confirmed by MD. Cyclic PTs a-Si↔Si-I, a-Si↔Si-IV, and Si-I↔Si-IV with non-repeatable nanostructure evolution have been discovered. These cyclic PTs offer additional carriers for plastic flow through shear transformation strain and TRIP.

3.8. Effect of temperature and strain rates on strain-induced PTs and CRs

There are relatively few old papers on the effect of these parameters, with different trends for different materials. For PT semiconductor-metal in $InSb$, growth in anvil rotation rate from $5 \cdot 10^{-3}$ to $5 \cdot 10^{-1} \frac{grad}{sec}$ for the rotation angle $\varphi = 10°$ decreases the PT pressure from 1.9 to 1.4 GPa [331]. An extrusion experiment revealed a reduction in PT pressure with increasing strain rate for $B1 \to B2$ PT in KCl and RbCl. In particular, a raise in strain rate by a factor of 6.15 for RbCl decreased the PT pressure from 0.75 to 0.62 GPa at strain of 0.1 [414]. At the same time, the anvil's rotation rate does not effect the kinetics of the polymer production [14, 16]. Also, the reduction anvil's rotation rate slows decomposition of some oxides [17]. For shock loading of $InSb$ single crystal in [415], the PT occurs at 0.95 GPa, much lower than 2.5 GPa at the static loading.

An increase in strain rate increases the $\sigma_y$, which promotes the PT due to higher shear stresses. On the other hand, there may not be enough time for time-dependent growth of the strain-induced nuclei, which suppresses PT progress. Increase in strain rate may reduce the role of dislocation plasticity vs. PTs (including amorphization) as a stress relaxation mechanism and substitute dislocation plasticity with twinning. One needs to consult the generalized Ashby's deformation map [412, 416, 417] to check whether for a given strain rate and temperature alternative to the dislocation glide mechanisms.

A priori, an increase in temperature should decrease the effect of plastic straining on PTs and CR because for reduction in the yield strength and increasing role of the recovery and relaxation processes, which initiates grain growth and reduction in dislocation density and number of dislocations in a pileup. Above recrystallization temperature, the effect of dislocation pileup should be absent. Indeed, plastic deformation of Pb in RDAC does not change visibly the PT pressure [418] because for Pb, room temperature is above the recrystallization temperature.

As it was reported in [8, 9] and discussed in Section 3.4 (see Fig. 20a,b), plastic straining very strongly accelerates the $PbO_2I \leftrightarrow PbO_2II$ PT at $T_h = 0.33$, quite strongly at $T_h = 0.47$, and slightly but still accelerates at $T_h = 0.75$. Even at high temperatures, dislocation pileups consisting of few dislocations can produce thermally activated nucleation before they annihilate [100, 102] (Section 6.5). During plastic compression in DAC of $Mg_2SiO_4$ [392], the thin $\beta$-spinel film was transformed from the olivine at the diamond surface at 19-35 GPa after holding at 575 $^oC$ ($T_h = 0.35$) for 10.5 hours. Since no PT was found in bulk, this indicates strain-induced PT due to large plastic shears caused by the contact friction. The most impressive plastic shear-induced promotion of PT at high temperature ($T_h \simeq 0.63$) is for PT from the orthoenstatite to clinoenstatite in $MgSiO_3$ [10], which occurs in the region of stability of the LPP, at 1 GPa and 920 $^0$C, i.e., 9.7 GPa below the $p_e$. The yield strength at this temperature is 2 times below that at room temperature. It is stated in [377] that the activation energy for the plastic strain-induced CR is zero, leading to independence of the reaction rates of temperature in the range [-196; 200 $^o$C]. At the same time, in [14], the CR rate slightly decreases with decrease in temperature.

While we do not intend to review here diffusion and diffusional PTs with change in composition, we would like to mention the concept of the *effective temperature* $T_{ef}$ under SPD [419, 420], which generalizes a similar concept introduced in [421] for diffusive PTs induced by severe irradiation. Thus, composition of phases after SPD at room temperature is equivalent to that under equilibrium conditions at some higher $T_{ef}$, but not at the higher pressure.

If temperature accelerates kinetics of PT or CR, fast SPD can indirectly promote them by increasing temperature, e.g., in shear bands. This is especially possible when strong TRIP or RIP is involved, leading to a positive mechanochemical feedback. This was concluded for CR in Ti-Si and Nb-Si mixtures in shear bands [388]. TRIP, causing self-blown-up heating in shear bands, was utilized to resolve the main puzzles in the mechanisms of the deep-focus earthquake [393]. In the opposite case when temperature suppresses PTs and CRs, e.g., by placing system in the region of stability of a LPP, plastic heating suppresses them as well.

The effect of the homologous temperature and strain rate $\dot{q}$ on plastic flow and microstructure was reviewed in [22, 23, 422]. An increase in temperature reduces steady hardness and torque and increases the steady grain size and strain needed to reach steady state. Increase in $\dot{q}$ from 0.0025 to 0.0625/s slightly reduces the steady grain size, more pronouncedly at high temperature. For dynamic HPT with up to 30,000 RPM and strain rate $> 10^3/s$ [423], hardness was higher for relatively small strains, but for large strain differences became negligible. To avoid significant heating, rotation increments were small ($< 40^o$) with intermediate cooling. Systematic studies of the HPT in the range 0.06-60 RPM and $\dot{q}$ from 0.004 to 20/s for Al, Cu, Fe, and Cu-30%Zn brass at 4 GPa and Ti at 2 GPa for 15 turns in [424] revealed that the steady hardness, shear stress, dislocation density, grain, and crystallite size are practically independent of $\dot{q}$. Small increase in crystallite and grain size and reduction in dislocation density for 20/s were related to some more essential heating. A possible explanation of the independence of all steady-state parameters of $\dot{q}$ was that $\dot{q}$ equally promotes grain fragmentation and dynamic recrystallization that compensate each other. The author's group performed the first in situ testing on Fe-7%Mn alloy up to 1,500 RPM and $\dot{q} = 2 \times 10^3/s$ in dynamic RDAC, but postprocessing is yet to be finished. Temperature increase due to the diamond should be much smaller than with metallic or ceramic anvils.

*3.9. Correlation between grain-size dependence of the yield strength and lowest pressure for strain-induced PTs*

Since increasing plastic strain in general reduces the minimum pressure for strain-induced PTs $p_\varepsilon^d$ (e.g., for Si-I→Si-III→Si-II PT and similar PTs for Ge [64, 332] and PT in InSb [64, 331]) and increases the yield strength $\sigma_y$, one can deduce that increasing $\sigma_y$ leads to reducing $p_\varepsilon^d$. Also, since increasing $\sigma_y$ is usually related to the reduction in the grain size (and increase in dislocation density), they also lead to the reduction in the $p_\varepsilon^d$. Of course, there are exceptions, e.g., when optimal plastic strain is needed for some PTs due to appearance of alternative phases [378] (Section 7.3) or suppressing effect of 2D disordering on PT, which also grows with plastic straining [74] (Section 7.4). Also, since usually $\sigma_y$ increases with the strain rate, and increasing strain rate reduces PT and CR pressure (e.g., for decomposition of some oxides [17] and PT in KCl and RbCl [414]), this again means that increase in $\sigma_y$ reduces $p_\varepsilon^d$. Again, there are exceptions, e.g., the yield of polymer is independent of the shear rate [14, 16]; steady hardness, shear strength, dislocation density, grain, and crystallite size are independent of the strain rate for 5 metals [424].

More recent in situ studies reveal the following. For $\alpha - \omega$ PT in ultra-pure Zr, $p_\varepsilon^d = 2.3$ GPa for annealed sample and $p_\varepsilon^d = 1.2$ GPa for strongly plastically predeformed by multiple rolling, i.e., nanograined, sample [80]. For $\alpha - \omega$ PT in commercially pure Zr, $p_\varepsilon^d$ reduced from 1.36 to 0.67 GPa, when crystallite size reduced from 65 to 48 nm (and dislocation density increased from $1.26 \times 10^{15}$ to $1.83 \times 10^{15} m^{-2}$) [82, 89, 425]. The linear reduction in $p_\varepsilon^d$ in RDAC with increasing plastic strain, corresponding to increasing dislocation density and microstrain, and decreasing crystallite size, was in situ found for PT olivine-spinel [83]. All these results mean that increase in $\sigma_y$ due to reduction in the grain/crystallite size reduces $p_\varepsilon^d$.

The most complete correlation between grain-size dependence of the yield strength and $p_\varepsilon^d$ was found for strain-induced PT in Si [81]; see Section 6.3 and Fig. 31a. Thus, in addition to the decrease in $p_\varepsilon^d$ with decreasing grain/particle size in the region of direct Hall-Petch effect, increase in $p_\varepsilon^d$ with further reducing grain/particle size in the region of the inverse Hall-Petch effect [263] was proofed in situ in DAC and RDAC for plastic strain-induced PT from Si-I to Si-II for 1 $\mu m$, 100 nm, and 30 nm particle sizes. This result is very nontrivial because for pressure-induced Si-I to Si-II PT, PT pressure grows with reduction in the particle size, and for 30 nm particles, Si-I does not transform to Si-II at all, transforming directly to Si-XI (Figs. 31a and 61). This result was predicted by dislocation pileup mechanism of strain-induced PT, and, consequently, strongly supports this mechanism and suggests methods to control the PT pressure for the strain-induced PTs by controlling grain size.

*3.10. Steady yield strength, microstructure, and minimum pressure for strain-induced PTs after SPD*

Here, we summarize some in situ results, already discussed for $\sigma_y$ (Section 2.1) and microstructure (Section 2.7) and will discuss some results for $p_\varepsilon^d$ (Section 7.5.1; also mentioned in Section 3.2). Generally, with increasing plastic deformation, $\sigma_y$, dislocation density and microstrain increase, but the crystallite size and $p_\varepsilon^d$ reduce, and their evolution depends on the plastic tensor strain $\boldsymbol{\varepsilon}_p$, including its mode, its entire path $\boldsymbol{\varepsilon}_p^{path}$, and pressure (except $p_\varepsilon^d$, which is the pressure itself). Due to infinite number of combinations, this makes finding some general rules hopeless. However, after some critical level of plastic deformation, all these parameters, the yield strength, dislocation density, crystallite size, and microstrain, as well as the minimum pressure for strain-induced PT reach steady values (not necessarily simultaneously), which are independent of $\boldsymbol{\varepsilon}_p$, including its mode and entire path $\boldsymbol{\varepsilon}_p^{path}$; dislocation density, crystallite size, and microstrain are also independent of pressure and pressure path; for monotonous and quasi-monotonous loading, materials also deform like isotropic and perfectly plastic with fixed surface of perfect plasticity $\varphi(\boldsymbol{\sigma}) = 0$.

For the yield strength and surface, this rule was suggested in [4, 47] and proved for numerous materials belonging to different classes, see Section 2. For microstructural parameters, it was formulated based on in situ experiments (i.e., under pressure) in [89] for Zr. However, steady hardness and microstructural parameters after SPD and pressure release were well-known for metals and ceramics [4, 20–22, 24–27, 47, 264–266]. Detailed review discussing state of the art of developing this rule for the yield strength and microstructure can be found in [90]. Rule for $p_\varepsilon^d$ was formulated in [80] for $\alpha - \omega$ PT in Zr based on in situ XRD study, while potentially, it could be deduced based on results in Fig. 19b and c for I→II and II→III PTs in PbTe from [358] and II→III PT in $C_{60}$ from [66] (Section 7.5.1). It was also concluded in [80] (Section 7.5.1) that $p_\varepsilon^d$ and the entire kinetic equation are independent of plastic strain $q_0$ below $p_\varepsilon^d$, i.e., before the PT. Also, strain-controlled kinetics is independent of $p-q$ path. The above rules drastically simplify the general theory and facilitate the development and experimental calibration of the specific models for plasticity, microstructure evolution, and strain-induced PTs.

Based on recent HPT experiments [424] with the strain rate from 0.004 to $20/s$ on Al, Cu, Fe, Ti, and Cu-30%Zn alloy, one can add that the *steady hardness, shear stress, dislocation density, grain, and crystallite size are practically independent of the strain rate.*

However, some sophistications were formulated in [82, 89] (Sections 2.8 and 7.5.3). Different steady states for all the above parameters were found after multiple rolling, torsion in RDAC, and compression in DAC with smooth and rough diamonds, respectively. Based on these experiments, each of these listed was claimed to be independent of $\boldsymbol{\varepsilon}_p$ and $\boldsymbol{\varepsilon}_p^{path}$, which is clearly contradictory; if it would be independent, then there should be a single value for the crystallite size, dislocation density, microstrain, yield strength, and minimum pressure for strain-induced PT. It was also known from ex situ experiments that the steady values of the grain/crystallite size and dislocation density are different after different SPD techniques (see Figure 13b,c).

This leads to formulation of a major new problem: *for which classes of plastic strain $\varepsilon_p$, $\varepsilon_p^{path}$, p, and pressure path $p^{path}$ material possesses one steady values of dislocation density, crystallite size, microstrain, yield strength ($\varphi(\sigma) = 0$ yield surface), and minimum pressure for strain-induced PT and for which of $\varepsilon_p$, $\varepsilon_p^{path}$, p, and $p^{path}$ classes there are jumps from one steady microstructure, yield strength, and minimum pressure for strain-induced PT to another?* This is an unsolved problem, and some steps toward its solution were suggested in [90].

We would like also to mention the following statement from [15]. Independent of the loading method (rotating Bridgman anvils, extrusion, shock wave, and extrusion through an angular gap with rotating dorn), the same reactions occur that do not start at uniaxial compression up to 8 GPa. Examples mentioned in [15] are the synthesis of amides from ammonium salts of saturated acids, polymerization of solid vinyl monomers, addition of $NH_3$ at the $-C=C$- bond (aminoacid synthesis), peroxide decay, esterification reaction, and synthesis of peptides from amino acids. Different loading methods produce different $\varepsilon_p$, including its mode, its entire path $\varepsilon_p^{path}$, p, and pressure path $p^{path}$, i.e., the above CRs are independent of them.

Note that SPD-induced composition and, in some cases, the volume fraction of a HPP reach a steady state, is independent of the initial state [35, 36], which is called *principle of equifinality*. As it is shown in Section 7.2, if evolution of volume fraction (or any other parameter) is described by the first-order differential equation with the difference of two terms responsible for growth and reducing, there is a stationary solution, which is independent of the initial conditions and straining path. This is one of the most probable sources of the principle of equifinality. The same is true for multiphase system (see Section 7.3) if all PTs can occur and there is no controlling parameter (e.g., pressure or dislocation density) that leads to multiple steady solutions. Otherwise, the steady state may depend on the initial conditions. For example [36], in Ti alloys, HPT leads to PT of the mixture of $\alpha$ and $\beta$ phases into $\omega$-Ti, and in most cases, the steady volume fraction of the $\omega$-Ti is independent of the initial proportion between $\alpha$ and $\beta$. However, most probable that the $\alpha$-Ti transforms into $\omega$-Ti through the intermediate $\beta$-phase only. Then, if initial annealing eliminates $\beta$ via eutectoid decomposition, PT to $\omega$ phase could be completely suppressed, i.e., equifinality is violated [36]. Similar properties of the evolution equations for microstructural parameters may lead to different stationary solutions (steady microstructures) for different straining paths.

### 3.11. First rules for the crystallite size and dislocation density evolution in phases during strain-induced $\alpha - \omega$ transformation in Zr and olivine-spinel PT

Radial distribution crystallite size $d$ and dislocation density $\rho_d$ in Zr in DAC with rough-DA and smooth-DA were measured in In [82, 425] using XRD like in Section 2.7 and [89], see Fig. 22a. They are quite heterogeneous and messy, along with measured heterogeneous pressure distribution and complex but unknown plastic strain tensor $\varepsilon_p$ and its path $\varepsilon_p^{path}$ distributions. This leaves a little hope that some general rules can be found. However, if all these points are plotted versus volume fraction of $\omega$-Zr, all points for each phase surprisingly belong to the same curve (Fig. 22b,c). This implies the existence of the unexpected and very informative rules for evolution of the crystallite size and the dislocation density during PT, formulated in the caption to Fig. 22b,c.

Since kinetics of $\alpha - \omega$ PT is known (see Eq. (40)), by approximating curves $d(c)$ and $\rho_d(c)$ in Fig. 22b,c, one can obtain the kinetics of evolution of $d$ and $\rho_d$ during strain-induced PT. These kinetic equations can be used for FEM simulations of microstructure evolution in DAC and RDAC, which is yet to be done. Generally, the problem of defining the initial dislocation density in the newly appeared phase, i.e., whether it is inherited from the parent phase or not, or what are the inheritance rules, is the key problem of the interaction between plasticity and PTs at different scales [93, 426, 427]. It never had a strict solution, and results in Fig. 22b,c give the first suggestion for Zr.

In addition to fundamental significance, results in Fig. 22b,c have essential applied potential. Traditionally, for producing almost single-phase $\omega$-Zr ($c = 0.82$) with a steady grain size of 200 nm, the HPT at 6 GPa and 5 anvil revolutions ($q = 80$) is necessary, see [29]. In [89], after the same rolling regime for the same Zr sample like in [82, 425] and then compression with rough-DA with $\Delta q = 1.84$ and up to 14 GPa, the steady $d_\omega = 37 - 60 nm$ and $\rho_{d\omega} = (0.75 - 1.30) \times 10^{15} m^{-2}$ were obtained. In Fig. 22b,c, in the course of the $\alpha - \omega$ PT in the pressure range of 0.67-3 GPa and $\Delta q = 0.48 - 1.22$ for rough-DA and $\Delta q = 0.43 - 0.85$ for smooth-DA, $d_\omega$ changes from 4 to 64 nm and $\rho_{d\omega}$ from 0.5 to 3.1 $\times 10^{15} m^{-2}$. It was concluded in [82, 425] that the plastic strain-induced PT is a much more effective method to refine microstructure and manipulate strength than the SPD at higher pressure alone. This is especially important for producing nanocomposites with optimal combination of strength (e.g., from strong but brittle $\omega$-Zr) and ductility (e.g., from weaker but more ductile $\alpha$-Zr).

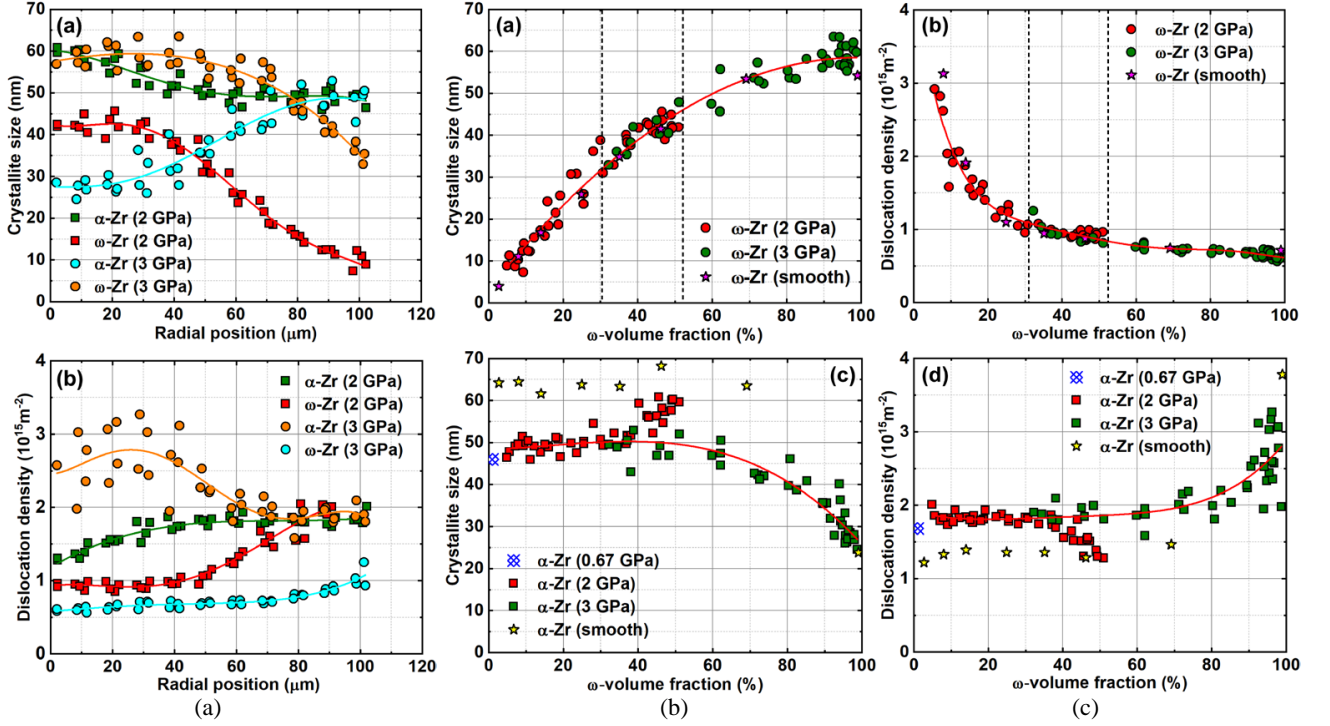

**Figure 22:** (a) Radial distributions of the crystallite size and the dislocation density in $\alpha$-Zr and $\omega$-Zr compressed with rough-DA to maximum pressures of 2 and 3 GPa. (b) Crystallite size and (c) dislocation density in $\omega$-Zr (top) and $\alpha$-Zr (bottom) from (a) plotted vs. phase fraction of $\omega$ phase at maximum pressures of 2 and 3 GPa with rough-DA and smooth-DA. Dash lines show region of the overlap of data for rough-DA from plots for 2 and 3 GPa. Plots in (b) and (c) lead to a remarkable rule for $\omega$-Zr: *there exist the unique curves for the crystallite size and dislocation density depending on volume fraction of $\omega$-Zr c only in the course of $\alpha - \omega$ PT, which are independent of pressure, plastic strain tensor $\varepsilon_p$ and its path $\varepsilon_p^{path}$, the same for rough- and smooth-DA and different initial $\alpha$-Zr microstructures.* The same rule is valid for $\alpha$-Zr, but different for rough-DA and smooth-DA and except for region $0.4 < c < 0.5$ with some scatter. Reproduced from [82].

Note that such a small crystalline or grain size at the initiating the PT is nontrivial, because at early stage of pressure-induced $\alpha - \omega$ PT in a nanograined Zr (and in multiple other metals), growth of small number of $\omega$ grains up to 10 $\mu m$ at room temperature within one hour was found in [428–430] and rationalized in [430].

Strong crystallite size and dislocation density variation during PT questions the applicability of the traditional mixture rule for the yield strength of a mixture with the *steady yield strengths* of phases used in FEM simulations (Section 8.3). One has to use $\sigma_{y\omega}(d_\omega, \rho_{d\omega}, p)$ and $\sigma_{y\alpha}(d_\alpha, \rho_{d\alpha}, p)$, and these dependencies may be non-monotonous in the regions of the direct and inverse Hall-Petch relationships. Such a small crystallite/grain size of the HPP at the beginning of PT may be a potential reason of the reduction in the yield strength during the PT for different materials discussed in [162, 431] (Section 3.12). The distribution of the yield strengths of phases could be *measured using XRD peak broadening or shifting* (see Section 2.5, [108, 208, 226–228, 233], and Fig. 9b,c). This could bring the modeling techniques for multiphase materials, especially with PTs, to a qualitatively new level.

A similar and even stronger rule was found *for olivine-spinel PT [83]: during the PT, the crystallite size of spinel is steady and independent of $\varepsilon_p$, its path $\varepsilon_p^{path}$, and volume fraction of spinel c.* Dislocation density had a significant scatter, preventing the claim of a similar conclusion. Note that to some extent, a similar rule for microstructure evolution was found in situ in [88] for pressure-induced $\alpha - \omega$ PT in Zr (see Section 4.3 and Fig. 23): *the microstrain, average crystallite size, and dislocation density in $\omega$-Zr for $c < 0.8$ are functions of the phase fraction of $\omega$-Zr only, which are independent of the pressure and plastic strain before the PT.*

It is necessary to mention important ex situ research on the grain size after SPD and PT for various oxides [432–436], which show that the grain size in the HPPs is smaller than that in the LPPs. However, obtained after load release for the final state only, they allow multiple interpretations (discussed in [85]). An intriguing conclusion was made in an ex situ study in [34] that the $\omega$ phase fraction in Ti increased with growing grain size and below the critical grain size, PT does not occur. This is opposite to the results for Zr [80, 82], olivine [83], and Si [81] (in the region of the direct Hall-Petch effect), showing that the smaller the crystallite/grain size is, the lower the PT pressure. The puzzle can be resolved by noting that the grain size in [34] was measured after PT. Consequently, it is not the grain size that affects the PT pressure, but the PT and phase fraction of $\omega$ phase determine the grain size. For such an interpretation, plot in [34] is consistent with Figure 22b. We are also reminded that during unloading after HPT of Ni at 8 GPa in [38], the dislocation density was decreased by 3 times and crystallite size was increased by more than twice, i.e., results of in situ and ex situ studies are quite different.

*3.12. Does the yield strength reduce during the PT?*

We have to distinguish data obtained with different methods.

*(a) Pressure gradient method.* Reduction in the yield strength after B1-B2 PT in NaCl by almost 50% was found in [162]. Such a drop in the strength versus jump in the volumetric strain during the PT belongs to the same straight line plotted using Bridgman's data [6] for KF, KCl, KBr, and KI. It was concluded that much smaller jump in the volumetric strain of $\sim 0.015$ for NaCl than for other halides ($\geq 0.1$) explained the reduction in the yield strength. Stress relaxation near the phase boundary was mentioned but not quantified.

It follows from the simplified equilibrium Eq. (4) that at steps in a radial pressure distribution observed in [62, 64, 66, 359] (see Section 3.2), the yield strength during the PT in the two-phase region is zero. This contradicts numerous stress-strain curves for TRIP steels during the PT [104]. As it will be shown in Section 8.3, Eq. (4) is not applicable in region with large gradients of $c$, i.e., implication about zero yield strength is wrong.

*(b) XRD with radial diffraction.* In [431] strength of NaCl increased from 0.22(6) to 0.36(6) GPa at 29 GPa in B1 phase. In the phase coexistence region between 28 and 32 GPa, the yield strength of B2 phase dropped sharply to 0.002(6) at 30 GPa than strongly increased to 1.8(6) GPa at 56 GPa. Note that the drop of the yield strength in B1 was not reported; just data for B2 had very low differential stress at the very beginning of the PT. The volume reduction during the PT was considered a cause for the stress relaxation. In fact, because of this, the differential stress within small B2 regions could be much smaller than $\sigma_y$ but sharply grow with small pressure/plastic strain increase. We see several possible interpretations of these results. (i) Due to small initial grain size of the HPP (e.g., 5 nm in $\omega$-Zr [82], in ringwoodite 4-10 nm [83] or 50-20 nm [43], or 10 nm for Si II, V, or XI [81]), HPP may be in the region of the inverse Hall-Petch relationship, which leads to a low yield strength. This is well documented for olivine-ringwoodite PT in [43]. This could also contribute to explanation of the reduction in strength during B1-B2 PT in NaCl [162] discussed above. (ii) Volume reduction and TRIP may lead to significant deviatoric stress relaxation, and measured differential stress may be significantly below the yield strength. (iii) If HPP consists of multiple crystallographic variants, this may lead to significant relaxation of deviatoric stresses immediately during the PT, without necessity of reaching the yield strength. The maximum driving force for the PT is achieved at zero deviatoric stress after completing the PT in nuclei [437]. For cases (ii) and (iii), experimental results do not imply that the yield strength of the HPP reduces during the PT. Generally, PT is considered a mechanism of stress relaxation, in addition to slip and twinning.

In a similar measurement, during the stishovite-$CaCl_2$-type PT in $SiO_2$, the differential stress drops from 4 GPa before the PT at 40 GPa to almost zero at 50 GPa, and then sharply increases to 5 GPa at 52-55 GPa after PT and grows to 7.5 GPa at 60 GPa [438]. Since this PT is related to vanishing tetragonal shear modulus (and, consequently, the Reuss bound for polycrystalline shear modulus) of the stishovite, this explains the drop of the differential stress and the yield strength to zero. Thus, statement about reduction of the yield strength during the PT in many cases is not supported by experiments.

*(c) XRD with axial diffraction.* Reduction in the deviatoric strain in Ge-I during its PT to Ge-II is claimed in [133], while it did not change in Ge-II. Deviatoric strain in each phase was determined based increase in the X-ray peak width (full-width at half-maximum of the XRD profile on $2\theta$-scale) [177, 227, 439]. Reduction in deviatoric strain was interpreted in terms volume decrease during the PT referring to the Eshelby inclusion model [440]. This result does not imply that the yield strength of Ge-I reduces during the PT, and additional stress relaxation may occur due to TRIP. As it is mentioned in Section 3.11, systematic measuring of the evolution of distribution of the dislocation densities, crystallite sizes, and the yield strengths in phases could bring the modeling techniques for multiphase materials, especially with PTs, at qualitatively new level.

*3.13. Summary and perspectives*

There are three main challenges in the discovery of new HPPs and transforming the discovery into industrial technologies: finding conditions for their synthesis at high pressures, retaining them at ambient conditions, and ways to reduce the PT pressure to a practically reasonable level. Based on the results summarized in this Section on HPT and in RDAC, SPD and controlling initial microstructure is a promising approach toward resolving these 3 problems. Thus, SPD (a) drastically reduces the PT and CR pressures; (b) leads to new PTs and CRs; (c) retains some metastable HPPs; (d) drastically accelerates transformation kinetics; (e) leads to nanostructured and amorphous materials, and (f) produces various important phenomena contributing to promoting the PTs/CRs. First, quite general rules were found for steady microstructure, yield strength, and minimum pressure for strain-induced PTs and its independence of the loading path. Due to existence of multiple steady states, a new general problem is formulated on finding loadings leading to transition from one steady state to another. Another rule states that during PT, all microstructural parameters depend on the volume fraction of the HPP only and are independent of the loading paths.

It was found that all promoting effects of SPD are more pronounced for nanograined materials, for which the grain size is in the region of Hall-Petch relationship for the yield strength. That is why it was suggested in [80–83, 85] first to prepare nanograined material with steady microstructure by SPD under normal pressure and apply plastic straining (even not necessary severe) under high pressure to promote PTs/CRs. For this case, the same minimum pressure for strain-induced PT/CR can be reached during compression in DAC like under torsion in RDAC, significantly increasing the community that works on strain-induced PTs/CRs. However, to complete the PT/CR under low pressure, torsion in RDAC is much more effective, because it pressure growth during torsion can be avoided or is much more limited than under compression in DAC. Also, PTs caused by the phonon instability are expected to be much more promoted by plastic straining than those caused by elastic instability. It was also found in situ in [82] that PT during plastic straining is a much more effective way to reduce the grain size and increase the dislocation density in HPP than plastic straining alone.

Various phenomena, such as PT induced by rotational plastic instability, shear banding, TRIP/RIP, and turbulent-like plastic flow, not only intensify plastic flow but also enhance strain-induced PTs and CRs. Some of them produce positive mechanochemical feedback, like TRIP/RIP, defect production, and PTs/CRs producing cascading structural changes [74, 388] and self-blown-up deformation-PT-TRIP-heating process in a shear band [393]. After better understanding, they can be utilized for strain- or defect-induced material synthesis.

Note that for some types of high-pressure loadings, e.g., indentation [441, 442] or uniaxial compression of nanospheres [443], plastic deformation does not produce visible reduction in PT pressure. This is not surprising from the point of view of dislocation pileup-based mechanism. Indeed, for single crystals, dislocation pileups do not appear due to lack of strong obstacles. For polycrystals, the high-pressure region is just below the indenter or other tools, but intense plastic flow is away from the high-pressure region, so they do not interact. At the same time, for small-scale strong material, like Si nanopillar, uniaxial compression produces very high stresses even in single crystals and lead to PTs [394] that were studied with in situ TEM. Similar experiments for a polycrystalline nanopillar may lead to *in situ detection of the dislocation pileup and corresponding nucleation mechanism or revealing new mechanisms for strain-induced PTs.*

## 4. Four-scale theoretical framework for plastic strain-induced PTs and CRs

*4.1. Thermodynamic conditions for stress-induced PTs and CRs in plastic material*

A general continuum thermodynamic and kinetic theory for stress- and temperature-induced PTs and CRs in elastoplastic materials was developed [92, 94] for small strains and [93] for large strains and reviewed in [101]. We will summarize the equations necessary for the current review only. The main geometric characteristic of martensitic or reconstructive PTs and CRs between crystalline phases is the transformation strain tensor $\boldsymbol{\varepsilon}_t$. It transforms a crystal cell of the parent phase into that of the product phase, both stress-free. For example, for cubic-tetragonal PT Si-I→Si-II [444] and cubic-monoclinic PT rhombohedral graphite→hexagonal diamond [445] tensors $\boldsymbol{\varepsilon}_t$ are

$$\boldsymbol{\varepsilon}_{tj}^{SiI-II} = \begin{pmatrix} 0.243 & 0 & 0 \\ 0 & 0.243 & 0 \\ 0 & 0 & -0.514 \end{pmatrix}; \quad \boldsymbol{\varepsilon}_{tj}^{G-D} = \begin{pmatrix} 0.024 & 0 & 0.105 \\ 0 & -0.034 & 0 \\ 0.105 & 0 & -0.35 \end{pmatrix}, \quad (12)$$

i.e., components can be quite large. For large-strain formulation, transformation between two lattices is described by transformation deformation gradient $\boldsymbol{F}_t = \boldsymbol{R}_t \cdot \boldsymbol{U}_t$, where $\boldsymbol{R}_t$ is the orthogonal lattice rotation tensor and $\boldsymbol{U}_t = \boldsymbol{I} + \boldsymbol{\varepsilon}_t$ is the transformational right stretch (Bain) tensor. To separate the effect of the pressure $p$ and deviatoric stress tensor $\boldsymbol{s}$, it is convenient to split $\boldsymbol{\varepsilon}_t$ into spherical $\varepsilon_{t0} = 1/3 \boldsymbol{\varepsilon}_t : \boldsymbol{I} = 1/3 \varepsilon_t^{ii} = 1/3(\varepsilon_t^{11} + \varepsilon_t^{22} + \varepsilon_t^{33})$ and the rest, i.e., deviatoric part $\boldsymbol{e}_t = \boldsymbol{\varepsilon}_t - \varepsilon_{t0} \boldsymbol{I}$ part (like in Eq. 1). For small strains, volumetric strain is $\varepsilon_{tv} = 3\varepsilon_{t0}$; for large strains, it is $\varepsilon_{tv} = J_t - 1$ with $J_t = Det \boldsymbol{F}_t = Det \boldsymbol{U}_t$. Even if transformation strain is not completely determined by the crystallography, e.g., for amorphous or liquid phases or completely reconstructive or diffusive PTs or CRs, the volumetric part of the transformation strain is determined via ratio of specific volumes or mass densities $\rho$ of phases, and the deviatoric part is proportional to deviatoric stress $\boldsymbol{s}$ [101, 446–449].

Additive decomposition of strain $\boldsymbol{\varepsilon}$ into elastic $\boldsymbol{\varepsilon}_e$, plastic $\boldsymbol{\varepsilon}_p$, and transformational $\boldsymbol{\varepsilon}_p$ parts is

$$\boldsymbol{\varepsilon} = \boldsymbol{\varepsilon}_e + \boldsymbol{\varepsilon}_p + \boldsymbol{\varepsilon}_t. \quad (13)$$

The net thermodynamic driving force $F$ for PT or CR for the appearance of the nucleus of HPP in the region $V_n$ bounded by surface $\Sigma_n$ at constant temperature $T$ is as follows [93, 94]:

$$F = \int_{V_n} X dV_n - \Delta \int_{\Sigma} \Gamma d\Sigma_n - \int_{V_n} K dV_n = \int_{V_n} \left[ \int_{\boldsymbol{\varepsilon}_1}^{\boldsymbol{\varepsilon}_2} \boldsymbol{\sigma} : d\boldsymbol{\varepsilon} - (\psi_2 - \psi_1) - \int_{\boldsymbol{\varepsilon}_{p1}}^{\boldsymbol{\varepsilon}_{p2}} \boldsymbol{\sigma} : d\boldsymbol{\varepsilon}_p \right] dV_n - \Delta \int_{\Sigma_n} \Gamma d\Sigma_n - \int_{V_n} K dV_n. \quad (14)$$

Here, $\psi = \psi^e + \psi^\theta$ is the Helmholtz free energy per unit volume, consisting of elastic and thermal parts, and $\Gamma$ is the interface energy per unit area, $K$ is the athermal threshold for PT per unit volume, subscripts 1 and 2 designate LPP and HPP, and $\Delta$ is difference of parameters after and before PT. In Eq. (14), $\boldsymbol{\sigma} : d\boldsymbol{\varepsilon} = \sigma_{ij} d\varepsilon_{ji} = \sigma_{11} d\varepsilon_{11} + \sigma_{22} d\varepsilon_{22} + \sigma_{33} d\varepsilon_{33} + 2\sigma_{12} d\varepsilon_{12} + 2\sigma_{13} d\varepsilon_{13} + 2\sigma_{23} d\varepsilon_{23}$ and $\boldsymbol{\sigma} : d\boldsymbol{\varepsilon}_p$ are the elemental total and plastic work (or plastic dissipation), where symmetry of stress and strain tensors is taken in account; the first two terms in the second part of Eq. (14) represent the total dissipation increment during PT due to processes in bulk; the first three terms are the dissipation increment due to PT only (i.e., total minus plastic dissipation) or the thermodynamic driving force for nucleation due to processes in bulk; $X$ designates its local value; the first four terms represent the thermodynamic driving force for nucleation. The PT or CR occurs when $F \geq 0$ only. All internal parameters, like geometry of nucleus, crystal lattice orientation in it, etc., are determined from the maximization of $F$. This was justified using the postulate of realizability [92, 93, 101, 450], which allows one to derive various extremum principles for dissipative systems and material instabilities. It is possible to show that for elastic materials without athermal dissipation ($\boldsymbol{\sigma} : d\boldsymbol{\varepsilon}_p = K = 0$), condition $F = 0$ reduces to the equality of the Gibbs energies of phases, and the maximization of $F$ is equivalent to the principle of the minimum of Gibbs energy. Including $K > 0$ even for elastic material already makes the Gibbs principle not applicable and requires more complex mathematical incremental and path dependent variational treatment [92, 93, 101, 450, 451]. In the simplest case, when $K$ and temperature are homogeneously distributed in a transforming region and elastic properties of phases are the same, Eq. (14) simplifies to

$$F = \int_{V_n} X dV_n - \Delta \int_{\Sigma} \Gamma d\Sigma_n - \int_{V_n} K dV_n = \int_{V_n} \int_{0}^{\boldsymbol{\varepsilon}_{t2}} \boldsymbol{\sigma} : d\boldsymbol{\varepsilon}_t dV_n - \left(\Delta \psi^{\theta}(T) + K\right) V_n - \Delta \int_{\Sigma} \Gamma d\Sigma_n. \qquad (15)$$

When the $V_n$ is a volume covered by a moving coherent phase interface $\Sigma$ during an infinitesimal time, Eq. (15) can be transformed to

$$F_{\Sigma} = X_{\Sigma} - K - 2\Gamma \kappa_{av} = \boldsymbol{\sigma} : \Delta \boldsymbol{\varepsilon}_t - \Delta \psi^{\theta}(T) - K - 2\Gamma \kappa_{av}; \quad X_{\Sigma} = \boldsymbol{\sigma} : \Delta \boldsymbol{\varepsilon}_t - \Delta \psi^{\theta}(T). \qquad (16)$$

Here, $\kappa_{av}$ is the mean interface curvature and $X_{\Sigma}$ is the thermodynamic driving force for the interface propagation excluding interfacial energy. The first term in the second part of Eqs. (15) and (16) is the transformation work in a nucleus or at phase interface, the main stress-related contributor to the net driving force. It looks very unexpected that Eqs. (15) and (16) do not explicitly involve plastic strain and have the same form as for an elastic material. However, stress variation within the nucleus during the PT/CR is affected by the evolution of the plastic strain field in the entire volume. An alternative approach [452–455] did not exclude plastic work from Eq. (14), which implies appearance of the plastic work in Eqs. (15) and (16), which could explain the promoting effect of plastic strain on PT and CR. However, exclusion of plastic work (i.e., utilizing dissipation increment due to PT/CR only in the expression for the thermodynamic driving force for PT/CR) was justified in [93, 94, 456–460] using a local description of PTs. It was also shown in [461] that not excluding plastic work leads to a conceptual contradiction, which disappears when it is excluded. Also, nanoscale [380, 381] and scale-free [462, 463] PFAs to interaction of PT and discrete dislocations also confirmed that the phase equilibrium condition at phase interfaces is described by the condition $X_{\Sigma} = 0$ (Section 6.6.3). Since $\boldsymbol{\sigma}$ explicitly includes contribution from dislocations, there was no need to include $K > 0$; the effect of $\Gamma$ was negligible. The main point is that the driving forces in Eqs. (15)-(16) involve the whole *stress tensor rather than the pressure only*; its effect on PT/CR is fully determined by the transformation strain tensor. Large-strain formulation is presented in [93, 101].

*4.2. Athermal threshold for PTs and CRs*

Athermal threshold in Eqs. (15)-(16) is similar to dry friction: PT and CR starts when the thermodynamic driving exceeds it. The athermal threshold produces the deviation of the actual stress or/and temperature for PT/CR from their thermodynamic equilibrium values and therefore stress/temperature hysteresis during cycling transformations and energy dissipation. There are several sources of the athermal threshold [92, 93, 101, 124, 464]: (a) Peierls barrier due to the discrete periodic structure of the crystal lattice, analogous to that for dislocation propagation; (b) nucleation barrier; (c) interaction of a propagating interface with a long-range stress field of various defects, e.g., dislocations, point defects (solute and impurity atoms, vacancies), grain, subgrain, and twin boundaries, stacking faults, and precipitates, and (d) emission of acoustic waves. For interface propagation, $K$ is also called athermal interfacial friction. The value of $K$ is expected to be dependent on the entire deformation-PT process and the evolving material microstructure. At the same time, a simple relationship

$$K = L\sigma_y\left(T, \boldsymbol{\varepsilon}_p, g_i\right)\varepsilon_{tv} \qquad (17)$$

was deduced in [92, 93] from the comparison of high-pressure experimental result in [64, 414, 465] on the proportionality between the pressure hysteresis and the hardness with the simple analytical solutions of the corresponding boundary-value problems. Here, $L$ is the proportionality factor and $g_i$ are microstructural parameters. The values $L$ for some PTs are given in [92, 93]. The proportionality between $K$ and $\sigma_y$ is physically reasonable because $K$ and $\sigma_y$ characterize resistance to the interface and dislocation motion, respectively, through the same defect microstructure. Since $\sigma_y$ increases with plastic strain before saturation, the immediate consequence of Eq. (17) is that plastic straining suppresses both direct and reverse stress/pressure-induced PTs. Thus, *plastic deformation of the HPP can be used to retain it under normal pressure*. For SPD, based on the rules described in Section 2.1, $\sigma_y$ and consequently $K$ attain their maximum and are independent of plastic strain and strain path.

A slightly different but similar relation between $K$ and $\sigma_y$ of an austenite for steels is suggested in [464, 466]. For shape memory alloys, the stress hysteresis is proportional to $K$. Since both the stress hysteresis and $\sigma_y$ linear depend on plastic strain [467–469], this also supports Eq. (17). Eq. (17) was utilised in analytical solutions in [92, 93, 101] and in FEM simulation in [96, 97, 99] for various pressure-, temperature-, and stress-induced PTs. However, as it is shown in [88] (Section 4.3), Eq. (17) cannot describe recent in situ experiments on pressure-induced $\alpha - \omega$ PT in Zr.

*4.3. Hydrostatic experiments: coupled microstructure evolution, $\alpha - \omega$ PT, athermal threshold, and plasticity for predeformed and annealed Zr*

A comprehensive in situ XRD study of the effect of the initial microstructure (a) on its evolution under hydrostatic loading prior, during, and after the $\alpha - \omega$ PT in commercially pure Zr and during unloading and (b) on $\alpha - \omega$ PT itself was performed in [88]. The goal here is to show complexity of the rationalizing even hydrostatic experiments and necessity of coupling PT and microstructure evolution. Two initial microstructures were produced by annealing and SPD by multiple cold rolling up to steady hardness.

*PT kinetics.* It was found that the $\alpha - \omega$ PT starts at lower pressure for pre-deformed sample (5.1 vs. 5.9 GPa for annealed sample) but completes at higher pressure (13.0 vs. 10.9 GPa), see Fig. 23. These results were interpreted in a way that plastic straining before PT promotes nucleation by producing new nucleating defects, but suppresses growth by producing more obstacles (dislocation forest, twin, subgrain, and grain boundaries, and point defects) for phase interface propagation. While the latter was taken into account through the athermal threshold $K$, stress concentration was not included in Eqs. (15)-(16). It is known that dislocations serve as nucleation sites for the initiation of phase transformation [100, 102, 103], which is confirmed by more recent phase-field simulations [381, 470–472]. Twin tips also produce stress concentrators promoting nucleation [473]. Consequently, an increase in plastic strain increases the number of dislocations and twins and promotes nucleation. To describe PT initiation, the following thermodynamic PT criteria were deduced from Eq. (15):

$$X^+ = p^+\varepsilon_{tv} - \Delta\psi^\theta = K^+ - 3z^+B^+\varepsilon_m^+\varepsilon_{tv}; \qquad X^- = p^-\varepsilon_{tv} - \Delta\psi^\theta = -K^- + 3z^-B^-\varepsilon_m^-\varepsilon_{tv}. \qquad (18)$$

Here, signs $\pm$ are for direct and reverse PTs, respectively, $B$ is the bulk modulus of transforming phase, $\varepsilon_m$ is the measured microstrain, $z > 1$ is the strain concentration factor. The new point in comparison with all previous treatments of stress-induced PTs [101] is the introducing local pressure concentrator $3zB\varepsilon_m$; also, all material parameters are different for the direct and reverse PTs. Still, one of the conclusions was that *Eq. (17) for K even along with advanced Eq. (18) cannot describe experiments in Fig. 23.*

Results in [88] on reduction in pressure for starting $\alpha - \omega$ PT in commercially pure Zr are consistent with those for extra pure Zr in [473] after plastic compression by 5% and 10% before PT. However, in [473] the kinetic curves $c(p)$ for different plastic strains did not intersect, and the reduction in PT initiation pressure is larger for plastic strain of 5% and then decreased for strain of 10%, which is difficult to rationalize physically. Note that the experiments in [473] have been done in multi-anvil apparatuses without a pressure-transmitting medium and microstructure was not recorded in situ, which may cause such results. The effect of 5 different dislocations and twin microstructures produced by pre-deformation and on the $\alpha \to \omega$ PT in Zr was studied in [474] experimentally and using the viscoplastic self-consistent crystal plasticity model [44, 475]. The main conclusion from [88] was that *to significantly reduce scatter in the characterization of pressure-induced PTs, one has to characterize at least the initial microstructure and desirably its evolution during the PT.*

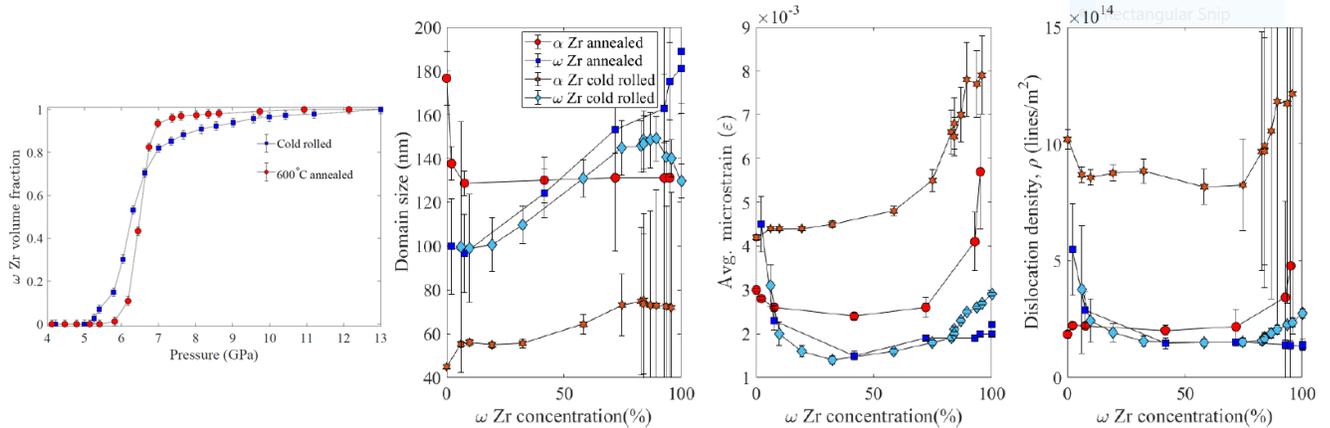

**Figure 23:** Phase fraction of $\omega$-Zr versus pressure for annealed at $600°C$ and cold rolled Zr samples (left) and average domain size, microstrain, and dislocation density in $\alpha$- and $\omega$-Zr for the annealed and cold rolled samples versus phase fraction of $\omega$-Zr (right) [88]. Closeness of microstructural parameters $\omega$-Zr for annealed and strongly predeformed samples implies the following important rule: *the microstrain, average crystallite size, and dislocation density in $\omega$-Zr for $c < 0.8$ are functions of the phase fraction of $\omega$ Zr only, which are independent of the pressure and $\varepsilon_p$ prior to PT.*

It is necessary to mention that results in [64, 414, 465] on the proportionality between the pressure hysteresis and the hardness were obtained from the force-displacement curves in piston-cylinder apparatuses without in situ XRD, with simplified pressure determination, and indium as a PTM. Therefore, some nonhydrostatic stresses were present and nucleation of a small amount of HPPs could not be detected, and contradictions of the results in [88] for Zr with those in [64, 414, 465] may be due to various reasons. Generally, results for all materials presented in [64] should be revisited using XRD with in situ synchrotron radiation.

*Microstructure evolution.* Nontrivial evolution of the microstructure during hydrostatic loading/unloading was revealed. The crystal domain size $d$ essentially decreases, and microstrain $\varepsilon_m$ and dislocation density $\rho_d$ increase during pressurizing for both single-phase $\alpha$ and $\omega$ phases. For the $\alpha$ phase, $d$ decreases from 60 to 45 nm before PT for the predeformed Zr and from 300 to 120 nm for the annealed Zr; $\rho_d$ grows linearly with pressure from 0.2 to $2.1 \times 10^{14}\, lines/m^2$ for the annealed Zr and from 6 to $1.1 \times 10^{15}\, lines/m^2$ for the predeformed Zr. These numbers for the predeformed sample are comparable to those in [89] after SPD, showing *potential for grain refinement and strengthening by hydrostatic pressure treatment.* Some advantages in comparison with SPD are that SPD may lead to a brittle $\omega$-Zr at low pressure, which may not be desirable. For $\omega$ phase $\rho_d$ grows to $5.6 \times 10^{14}\, lines/m^2$ for the predeformed Zr at 15.9 GPa and to $3.03 \times 10^{14}\, lines/m^2$ at 15.6 GPa.

On pressure release, average domain size in the recovered $\omega$ phase remains nearly the same as at the highest measured pressure, i.e., 150 nm and 80 nm for the annealed and the cold rolled samples, respectively, versus 300 nm and 60 nm in $\alpha$-Zr before pressure cycle for the annealed and the predeformed Zr, respectively. Thus, for predeformed Zr pressure cycle increases and $d$ is therefore useless. In the same cycle for predeformed sample, the final $\rho_d = 3 \times 10^{14}\, lines/m^2$ (2 times lower than in the starting $\alpha$-Zr) and for annealed sample $\rho_d = 3.4 \times 10^{14}\, lines/m^2$ (an order of magnitude larger than in the starting $\alpha$-Zr and even larger than for predeformed final $\omega$-Zr, but lower than for the initial predeformed $\alpha$-Zr). Surprisingly, for the annealed sample, the final $\rho_d$ and $\varepsilon_m$ in the $\omega$-Zr are larger than for the SPD-treated sample. Thus, irreversibility in the changes in microstructure during pressure cycle is of practical interest for producing nanograined materials. This cycle, as an alternative to SPD *improves microstructure for the annealed sample and is not of use for the predeformed sample.* The irreversibility of the effect of pressure on the microstructure implies the necessity to add pressure path $p^{path}$ as a factor controlling microstructure and should be included in the list of independent parameters for its description and related description of the yield strength and yield surface. Also, a strong change in the microstructure under hydrostatic loading shows significant advantage of the in situ versus post-mortem studies.

Fig. 23 presents comparison of evolution of the microstructural parameters for both samples with changing volume fraction of $\omega$-Zr. The main and unexpected result is that *for pressure-induced PT, the microstrain, average crystallite size, and dislocation density in $\omega$-Zr for $c < 0.8$ are functions of the phase fraction of $\omega$-Zr only, which are independent of the pressure and plastic strain before the PT*, to some extend similar to what was observed for strain-induced PTs. Since the evolution of microstructural parameters in $\alpha$-Zr significantly differs for annealed and predeformed samples, mostly due to different initial values; this makes impression that the the microstructure is not inherited during PT and the PT wipes out the entire memory about the microstructure in $\alpha$-Zr. However, for $c > 0.8$, some memory re-appears by some unknown mechanism and a cold-rolled sample has finer domain size and larger dislocation density and microstrain than the annealed Zr until the end of PT and in single-phase $\omega$-Zr. It looks like 20 % of retaining $\alpha$-Zr transforms differently somehow and keeps the memory of the initial microstructure.

*Plasticity theory under hydrostatic loading.* It is generally accepted in the macroscopic theory of plasticity [4, 165, 476] that plastic deformations do not occur at hydrostatic loading of void-free materials. This neglects the well-documented physical causes for micro-plastic deformation at hydrostatic conditions, like high local non-hydrostatic internal stresses due to various defects (dislocations, twins, subgrain and grain boundaries, etc.) and nonuniform and anisotropic elastic moduli. Also, elastic moduli, dislocation core energy, and structure vary with pressure, which causes to redistribution of dislocation configurations, increase in dislocation density, grain refinement, and others. Confronting this paradigm, it was suggested in [88] to describe macroscopic plasticity under hydrostatic loading in terms of measured $\rho_d$ and $d$ as internal variables instead of a change of geometry (plastic strain), which is negligible. Then the yield strength is varied according to Eq. (6), which gives additional pressure-dependent contributions to $\sigma_y$. The simplest estimate of plastic strain was based on Orowan's equation for plastic shear $\gamma_p = \rho_d b l$, where $l$ is the averaged distance travelled by dislocations [262]. A rough estimate based on an experiment in [88] gave $\gamma_p = 1.21\%$ for the annealed Zr and 0.75% for the predeformed samples.

*EOS.* It was found in [88] that the EOS curves for lattice parameters $c$ and $a$ and volume for the $\alpha$-Zr for predeformed sample are clearly above those for the annealed Zr, while for $\omega$-Zr they are very close. Based on the obtained experimental data, it was deduced that the difference in EOS for $\alpha$-Zr for the predeformed and annealed samples can be related to different $\varepsilon_m$. Microstrain characterizes internal elastic strains and stresses with significant non-hydrostatic components. Since the difference between the EOS under hydrostatic and non-hydrostatic stresses is well-documented [222, 229, 477] (which is true even for microstructure-free materials [478]), this explains different EOS for the predeformed and annealed $\alpha$-Zr. For predeformed and annealed $\omega$-Zr, microstrains are very close, which explains the closeness of EOS. Since $\rho_d$ and $d$ for the predeformed and annealed $\omega$-Zr are different, they were excluded from the main parameters affecting EOS.

Thus, hydrostatic experiments were considered as "clean" ones under well-defined pressure and without devi-atoric stresses and plasticity. However, these conditions are true for the liquid/gaseous PTM before solidifying rather than for a solid sample. The results in [88] imply the *necessity of developing experiment-based predictive constitutive equations for the coupled EOS, pressure-induced PTs, microstructure evolution, and plasticity.*

Note that in 4:1 methanol-ethanol PTM, $\alpha - \omega$ PT in the coarse-grained Ti occurs in the range 10.2-14.7 GPa in [325], in [327] strongly pre-deformed nanocrystalline Ti transforms in the range 10.0-13.0 GPa, i.e., preliminary plastic deformation slightly promotes nucleation and more essentially growth.

*4.4. Thermodynamic theory for stress-induced PTs/CRs fails to describe strong reduction in PT pressure*

This was shown in [479] utilizing a simplified version of Eqs. (15) or (16) with neglected interfacial energy and constant stress tensor during the PT. A simplified expression for transformation work $W = \boldsymbol{\sigma} : \Delta \boldsymbol{\varepsilon}_t = \boldsymbol{\sigma} : \boldsymbol{\varepsilon}_t$ is $W = -p\varepsilon_{tv} + \tau \gamma_t$, where deviatoric part of the transformation work is express as a product of shear stress $\tau$ and transformation shear $\gamma_t$ (equal to double of shear component of the transformation strain tensor), neglecting other components. Thus, the PT criterion is $-p\varepsilon_{tv} + \tau \gamma_t = \Delta \psi^\theta + K$. Under hydrostatic conditions, $W = -p_h^d \varepsilon_{tv} = \Delta \psi^\theta + K$. Under shear stresses, which are limited by the macroscopic yield strength $\tau_y$, the PT pressure $p_\tau$ is determined from $W = -p_\tau \varepsilon_{tv} + \tau_y \gamma_t = -p_h^d \varepsilon_{tv}$, implying $p_h^d - p_\tau = \tau_y \gamma_t / |\varepsilon_{tv}|$. We took into account that $\varepsilon_{tv} < 0$.

Take the generic values $\gamma_t = 2|\varepsilon_{tv}|$ and $\tau_y = 1$ GPa. Then, $p_h^d - p_\tau = 2$ GPa only. Even if we increase $\gamma_t = 4|\varepsilon_{tv}|$ and $\tau_y = 2$ GPa, then $p_h^d - p_\tau = 8$ GPa only. Table 1 shows much larger numbers for PTs $hBN \to wBN$, graphite to cubic and hexagonal diamonds, Si-I→Si-II and Si-I→Si-III. To be specific, for Si-I→Si-II [81], based on Eq. (12), $\varepsilon_{tv} = -0.249$ and very large $\gamma_t = 0.243 + 0.514 = 0.757$, $\gamma_t / |\varepsilon_{tv}| = 3.04$. Taking $\tau_y = 2.5$ GPa, we obtain $p_h^d - p_\tau = 7.60$ GPa only, while in experiment for 100 nm particles it is 15.9 GPa [81]. The yield strength should be doubled, and maximum shear stresses should be perfectly aligned with direction of $\gamma_t$ to approach the experimental PT pressure reduction.

For PT Si-I→Si-III, $\boldsymbol{\varepsilon}_t = \{-0.002; -0.136; 0.058\}$ [480] with the volumetric strain of $\varepsilon_{tv} = -0.0885$ and $\gamma_t = 0.136 + 0.058 = 0.194$. Then $\gamma_t / |\varepsilon_{tv}| = 2.19$, and for the same $\tau_y = 2.5$ GPa, one obtains $p_h^d - p_\tau = 5.48$ GPa only. Since PT to Si-III was not observed under hydrostatic pressure at all, such reduction in the PT pressure by shear stress qualitatively contradicts obtaining Si-III at 0.6 GPa in [81].

Note that despite the estimates in [479], there are various attempts to explain reduction in PT pressure because of plastic flow by taking into account deviatoric or shear stress in the conventional thermodynamic treatment. For example, the thermodynamic framework in [481] was used to analyze the effect of deviatoric stresses on the $\alpha - \omega$ and $\omega - \beta$ phase equilibrium lines. However, the main contributor to the effect of deviatoric stresses, deviatoric transformation strain, was not included. Similarly, in [482], devoted to significant reduction in PT pressure for PTs in BN in RDAC, transformation shear strain is neglected and the reduction in PT pressure is due to change in elastic shear moduli and relaxation of the energy of shear stresses (while relaxation cannot be expected for PT to stronger HPPs of BN). The maximum reduction in PT pressure in [482] was $\Delta p = \tau_y^2 / (2\mu |\varepsilon_{tv}|) = 0.5 \tau_y \gamma_e / |\varepsilon_{tv}|$, i.e., similar to the above equation for $p_h^d - p_\tau$ but with the half the elastic shear strain $\gamma_e = \tau_y / \mu$ instead of transformation shear strain $\gamma_t$. For polycrystalline hBN (in which PT occurs), $\tau_y \simeq 0.5 \sigma_y = 0.15$ GPa [386] and $\mu = 20$ GPa, i.e., $0.5 \gamma_e \simeq 0.004$; even if we take $\mu = 6$ GPa [482], we will get $0.5 \gamma_e \simeq 0.013$. For PT from hBN to wBN, $\boldsymbol{\varepsilon}_t = \{0.018; 0.018; -0.365\}$ [483, 484] with $\varepsilon_{tv} = -0.342$ and $\gamma_t = 0.365 + 0.018 = 0.38$, i.e., it is at least 29 times larger than $0.5 \gamma_e$. Should the authors of [482] insert the numbers, they would obtain $\Delta p = 0.006$ GPa. Even funnier is the estimate of the radius of the critical nucleus in Eq. (11) in [482] for homogeneous nucleation under shear stresses to explain reduction of PT pressure in RDAC, because (in addition to wrong volume $V$ in denominator) homogeneous nucleation requires a very large thermodynamic driving force [485]; that is why heterogeneous nucleation at defects is usually considered. In [445, 483, 484], transformation work for the full stress tensor was properly included in the expression for the driving force for PT in BN and carbon and used for a qualitative analysis; however, numbers also were not inserted. The effect of nonhydrostatic stresses on $\alpha - \omega$ PT in Zr was quantitatively analysed analytically and with PFA in [486]. It was concluded that their effect is insufficient to explain experimentally observed reduction in PT pressure; consideration of the plastic strain-induced PT is required.

After learning that Santa Claus does not exist, the author stopped believing in miracles. He could not believe that thermodynamics can fail and assumed that some mechanisms of local stress magnification should be first involved, and then thermodynamics will work again. This idea was realized in [100, 102], see below.

*4.5. Sketch of four-scale modeling for strain-induced PT*

Even before estimates presented in Section 4.4 were published, it became clear that PTs and CRs caused by plastic flow must be considered as a separate class of plastic strain-induced PTs/CRs which occur by nucleation at new defects permanently generated during plastic flow [100, 102]. The first simplest analytical three-scale theory for strain-induced PTs/CRs was presented in [100, 102], a sketch of which, after intensive extension and adding the fourth atomistic scale, is presented in Fig. 24. The main physics is related to nanoscale processes of nucleation and growth of HPP at the defect with strongest stress concentrator – dislocation pileup with large number of dislocations – because other defects cannot describe experimentally observed drastic reduction in PT/CR pressure by plastic straining. In addition to an analytical approach, nanoscale PFA, MD, and CAC (i.e., MD & FEM) approaches were developed and applied to study different aspects of interaction of dislocation pileups and other generated discrete dislocations, and PT to HPP, see Section 6. These approaches supplement and enrich each other. Thus, nanoscale PFA (Section 6.6) gives general extended thermodynamic framework for PTs, dislocations, and their interaction, which includes energy barriers between phases (in contrast to continuum approaches), crystallographic variants, and interfacial energies. It allows to directly model barrierless nucleation and evolution of dislocations and HPP without any a priori assumptions and hands waving. MD simulations in Section 6.8 provide not only atomistic features of dislocation pileup-induced PT, but also interaction between pileup and grain and phase boundaries, including partial or complete passage of the dislocations through grain boundaries, their absorption, rearrangement of the interfaces, and the tip of dislocation pileups. The CAC approach, combining MD simulations at the place of interest (e.g., tip of dislocation pileup interacting with grain boundary) and coarse-grained FEM simulations elsewhere, allows increasing sample size. Simulation results lead to some advancing of the analytical model in Section 6.3. Some experimental confirmations of the consequences of the dislocation pileup mechanism are presented, especially in Sections 6.3, 6.4, and 6.10.

Conceptual nanoscale results, along with some micromechanical and thermodynamic treatments, are scaled up for the development of the microscale kinetic equation for strain-induced PTs/CRs between two and multiple phases in Sections 7.1- 7.3. The kinetics is experimentally confirmed for $\alpha - \omega$ PT in Zr in Section 7.5 and, after developing a macroscale model in Section 8.3 and corresponding CEA-FEM methods, generalized for the same PT in Section 7.6. Alternative coupled kinetics of hBN→wBN PT and 2D disordering in BN is developed in Section 7.4. Also, scale-free PFA was developed to increase sample size to the microscale and applied to simulations of interaction between evolving dislocation pileups and HPP in a polycrystalline sample in Section 6.7.

Nanoscale results are also scaled down in Section 5 using atomistic simulations combined with PFA for defect-free crystals. Since nucleation at the tip of the dislocation pileup is barrierless, it is related to the loss of stability of the crystal lattice. Also, in the nanoscale region near the tip of the pileup, the crystal is practically defect-free. That is why DFT and MD are utilized in Section 5.2 to obtain a general lattice instability criterion for Si-I→Si-II PT under general stress tensor. Since this requires too many variants of calculations, this criterion was first derived analytically within PFA and then verified by atomistic simulations. Also, $5^{th}$-degree elastic energy for Si-I was found in Section 5.3, which describes DFT results for the energy and stress-strain curves under complex loading well, which can also be used in the macroscopic models.

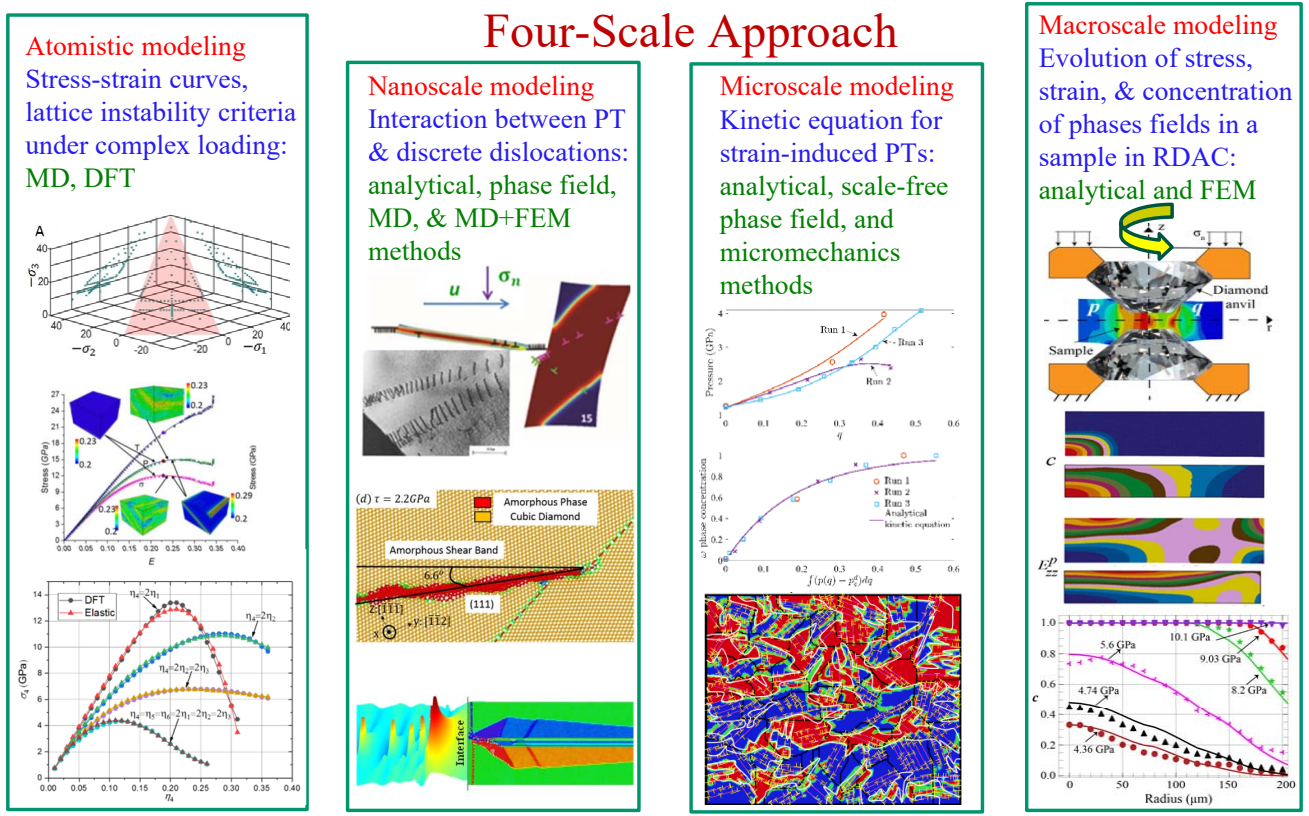

**Figure 24:** Sketch of four-scale modeling

The kinetic equations for strain-induced PTs are part of a macroscale large-strain model for coupled plastic flow and PT presented in Section 8. Simple analytical solutions in Sections 8.1 and 8.2 give some intuition, but quantitative results for DAC and RDAC are obtained with FEM in Section 8.3. They are used for studying the effect of various parameters on the coupled deformation-PT processes, reproducing and interpreting multiple experimentally-observed phenomena, predicting new effects, and developing coupled FEM-experimental methods for calibrating all constitutive parameters and determining all tensorial stress-strain fields and volume fraction of phases. Combination of the results from all four scales gives a multifaceted and multidisciplinary view of the SPD and strain-induced PTs/CRs, and suggests novel methods for controlling the PTs and economic synthetic pathways to new or known HPPs.

Note that while an absolutely major part of the modeling and simulation works reviewed here are devoted to PTs, many results are equally applicable to CRs. For CRs, plastic flow drastically accelerates diffusion, fragmentation, and mixing of constituents [14, 16, 377]; thus, diffusion and mixing is not a bottleneck process that determines a time scale for strain-induced CRs. As mentioned in [14], strain-induced CRs occur at the stress concentration sites as well, i.e., one of the nanoscale mechanisms of promoting of CRs in crystalline material may be related to dislocation pileups. The structure of microscale kinetic equations, while motivated by dislocation pileup-based mechanism, can be repeated in a much more general framework. That is why microscale kinetics and macroscale treatment of the sample behavior should be formally the same for PTs and many CRs; when we will write about PTs in this respect, this should be *applicable to the CRs* as well in many cases. Of course, such an approach looks like a strong oversimplification for chemists. However, this is similar to how firefighters distinguish a violin from a piano – a piano burns longer, which is sufficient for their purposes.

## 5. Crystal instability conditions under stress tensor

PTs in solids are mostly characterized by equilibrium pressure-temperature phase diagrams [305], each line of which is obtained from the equality of the Gibbs energy of the corresponding phases or zero driving force for the PT. The phase equilibrium conditions in elastic materials have been extended for the general stress tensor in geological [306–309], material science [310–313], and continuum mechanics [314] literature, and summarized in a review [315] and book [316]. However, in experiments, especially at room and low temperatures, PTs occur under significant deviation from equilibrium [77, 317–319], i.e., with large hysteresis. For example, at room temperature, the graphite-diamond equilibrium pressure is 2.45 GPa; however, due to hysteresis, the PT is observed at 70 GPa [318] for crystalline precursor and at 80 GPa [320] or 100 GP and 400 $^oC$ [321] for a glassy carbon. While the cBN is stable at ambient conditions [319], disordered low-pressure hBN does not transform even at 52.8 GPa [77]. At continuum level, this deviation is formalized through introducing an athermal threshold $K$ in Eqs. (14)-(16) for the thermodynamic driving force $X$. Some stable phases, e.g., ringwoodite [83] or Si II in 30 nm particles [81] do not appear under hydrostatic loading, because hysteresis is so high that other phases appear first. These PTs are nucleation-controlled, because the system has to overcome a Gibbs energy barrier between the parent and product phases. When thermal fluctuations can be neglected, e.g., at low temperature and short times, the PT condition is related to the lattice instability and barrierless nucleation. Even the theory of the temperature-induced martensitic PTs is based on the barrierless nucleation due to stress field of some dislocation configurations [103, 487, 488].

Lattice instability under nonhydrostatic stresses is of great importance for understanding and describing various phenomena, like crystal-crystal, crystal↔melt, and crystal↔amorphous PTs, and defect nucleation, like twins, dislocations, voids, and cracks. Instability is usually described in terms of the order parameters $\eta_i$ responsible for one or several of these processes. For PT, the Gibbs energy $G(\boldsymbol{\sigma}, T, \eta)$ at fixed $\boldsymbol{\sigma}$ and $T$ has a two-well structure, with local minima corresponding to the parent (often taken as $\eta = 0$) and product ($\eta = 1$) phases and energy barrier between them for $0 < \eta < 1$. When $\boldsymbol{\sigma}$ and $T$ vary, the energy barrier varies as well. Lattice instability corresponds to the disappearance of the Gibbs energy barrier.

We will focus on two main types of lattice instability: elastic and phonon instabilities. Instability with respect to some types of uniform deformation is called the *elastic instability*; the order parameter is the elastic strain tensor or some of its components or their combinations. Instability with respect to intra-cell atomic displacements or relative shifts between multilattices is described by *phonon instabilities*.

*5.1. Elastic instability*

*Elastic lattice instability* starts when $det\boldsymbol{C} = 0$ for the first time, with $\boldsymbol{C}$ for fourth-rank tensor of elastic moduli [489, 490]. Since $det\boldsymbol{C} = 0$ is 6 × 6 matrix, in general $det\boldsymbol{C} = 0$ results in 6 independent equations stating that some elastic moduli or their combinations tend to zero. Symmetry of the lattice reduces number of independent equations, e.g., up to 3 for cubic system. Examples for different crystal lattices can be found, e.g., in [491, 492]. Elastic instability results in softening of an acoustic mode in the vicinity of the Γ-point.

For the finite strains or prestressed crystals, the same criterion $det\boldsymbol{B} = 0$ is valid for the effective elastic moduli tensor $\boldsymbol{B}$ (the generalized Born criterion) [493–507]. Phonon instability (soft-mode) criterion [490, 503, 504, 506–508] is expressed as $[\omega(\boldsymbol{q}s)]^2 < 0$, i.e., phonon frequency $\omega$ for some wavevectors $\boldsymbol{q}$ and polarization $s$ became imaginary. Both elastic and phonon instability may interact and compete, which also has to be taken into account.

There was various confusion in the general determination of the effective elastic moduli for nonlinear elasticity and finite strains, analyzed in [478, 509]. In particular, a quite general and correct description of the elastic moduli with respect to the reference configuration under arbitrary stress tensor was developed in [493–497]. In modern continuum formulation [478], these moduli $\boldsymbol{B} = \{B_{ijkl}\}$ are defined by the relationship

$$\overset{\triangledown}{\boldsymbol{\sigma}}_J = \boldsymbol{B} : \boldsymbol{d}, \tag{19}$$

or, in the incremental component form

$$\sigma_{ij}(t+\Delta t) = \sigma_{ij}(t) + B_{ijkl}\epsilon_{lk} + \frac{1}{2}(\sigma_{kj}\delta_{li} - \sigma_{jl}\delta_{ik} - \sigma_{il}\delta_{jk} + \sigma_{ik}\delta_{jl})\omega_{lk};$$
$$B_{ijkl} = \bar{C}_{ijkl} - \sigma_{ij}\delta_{kl} + \frac{1}{2}(\sigma_{il}\delta_{jk} + \sigma_{ik}\delta_{jl} + \sigma_{kj}\delta_{li} + \sigma_{lj}\delta_{ki}). \qquad (20)$$

Here $\overset{\triangledown}{\boldsymbol{\sigma}}_J = \dot{\boldsymbol{\sigma}} - \boldsymbol{w}\cdot\boldsymbol{\sigma} - \boldsymbol{\sigma}\cdot\boldsymbol{w}^T$ is the Jaumann derivative of the Cauchy stress, $\boldsymbol{d}$ is deformation rate, $\boldsymbol{w}$ is the antisymmetric spin tensor, $C_{ijkl} = \frac{\partial^2\psi}{\partial E^e_{ij}\partial E^e_{kl}}|_{E^e_{cd}=0}$ and $\bar{C}_{ijkl} = J_e^{-1}F^e_{lt}F^e_{ks}F^e_{in}F^e_{jm}C_{mnst}$ are the elastic moduli in the reference (undeformed) and current (deformed) configurations, respectively, $\boldsymbol{E}_e$ is the Lagrangian elastic strain, $\psi(\boldsymbol{E}_e)$ is the elastic energy, $J_e = det\boldsymbol{F}_e$, $\epsilon_{lk} = d_{lk}\Delta t$ and $\omega_{lk} = w_{lk}\Delta t$ are small strain and rotation increments with respect to the state at time $t$. Thus, effective elastic moduli $\boldsymbol{B}$ connect small strain increments from any prestressed state with small increment of the Cauchy stress when lattice rotations are absent. The moduli connecting $\overset{\triangledown}{\boldsymbol{\sigma}}_J$ and $\boldsymbol{d}$ are broadly used in computational algorithms in various FEM codes as an input not only for elastic, but also various inelastic material models [510].

The tensor $\boldsymbol{B}$ was introduced in [495–497], is often called the "Wallace tensor," and is broadly used to evaluate elastic instability from the condition $det\boldsymbol{B} = 0$ [500, 506, 507, 511, 512]. Tensor $B_{ijkl}$ is symmetric with respect to permutations $i \leftrightarrow j$ and $k \leftrightarrow l$, but is generally not symmetric in $(i,j) \leftrightarrow (k,l)$. Also, symmetry of $B_{ijkl}$ is determined not by symmetry of unloaded lattice, but by symmetry of the lattice in the deformed state, i.e., by applied stress tensor. For example, cubic lattice under general stress state has triclinic symmetry, and all elastic constants are not zero. A sufficient but not necessary condition for elastic instability is $B_{ii} = 0$, i.e., instability occurs at reaching the maximum Cauchy stress at stress-strain curve for direct and minimum for the reverse PT (Fig. 25a), respectively. It is much simpler, and gives a correct result if supported by another practical way to detect instability. Namely, in MD simulations, one has to produce complex quasi-static loading by applying true stresses until system cannot be equilibrated [513]. After this was proved for PT Si I→Si II in MD in [513], an instability approach based on the extrema of the stress-strain curves was applied for this PT in [444] (Fig. 25a).

A method based on the extreme point of the stress-strain curves may also work even if corresponding $B_{ii} \neq 0$. In particular, in DFT simulations under hydrostatic loading, each of curves can be continued for higher $|\sigma_3|$ (Fig. 25a), i.e., dashed curve for Si I can be continued to higher strains for which Si II is more stable (has lower $|\sigma_3|$) and curve for Si II can be continued to lower strains for which Si I is more stable. Choice of the lowest curves gives the resultant stress-strain curve with a jump in $B_{33}$ instead of $B_{33} = 0$; strict instability criterion $det\boldsymbol{B} = 0$ in this case corresponds to the condition of the tetragonal instability $B_{11} < B_{12}$. Similarly, instability for Si II to Si I PT in MD simulations in [513] starts with a jump in $B_{33}$ rather than at $B_{33} = 0$, and a conclusion was made that elastic instability criterion does not work. This statement was not justified, because instability may occur when any of the elastic conditions, $B_{11} < |B_{12}|$, $B_{44} < 0$, $B_{66} < 0$, or $(B_{11} + B_{12})B_{33} < 2B_{13}^2$ [491, 492], is met, resulting in the jump in $B_{33}$ in the stress-strain curve. Note the approach in [513] on detecting the instability by reaching applied true stresses for which system cannot be equilibrated in MD simulations, is applicable for the phonon instabilities as well, provided that interatomic potential can reproduce them.

*5.2. Combined phase-field and atomistic approaches*

While studies of the lattice instabilities for various material classes under hydrostatic and two-parametric nonhydrostatic loadings have been performed using MD and DFT [507, 514–518], determining the instability criteria under the full stress tensor seemed daunting because of the large number of combinations. In the first papers on this topic [444, 513], an unexpected help came from the crystal lattice instability criterion formulated within the large-strain PFA to martensitic PTs [444, 513, 519, 520], after essential iterative development of the PFA in comparison with results of the atomistic simulations. The order parameters $\eta_i$ in this approach are not related to the total strain tensor and are internal variable, i.e., they describe phonon instability. Using the thermodynamic procedure, the following instability criterion was derived:

$$W_{ins} = 0.5\boldsymbol{\sigma}{:}\boldsymbol{F}_e^{T-1} \cdot \left.\frac{d^2\bar{\boldsymbol{U}}_t}{d\eta^2}\right|_{\eta=0} \cdot \boldsymbol{F}_e^T = 0.5\sigma_{lk}F_{ik}^{e-1}\left.\frac{d^2\bar{U}_{ij}^t}{d\eta^2}\right|_{\eta=0} F_{lj}^e \geq A. \tag{21}$$

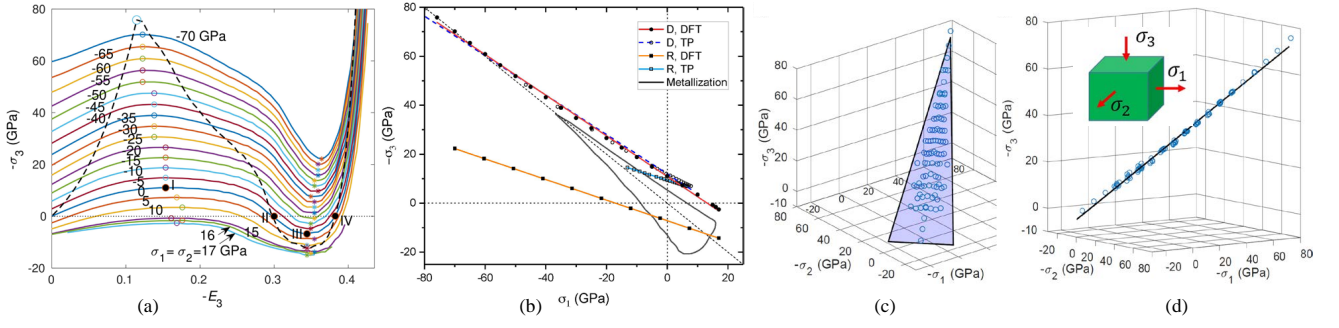

**Figure 25:** DFT results from [444] under three stresses normal to cubic faces of Si I. (a) Relationship between the true (Cauchy) stress $\sigma_3$ and Lagrangian strain $E_3$ for compression/tension along the cubic direction for different lateral stresses $\sigma_1 = \sigma_2$ shown near the curves for Si I↔Si II PTs. Hollow and star symbols designate elastic instability points for forward and reversed PTs, respectively. Dashed line is the stress-strain curve for hydrostatic loading $\sigma_1 = \sigma_2 = \sigma_3$. Solid symbols mark elastic instability and zero-stress points for the uniaxial stressing with $\sigma_1 = \sigma_2 = 0$. (b) Elastic instability lines in $\sigma_3 - \sigma_1 = \sigma_2$ plane for direct (D) Si I→Si II and reverse (R) Si II→Si I PTs from DFT and MD with the Tersoff potential from [513, 545], and metallization curve from DFT. Hydrostatic condition is shown with a dashed diagonal line. (c) Criterion for Si I→Si II PT for triaxial normal stresses. Plane corresponds to a constant value of the modified transformation work in Eq. (22) and points are the DFT results. (d) Rotated figure in (c) until plane is visible as a line to visualize a good correspondence between PFA prediction and DFT results.

Here, $\bar{\boldsymbol{U}}_t(\eta) = \boldsymbol{I} + \bar{\boldsymbol{\varepsilon}}_t(\eta)$ and $\bar{\boldsymbol{\varepsilon}}_t(\eta)$ are the interpolation of the transformation deformation gradient and transformation strain for the PT process, respectively, $\bar{\boldsymbol{\varepsilon}}_t(0) = \boldsymbol{0}$, $\bar{\boldsymbol{\varepsilon}}_t(1) = \boldsymbol{\varepsilon}_t$, $\boldsymbol{\varepsilon}_t$ is the transformation strain for the complete PT, and $A$ is the height of the double-well barrier. For cubic-to-tetragonal transformation Si I↔Si II, considered in [444, 513, 519, 520] as an example, $\boldsymbol{\varepsilon}_t = \{\varepsilon_{t1}, \varepsilon_{t1}, \varepsilon_{t3}\}$, $\left.\frac{d^2\bar{\boldsymbol{U}}_t}{d\eta^2}\right|_{\eta=0} = 2\{b_1\varepsilon_{t1}, b_1\varepsilon_{t1}, b_3\varepsilon_{t3}\}$, where $b_i$ are the parameters in the interpolation function of $\bar{\boldsymbol{\varepsilon}}_t(\eta)$. In Eq. (21), the terms with the jump of elastic moduli are eliminated by the choice of the proper interpolation function; otherwise, they give strong nonlinearity not present in the results of the atomistic simulations.

Under all normal $\sigma_i$ and shear stresses $\tau_{ij}$ applied and rigid-body rotations eliminated by imposing a constraint $F_{12} = F_{13} = F_{23} = 0$, the PT criterion (21) (with sign =) can be presented in more explicit form:

$$W_{ins} = b_3\sigma_3\varepsilon_{t3} + b_1(\sigma_1 + \sigma_2)\varepsilon_{t1} + \frac{b_1\varepsilon_{t1} - b_3\varepsilon_{t3}}{F_{11}^e F_{22}^e}\left[\tau_{32}F_{32}^e F_{11}^e + \tau_{31}(F_{31}^e F_{22}^e - F_{32}^e F_{21}^e)\right] = A. \tag{22}$$

Here, $\sigma_3$ is in the tetragonal direction. Since $\sigma_3\varepsilon_{t3} + (\sigma_1+\sigma_2)\varepsilon_{t1}$ is the transformation work, $W_{ins}$ and the criterion (22) are called the modified transformation work and the modified transformation work criterion [513]. Since the shear stresses do not contribute to the transformation work for a cubic to tetragonal PT, the terms proportional to the $\tau_{ij}$ are due to geometric nonlinearity and do not include additional material coefficients. Shear stresses modify the crystal geometry, which affects transformation work of the normal stresses. For $\tau_{ij} = 0$, the PT criterion (22) is linear in stresses and should represent a plane in 3D $\sigma_i$ space. This was confirmed both with MD [513] and DFT [444], and results of both are very close (Fig. 25b-d). For the reverse, linear PT criterion is also valid, slopes for MD and DFT are close, but lines are significantly shifted (Fig. 25b). DFT predicts that Si II is metastable at zero stresses, i.e., does not have elastic and phonon instabilities [515]; however, Si was never recovered in experiments and transforms to Si III and Si XII or amorphous Si [81, 441, 442]. Only recently, loading path was found in dynamic [521] and RDAC [81] experiments with retained Si II.

A strong effect of the nonhydrostatic stresses is evident. While under hydrostatic conditions the PT pressure is 75.8 GPa, under uniaxial loading $\sigma_3 = -11.03$ GPa, i.e., the pressure is 3.68 GPa, which is 20.6 lower than under hydrostatic conditions. Strong effect of nonhydrostatic stresses explains large scatter in a PT pressure under quasi-hydrostatic conditions. The problem is that one cannot apply uniaxial 11 GPa because for real (defective) material $\sigma_y$ is much lower. Only for initially (almost) defect-free nanopillars uniaxial stresses exceeding 10 GPa can be applied for some directions [394, 522]. However, such strong nonhydrostatic stress states in which nonhydrostatic stresses are limited by the theoretical strength instead of macroscopic $\sigma_y$ (which is one-to-two orders of magnitude smaller) can be realized in a nanosized volume of polycrystalline materials at the tip of defects with strong stress concentration, like dislocation pileup with large numbers of dislocations (Section 6).

*Effect of shear stresses.* According to the criterion (22), shear stress $\tau_{21}$ does not affect it, which was confirmed by DFT PTs (Fig. 26a) within the relative error of 6%. Some reduction in the instability stress $\sigma_3$ with increasing $\tau_{31}$ is observed in Fig. 26b. The relative error of the instability stress $\sigma_3$ compared to the analytical prediction (22) for majority of combinations of $\tau_{31}$ and $(\sigma_1 + \sigma_2)/2$ is between $+4\%$ and $-6\%$. For neglected shear stresses in Eq. (22), the deviation of $\sigma_3$ from the linear expression vs. $\sigma_2$ and $\sigma_3$ is within 10% for $\tau_{31} < 8$GPa.

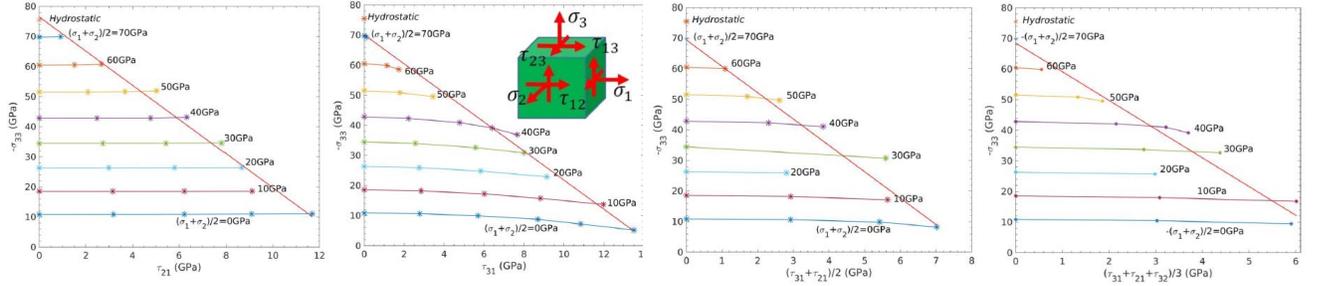

**Figure 26:** The effect of shear stresses on the tetragonal instability stress $\sigma_3$ for different $\sigma_1 = \sigma_2$ from DFT [444]. All symbols correspond to the maximum stress $\sigma_3$ in a $\sigma_3 - E_3$ curve. Symbols with the largest shear stresses roughly correspond to the shear instability. Straight inclined lines represent linear approximations of the relationship between shear stresses for shear instability and $\sigma_3$.

The combined loading with two and three shear stresses affect the instability stress $\sigma_3$ less than $\tau_{31}$ alone (Fig. 26c,d) (a) due to a lower averaged shear stress that causes shear instability and (b) a minor contribution of $\tau_{21}$. A relative error is within $\pm 4\%$. The $\tau_{31}$-strain curves for different combinations of the normal stresses, corresponding, below, and after tetragonal instability were calculated in [444]. Straight lines describing the shear instability (corresponding to the maximum at the shear stress-strain curves) are shown in Fig. 26b-d. These shear instability lines are not described by the criterion (22) but just limit the range of stresses for which PT Si I to Si II occurs. Under dominating shear stresses, e.g., in a shear bands, Si I transforms to Si IV and then amorphous Si or directly to amorphous Si [403], see Section 6.8.1.

Note that for applications of Eq. (22) one needs to know both stresses and elastic strains, i.e., the nonlinear elasticity rule, and include the elasticity rule (given in [523]) in Eq. (22), which makes it strongly nonlinear in elastic strains. Use of the linear version of Eq. (22) without shear stresses with an error of 10% is quite practical for many cases. It is amazing that the instability criterion (22) for the general stress tensor has just 2 material parameters.

### 5.3. $5^{th}$-degree elastic energy for Si-I

Nonlinear elastic properties of single crystals at finite strains determine material response to extreme stresses under dynamic (e.g., in shock waves) and static high pressures, in defect-free crystals and nanoregions, and near interfaces with significant lattice mismatch, where plastic stress relaxation is suppressed. They also determine elastic instabilities of the crystals. We discuss here some selected DFT results on the determination of the higher-order elastic energy. The third-order [524–526] and in few cases fourth-order elastic constants [527, 528] were determined for different crystals but at small strains (e.g., 0.02-0.03). Therefore, as written in [528], fourth-order elastic constants "should be treated as an estimation only," e.g., for diamond and Si. Indeed, third-and fourth-order elastic constants for diamond from [528] were found to be inconsistent with the equation of state of diamond [143], and have been modified to fit FEM modeling in [143] to the experiment in [132].

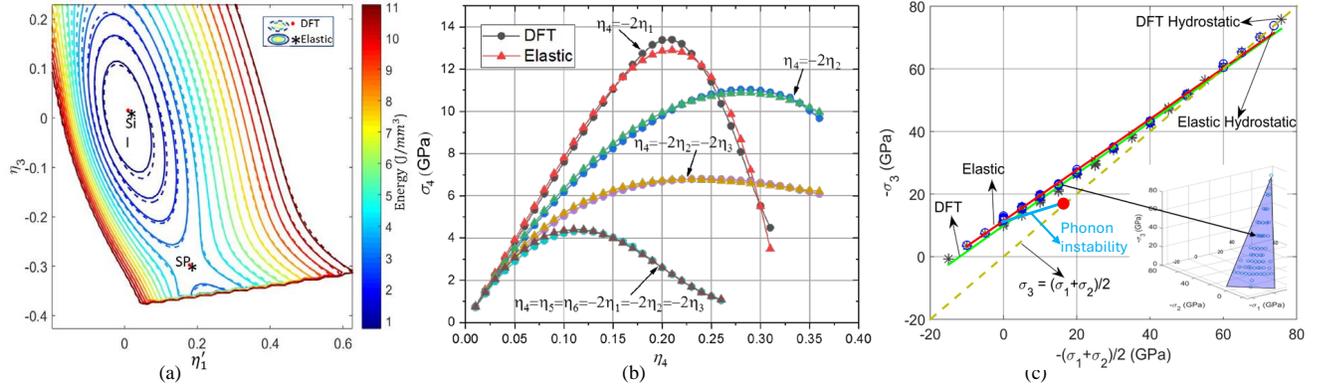

**Figure 27:** (a) A good correspondence between contour lines for the fifth-degree elastic energy and DFT simulations for Si I in $\eta_1' = \eta_2' - \eta_3$ plane ($\eta_1' = \eta_2'$ are rotated by $45^0$ around axis 3 coordinate system, as in DFT unit cell). (b) Comparison between true shear stress $\sigma_4$–strain $\eta_4$ curves from fifth-degree elastic potential and DFT simulations for various combinations of normal and shear strains; all non-mentioned strains are zero. (c) Comparison of the elastic instability condition $\sigma_3$ versus $0.5(\sigma_1 + \sigma_2)$ for Si I $\to$ Si II PT obtained from the analytical elastic energy (circles) and DFT simulations (*). Red point added here corresponds to the phonon instability from [537]. Blue line connects this point and elastic instability point for $\sigma_1 = \sigma_2 = 0$, for which phonons are stable. The inset shows the same data from the elastic energy in $3D$ space $\sigma_i$, which are also very close to the instability plane from the DFT simulations in [444]. Reproduced from [523].

Since lattice instability may occur at finite strains (e.g., at 0.2 for Si [444] (Figs. 26a and 27a,b) or 0.3-0.4 for $B_4C$ [407, 408]) and extrapolation to large strain may contain essential error, even such higher-order elastic energy cannot describe it. Stress-strain curves at finite strains are obtained for some specific loadings in [79, 407, 408, 444, 507, 516–518], which is, however, insufficient for completing an expression for elastic energy for the full strain tensor. Consequently, a higher-order elastic constant must be determined for a finite strain that includes the lattice instability. Since a big data was generated in [444] for Si under all six components of the deformation gradient at strains exceeding instability strains, fifth-degree elastic energy in terms of Lagrangian strain was approximated by minimizing error relatively to DFT results in [523]; the additional required DFT results have been included. Thus, 3 second-, 6 third-, 11 fourth-, and 18 fifth-order constants were found. A very good correspondence between elastic energy prediction and DFT results was obtained in terms of energy, stress-strain curves at various complex loadings, and criterion for Si I$\to$Si II PT under 3 normal stresses (Fig. 27). Obtained analytical expression can improve continuum modeling of Si behavior under extreme static and dynamic loadings, especially for PFA [520, 529–532], for which second-order elasticity was used.

Note that experimental data for the second-order elastic constants for Si I from 0 K up to the melting point [533, 534] and results of reactive MD simulations [534] show weak temperature dependence. For all elastic constants up to fifth order, it can be calculated with ab initio MD. Recent developments, including determination of the fourth-order elastic constant for Si I and nonlinear elasticity of amorphous Si, can be found in [535, 536].

As it was mentioned above, the instability stress in [444, 513, 523] was determined as the maximum stress at the stress-strain curves for the direct and minimum for the reverse PTs (Fig. 25a), which is not strict. While it was supported for many cases by the impossibility to equilibrate higher (lower) stresses in MD simulations, the Tersoff potential used in MD does not reproduce DFT results well for the reverse PT; shear stresses did not exceed 3 GPa (i.e., much lower than in Fig. 26 from DFT) after which material was disordered. Application of a general elastic instability criterion $det\boldsymbol{B} = 0$ should give more precise results, especially under shears, which lower crystal symmetry and give more conditions from these equations. Elastic potential derived in [523] allows to do this. This may reveal not only instabilities leading to Si II, but also to other phases to which Si I may directly transform (e.g., Si III and XI [64, 81]), fracture, slip, and twinning. While analytical expression for $\boldsymbol{B}$ and, especially, $det\boldsymbol{B}$ is extremely nonlinear, bulky, and may require independent approximations, a solution is definitely possible. The problem is how to visualize in 6D space and represent it in the way useful for further applications. Fourth-order elasticity for diamond, along with instability criterion $det\boldsymbol{B} = 0$, was utilized in FEM simulations and analysis of the effect of diamond anvil geometry and friction condition on anvil performance in [217] (Section 2.4).

*Primitive account for phonon instability.* The most serious problem is that phonon instability was not taken into account in DFT simulations [444], which could overestimate the PT stresses. Indeed, as was reported in [537] in the supplement, the minimum phonon instability pressure under hydrostatic loading is 18.3 GPa, much lower than the elastic instability pressure of 75.8 GPa. Since there is no phonon instability in Si I for uniaxial loading (Zarkevich and Levitas, unpublished), we added a red point for the phonon instability in Fig. 27c and connected the instability points for the uniaxial and hydrostatic loadings by a straight line. Thus, the instability condition can be approximated as

$$\sigma_3 = -11.45 + 0.4066(\sigma_1 + \sigma_2) \quad \text{for } \sigma_1 + \sigma_2 \geq 0 \text{ (elastic instability)}$$
$$\sigma_3 = -11.45 + 0.1877(\sigma_1 + \sigma_2) \quad \text{for } -36 \text{GPa} \leq \sigma_1 + \sigma_2 < 0 \text{ (phonon instability)}. \tag{23}$$

Similar adjustment can be made in Eq. (22).

*Metallization.* The electronic structure in Si I was studied under different combinations of normal stresses in [444]. For each $\sigma_1 = \sigma_2$, the band gap reaches its maximum for some $E_3$, corresponding to stresses close to $\sigma_1 = \sigma_2 = \sigma_3 \approx -10$GPa. Some deviations from this strain in both directions reduce the band gap to zero, resulting in the metallization curve presented in Fig. 26b. The effect of non-hydrostatic stresses on the metallization is very strong. The metallization precedes the Si I to Si II PT for all stresses, i.e., a sufficiently deformed Si I under stress is metallic. This transition is not visible in the stress-strain curves (Fig. 26a). The metallization line is closed in the $(\sigma_3, \sigma_1)$ plane. Note that for simple shear, phonon instability occurs at a shear strain of 0.22 prior to the elastic instability, leading to the second-order PT [538].

Recently, a DFT study of phonon instability in the 6D strain (rather than stress) tensor space was performed for diamond and Si [539–541]. A small strain tensor was probably utilized, which of course must be substituted with the finite strain measure, like Lagrangian strain. Since it was not guided by PFA or other theoretical predictions, analytical description was lacking, and various sections of the instability surface were presented. In [540], conditions for metallization/demetallization and reversible indirect-to-direct bandgap transitions in diamond within the region of it phonon stability were found. Machine learning was utilized to fit the strain states against the values of the bandgap determined with DFT. This model was used in FEM simulations of the distribution of bandgap in the bent diamond nanoneedles. Conditions for diamond-graphite PT (which was observed experimentally in [542, 543]) due to phonon instability were determined. Since tensile stains up to 9% can be reached experimentally without fracture in the bent nanoneedles [544], many strain states of interest can be reached. Similar DFT and machine learning approaches to Si and diamond have been realized for the elastic strain engineering of bandgap [539] and thermal conductivity of diamond [541].

*Homogenous hysteresis-free $1^{st}$-order PTs with stable intermediate phases.* MD results for direct and reverse PTs are close, and, starting with tensile $\sigma_1 = 10.2$ GPa, coincide. This leads to very interesting phenomena, observed with MD [545] and corresponding PFA simulations [520, 529]. Coincidence of the instability conditions for direct and reverse PTs result in unique homogeneous and hysteresis-free $1^{st}$-order PTs, for which each intermediate phase (crystal lattice) along the PT path is in neutral thermodynamic equilibrium. It can be arrested and researched for any fixed strain in one direction. There is a continuum (an infinite number) of such intermediate phases without an energy barrier between them. Usually, after instability, HPP nucleates, producing a two-phase mixture separated by interfaces, which motion causes an energy dissipation. Here, no interfaces and energy dissi-pation appear. Even a regular two-phase system, while approaching these stress states, continuously transforms to uniform intermediate phases by interface broadening and remains uniform in a broad range of stresses. This also allows for the stabilization of unique heterogeneous intermediate structures, which may possess unexpected properties. Indeed, PFAs in [546–550] showed that the ratio of the widths of two different phase interfaces, surfaces, or dislocation cores strongly affects nano- and macroscale PT properties for various PTs and their interaction with dislocations (see also review [551]). Zero hysteresis and homogeneous transformations are the optimal property for various PT applications (e.g., for shape memory alloys for actuation or medical applications [552–554] or caloric materials [554–556]), which strongly reduce damage and energy dissipation. Multiaxial stresses may control all the above, producing new properties and phenomena. Further increase in $\sigma_1$ ends with the unusual $2^{nd}$-order PT with a large jump in elastic moduli, without change in symmetry, and intracell atomic displacements.

Unfortunately, further DFT calculations [444] did not confirm the coincidence of the conditions for the direct and reverse PTs (Fig. 25b). Still, since predicted properties and phenomena are very interesting and of significant applied interest, such materials could be potentially designed by proper multicomponent alloying.

*Why do results for elastic instability in MD and DFT and phonon instability in PFA coincide?* The irony of the above results in [444, 513] is that the *phonon* instability criterion (22) obtained with PFA describes the *elastic* instability from MD and DFT for direct and reverse Si I↔Si II PTs. The question "Why do results for elastic instability in MD and phonon instability in PFA coincide?" was formulated in [513] and attempted to be partially answered in [557]. Thus, the simplest finite-strain model with quadratic elastic energy in terms of Lagrangian strain results in a strongly nonlinear equation for the true stress $\boldsymbol{\sigma}$. This leads to a qualitative reproducing of all of the features in [513] (and most of features in [444]) connected to PTs and lattice instability under 3 normal stresses $\sigma_i$. Namely, (a) criteria for elastic lattice instability for direct and reverse PT between Si I and Si II are linear functions of the true stresses $\sigma_i$; (b) these criteria depend on $\sigma_2 + \sigma_1$ instead of on $\sigma_1$ and $\sigma_2$ independently; (c) the effect of the shear stress $\tau_{13} < 3$ GPa is negligible; (d) the instability stresses $\sigma_3$ for direct and reverse PTs coincide above some tensile stress $\sigma_2 = \sigma_1$; (e) a jump in strain $E_3$ from Si I to Si II reduces with increasing tensile stress $\sigma_2 = \sigma_1$ and tends to zero, which leads to change from the first-order PT to the second-order PT. Thus, surprisingly, all the above phenomena are consequences of the finite elastic strains and do not require specific physical mechanisms. Then, the coincidence of the elastic and phonon instability for loading by normal true stresses $\sigma_i$ can be explained by linearity in stresses in both of them. Indeed, MD and DFT results show the linearity of PT criteria in stresses $\sigma_i$. Because of this, in the PFA, the nonlinear terms related to the jump in elastic moduli were eliminated, and two material parameters in the linear criterion were calibrated utilizing MD (or DFT) results.

*5.4. Stress-measure dependence of phase transformation criterion*

The elastic instability criterion $det\boldsymbol{B} = 0$ and its applications in [500, 506, 507, 511, 512] make the impression that the problem is solved unambiguously. However, this is not the case and is a consequence that the formulation is presented in terms of true stress $\boldsymbol{\sigma}$ by default. However, other stress measures are often used for different purposes, like Kirchhoff stress $\boldsymbol{\tau} = J\boldsymbol{T}$, the first Piola–Kirchhoff stress $\boldsymbol{P}$ (force per unit undeformed area), and the second Piola–Kirchhoff stress $\boldsymbol{T}$ (stress tensor work-conjugate (i.e., producing work) to the Lagrangian $\boldsymbol{E}$). A more general elastic instability criteria [498, 558] are expressed in terms of arbitrary measures of stress and work-conjugate strain; they clearly depend on the chosen (prescribed) stress/strain measure [498, 505]. For a uniform solution, when stability is evaluated under certain prescribed stress measure, this measure determines the choice of the unique criterion. However, for *heterogeneous* solutions of a boundary-value problem, stress tensor can only be given at the boundary of a sample, stress measure for each material point is *undefined*, and elastic instability cannot be uniquely determined. Alternative approach based on loss of positive definiteness of acoustic tensor [559] (which is a finite-strain generalization of Hadamard's work [560]) also depends on the selection of strain and stress measures.

At the same time, it was shown analytically in [519, 520] that within the PFA, the PT criterion describing the phonon instability (Section 5.2) is independent of the prescribed stress measure. Paper [561] sheds additional light on this issue. PFA model from [520, 529] calibrated by MD results from [513] was used. Fig. 28a shows equilibrium stress-strain curves for different stresses for a certain uniform deformation of Si I. For all four stress measures, PT instability described by criterion (22) starts at the same strain, visualising an analytical result in [519, 520] that the PT instability is independent of the prescribed stress measure.

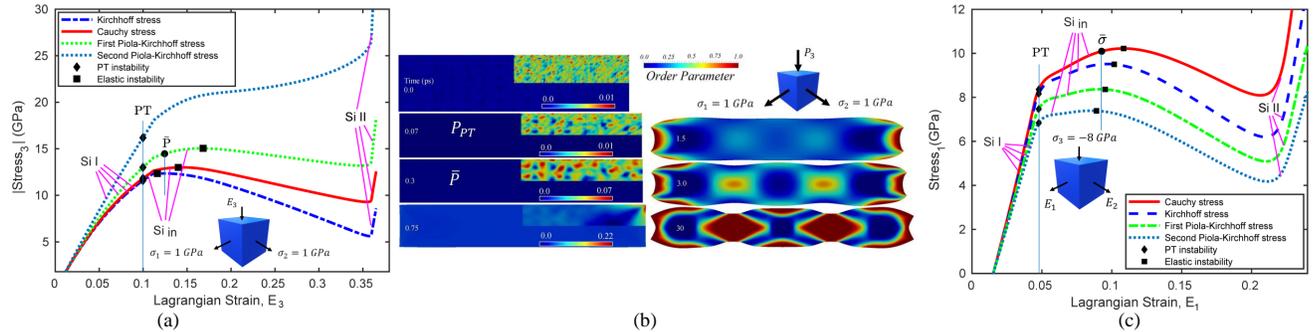

**Figure 28:** (a) Equilibrium stress-strain curves for uniform deformation of Si I with prescribed $E_3$ at $\sigma_1 = \sigma_2 = 1$ GPa. Hierarchy of phonon and elastic instability points are designated by symbols. Intermediate phase $Si_{in}$ between points of phonon (PT) and elastic instability appears via third-order PT. (b) Nanostructure evolution for loading by fixed first Piola-Kirchhoff stresses $\bar{P}$ (shown in (a)) at the top face with the magnitude between the peaks for the Kirchhoff and the Cauchy stresses, along with $\sigma_1 = \sigma_2 = 1$ GPa. (c) Equilibrium stress-strain curves for uniform deformation under fixed $\sigma_3 = -8$ GPa and growing tensile strains $E_1 = E_2$.

In a small strain formulation, the equilibrium stress-strain curve has a reducing branch until complete PT, and initiation of the PT leads at constant stress to its completion through a non-equilibrium process. However, for finite strain in Fig. 28a, after PT instability, each stress measure keeps increasing until maximum (except $T$, which increases monotonously). In this region, order parameter $\eta$ reaches equilibrium value $0 < \eta < 1$ for each $E_3$, i.e., describes some intermediate phase $Si_{in}$ between Si I and Si II. It is shown that fulfilment of criterion (22) leads to the continuous $3^{rd}$-order PT. For strain-controlled uniform deformation and perturbations, stable uniform behavior is observed till end of PT and further loading of Si II. For any prescribed stress measure (instead of $E_3$), a non-equilibrium $1^{st}$-order PT $Si_{in} \to$ Si II occurs after this stress reaches maximum due to elastic instability.

The equilibrium $3^{rd}$-order PT between the phonon and elastic instability stresses is completely reversible without hysteresis and dissipation for any prescribed stress; for prescribed $T$ it occurs till completing (Fig. 28a). These features are very important for shape memory alloys [552–554] and caloric materials [554–556]; see also [545].

For the loading of the sample shown in Fig. 28a by fixed $P$ with the magnitude between the peaks for $\tau$ and $\sigma$ stresses with initial random heterogeneous field $\eta$ in the range $(0; 0.01]$, heterogeneous nanostructure evolves and ends with the complex stationary two-phase configuration (Fig. 28b). This is unexpected, because the prescribed $P$ is below the elastic instability stress and for homogeneous fluctuations, the system remains in equilibrium. However, since strain exceeds the elastic instability for the Kirchhoff stress, heterogeneous fluctuations cause the initiation and significant progress of the PT, probably due to locally exceeding elastic instability in some region.

For the loading shown in Fig. 28c with a biaxial tensile component, a completely different sequence of the stress-strain curves takes place, with the lowest for $T$ and highest for $\sigma$ stress. Loading with the Cauchy stress above the strain, exceeding the elastic instability for the lowest curve for $T$, along with heterogenous fluctuations, was sufficient to initiate PT and complete it in the major part of the sample. Thus, for *tensile* loadings (including determination of the theoretical strength in tension), the most popular results based on the *Cauchy stress* *overestimate* the reality.

While these examples make clear that for any loading, the first-order PT is determined by stress measure producing the lowest stress-strain curve (which is different for different loading modes), still problem remains. In general [498, 505], there is an infinite number of work-conjugate measures of stress and strain couples, and choice of the measures that give the lowest stress is not a trivial task. On the positive side, in simulations, such a choice is performed automatically. Also, differences in instability strains between the most popular stresses (like here) and some exotic stresses may not be essential.

*Finite lattice rotations.* Another major problem is taking into account finite lattice rotations in the determination of instability. As it follows from Eq. (20), the elastic instability criterion $det\boldsymbol{B} = 0$ neglects rotations. In [558], under fixed load, rotations were varied arbitrarily, which led to strong constrains on material stability. It was shown in [519] that under prescribed $\boldsymbol{P}$, rotation tensor is defined unambiguously, i.e., it cannot vary independently. Theory in [519, 520] takes into account rotations, but was confirmed in atomistic simulations [444, 513] for diagonal (shear-free) transformation strain only, for which rotations play a secondary role. The author's team's preliminary experience with shear-containing transformation strain shows that lattice rotations are still a big trouble.

*5.5. Summary and perspectives*

Lattice instability for perfect crystals under a general stress tensor plays a key role in determining of the upper bound of how deviatoric stresses can reduce the PT pressure. Due to lack of defects, shear stresses are limited by theoretical strength and are one to two orders of magnitude larger than $\sigma_y$ for real defective materials. Such conditions can be realized in the nanosized samples, like nanowires, nanopillars, nanoparticles, nanofilms, or in some grains of nanograined materials. In real defective materials, they can be realized in the nanosized defect-free regions near strong stress concentrators, like tip of dislocation pileups, twins, disclinations, etc. Indeed, an averaged distance between two dislocations is $\Delta l = 1/\sqrt{\rho_d}$, where $\rho_d$ is the dislocation density. For severely pre-deformed material, $\rho_d \simeq 10^{15}/m^2$ (Fig. 12), then $\Delta l = 32$ nm, which gives an approximate size of dislocation-free region. For comparison, for annealed metals, $\rho_d \simeq 10^{10}/m^2$ and $\Delta l = 10\,\mu m$, but this is less relevant for our topic. Nucleation of the HPP phase in such nanoregions requires fulfillment of corresponding lattice instability conditions, which determines their importance for our studies. Due to a large number of combinations, determination of the general lattice instability condition is a very computationally expensive problem, with very few general results for few materials. Utilizing PFA, which gives, under certain assumptions, an analytical expression for the lattice instability condition due to phonon instability [444, 513], could be a significant help. This help is not altruistic, because it leads in turn to the development and specification of the PFA [444, 513, 520]. Machine learning is another way to reduce computational costs [539–541]. For elastic instability, one of the simplest ways is to find an analytical expression for the higher-order elastic energy at finite strain, based on atomistic simulations and supplemented/verified by experiments, and then use condition $det\boldsymbol{B} = 0$ with the Wallace tensor (Eq. (20)) to determine all possible instabilities. This approach was not yet implemented for any material. As we will see at the end of Section 6.1, phonon instability should play a more important part in the promotion of barrierless nucleation at the tip of dislocation pileup than the elastic instability. Thus, this is a very exciting and promising field of research, with various unsolved formal and physical problems.

# 6. Nanoscale mechanisms of strain-induced phase transformations

*6.1. Nucleation at the tip of the dislocation pileup: analytical treatment*

To explain the reduction in PT/CR pressure by a factor of 10-100 (Section 3.1), a defect with the strongest stress concentration is required. This is the dislocation pileup, which was used to model temperature-induced martensitic PT [488], slip transfer through grain boundary [562], and deformation twinning [417]. The model in [100, 102] considered nucleation of HPP at the tip of the strain-induced dislocation pileup [100, 102] and initiated works on nanoscale mechanisms of strain-induced PTs/CRs, which explained some of the phenomena described in Section 3 and identified controlling parameters. This does not exclude that other defects (grain and subgrain boundaries, disclinations, twins, and stacking faults) are sufficient for some systems with the smaller reduction in PT pressure. Still, some points from the approach outlined below are applicable.

An analytical solution on the nucleation of an HPP at the tip of the edge dislocation pileup of length $l$ against an obstacle (e.g., grain boundary) under action of applied normal $\sigma_1$ and $\sigma_2$ and shear $\tau$ stresses was outlined in [563] and with more details in [100, 102] (Fig. 29a). Plane stress linear elastic isotropic problem formulation with small strains is assumed. A nucleus represents a pill-box with sizes $2L \times 2c \times b$, inclined under angle $\phi$ to the pileup. The stress field of the superdislocation was used [562]

$$\sigma = \frac{l\tau}{2}\frac{\sin\phi}{r}; \qquad \tau_l = \frac{l\tau}{2}\frac{\cos\phi}{r}; \qquad l\tau = \frac{\mu N |\boldsymbol{b}|}{\pi(1-\nu)}, \tag{24}$$

where $N$ is the number of dislocations in a pileup, $r$ is the distance from the tip, $\tau_l$ and $\sigma$ are the shear and normal stresses along sides $L$ and $c$, $\boldsymbol{b}$ is the Burgers vector, and $\nu$ is the Poisson's ratio. Since stresses are proportional to $\mu N/r$, they may exceed the lattice stability limit $\sim 0.1\mu$ in the finite nm-size region.

Stresses in the nucleus consist of the sum of external stresses, internal stresses due to dislocation pileup, and transformation strains determined using the Eshelby solution for ellipsoidal inclusion. Transformation strain $\boldsymbol{\varepsilon}_t$ in nucleus is an invariant-plane strain with shears of $0.5\gamma_t$ along the sides (due to symmetry of $\boldsymbol{\varepsilon}_t$), and normal strain $\varepsilon_n = \varepsilon_{tv}$ (because of absence of other normal transformation strains) along side $c$. After calculations, the following expression for the thermodynamic driving force $F$ for the appearance of the nucleus was obtained

$$\begin{aligned}F &= AL^2 + BL; \qquad B = l\tau bn \ln\frac{2}{n}\sqrt{(\varepsilon^2+\gamma_t^2)} - 4\Gamma b; \quad \tan\phi = \frac{\varepsilon}{\gamma_t}; \quad s = \frac{\mu\pi}{8(1-\nu)}; \\ A &= (\sigma_2\varepsilon_{tv} + \tau\gamma_t)4bn - (\Delta\psi^\theta + K)4bn - 4bn^2 s(\varepsilon^2+\gamma_t^2) - 8\Gamma n; \quad n = c/L \ll 1\end{aligned} \tag{25}$$

with $\varepsilon = |\varepsilon_{tv}|$. Angle $\phi$ was determined from the maximization of the $F$. Since the applied shear stress $\tau$ causes dislocation motion in the pileup plane and direction, it is equal to the critical resolved shear stress for dislocation motion, and can be approximated by the macroscopic yield strength in shear $\tau_y$.

*6.2. Barrierless Nucleation*

If $B > 0$, which is the case for large $N$ and strong enough stress concentration and, consequently, $\tau l$, one has $F > 0$ for relatively small $L$, i.e., nucleation is barrierless. The condition $F = 0$ results in the thermodynamic equilibrium length $L_e = -B/A$ of the nucleus, which for $\sigma_1 = \sigma_2 = -p$, can be resolved for the minimum pressure for direct strain-induced PT,

$$p_\varepsilon^d = \frac{\Gamma}{\varepsilon}\left(\frac{2}{b} + \frac{1}{nL}\right) + \frac{\Delta\psi^\theta + K + ns(\varepsilon^2+\gamma_t^2)}{\varepsilon} - \tau\left(\frac{\gamma_t}{\varepsilon} + \frac{1}{4}\sqrt{1+\left(\frac{\gamma_t}{\varepsilon}\right)^2}\frac{l}{L}\ln\frac{2}{n}\right). \tag{26}$$

The reduction in $p_\varepsilon^d$ due to shear stresses is especially large for PT with large $\gamma_t/\varepsilon$. The best example of is martensitic PT, e.g., in $NiTi$ shape memory alloy with $\gamma_t = 0.13$ and $\varepsilon = 0.0034$ [564], $\gamma_t/\varepsilon = 38$. However, for such materials, due to small transformation work $p\varepsilon$ of pressure, there is not much sense to apply pressure at all.

Assuming $n = 0.1$, the criterion for barrierless nucleation, $B > 0$, can be expressed as

$$l\tau > 13.35\frac{\Gamma}{\sqrt{(\varepsilon^2+\gamma_t^2)}}; \qquad N > 13.35\frac{\Gamma(\pi(1-\nu))}{\mu|\boldsymbol{b}|\sqrt{(\varepsilon^2+\gamma_t^2)}}. \tag{27}$$

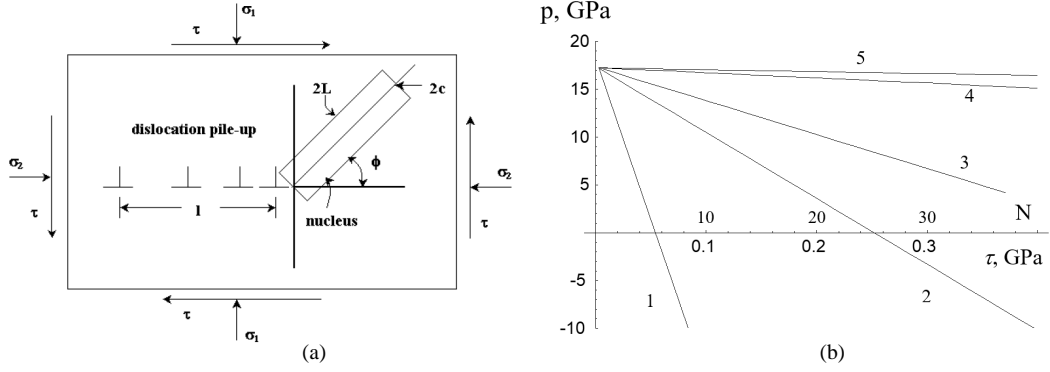

**Figure 29:** (a) The schematic of the appearance of pill-box shaped nucleus at the tip of the dislocation pileup against an obstacle [102]. (b) The PT pressure versus shear stress (or number of dislocations $N$) for the barrierless nucleation for the pileup length $l = 1\,\mu m$ and different nucleus length $2L$: $1 - 10$ nm; $2 - 50$ nm; $3 - 100$ nm; $4 - 1000$ nm; $5 - 2L = \infty$. Adapted from [102] with some corrections of parameters in the caption.

For the chosen typical values $\Gamma = 0.1$ N/m, $|\boldsymbol{b}| = 3 \cdot 10^{-10}$ m, $\varepsilon = 0.1$, $\gamma_t = 0.2$, and $\nu = 0.3$, Eq. (27) simplifies to $l\tau > 5.97 Nm^{-1}$ and $N > 43.76/(\mu\,(GPa))$. These conditions can be met straightforwardly for reasonable values of $\tau = 0.2$ GPa, $l = 30$ nm, and $N = 1$ for $\mu \leq 43.76$ GPa; i. e., even a single dislocation can lead to barrierless nucleation for some materials. E.g., for iron $\mu = 82$ GPa and for $\alpha-$Zr $\mu = 36$ GPa, only one or two dislocations are sufficient for barrierless nucleation. However, many more dislocations are required to produce a nucleus of a reasonable size of at least several tens of nanometers for nanograined materials and up to micrometers for large-grained materials.

Let us accept $\Delta\psi^\theta = 1$ GPa, $K = 0.5$ GPa, and $\varepsilon = 0.1$. This results in $p_{eq} = \Delta\psi^\theta/\varepsilon = 10$ GPa and pressures for direct-reverse PT (when the effect of $\Gamma$, shear stresses, and internal stresses is neglected), $p = (\Delta\psi^\theta \pm K)/\varepsilon$, of 15 and 5 GPa. If in addition to the above material parameters, we accept $\gamma_t = 0.2$, $\mu = 80$ GPa, and $b = 10^{-6}$ m, then Eqs. (26) simplify to

$$p_\varepsilon^d(GPa) = 17.24 + 10/L(nm) - (2 + 1.675 l/L(nm))\,\tau(GPa); \qquad (28)$$

$$p_\varepsilon^d(GPa) = 17.24 + 10/L(nm) - 21.83 N/l(nm) - 18.28 N/L(nm). \qquad (29)$$

In Eq. (29), $\tau$ is expressed in terms of $N$ using Eq. (24), i.e., $\tau(GPa) = 10.91 N/l(nm)$ or, for $l = 1\,\mu m$, $\tau(GPa) = 0.0109 N$. Plots of the Eqs. (28) and (29) for several $2L$ vs. $N$ and for $l = 1\,\mu m$ vs. $\tau$ are shown in Fig. 29b. Analysis in terms of $N$ is simpler since it is independent of $l$. Thus, 25 dislocations in a pileup can reduce pressure to produce 50 nm nucleus down to zero; 6 dislocations can do the same to produce 10 nm nucleus. Both are realistic numbers for various materials, e.g., for $\alpha - \omega$ PT in Zr [82], Si I to Si II and Si III PT [81], and PTs from graphite to cubic and hexagonal diamonds [79], for which PTs occur at sub-GPa and pressure was reduced by 5 - 70 GPa. The obtained results also show that the strain-induced PT below the $p_e = 10$ $GPa$ and even the reverse PT pressure under hydrostatic loading of 5 GPa *can be explained by the nucleation at the dislocation pileup*. Note the PFA simulations in [380] also gives the linear relation between $p_\varepsilon^d$ and $N$ with the slope of 2.3 GPa/dislocation (Section 6.6.2 and Fig. 29b), which corresponds $L = 8$ nm in Eq. (29), which is reasonable. It is clear from Eqs. (26) and (29) that increase in nucleus length $L$ requires increase in pressure, $N$, and $\tau l$. Note that while we consider plastic strain-induced PTs, nanoscale model does not include plastic strain and treats them as stress-induced PTs, but in the presence of the strong stress concentrator at the tip of the plastic strain-induced dislocation pileup.

It was mentioned in [100, 102] that the barrierless nucleation does not need a waiting time for proper thermal fluctuations, which explains the plastic strain-controlled instead of time-dependent kinetics. Indeed, after very fast barrierless nucleation, growth to the thermodynamically equilibrium length $L_{eq}$ should occur fast. If the observation time is larger than the growth time, this PT looks instantaneous. Without deformation, new pileups and nuclei do not emerge, and growth of the nuclei is thermodynamically prohibited. Until recent experiments [82], such an explanation was considered satisfactory. However, in situ study of $\alpha - \omega$ PT in Zr under compression in DAC with rough diamonds revealed essential increase of the volume fraction of $\omega$-Zr by up to 35% within 1 hour at the fixed sample thickness. Some rationale for time-dependence was given in [82] (Section 7.5.3.

Since edge dislocations produce not only shear but also normal stress concentration, Eq. (26) shows the reduction in PT pressure even for $\gamma_t = 0$, i.e. when $\tau$ does not contribute to the transformation work and does not affect stress-induced PTs. Isostructural, fcc-fcc electronic PTs in $Ce$ and its alloys (e.g., CeP) give a chance to check this theoretical prediction. It was obtained in [64] that under hydrostatic loading of Ce at room temperature, pressures for direct $p_h^d = 0.80$ GPa and reverse PT $p_h^r = 0.57$ GPa, giving a hysteresis $H_p = 0.23$ GPa. Under shear, pressures for strain-induced direct $p_\varepsilon^d = 0.675$ GPa and reverse PT $p_\varepsilon^r = 0.65$ GPa, and hysteresis $H_\varepsilon = 0.025$ GPa. Thus, these results are consistent with the prediction of the model.

As mentioned in [478], since in Eq. (24) $\sigma_{ij} \sim \mu(p)$, an essential difference is expected between the lattice instability at the tip due to elastic or phonon instability. For elastic shear instability, $\mu \to 0$, leading to minor stress concentration. *Since the phonon instability is not connected with the reduction in $\mu$, the drop in PT pressure can be essentially greater.* This difference may partially explain a broad range of the pressure reduction due to plastic straining for different materials and loadings. This prediction is to be checked experimentally, e.g., for the stishovite-CaCl$_2$-type PT in $SiO_2$, which is related to vanishing tetragonal shear modulus [438]. Note that the type of instability depends on the stress state, as we discussed for Si-I→Si-II PT in Eq. (23).

*6.3. Grain-size dependence of the transformation pressure for strain-induced PTs*

Similar analysis was performed in [100, 102] in terms of $\tau l$ instead of $N$. While $\tau = \tau_y$, length of dislocation pileup $l$ depends significantly on microstructure and is usually assumed to be a fraction (from quarter to half) of the grain size. Since $\tau_y$ and $l$ are limited, one expects a minimum possible pressure, $p_\varepsilon^d$, below which strain-induced PT cannot occur. An increase in $\tau_y$ by increasing $q$ and $\dot{q}$ implies a reduction in $p_\varepsilon^d$, which is indeed observed in experiments [17, 80, 83, 331, 332, 414].

Another straightforward conclusion in [100, 102] was that the larger $l$ is, the stronger stress concentration and reduction in PT pressure are. Since grain size significantly reduces during SPD, one suggested way to intensify PT was related to the increase in grain size by annealing and recrystallization after compression of the sample at pressures below $p_\varepsilon^d$. In particular, reduction in the grain size of HMX organic explosive from 300 to 4 $\mu m$ increased CR and detonation start pressure in shock by 2.5 and 1.7 times, respectively [565]. Similarly, reduction in the grain size of RDX organic explosive from 135 to 6 $\mu m$ increased detonation start pressure in shock from 8.5 to 12.0 GPa. Analysis in [565] was related to dislocation motion, but it is not clear whether initiation of reaction and detonation can be related to strain-induced, stress-induced, or combined CR.

While many predictions of our analytical model were later qualitatively confirmed by the nanoscale [380, 381, 479] and scale-free [462, 463] PFA simulations, MD [566] and CAC [567] simulations, and above mentioned experiments [64, 79, 81, 82], for micron and submicron grain sizes, the implication that the increase in the grain size reduces $p_\varepsilon^d$, contradicted recent experiments. Thus, for $\alpha - \omega$ PT in ultra-pure Zr, $p_\varepsilon^d = 2.3$ GPa for annealed sample and $p_\varepsilon^d = 1.2$ GPa for strongly plastically predeformed by multiple rolling, i.e., nanograined, sample [80]. For $\alpha - \omega$ PT in commercially pure Zr, $p_\varepsilon^d$ reduced from 1.36 to 0.67 GPa; when measured with XRD, crystallite size reduced from 65 to 48 nm (and dislocation density increased from $1.26 \times 10^{15}$ to $1.83 \times 10^{15} m^{-2}$) [82, 89, 425].

This contradiction was resolved in [82, 425]. It follows from nanoscale PFA [380, 381, 479], MD [566], and CAC [567, 569] simulations, that, for large shear stresses, almost all dislocations are located at the grain boundary or other obstacle, making a step or superdislocation (Fig. 30). Consequently, length of the dislocation pileup $l \simeq N|\boldsymbol{b}| \ll d$ is not limited by the grain size $d$. This eliminates the suppressive effect of the reducing grain size on the strain-induced PTs. On the other side, shear stress $\tau = \tau_y = \tau_\infty + kd^{-0.5}$ increases with the grain size reduction according to the Hall-Petch relationship [263], where $\tau_\infty$ and $k$ are parameters. Thus, reduction in the grain size increases stress concentration in Eq. (24) and reduces PT pressure in Eq. (26), which explains experiments for Zr.

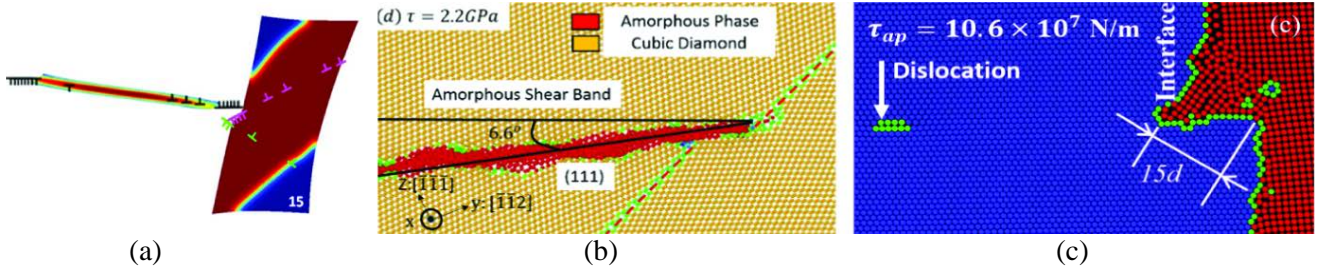

**Figure 30:** (a) Dislocation pileup in the left grain produces a superdislocation (step) at the grain boundary that induces a cubic-tetragonal PT and dislocation nucleation and glide in the right grain. The nanoscale PFA simulations from [380] for a model with $p_\varepsilon^d = 0$ GPa, $p_h^d = 15.9$ GPa, $p_e = 3$ GPa and applied $\gamma = 0.35$. For the applied pressure of 1.59 GPa, $c = 0.77$. (b) MD simulations from [566] show that dislocation pileup in the right grain produces a superdislocation at the $\Sigma 19$ grain boundary in Si-I and amorphous band. (c) Step at the phase interface consisting of 15 dislocations induces square-hexagonal PT in the CAC approach from [567]. The HPPs are red here and in other figures below.

A more focused and broader approach was realized for strain-induced PTs in Si [81]. It included an analytical treatment that for very small grains, the inverse Hall-Petch relation [263] is applicable, and $\tau_y$ reduces with the reducing grain size. Also, grain boundary sliding competes with the generation of the dislocation pileups, reducing $N$ in Eq. (29). Thus, in addition to the decrease in $p_\varepsilon^d$ with with decreasing grain size in the region of direct Hall-Petch relation, an increase in $p_\varepsilon^d$ with with reducing grain size in the region of the inverse Hall-Petch effect was predicted. This prediction was confirmed in situ experimentally in DAC and RDAC for PT Si-I→Si-II for 1 $\mu m$, 100 nm, and 30 nm particles (Fig. 31a), thus leading to the main rule in the field – correlation between the direct and inverse Hall-Petch effect of particle size on $\sigma_y$ and $p_\varepsilon^d$. The result is very nontrivial because for pressure-induced Si-I to Si-II PT, PT pressure grows with reduction in the particle size, and for 30 nm particles, Si-I does not transform to Si-II at all, transforming directly to Si-XI (Fig. 61). Also, this result strongly supports dislocation pileup mechanism of strain-induced PT and suggests methods to control $p_\varepsilon^d$ by controlling grain size. In addition, since grain size at the initiation of PT in most of the cases was not measured, this could lead to significant scatter in PT pressures reported by different groups.

Note that the PFA solution on coupled PT and dislocation evolution in a model bicrystal under compression and shear shows the promotion of PT with increasing grain size from 25 and 50 nm, in agreement with the experiments in Fig. 31a, see [381] and Section 6.6.1. The linear reduction in PT pressure in RDAC with increasing plastic strain, corresponding increasing dislocation density and microstrain, and decreasing crystallite size was in situ found for PT olivine-spinel [83]. Some ex-situ [432–436] and in-situ studies [568] of the grain size after PTs in various oxides produces by HPT will be analysed in Section 3.11. Other notable HPT studies include [570, 571] and review [572]. With increasing temperature, the promoting effect of the plastic straining on PTs should be reduced based on the suggested mechanism, which is observed in experiments (Section 3.8). The reasons include increase in the grain size, reduction in $\sigma_y$, increased absorbtion of dislocations by the grain boundaries, recrystallization leading to annihilation of dislocations, and activation of diffusion-controlled creep mechanisms competing with dislocation pileup formation. Above recrystallization temperature, formation of dislocation pileups and the strain-induced reduction of the PT pressure are not expected. This is consistent with the results in [573] on the fcc-hcp PT in lead, which show minor difference between minimum PT pressure of 12.8 GPa in RDAC and hydrostatic compression. Currently, the dislocation pileup-based mechanism is the only one used in the literature for interpretation of the reduction in PT pressure by plastic straining.

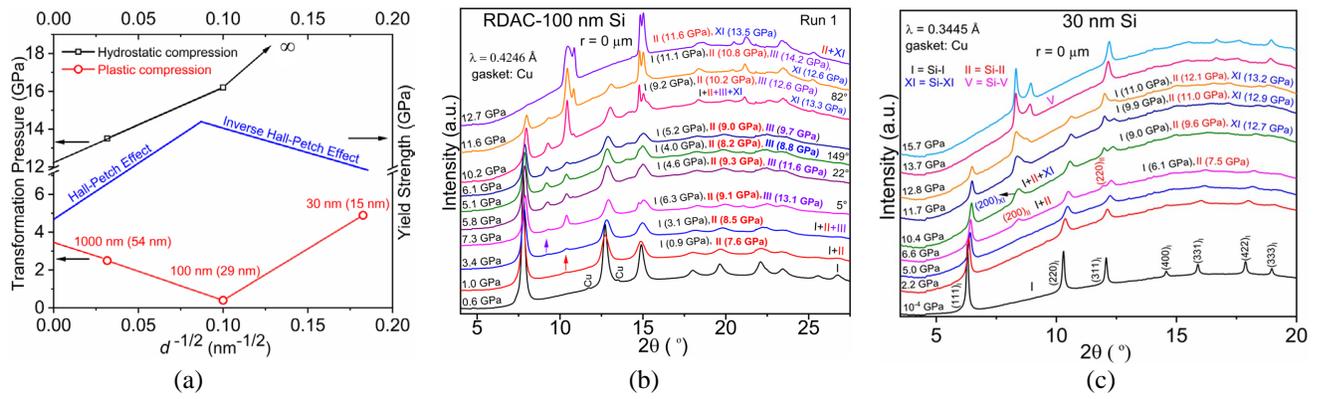

**Figure 31:** (a) Experimental results on particle-size dependence of the minimum pressure for Si-I→Si-II PT under hydrostatic compression and plastic deformation, and a sketch of particle-size dependence of $\sigma_y$ for the direct and inverse Hall-Petch relationships. The correlation between the particle size's direct and inverse Hall-Petch effects on $\sigma_y$ and $p_\varepsilon^d$ is obtained in agreement with the dislocation pileup-based nucleation mechanism. (b) and (c) XRD patterns for compression of 100 nm and 30 nm Si, respectively. Numbers on the left side are pressures averaged over the phase mixture. Numbers in parentheses near the Si phases show pressure within these phases. Bold symbols are used for large pressure values in Si-II and Si-III in comparison with Si-I. Reproduced from [81].

Note that dislocation pileup produces compressive (for $\phi > 0$ and side with extra atomic planes) and tensile (for $\phi < 0$ and side with missing atomic planes) pressures of the same magnitude, simultaneously promoting both direct PT in the LPP and reverse in the HPP. This is included in the microscale model in [100, 102, 378] and Section 7.

### 6.4. In situ experimental proof of the strong pressure concentration in strain-induced nuclei

While there are various experimental results mentioned above consistent with the dislocation pileup-based mechanism for the initiation of the strain-induced PTs, there is a major problem. Although there are numerous TEM images of dislocation pileups in materials with micron and larger-grain sizes, dislocation pileups have never been observed for nanograined materials. This is not surprising for us, because the energy of a superdislocation is proportional to $(N|\boldsymbol{b}|)^2$ and is very high, and it is equilibrated by large external shear stresses. After removing shear stresses, large repulsive forces lead to a significant increase in dislocation spacing. For nanograined materials, dislocations may reach the opposite side of a grain boundary and be absorbed, which we observe in simulations [380, 381]. CAC simulations [567, 569] show that even during loading, dislocation pileup may split into two, reducing its energy; however, this happened after a HPP has already nucleated. That is why there are low chances for postmortem observation of dislocation pileups with large $N$. We would like to mention that some relatively novel precession electron diffraction techniques show in nanograined Ti a large dislocation density with significant localizations [574]; however, dislocation pileups could not be separated. We are also not aware of any methods to detect dislocation pileups during in situ experiments with DAC or RDAC, because XRD beam for axial diffraction averages all elastic strains over the sample thickness, and there is no way to focus on dislocation pileups.

We suggested an idea that based on our analytical [100, 102] and PFA [380, 381, 479] results, which show strong stress concentration in the small regions of the HPP immediately after their appearance near the tip of the dislocation pileups. Since XRD allows us to measure lattice parameters in each phase separately, we can expect that as soon as XRD signal of good enough quality can be collected (which is possible at $c \simeq 0.03-0.05$), it should show significantly higher pressure in the HPP than in the LPP. However, this effect was not confirmed for $\alpha - \omega$ PT in Zr [80], olivine-spinel PT [83], and PT from hexagonal to wurtzitic BN (unpublished yet data), probably due to effective stress relaxation immediately after nucleation.

The first success was achieved for PTs in Si [81]. During plastic straining of 100 nm Si, pressure in small Si-II regions was from 5 to 7 GPa higher than in the Si-I (Fig. 31b), despite the fact that for pressure-induced PT, it was by 2.7 GPa lower due to $\varepsilon_{tv} = -0.22$. Thus, relative increase in pressure was as high as 7.7 to 9.7 GPa! Such a strong stress concentration could definitely be caused by dislocation pileup, and we do not know of any other defects that can cause it. This is consistent with the result that for 30 nm Si particles, due to lower dislocation role in the region of the inverse Hall-Petch relation, the pressure difference is 1.4 GPa only (Fig. 31c). With increasing volume fraction $c$ of Si-II for 100 nm Si, the difference in pressure between Si-I and Si-II reduces because stresses are averaged over the larger volume of Si II, where stresses reduce away from the tip of the dislocation pileups. A similar increase in pressure from Si-I to Si-II was observed in micron Si particles and from Si-I to Si-III in 100 nm Si (Fig. 31b). After an upgrade of synchrotron radiation facilities at APS, good enough XRD patterns could be collected at a much smaller volume fraction of HPPs, which will lead to detection of even stronger stress concentration. Obtained in [81] results brings an understanding of the mechanisms of strain-induced PT at the new level.

Note that for strong materials like Si, *some high-pressure PTs can be studied under uniaxial compression of nanopillars with in situ TEM* [394, 522]. In [394], single crystal, $<111>$-oriented Si I nanopillar was subjected to uniaxial compression with stresses reaching 18 GPa, and PT to Si-IV in a shear band, nucleation and accumulation of dislocations leading to amorphization in the shear band was studied with in situ TEM. Similar experiments for a polycrystalline nanopillar with the grain size between 30 and 100 nm may lead to *detection of the dislocation pileup and corresponding nucleation mechanism or revealing new mechanisms for strain-induced PTs.*

*6.5. Thermally activated nucleation*

For $B < 0$ in Eq.(27), the PT can occur at $A > 0$ only. Because $F < 0$ for $L < -B/A$, thermal fluctuations are needed for nucleation. Theory for thermally activated nucleation at the dislocation pileup developed in [100, 102] led to the following conclusions in (Fig. 32). For the chosen model material parameters, the thermally activated regime occurs at low shear stress ($\tau \leq 0.03 GPa$) and 1-2 dislocations. While the effect of $\tau$ and $N$ is much stronger for a thermally activated nucleation than for the barrierless one, since $\tau$ and $N$ are limited by the condition $B \leq 0$, the minimum PT pressure reduced by $\sim$3 GPa at 1,000 K and $\sim$8.5 GPa at 300 K.

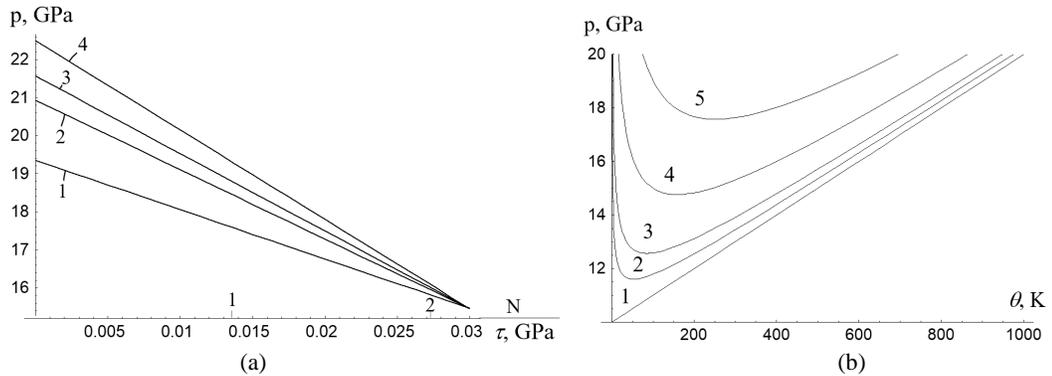

**Figure 32:** PT pressure versus shear stress (number of dislocations) (a) and temperature (designated here as $\theta$) (b) for thermally activated nucleation at dislocation pileup [102]. (a) Results for the pileup length $l = 20$ nm and different temperatures: 1 – 1000 K; 2 – 500 K; 3 – 400 K; 4 – 300 K. (b) Results for different $\tau_i l = \tau l$, expressed in $GPa$ nm: 1 – the equilibrium PT line; 2 – $\tau_i l = 5.6$; 3 – $\tau_i l = 5.2$; 4 – $\tau_i l = 4.0$; 5 – $\tau_i l = 2.0$.

Since $N$ is small, the effect of pre-existing dislocations was also approximately included in the model in terms of corresponding internal shear stress $\tau_i$. Results in Fig. 32b demonstrate that increase in $\tau l = \tau_i l$ reduces overshooting of the PT start line with respect to the phase equilibrium line. These features correspond to experimental results in [8, 9] for the PTs $PbO_2 I \leftrightarrow PbO_2 II$, $MnF_2 I \leftrightarrow MnF_2 II$, quartz $\leftrightarrow$ coesite PT in $SiO_2$, litharge $\leftrightarrow$ massicot PT in $PbO$, calcite $\leftrightarrow$ aragonite PT in $CaCO_3$, as well as reaction 2 jadeite $\leftrightarrow$ nepheline + albite (Section 3.4). One of the conditions for such a correspondence, not only for direct but also for the reverse PT, is that $K$ and the energy of the internal stresses due to $\varepsilon_t$ are negligible. Also, internal and external shear stresses should be small enough not to affect the $p-T$ phase equilibrium line. For such materials, experiments on the strain-induced PT can indeed determine the $p-T$ phase equilibrium line by reducing kinetic hysteresis. The opposite statement for the case with the barrierless nucleation will be analysed in Section 7.

Note that Gilman's mechanism [335–337], that requires large elastic shears, could be realized near the tip of a dislocation pileup. Similar consideration for both barrierless and thermally activated nucleation can be performed for the pileup of screw dislocations. However, they are less potent [562] because (a) they do not generate normal stresses and (b) the shear stress is reduced by a factor of $1/(1-\nu)$. That is why they may lead more often to the thermally activated nucleation than the edge dislocations.

### 6.6. Nanoscale phase-field approach simulations for a bicrystal

Despite the conceptual success of the analytical solution in explaining various phenomena, the problem formulation and solution itself were clearly oversimplified. Indeed, using linear elasticity, infinite stresses at the tip of a pileup, small strain formulation, single nucleus in the infinite space (i.e., just a nucleation), and lack of stress relaxation due to dislocations or twins near the tip of the dislocation pileup questioned whether solution and conclusions are reliable. To resolve most of these issues and bring the study to a qualitatively higher level, the first PFA for the coupled dislocation evolution and PT was developed [426, 427, 472, 479], which synergistically combined the most developed large-strain theories for martensitic PTs [519, 575–577] and dislocations [547, 578, 579] at that time, including their nontrivial interactions. It was implemented in a FEM COMSOL Multiphysics code. Numerous plane-strain problems on coupled evolution of straight edge dislocations leading to the pileups and the HPP in a bicrystal under compressive stress and shear strain (Fig. 33a) have been solved in different formulations [380, 381, 479]. The slip systems and dislocations are inherited during the PT. The resistance to dislocation propagation is due to elastic interaction between dislocations and other internal stresses only; there is no athermal resistance due to Peierls barrier. For all problems, initially dislocation-free samples have been considered.

#### 6.6.1. Solution for periodic boundary conditions

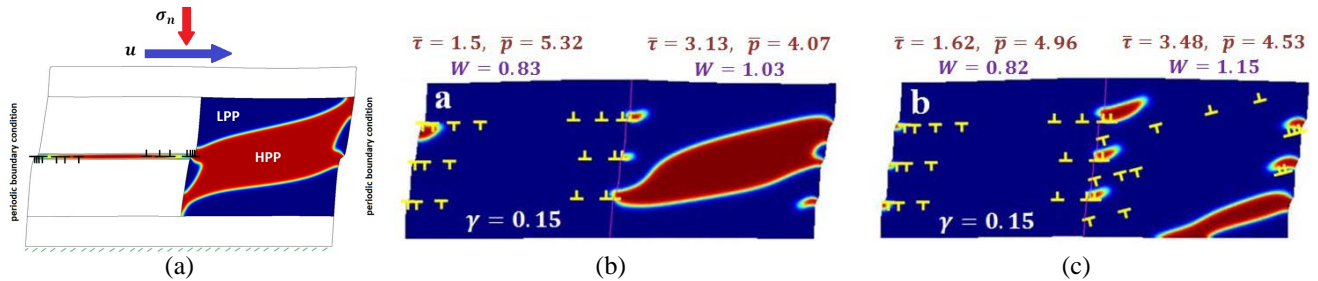

**Figure 33:** (a) A bicrystal under normal stress $\sigma_n = 3.05$ GPa (leading to an initial pressure averaged over the sample $\bar{p} = 2.0$ GPa) and shear strain with periodic boundary conditions on the lateral sides and two rectangles above and below the sample, for which the elastic problem without PFA was solved only to mimic the effect of the surrounding grains. Dislocations are in the left grain, and PT occurs in the right grain only. A stationary nanostructure for $\bar{p}_s = 0.81$ GPa and applied averaged shear $\bar{\gamma} = 0.2$ [381]. (b) Stationary configuration of HPP and dislocations under compression and $\gamma = 0.15$ [479]. Three slip systems are included in the left grain only, and dislocations are pinned before PT starts, i.e., not evolved during the PT. (c) The same as in (b), but with three slip systems in the right grain inclined under $15^0$, which lead to strong suppression of HPP [479]. Numbers on the top of the grains in (b) and (c) show shear stress $\bar{\tau}$, pressure $\bar{p}$, and transformation work $W$ averaged over each grain, all in GPa.

The following material properties are used. Shear modulus $\mu = 71.5$ GPa and bulk modulus $B = 112.6$ GPa are the same for both phases. Transformation strains are $\varepsilon_{tx} = \varepsilon_{ty} = -0.05$ and $\varepsilon_{txy} = 0.1$ (i.e., $\varepsilon = 0.1$, $\gamma_t = 0.2$); the phase interface energy and width are $0.34 \, J/m^2$ and 1.4 nm, respectively. Phase equilibrium pressure $p_e = 10$ GPa, instability pressure for the LPP $p_{in}^d = 20$ GPa and the HPP $p_{in}^r = -10$ GPa. For the hydrostatic loading and with a single dislocation present, the nucleus appears at $p_h^d = 15.75$ GPa. For slip, the critical resolved shear stress for the lattice instability and nucleation of dislocation in LPP $\tau_A^c = 0.5$ GPa and HPP $\tau_M^c = 1.5$ GPa. Since multi-well energy for dislocations is proportional $\tau^c$, it increases by a factor 3 during the PT, which produces a thermodynamic driving force for dislocation motion from the HPP to the LPP.

*Dislocation pileup-induced PT in the elastic nanograin.* For the problem in Fig. 33a, shearing leads to consecutive nucleation of dislocations of opposite signs from both grain boundaries and moving them toward each other. For $\gamma = 0.2$, 7 stationary dislocations appear, 3 of which create steps and 4 create pileup at both grain boundaries. After this, PT evolution starts in the right grain, see Fig 34a. Strong stress concentration near the tips of pileups cause the appearance of two HPP nuclei that initially grow independent of each other. Their geometry is close to that assumed in the analytical model in [100, 102] (Fig. 29a). Due to the small distance between HPP regions, they interact and coalesce, leading to stationary structure with $c_s = 0.51$.

Before PT, the averaged shear stresses are $\bar{\tau} = 5.86$ GPa and 8.77 GPa in the left and right grains, respectively, with 7.31 GPa averaged over the sample. These stresses are much higher than $\tau_A^c = 0.5 GPa$ due to back stresses from the grain boundaries and elastic right grain. They produce more than an order of magnitude larger transformation work than expected from the macroscopic critical resolved shear stresses, which is important not only for nucleation, but also for stabilization of the HPP interface, see below. During PT: (a) $\bar{\tau}$ is decreased to 4.75 and 5 GPa in the left and right grains, respectively, and 4.9 GPa averaged over the bicrystal; (b) $\bar{p}$ is reduced from 2 to 1.69 GPa in the left grain, and 0.06 GPa in the right grain (due to the compressive $\varepsilon_v$), with 0.81 GPa in the bicrystal. Note that in experiments, the averaged pressure PT pressure is measured and reported after PT increment. Consequently, the PT pressure dropped from 15.75 GPa under hydrostatic loading to 0.81 GPa under relatively small shear, i.e., by almost 20 times, with a significant $c_s = 0.51$. Analysis of the local stress evolution in [381] shows some surprising results. The zero-pressure contour lines surround the HPP during the entire process and practically coincide with the upper and lower interfaces for the stationary solution. Shear stress at the corresponding contour lines surrounding the nuclei reduces from $> 9$ to $< 5$ GPa for the stationary solution.

*Competition between PT and slip.* When for the same problem, dislocation evolution in the transforming grain is included, the PT is significantly suppressed (Fig. 34b and c), and $c_s = 0.166$. Evolution of phase and dislocation structures strongly depend on the relative mobility of the dislocations and phase interfaces, expressed in terms of the ratio $L_\xi/L_\eta$ of the kinetic coefficient for dislocation evolution $L_\xi$ and PT $L_\eta$. However, *stationary dislocation and, especially, phase structures are very similar*, with the same $c_s = 0.166$ for $L_\xi/L_\eta$ in the range of $0.1 - 80$. This, however, could be a consequence of the absence of other mechanisms of their pinning, e.g., by point defects, stacking faults, etc. Relatively low dislocation mobility, $L_\xi/L_\eta = 0.1$, PT initially proceeds almost like without plasticity (Fig. 34b), because just one dislocation appears near the tip of a pileup. With increasing number of dislocations, the reverse PT starts. Dislocations relax shear stresses and (slightly) relax pressure; the HPP region gets smaller and then splits in two. Two dislocation pileups are formed in the transforming grain, and compressive stresses due to extra atomic planes in these pileups stabilize both HPP regions. The transformed regions are confined by dislocations, and dislocation pileups coincide with the phase interfaces, making them semicoherent. Semicoherency dislocations produce an athermal friction, which leads to interface arrest, which was quantified with PFA simulations in [472, 547].

For $L_\xi/L_\eta = 4$ (Fig. 34c), stresses relax and suppress the PT from the beginning. Nucleation at the tip of dislocation pileup at the right grain boundary occurs practically like for elastic grain because there are no slip systems. At the left side of the grain, tip of the pileup generates dislocations instead of the HPP. Only after dislocations in the right grain piled up near the left grain boundary and generated high compressive stresses, a small nucleus of the HPP appeared, but did not grow. Consequently, dislocation pileups in the transforming grains provide an *additional mechanism of nucleation and (mostly) stabilization of a HPP* by compressive stresses.

Consequently, *plasticity provides two competitive effects: it generates dislocation pileups with strong concentration of all stresses promoting nucleation and growth of a HPP, but it also suppresses the PT by relaxing the stress concentrators.* Resultant phase and dislocation structures depend on crystal lattices of the phases, grain orientations, slip systems, applied stresses, etc., and can be controlled (see below).

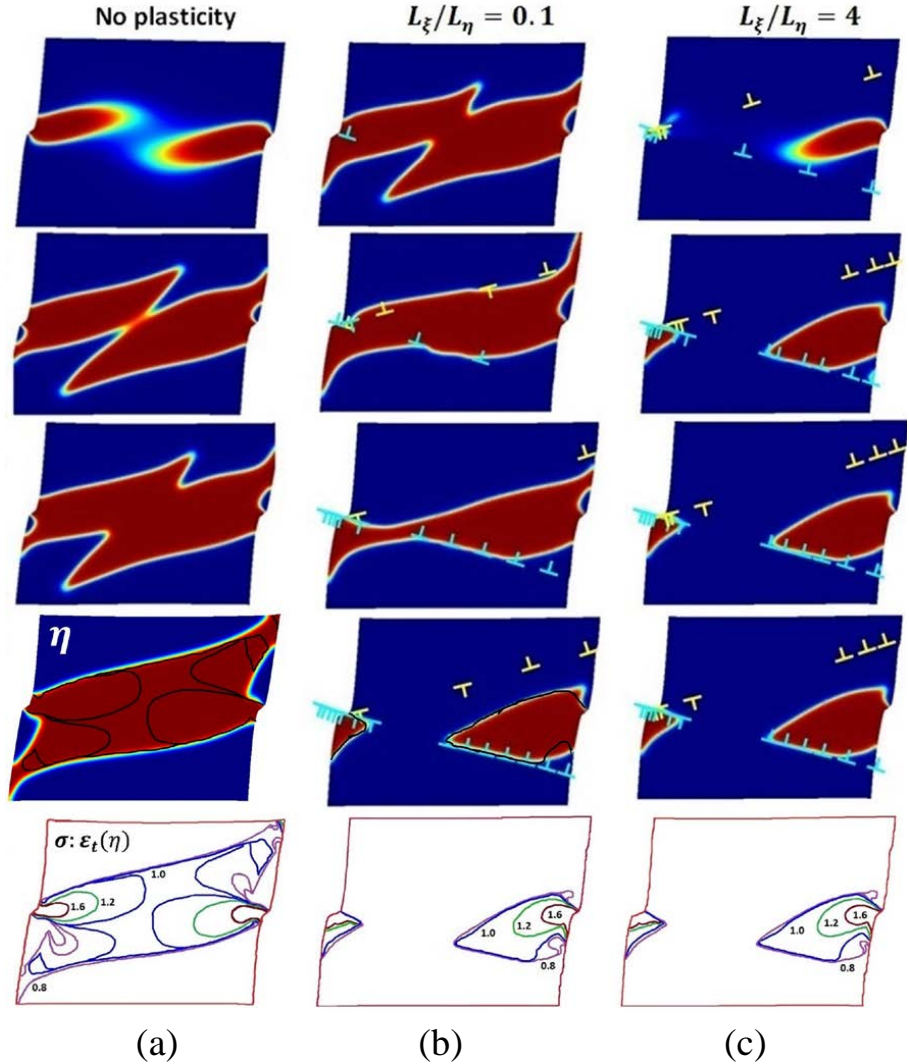

**Figure 34:** Nucleation and evolution of the $HPP$ (a) and $HPP$ and dislocations (b and c) in the right grain until stationary nanostructure at $\gamma = 0.2$ [381]. Stationary state for (a) is also shown in Fig. 33a. Ratio of the kinetic coefficient for dislocation evolution $L_\xi$ to that for the PT $L_\eta$ is 0.1 in (b) and 4 in (c). Averaged over both grains pressure $\bar{p}$ is 2 GPa before PT and $\bar{p}_s = 1.36$ GPa (b) and $\bar{p}_s = 0.9$ (c) in the transformed grain. The lowest row is the contour lines for the transformation work $\boldsymbol{\sigma}:\boldsymbol{\varepsilon}_t(\eta)$. The black contour lines of the equilibrium transformation work $\boldsymbol{\sigma}:\boldsymbol{\varepsilon}_t(\eta) = 1$ are also plotted on the HPP stationary states in (a) and (b). They coincide with the phase interfaces almost everywhere.

Qualitatively similar comparative results were obtained in [479] with three horizontal glide systems in the left grain and three inclined under $15^0$ glide systems in the right grain, without (Fig. 33b) and with dislocations in the right grain (Fig. 33c). For both cases, for $\gamma = 0.15$, a tiny amount of HPP was obtained in the left grain, because (a) glide systems are parallel to the applied shear, relaxing shear stress, and (b) due to small vertical distance between slip systems ($\sim 4$ nm) within which pressure from dislocations varies from strongly positive to strongly negative. In the right grain $c_s = 0.347$ without slip and just 0.126 with slip, despite higher $\bar{p}_s$ and $\bar{\tau}_s$ with slip (Fig. 33b and c). However, a situation is far from being hopeless. Increased $\gamma = 0.2$, while not promoting PT in the left grain, leads to $c_s = 0.55$ for the elastic right grain and $\bar{c}_s = 0.45$ with plasticity (Fig. 35 [381]). Thus, suppression of PT due to plasticity is not significant. Fields of pressure and shear stress are also shown in Fig. 35. The upper and lower phase interfaces for the elastic right grain are close to the constant $p = 4$ and $\tau = 3$ GPa lines. With slip in both grains, the upper phase interface is arrested by the dislocations but the lower boundary approximately follows $p = 4$ GPa line. An even better result for the right grain is given in [381] for $\gamma = 0.15$: without and with plasticity, $c_s = 0.439$, $\bar{p}_s = 5.66$ GPa, and $\bar{\tau}_s = 2.46$ GPa. The reason is that just two dislocations of opposite signs appear in each inclined slip system; however, they disappear while approaching the stationary solution. Thus, for such loading parameters, *PT wins competition over dislocations*, and many other similar optimal loadings are possible for which *shear promotes PT instead of plasticity*.

*Grain size effect.* When the grain sizes were increased from $25 \times 20 nm^2$ to $50 \times 40 nm^2$, but with the same 3 slip systems in each grain, $c_s$ in the right grain increased from 0.439 to 0.485 with slightly lower $\bar{p}_s$ and $\bar{\tau}_s$, and in the left grain from 0.02 to 0.06, with significantly larger $\bar{\tau}_s$ and slightly lower $\bar{p}_s$. The reason is in the larger spacing between slip systems, and more dislocation-free space. Note that for PT Si-I to Si-II, an increase in the crystallite size in this range (in the region of the inverse Hall-Petch effect) also leads to the promotion of the PT, see Fig. 31a [81]. However, many other reasons not included in the current model (like grain boundary sliding) may contribute to this experimental effect.

*Reversibility of PT.* During complete unloading, the complete reverse PT occurs, dislocations disappear, and geometry returns to its initial state, i.e., all processes are entirely reversible. Even if dislocation evolution equations are not solved, i.e., they are pinned before unloading begins, the complete reversal to the LPP still occurs. Thus, existing dislocations are not sufficient to provide a proper athermal resistance at such large deviation from $p_e = 10$ GPa. The problem with simulating the metastability of the HPPs is to be solved. An athermal resistance to the interface motion due to point defects, stacking faults, and other defects should be included in the model [580, 581].

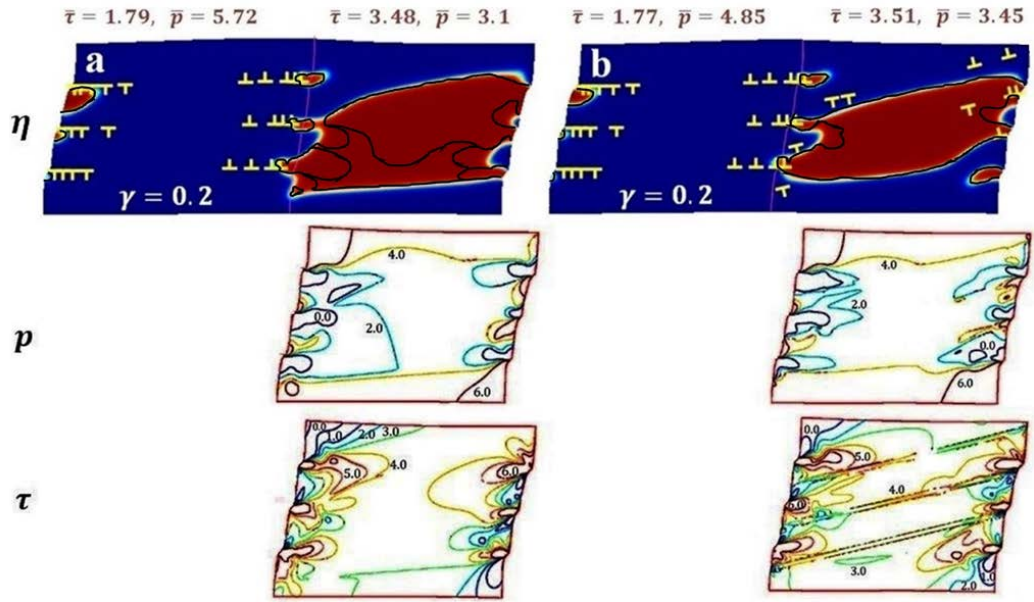

**Figure 35:** Stationary distributions of the HPP, dislocations, and pressure and shear stress contour lines for the same sample, problem formulation, and slip systems like in Fig. 33b and c but for $\gamma = 0.2$ [381]. (a) For elastic right grain. (b) With plasticity in the right grain. The black contour lines of the thermodynamically equilibrium transformation work $\boldsymbol{\sigma}{:}\boldsymbol{\varepsilon}_t(\eta) = 1$ are shown on the HPP regions. They coincide with the phase interfaces almost everywhere.

### 6.6.2. Solutions for equal prescribed normal stresses and superposed shear [380]

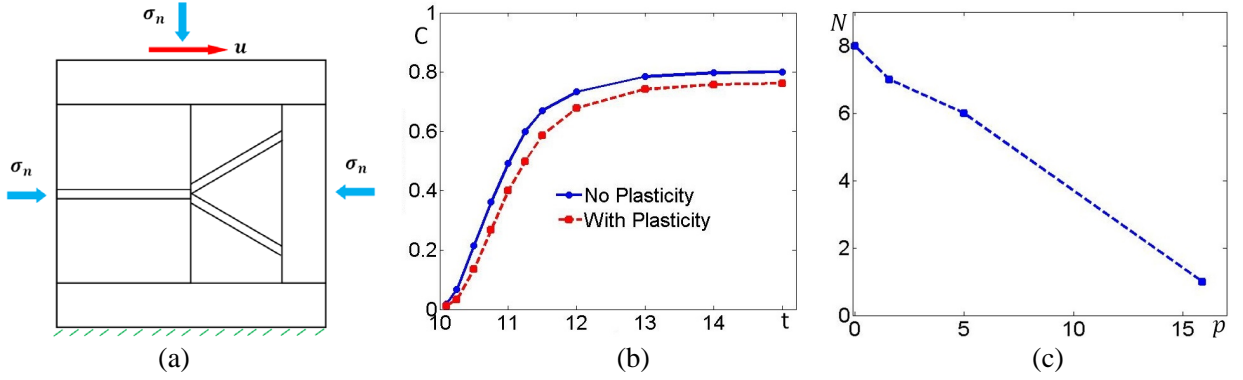

**Figure 36:** (a) Schematics of the bicrystal with 3 elastic layers under equal normal stresses and shear displacement. (b) Volume fraction of the HPP vs. time for $p = 1.59$ and $\gamma = 0.35$. (c) The minimum number of dislocations in a pileup to nucleate HPP as a function of the applied pressure. Reproduced from [380].

Geometry of the problem is shown in Fig. 36a. The material parameters are: $\mu = 71.5$ GPa and $B = 112.6$ GPa for both phases; $\varepsilon_{tx} = \varepsilon_{ty} = -0.1$; $\varepsilon_{txy} = 0.15$ (i.e., $\varepsilon = 0.2$, $\gamma = 0.3$); interface energy is $1.12 J/m^2$ and width is 1.0 nm; $\tau_A^c = 1.2$ GPa and $\tau_M^c = 3.6$ GPa; $p_e = 3$ GPa, and $p_{cl} = 17.6$ GPa; $p_h = 15.9$ GPa. Three external pressures, 0, 1.59 ($= 0.1 p_h$), and 5 GPa ($= 0.32 p_h$) were considered. For $p = 0$ and $\gamma = 0.3$, with plasticity in the right grain, PT did not start when 4 dislocations produced a pileup, but initiated with 8 dislocations. This result motivated our DAC and RDAC experiments at sub-GPa pressure for PTs from graphite to diamond [79], $\alpha - \omega$ PT in Zr [82], and Si-I to Si-II and Si-III PT [81], for which PT pressure was reduced by 5 - 70 GPa.

For $p = 1.59$ GPa and $\gamma = 0.35$, the evolution of dislocations and the HPP is shown in Fig. 37. Corresponding evolution of $c$ is given in Fig. 36b; evolution of $\bar{p}$ and $\bar{\tau}$ versus time and $c$ is shown in Fig. 38. In contrast to the periodic boundary conditions, plasticity in the right grain reduces $c_s$ from 0.8 to 0.75 only, i.e., its effect is weak. Due to dislocation pileup, $\bar{p} \simeq 4.3$ GPa at the beginning of the PT and, due to volume reduction, it drops to $\bar{p}_s \simeq -1.3$ GPa; $\bar{\tau}$ reduces from $\sim 7.3$ GPa to $\sim 3$ GPa. Still, HPP remains metastable. Both $\bar{p}$ and $\bar{\tau}$ reduce linearly with increasing $c$. Surprisingly, for $p = 5$ GPa and $\gamma = 0.3$, $c_s = 0.8$ with and without plasticity, i.e., is practically the same as for $p = 1.59$ GPa. The initial and stationary pressures are higher by $\sim 2.5$ GPa than for $p = 1.59$ GPa and $\bar{p}_s \sim 1$ GPa; $\bar{\tau}_s$ is lower by $\sim 1$-$1.3$ GPa.

The minimum number of dislocations in the pileup to initiate the PT versus the applied pressure is shown in Fig. 36c This curve can be approximated by $p_\varepsilon^d (GPa) = 18.2 - 2.3N$. Despite the significant difference in formulations, the slope corresponds to that in an analytical solution Eq. (29) for $L = 8$ nm, which is reasonable.

Results for different $L_\eta/L_\xi$ are quite different than for the periodic boundary conditions. Results above were for $L_\eta/L_\xi = 0.26$; for $L_\eta/L_\xi = 0.052$, HPP does not nucleate even at $p = 5$ GPa and $\gamma = 0.3$ because only dislocations are produced. Because of lack of nucleation, a surprising result is obtained that the *stationary solution depends on $L_\eta/L_\xi$, while this parameter even is not included in the stationary Ginzburg-Landau equations*. Similarly, for large enough $L_\eta/L_\xi$ when dislocations in the right grain do not nucleate, another stationary solution appears with a much larger HPP region (like in Fig. 34). This jump from one steady solution to another may give a hint on how to *rationalize jump from one steady microstructure to another for different SPD processing or classes of $\boldsymbol{\varepsilon}_p$, $\boldsymbol{\varepsilon}_p^{path}$,*

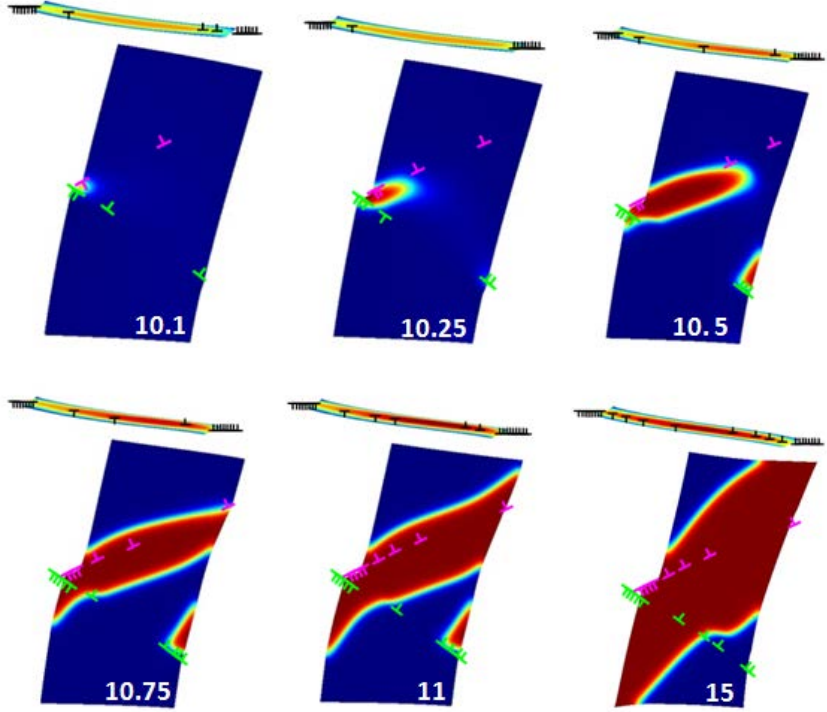

**Figure 37:** Evolution of the HPP and dislocations for $p = 1.59$ GPa and $\gamma = 0.35$ until the stationary solution. Dislocation evolution in the left grain is presented at the top [380].

$p$, and $p^{path}$ (Sections 2.1 and 3.10). For $L_\eta/L_\xi = 1$, HPP and dislocations appear and evolve simultaneously, and the steady solution does not essentially differ from that for $L_\eta/L_\xi = 0.26$. In contrast to solution with periodic boundary conditions, *no grain size effect* was found for problem formulation in Fig. 36. This may be related to just one slip system in the left grain and no plasticity in the right grain, because the main reason for the grain size effect in Section 6.6.1 was the larger spacing between slip systems and more dislocation-free space.

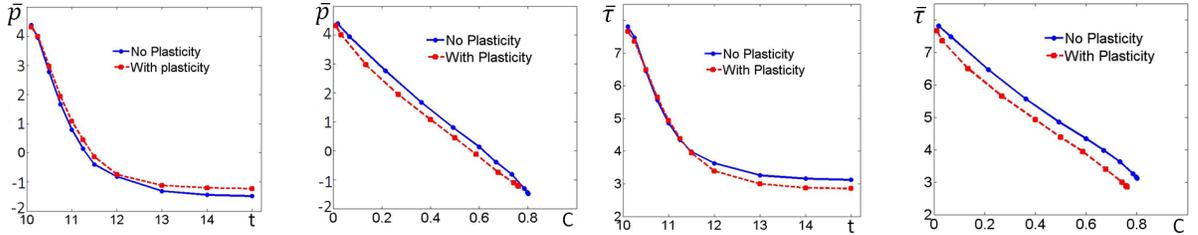

**Figure 38:** Pressure and shear stress averaged over right grain vs. time and the volume fraction of HPP for $p = 1.59$ and $\gamma = 0.35$ with plasticity and for elastic right grain. The effect of plasticity is weak. With plasticity, pressure is higher for the same time but lower for the same $c$ [380].

### 6.6.3. Local and averaged transformation work-based analysis of the equilibrium configuration

It was suggested in [380, 381] to interpret the steady geometry of the HPP region and $c_s$ using a simplified thermodynamic equilibrium conditions across a steady phase interface and the transformation work averaged over the transformed grain and also HPP region. Simplification consists of neglecting the interfacial energy and difference in elastic properties of phases, and using a small-strain approximation for evaluating the transformation work $w_t = \boldsymbol{\sigma}\!:\!\boldsymbol{\varepsilon}_t$, i.e., using Eq. (16)[92–94, 461]:

$$X_\Sigma = \boldsymbol{\sigma}\!:\!\boldsymbol{\varepsilon}_t - \Delta\psi^\theta = 0 \quad \Rightarrow \quad \boldsymbol{\sigma}\!:\!\boldsymbol{\varepsilon}_t = \Delta\psi^\theta. \tag{30}$$

However, there was an alternative competing approach based on using the Eshelby driving force (i.e., total work, including plastic work, instead of $w_t$) [452–455], which was analyzed in detail and criticized in [461]. PFA offers a new opportunity to check which expression for $X_\Sigma$ and corresponding phase equilibrium condition are valid. The $w_t$ contours for the steady solution are shown in Figs. 34 and 35 for cases without and with plasticity in the transforming grain. Note that $w_t$ within an HPP should not satisfy the phase equilibrium or PT ($\boldsymbol{\sigma}{:}\boldsymbol{\varepsilon}_t > \Delta\psi^\theta$) criteria. The values of $w_t$ within a HPP, if much smaller than $\Delta\psi^\theta$, show the potential for nucleation of an LPP only. If the instability condition for the reverse PT is not met within the HPP, the HPP remains metastable. Pressure within HPP and averaged over the grain can be much lower than $p_e$ and be even tensile, without causing the reverse PT. Similarly, $\boldsymbol{\sigma}{:}\boldsymbol{\varepsilon}_t > \Delta\psi^\theta$ in the LPP does not mean nucleation of the HPP because nucleation requires much larger transformation wok than phase equilibrium. It is evident from Figs. 34 and 35 that the upper and lower phase interfaces coincide with lines $\boldsymbol{\sigma}{:}\boldsymbol{\varepsilon}_t(\eta) = 1$, meaning that the *phase equilibrium condition is met at these interfaces.* This condition is not met at some small portions of the interfaces due to contribution of the interfacial energy at large curvatures or strong stress concentrators. *The main input to $w_t$ at the steady interfaces is due to the shear stresses.* In particular, for elastic right grain in Figs. 34a, the pressure is close to zero at the interfaces. Eq. (30) is also confirmed for stationary interfaces for problems solved in [380] with boundary conditions shown in Fig. 36. In contrast to [381, 479], where dislocations in many cases were localized at interfaces (Fig. 34b and c), in [380], interfaces were practically dislocation free. This, however, is in agreement with some experiments [21, 33] in which the most of the interfaces were coherent after HPT. The validity of Eq. (30) has several important consequences:

1. It proves the equivalence of the definition of $X_\Sigma$ for interface propagation in plastic materials for the sharp interface formulation [92–94, 461] and the finite-width formulation within PFA.

2. Plastic work does not contribute to $X_\Sigma$, in agreement with the sharp-interface approaches in [92–97, 99, 461] and in contrast to the approach based on the thermodynamic Eshelby driving force in [452–455]. This gives an additional strong argument in a long discussion on the correct definition of $X_\Sigma$.

3. In the sharp-interface approach, the interface equilibrium condition is more general than in Eq. (30), namely $X_\Sigma = K$; but we proved this equation with $K = 0$. This, however, is not contradictory. In the sharp-interface approach, internal stresses due to dislocations are not included. Since interaction between phase interface and stress field of dislocations (and other defects) is one of the main sources of the interface friction [103, 464], the effect of the stress field of dislocations is effectively included in $K$. In the current PFA, internal stress field of dislocations, and, consequently, the athermal threshold, are automatically included in stress $\boldsymbol{\sigma}$ through solution of the elastic problem, which is more precise than using the effective empirical parameter $K$. Note that $K$ produced by misfit interfacial dislocations have been studied with the same PFA in [472, 547].

4. Strain-induced PTs are described locally and in averaged sense (see Eq. (33) below) by traditional phase equilibrium equations using stress tensor, like stress-induced PTs but with taking into account evolution and stress field of dislocations. The results in Eq. (30) and Eq. (33) seemingly confront our statement in [100, 102] (Section 7.1) that the phase equilibrium conditions do not participate in the microscopic (averaged) kinetics of strain-induced PTs and cannot be extracted from the strain-induced experiments. In fact, there is no contradiction here. Even recent experimental and experimental-computational approaches that determine the kinetic equation [80, 82, 108, 109] cannot find $p_e$ or $\boldsymbol{\sigma}_e$. The main difference is in more detailed treatment at smaller scale in PFA than in the averaged treatment in [80, 82, 100, 102, 108, 109]. Another point is that Eqs. (30) and (33) do not include other contributions to the athermal friction (due to point defects, stacking fault, subgrain boundaries, etc.), but dislocations.

Let us introduce the transformation work $W_t$ and $W_M$ averaged over the entire transforming grain area and

HPP, respectively, and $W_t^*$ for a completely transformed grain, assuming the same stress distribution:

$$W_t = <\boldsymbol{\sigma}\!:\!\boldsymbol{\varepsilon}_t(\eta)>_0; \quad W_M = <\boldsymbol{\sigma}\!:\!\boldsymbol{\varepsilon}_t>_M; \quad W_t^* = <\boldsymbol{\sigma}\!:\!\boldsymbol{\varepsilon}_t>_0, \tag{31}$$

where $<...>_0$ and $<...>_M$ denote averaging over the total grain area and HPP, respectively; $\boldsymbol{\varepsilon}_t(\eta)$ is constant within HPP and zero within the LPP, and for fully transformed grain $\boldsymbol{\varepsilon}_t$ is constant within the entire grain. Then

$$W_t = <\boldsymbol{\sigma}\!:\!\boldsymbol{\varepsilon}_t>_M \bar{c} = <\boldsymbol{\sigma}>_M \!:\!\boldsymbol{\varepsilon}_t \bar{c}; \qquad W_t^* = <\boldsymbol{\sigma}>_0 \!:\!\boldsymbol{\varepsilon}_t. \tag{32}$$

For periodic boundary conditions and based on solution of 4 problems with and without plasticity, the following important results were obtained in [381]:

$$<\boldsymbol{\sigma}>_0 \!:\!\boldsymbol{\varepsilon}_t \simeq <\boldsymbol{\sigma}>_M \!:\!\boldsymbol{\varepsilon}_t \simeq \Delta\psi \quad\Rightarrow\quad <\boldsymbol{\sigma}>_0 \simeq <\boldsymbol{\sigma}>_M . \tag{33}$$

Eq. (33) means that for a stationary state, the thermodynamic phase equilibrium criterion is valid (with an error within 6%) if the transformation work is evaluated in terms of stresses averaged either over the entire grain or HPP, which are approximately equal. An unexpected Eq. (33) should help one to scale up results of a nanoscale simulation to the microscale, namely, to derive an advanced strain-controlled kinetic equation for PT.

For the boundary conditions used in [380], with and without plasticity, (Section 6.6.2 and Fig. 36), the same results are valid but with larger error: for $p = 1.59$ GPa and $\gamma = 0.35$ within 9% and for $p = 5$ GPa and $\gamma = 0.3$ within 20%. Such a difference is related to the effect of boundary conditions and their effect on the definition of the averaged transformation work. In the averaging theorem for elastoplastic materials [582–585], including PTs [452, 586, 587], all averaged parameters–stress and total strain (or deformation gradient) tensors, and total stress work– are defined through tractions and displacements at the surface. They are strictly transformed to volume-averaged for a specific boundary condition only, like uniform stress tractions, linear displacements, or periodic boundary conditions. Even for such conditions, for contributions to the total work–elastic, plastic, and transformation work–simple definitions, like in Eq. (31), cannot be justified, i.e., they are approximate. Thus, error comes from the larger violation of the "legal" boundary conditions than for periodic ones.

It was also found in [380] that the contribution to the stationary transformation work from deviatoric (shear) stresses is essentially larger than that due to pressure. For $p = 1.59$ GPa, pressure is even negative due to volume reduction (Fig. 38). This seemingly confronts our earlier statement in [479] (Section 4.4), where input of the external shear stress to the transformation work was small. The difference here is in focusing on the transforming grain only, which includes a strong stress concentrator due to dislocation pileup, higher $\tau_y$ due to grain size effect and reaching theoretical strength near the tip of the pileup, and large pressure drop due to volume reduction.

Let us reexamine the evaluation of the effect of shear stress on PT condition made in [479] and Section 4.4. First, we evaluate the phase equilibrium condition Eq. (33) $W_t^* = \Delta\psi = 0.1$ GPa, giving $p_e = 10$ GPa instead of nucleation condition with $p_h^d = 15$ GPa. Second, instead of macroscopic $\tau_y = 1$ GPa, we take $\bar{\tau} = 3.5$ GPa from Fig. 33c. Fig. 6 in [479]. Then for $\varepsilon = 0.1$ and $\gamma_t = 0.2$, condition $W_t^* = \bar{p}\varepsilon + \bar{\tau}\gamma_t = 1$ is met at $\bar{p} = 3$ GPa. While this is an impressive pressure reduction in comparison with $p_h^d = 15$ GPa (and $p_e = 10$ GPa), it is just by 5 instead of 50-100 times in Table 1. However, in experiments, the record pressures for the nucleation of the HPP are reported rather than pressure for a stationary configuration with essential $c$. They are determined by the lattice instability or barrierless nucleation conditions, which are different from the thermodynamic equilibrium conditions. In the very small volumes, $\bar{\tau}$ tends to the theoretical strength. In Fig. 38 for $p = 1.59$ and $\gamma = 0.35$, $\bar{p} = 4.2$ GPa and $\bar{\tau} = 7.3$ GPa at the beginning of PT Even the thermodynamic contribution of $\bar{\tau} = 7.3$ is equivalent to the pressure of $\bar{\tau}\gamma_t/\varepsilon = 14.6$ GPa, resulting in a total pressure of 18.8 GPa, close to the instability pressure for the LPP $p_{in}^d = 20$ GPa. The Effect of even higher local shear stresses is much larger.

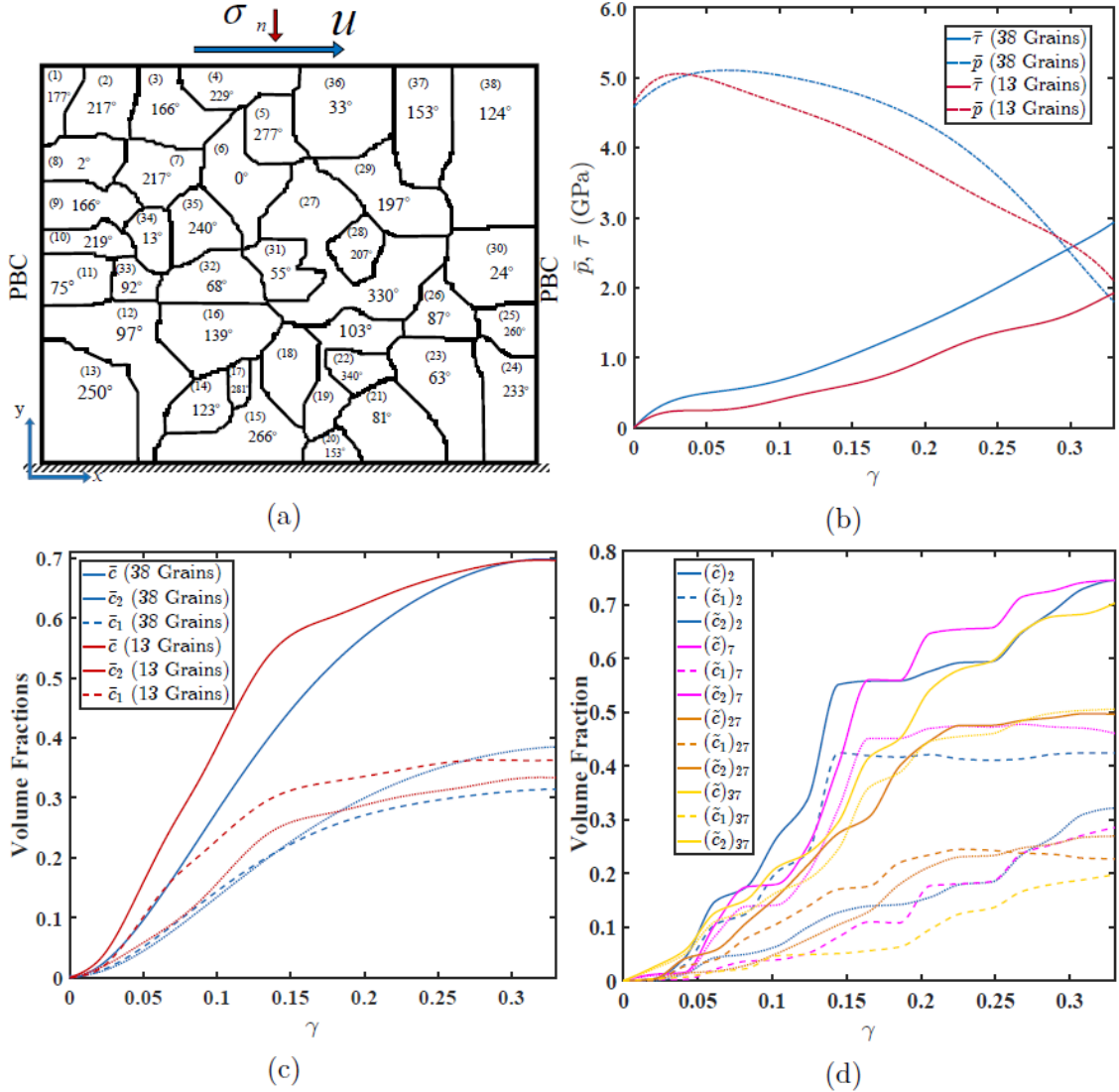

**Figure 39:** (a) A schematic of the grain structures of a polycrystalline sample with 38 grains loaded by normal stress $\sigma_n = 6.05$ GPa and shearing rate $\dot{\gamma} = 0.004s^{-1}$. Within each grain, the grain # is shown in parentheses and orientation of each is given by corresponding angle. (b) Evolution of the averaged pressure $\bar{p}$ and shear stress $\bar{\tau}$ as a function of the prescribed $\gamma$ for samples with 13 [462] and 38 grains. (c) Evolution of the volume fraction of two martensitic variants, $\tilde{c}_1$ and $\tilde{c}_2$, and the HPP $\tilde{c} = \tilde{c}_1 + \tilde{c}_2$, averaged over the sample for a polycrystal with 13 [462] and 38 grains. (d) Evolution of the volume fraction of two martensitic variants, $\tilde{c}_1$ and $\tilde{c}_2$, and the HPP $\tilde{c}$, averaged over each grain, in several grains, as a function of $\gamma$. Subscript shows the grain number in (a). Reproduced from [463].

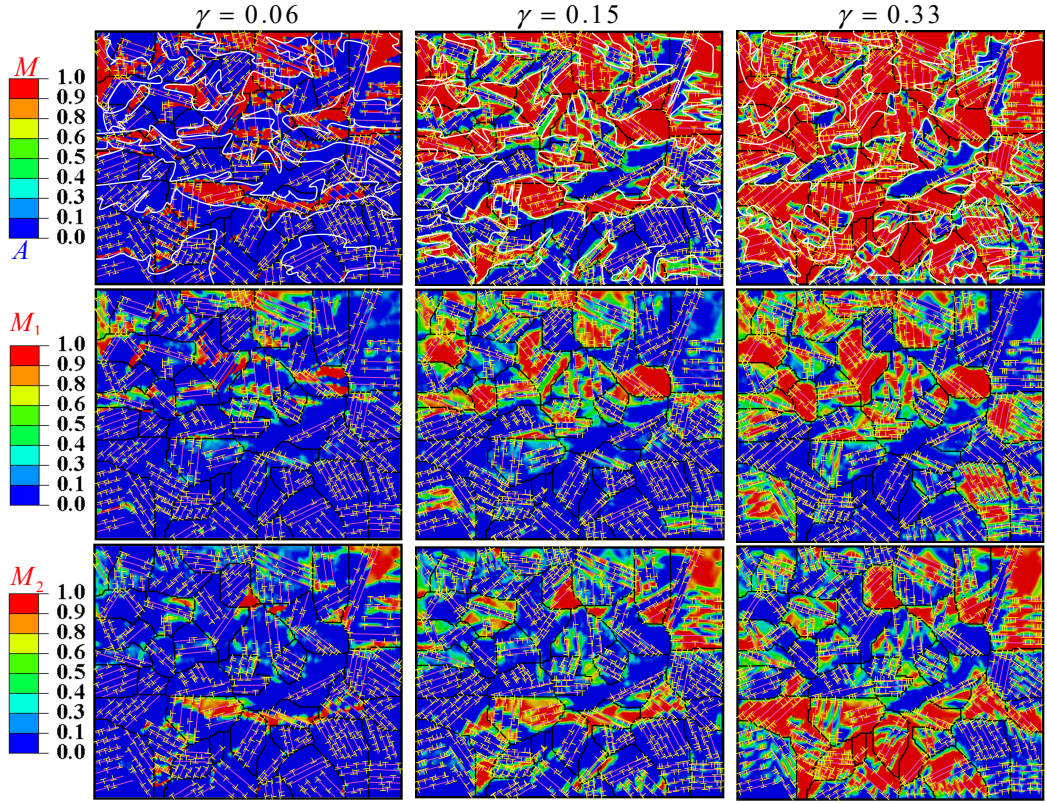

**Figure 40:** Evolution of the dislocation structure, the HPP ($M$), and two martensitic variants of the HPP ($M_2$ and $M_2$) in a sample loaded by normal stress $\sigma_n$ =6.05 GPa and $\dot{\gamma} = 0.004s^{-1}$ like in Fig. 39a, under three different prescribed shears [463]. Along the white line in the top raw, the thermodynamic phase equilibrium criterion Eq. (30) is satisfied.

### 6.7. Scale-free phase-field simulations for a polycrystal

In the above nanoscale PFA, the phase interface widths and the dislocation cores are ∼1 nm. Within the interface widths and dislocation cores, corresponding order parameters change by unity, producing very heterogeneous fields. To receive the mesh-independent solutions, these width and core should be discretized by 4-5 finite elements. This restricts the nanoscale PFA to sample sizes <0.1-1 $\mu m$. To resolve this issue for PTs, very different scale-free PFA approach for the martensitic PT was developed in [588–590]. In this model, the volume fraction of the martensite (HPP), $c$, is the order parameter responsible for the material instability and strain softening, leading to the transformation strain localization and formation of a discrete martensitic microstructure, similar to that in the nanoscale PFA. The volume fractions of martensitic variants, $c_i$, are not the order parameters but internal variables, not producing material instabilities. Therefore, there are no interfaces between different martensitic variants. In contrast to the nanoscale PFA, gradient energy term is excluded, i.e., the model is local. Still, simulations show that solution is practically mesh-independent. Expression for the local energy vs. order parameters is based on the mixture theory with the simplest interaction term. All these lead to significantly more computationally economic model than the nanoscale PFA and allows one to treat much larger samples.

Scale-free description of dislocations in [462, 463] was realized through solution of the contact problem. By definition [562], a dislocation is produced by a relative glide of two sides of a cut in a continuum, by a Burgers vector $\boldsymbol{b}$, keeping the traction and the normal displacement across the cut continuous. Similarly, multiple continuously distributed dislocations can be obtained by larger relative sliding $\boldsymbol{u}_s$ along the slip surface. The same continuity conditions of the traction and normal displacement across the contact surface are met for the relative slip between two contacting deformable bodies [591, 592]. This was the rationale in [462] to suggest a model in which continuous distribution of dislocations along selected discrete slip (contact) surfaces was determined through solution of the contact problem. Sliding rule in contact problem mimicked slip rule for a single crystal. Similar contact formulation is valid for shear bands [593], thin twines [594], or shear cracks [595]. ABAQUS FEM code was used to implement developed models [592]. While scale-free model is much simpler than the nanoscale PFA models in [380, 381, 426, 427, 479], it reproduced well the stress field of a single dislocation and reasonably well the evolving dislocations and HPP obtained with nanoscale PFA in Fig. 34. Due to lack of a internal length, the model is scale-free and can be used for a sample size from tens of nm to km, e.g., for modeling strain-induced PTs in geophysical applications [83, 393].

Various problems on PT and dislocation evolution in a bicrystal and polycrystal under compression and shear are solved in [462, 463]. The material properties used for problems shown in Figs. 39 and 40 are: $\varepsilon_{tx} = \varepsilon_{ty} = -0.05$ and $\varepsilon_{txy} = 0.1$ the Young's modulus $E = 177$ GPa; $\nu = 0.24$; $p_e = 10$ GPa, $p_{in}^d = 20$ GPa, $p_h^d = 14.7 GPa$, and $\tau^c = 0.3$ GPa for LPP and both variants of the HPP were used; $\{112\}\langle 111\rangle$ slip system is chosen in the LPP, which is inherited by each martensitic variant. Plane strain formulation, edge dislocations, and two martensitic variants have been considered. It is necessary to underscore the basic difference between $\tau^c$ for the scale-free model, which characterizes athermal resistance to sliding, and corresponding parameter in the nanoscale model in Section 6.6. It characterizing a barrier for dislocation nucleation, which then propagates without athermal resistance due to Peierls barrier.

Compression and shear loading for a polycrystalline sample with stochastically oriented 38 grains and periodic boundary conditions on the lateral sides is shown in Fig. 39a. Evolution of the dislocations, the HPP ($M$), and two martensitic variants of the HPP ($M_2$ and $M_2$) for 3 shears is shown in Fig. 40. Evolution of various averaged parameters is presented in Fig. 39b-d, in comparison with results for 13 grains from [462].

As can be seen from Figs. 39d and 40 for the HPP $M$, the PT progress is very heterogeneous from grain to grain. The same is true for each martensitic variant. The HPP mostly nucleates near the dislocation pileup tips in the adjoining grains and close to the extra atomic planes of dislocations within the same grains. The experimentally detectable amount of the HPP in the polycrystal ($\bar{c} \leq 0.05$) appears at pressure of 5 GPa (for both samples with 38 and 13 grains) and $\gamma =$0.02-0.03. Kinetic curves in Figs. 39d exhibit multiple plateaus followed by stages with intense growth. Plateaus correspond to the dislocation accumulation until they join the pileups and increase the stress concentrations, leading to the PT progress.

Pressure drops to 1.8 GPa at $\gamma = 0.33$ for both samples. Despite the perfectly plastic model with $\tau_c = 0.3$ GPa for a single crystal, $\bar{\tau}$ increases to 2.9 and 1.9 GPa for 38 and 13 grains, respectively. This work-hardening is due to back stresses from the dislocation pileups with increasing number of dislocations at grain boundaries. An increase in $\bar{\tau}$ (and, consequently, $\tau_y$) with increasing number of grains, and, consequently, with reduction in the grain size, reproduces experimental Hall-Petch effect [263]. Such an evolution in $\bar{p}$ and $\bar{\tau}$ implies that the contribution of the $\bar{\tau}$ to the transformation work growth during the loading, and exceeds pressure contribution by more than 2 times for $\gamma = 0.33$ and 38 grains.

According to Fig. 39c, the PT progress is faster for a sample with larger grains, despite the smaller $\bar{p}$ and $\bar{\tau}$. This was interpreted in term of longer dislocation pileups promoting nucleation and a smaller grain-boundary area producing a smaller overall resistance to the interface propagation for larger grains. At the same time, for both grain sizes, $\bar{c}_s = 0.7$, was reached and did not change at further shearing. Multiple pockets of the LPP remain in practically all grains (Fig. 40). Steady two-phase mixture at ambient pressure after SPD with HPT was broadly reported [29, 34, 35, 432]; this steady state was independent of pressure-plastic strain loading paths in [35]. However, it was mentioned in [29] that $\alpha - \omega$ PT in Zr at 6 GPa and 20 turns could be complete, but reverse PT occurs during the sample preparation for XRD. Indeed, complete $\alpha - \omega$ PT in Zr was obtained DAC and RDAC in situ in [80, 82, 89, 108, 109] at much smaller $p$ and $q$, and reverse PT was not observed at torsion even at 0.2 GPa [80]. Nevertheless, the existence of the two-phase stationary state means that further straining is pointless. This is one more confirmation (see also [84, 100, 102, 378]) that plastic strain should be optimal. Note that the microscale kinetic model (see [100, 102] and Section 7.1) predicts steady, loading path-independent $c_s(p)$ when $p_\varepsilon^d < p_\varepsilon^r$.

It follows from Fig. 40 for $M$ that despite the very sophisticated geometry, microstructure, and stress state, the *local phase equilibrium Eq. (30) is met* for most of interfaces, excluding small regions with large curvature and strong stress concentration due to dislocation pileups. Thus, despite drastic difference in equations, the *scale-free PFA, similar to the nanoscale PFA, produces phase interfaces consistent with the sharp interface approach.* Note that a finite-strain generalization of Eq. (30), which includes jump in elastic moduli and complex athermal friction, was also confirmed for 3D problem but without dislocations in [531]. What is also important that *for stresses averaged over the entire polycrystal or HPP, and $\boldsymbol{\varepsilon}_t$ averaged over the HPP, Eqs. (33) for stresses, transformation work, and averaged phase equilibrium condition are met.* This is very surprising results that have a potential for the development of the coarse-grained theory for the interaction PT and discrete. In combination with the plots 39b-d, they may lead to the development of a more advanced microscale kinetic equation than those suggested in [100, 102, 108, 109] and Section 7.1.

Note that Eqs. (33) are not applicable if averaging is performed over the each grain separately. While the boundary conditions for a polycrystal in Fig. 39a satisfy the averaging theorems [452, 582–587], this is not the case for each grain, because they are determined by interaction with the surrounding grains.

## 6.8. Atomistic and concurrent atomistic-continuum simulations of strain-induced nucleation

The above continuum simulations considered grain boundaries as structure-less fixed lines of jump in crystal orientation, not penetrable for dislocations. MD simulations allows one to reveal not only atomistic features of dislocation pileup-induced PT, but also interaction between pileup and obstacles (grain and phase boundaries), namely partial or complete passage of the dislocations through grain boundaries, absorption of dislocations, rearrangement of the interfaces and the tip of dislocation pileups, see examples below.

### 6.8.1. Amorphization in Si

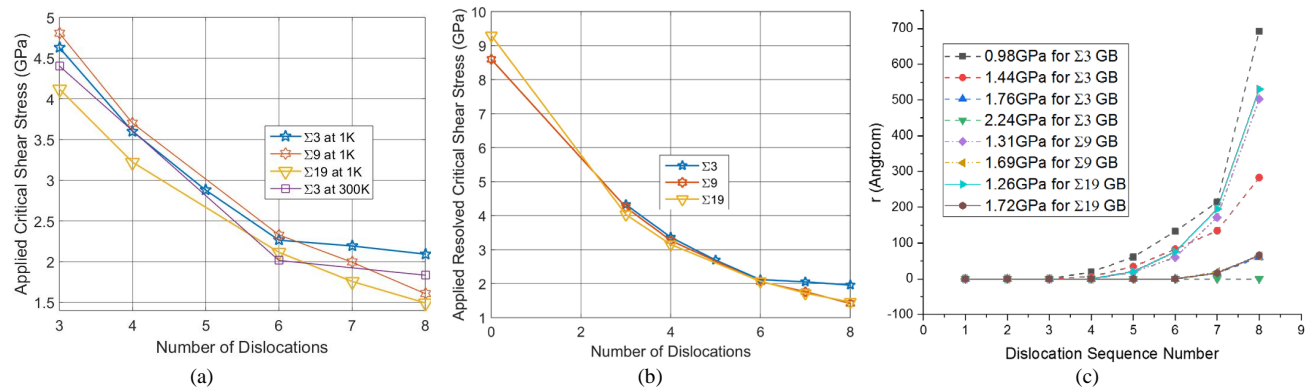

**Figure 41:** (a) and (b) The applied critical shear stress $\tau_{ap}^c$ and the applied resolved critical shear stress $\tau_r^c$ on the amorphization plane, respectively, required for the formation of an amorphous shear band at the tip of dislocation pileup against 3 different grain boundaries in Si vs. the number of dislocations in a pileup $N$. The shear stresses for lattice instability for (111) and ($1\bar{1}0$) amorphization planes in a dislocation-free crystal are shown for $N = 0$. (c) The distance of 8 sequential dislocations from the grain boundary for different $\tau_{ap}^c$ and 3 grain boundaries. Reproduced from [566].

Si-I undergoes amorphization under compression in DAC of nanocrystalline and nanoporous Si [596], shock loading [395–397], uniaxial compression of a nanopillar [369, 522], nanoindentation [441, 442], scratching and machining [398–401]. In addition to the thermodynamic reasons caused by the elimination of dislocation (and other defects)t energy, amorphization is also considered an additional carrier for plastic flow [402].

The first MD simulation of the PT at the tip of dislocation pileup under shear was presented for amorphization in Si-I [566]. Shuffle $60^o$ dislocation pileups with the Burgers vector $\boldsymbol{b} = a_o/2 <110>$ against the Σ3 (coherent twin boundary), Σ9, and Σ19 grain boundaries were considered. The Stillinger-Weber (SW) interatomic potential [597] and LAMMPS code [598] were utilized. Formation of a step at the grain boundary leading to an amorphous band, propagating away from the tip of a dislocation pileup, was observed for all grain boundaries, see Fig. 30b. The $\tau_{ap}^c$ required for the formation of an amorphous band strongly reduces with increasing number of dislocations in a pileup $N$, initially almost linearly, then slower (Fig. 41a). This dependence is rationalized based on the concept that amorphization starts due to reaching the local lattice instability conditions, which for Si-I depends on all six components of $\boldsymbol{\sigma}$ [444, 513, 523]. Namely, it was postulated that the amorphization along a chosen plane starts when the local resolved shear stress at the tip of the dislocation pileup, averaged over the small disordered embryo, reaches the shear instability stress $\tau_{in}$ for the same plane for a perfect crystal. Utilizing this criterion and simple calculations, results in Fig. 41a, presented in terms of $\tau_r^c$ on the amorphization plane in Fig. 41b. One can see curves for all 3 grain boundaries practically coincide, confirming correctness of the suggested criterion.

Note that the thermodynamic amorphization criterion based on the extended Patel-Cohen approach to martensitic PTs [599] and classical nucleation theory was suggested in [397]. Since $\gamma_t$ for amorphization is not defined by crystallography, it is assumed to be the same as for twinning for cubic lattices, i.e., $\gamma_t = 0.707$. It was also mentioned that with increasing shock amplitude and defect density, mechanism of barrierless nucleation suggested in [102, 479] is very likely. As the next step [395, 404], a more general criterion, utilizing a modified transformation work from [513], was introduced by adding an interfacial energy and energy of defects to the expression in [513]. It is necessary to mention that in [513], a criterion for lattice instability of the perfect crystal was suggested, but in [395, 404], it was a thermodynamic criterion for a PT, i.e., they have a different physical sense. Thus, criterion in [395, 404] was motivated by expression in [513] but cannot be considered as a generalization.

Fig. 41c shows how dislocations approach different grain boundaries with increasing applied shear stress, producing a step. After nucleation of a disordered embryos near all the grain boundaries, different mechanisms of the amorphous shear band propagation for different grain boundaries were revealed. For the Σ3 boundary, an amorphous band propagates (without any precursors) along the (112) plane and generates partial and complete dislocations from the crystalline-amorphous interface. For the Σ9 boundary, propagation of an amorphous band along the (110) plane is preceded by nucleation of partial dislocations producing a stacking fault. For the Σ19 boundary, a single-layer stacking fault along the (111) plane transforms into a two-layer band with the atomic bonds rotated by $30^o$ with respect to those in the surrounding crystal, following by amorphization along the (111) plane. Amorphization of similar Si bicrystals under shear and different temperatures but without initial dislocation pileup was considered in [600].

At the same time, a shuffle screw dislocations transmit through all the same grain boundaries [601]. When a dislocation interacts with a heptagon site of all these grain boundaries, it transmits directly. However, while interacting with a pentagon site, it initially cross slips to a plane on the heptagon site before passing the grain boundary. The barrier for transmission was evaluated with the climbing image nudged elastic band method. For both Σ9 and Σ19 grain boundaries, the energy barrier from the heptagon sites is much larger than the pentagon sites. Still, since the energy barrier for all the grain boundaries equal or slightly larger than the Peierls barrier, *perfect screw shuffle dislocations cannot pile up against these grain boundaries and cannot promote the PT*. This is very important limitation that should be considered for all materials. Moreover, moving at a constant velocity dislocations require almost two times lower applied shear stress that the initially static dislocations in order to transmit through the Σ9 and Σ19 grain boundaries on heptagon site.

The Peierls barrier plays an important part in the formation of the dislocation pileup and subsequent nucleation of the HPP. It is absent in the analytical model (Section 6.1) and nanoscale PFA (Section 6.6), but naturally appears in MD simulations, and is introduced in the scale-free PFA (Section 6.7). In this regard, an important phenomenon was predicted analytically in [602, 603], based on simulations on simple 2D lattice in [604], and with MD for such a high-Peierls-barrier materials as Si in [605] that the stationary dislocation can propagate though the crystal at stresses significantly below the static Peierls stress $\tau_p^s$. The dynamic Peierls stress in [602, 603] was introduced as a function of dislocation velocity (i.e., a dissipation stress); the dynamic Peierls stress, $\tau_p^d$, was defined in [605] as the minimum applied shear stress $\tau_{ap}$ at which a stationary motion is possible. A stationary motion of shuffle screw and $60^o$ dislocations in Si was studied in [605] using the Stillinger-Weber potential [597] and the continuum equation of motion, utilizing inputs from the MD. The obtained $\tau_p^d$, $0.33 GPa$ for a screw dislocation and $0.21 GPa$ for a $60^o$ dislocation was found to be 5-7 times lower than the $\tau_p^s$ of $1.71 GPa$ and $1.46 GPa$, respectively. The critical initial velocity vs. $\tau_{ap}$ above which a stationary dislocation propagation is possible was also determined. Two reasons for the possibility of dynamic dislocation motion below the $\tau_p^s$ were suggested: (a) the periodic lattice resistance stress is lower than the $\tau_p^s$ almost everywhere, and changes the sign, which helps the dislocation motion; (b) high kinetic energy, allowing to overcome the $\tau_p^s$. Such fast dislocation motion below the $\tau_p^s$ and their accumulation against the grain boundaries create a potential for a dynamic intensification of plastic straining and strain-induced PTs, e.g., by ball milling, short peening, dynamic HPT, and shock loading. Similar approaches can be developed for partial dislocations, twins, and phase interfaces. Concept of the velocity dependent dynamic Peierls stress could be useful for developing the meso- and macro-scale viscoplastic models.

*6.8.2. Square to hexagonal phase transformation in 2D materials*

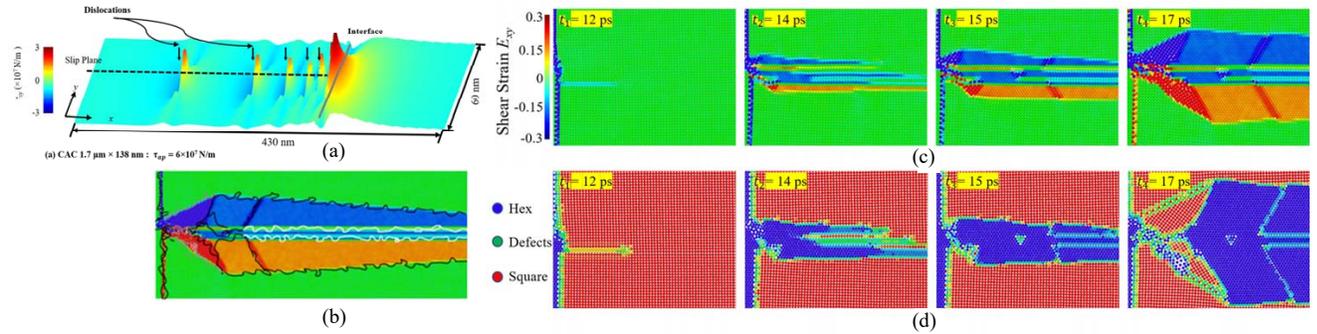

**Figure 42:** Results of CAC simulation of 2D pre-sheared two-phase sample during compression [567]. (a) Phase interface separates hexagonal (HPP) phase on the left with dislocation pileup and square (LPP) phase on the right, which transforms to HPP. The distribution of internal shear stresses due to pileup is presented. (b) Stationary field of the shear strain in the rightregion with the same colors like in (c). Green corresponds to the initial LPP; blue and red on the right side are twin-related variants of the HPP; dark blue and red regions near tip of the pileup are LPP regions producing twin interfaces with the initial LPP and obtained by the reverse PT from the HPP due to splitting of the dislocation pileup in several smaller pileups and reduction in stress concentration (Fig. 43f); dark blue and red regions within HPP are stacking faults. The phase interface equilibrium conditions (Eq. (34)) (using deformation gradients from both sides of each interface) are met along the black lines . (c) and (d) Evolution of the shear strain and phases (based on coordination number), respectively, in the right region during the PT from LPP to twinned HPP followed by reverse PT near the tip of dislocation pileup producing twins in the LPP (twinning by direct-reverse PT).

Since MD simulations are restricted to the small sample size, CAC simulation method was developed [606–611] and applied to combine study of the interaction of dislocations and grain/phase boundary, nucleation and evolution of phase interfaces with atomistic resolution, and continuum treatment of long dislocation pileups away from the grain and phase boundaries [567, 569]. 2D two-phase system in Fig. 42a with 16 dislocations in the HPP was considered undergoing square (LPP) to hexagonal (HPP) PT with two twin-related variants. Right part of a system (including some material on the left side of the interface) was atomistic, and the left part was a course-grained continuum. A similar system of a smaller size and 8 dislocations was treated with MD. Lateral boundaries were stress-free. System is first subjected to shear, and then compression caused the PT. Such a sequence, opposite to what is done in RDAC experiments and in the above Sections, is chosen to avoid the phase interface propagation without dislocation pileup. Dislocation passage through the interface is suppressed. Shear stresses pushed dislocations to the phase interface, producing a step (superdislocation) (Fig. 30c) and pileup behind the tip. Stress concentrator at the tip of a pileup initiates a stacking fault and then twinned HPP with a horizontal twin boundary, which grows as a wedge to the right (Fig. 42b,c). At the same time, the rearrangement of the local atomic structure near the tip of the dislocation pileup leads to splitting of a single pileup into multiple smaller pileups, reducing stress concentration (Fig. 43f). This, in turn, leads to the reverse PT, but not to the original LPP lattice, but to the twinned LPP lattice producing twin boundaries with the initial LPP phase (Fig. 42b,c). This is a very interesting case of twinning in the LPP though the direct and then reverse PTs under increasing loading. The reverse PT is localized, keeping large amount of the HPP despite the strongly reduced stress concentration, which has already played its part in the nucleation of the HPP.

As a finite strain generalization of the thermodynamic equilibrium condition Eq. (30) found for nanoscale [380, 381] and scale-free [462, 463, 531] PFAs, the following phase equilibrium condition for different interfaces was considered:

$$X = \boldsymbol{P}^T : (\boldsymbol{F}_{t2} - \boldsymbol{F}_{t1}) - \Delta \psi = 0. \tag{34}$$

For twin boundaries and stacking faults $\Delta \psi = 0$. Fig. 42b demonstrates the *fulfilment of the thermodynamic equilibrium condition (34) for all interfaces in the stationary configuration: different twin-related regions of LPP - variants of the HPP, twins in LPP and HPP, initial vertical LPP-HPP interface, and even stacking faults within HPP twins.*

Among all introduced dislocations in the HPP, $N = 16$ for CAC and $N = 8$ for MD, $N_a$ dislocations produce a step (superdislocation) or tip and $N_b$ dislocations are located behind the pileup tip (Fig. 43c); $N = N_a + N_b$. The calculated shear stress distribution ahead of the pileup was fitted in [567] as a sum of these two contributions, each of which generalizes known analytical expressions [562] for superdislocation and dislocation pileup, respectively:

$$\tau_{xy} = \frac{z\mu N_a |\boldsymbol{b}|}{2\pi(1-\nu)(r-r_0)} + \frac{K_b}{\sqrt{\pi(r-r_0)}} + \tau_0. \tag{35}$$

Here, $r_0$ characterizes the location of the maximum internal stress at the pileup tip, which was found to be in the middle of a step; $\tau_0$ characterizes initial internal stress in dislocation-free system due to vertical phase interface; $K_b$ is the stress intensity factor for the dislocations behind the pileup tip. The correction factor $z$ collectively takes into account the effect of finite strain, geometric changes, system and step size, nonlinear anisotropic elasticity, lack of strict additivity of stresses in a nonlinear material, and two-phase system, which are not present in the analytical model. It can be described by the relationship

$$z = 0.1656 e^{0.2173 N_a} + 0.8777 \tag{36}$$

shown in Fig. 43d, which is independent of $N_b$. This is very important result, which shows that the correction factor for the step to Eq. (24) used for analytical treatment is $\sim 2$ for 9 dislocations and $\sim 6$ for 16 dislocations. While it was not checked in [567], but based on Eq. (24), pressure should be subjected to the same correction, which means that the *promoting effect of dislocation pileup with large number of dislocations is essentially larger* than presented in [100, 102] and Section 6.1.

Fig. 43e shows nontrivial contribution of the number of dislocations behind the tip to the $K_b$. Note that in contrast to the analytical solution, dislocation at the obstacle is excluded from $N_b$ and included in $N_a$. That is why 1 dislocation produces negligible contribution to the stress singularity. $K_b$ increases almost linearly with $N_b$, from $\sim 1$ for 2 dislocations to $\sim 12$ for 8 dislocations. With increasing shear stress $\tau_{ap}$, $N_a$ slightly nonlinear increases and $N_b$ decreases keeping $N = const$ (Fig. 43c), and the effect of the superdislocation is getting dominating. The reduction in the compressive stress $\sigma_c$ to start the PT with the increasing $N_a$ and corresponding stress intensity factors $K_a$ (which combines all terms in Eq. (35) $\sim 1/(r-r_0)$), from $22 \times 10^7$ N/m for a single dislocation to $4 \times 10^7$ N/m for 14 dislocations is shown in Fig. 43b. The fastest reduction occurs for $N < 8$, then it essentially slows. Presenting a similar reduction in $\sigma_c$ with reduction in $N_b$ and corresponding $K_b$ looks confusing and non-informative; it is caused by the constraint $N = N_a + N_b$ and increase in $N_a$.

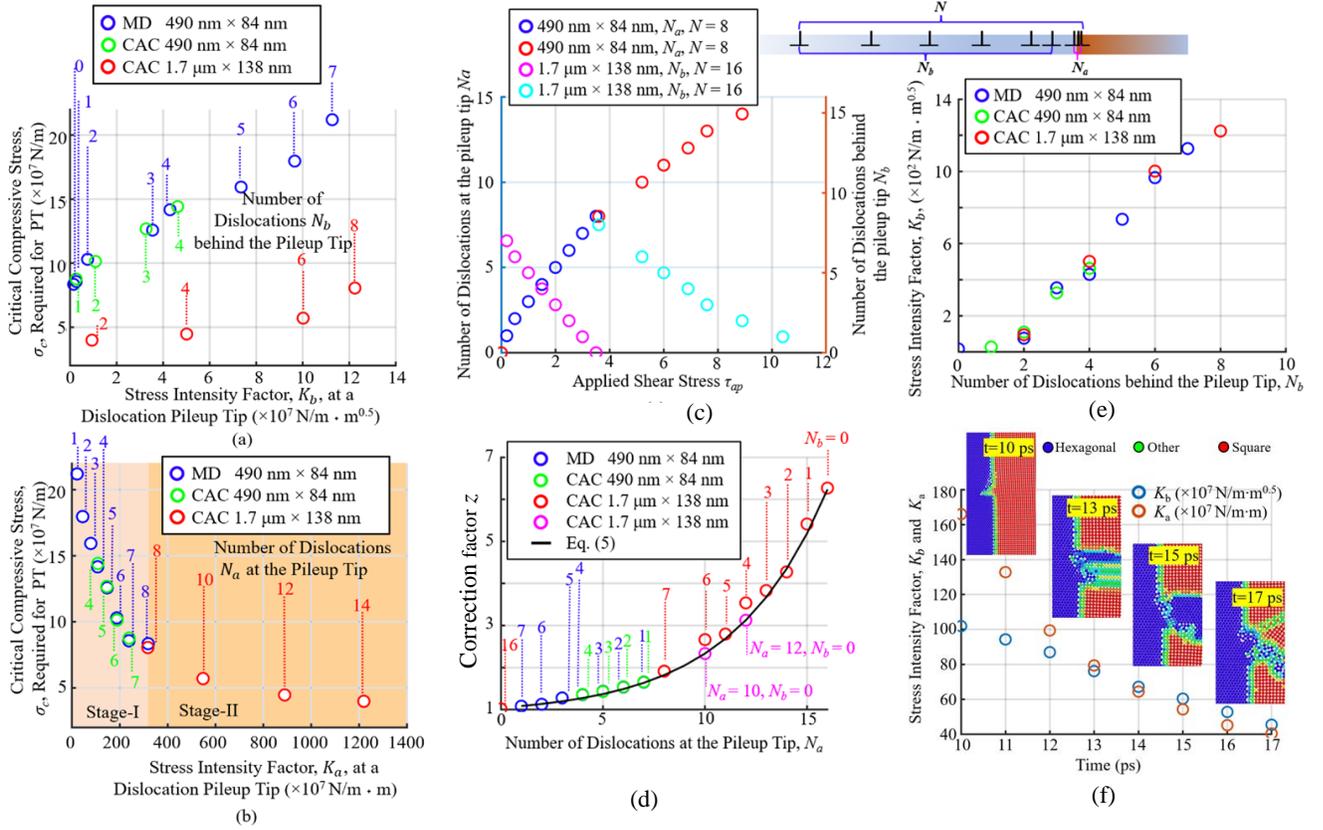

**Figure 43:** (a) and (b) MD and CAC results for the compressive stress $\sigma_c$ to start the PT vs. the stress intensity factors $K_b$ behind the pileup tip and $K_a$ ahead of the pileup tip, respectively. (c) The number of dislocations behind the pileup tip $N_b$ and at the pileup tip $N_a$ vs. the applied shear stress $\tau_{ap}$. (d) The coefficient $z$ in Eq. (35) vs. the number of the dislocations at the pileup tip $N_a$. Line is described by Eq. (36). (e) The stress intensity factor, $K_b$, vs. the number of the dislocations, $N_b$, behind a pileup tip. (f) The evolution of the stress intensity factors, $K_b$ and $K_a$, with the insets demonstrating the snapshots of the rearrangement of the local atomic structure near the tip of the dislocation pileup during the direct and then reverse PTs. Reproduced from [567].

The above results were obtained before the PT initiation. During the PT, initially, $K_a$ is 1.6 times larger than $K_b$ (Fig. 43f) and both decrease with compression. This happens because a superdislocation disassociates into multiple superdislocations with smaller number of dislocations in each, leading to local reverse PTs. All these results are useful for the development of a more advanced analytical model than in [100, 102] for the nucleation at the tip of dislocation pileup.

### 6.9. Promotion of PTs/CRs in dislocation-free crystalline and amorphous materials

If crystalline materials are (almost) dislocation-free, e.g., due to small size in experiment or simulation conditions, dislocation pileup mechanism apparently does not work. In this case, shearing under pressure or other type of loadings leads to crystal lattice instability resulting in PTs to other crystalline or amorphous phases, fracture, slip, or twinning (Section 5), i.e., the same as at the tip of dislocation pileup but in a larger volume. For example, PT Si-I↔Si-II in a defect-free single crystal under general stress tensor was studied in [513, 545] with MD and [444] with DFT; this study shows that deviatoric stresses essentially reduce the PT pressure. Essential (but much smaller than in experiment) reduction in PT pressure from graphite to diamond due to shear was demonstrated with DFT and MD in [79]. Much closer to experimental reduction in the PT pressure for nanocrystalline graphite to diamond down to 2.3 GPa at 950 K due to shearing was obtained [612]. In addition to promotion of heterogeneous nucleation at the grain boundary and in the absence of dislocations, large elastic strains of the magnitude similar to that at the tip of dislocation pileup appear in the constraint region near the grain boundary, causing this PT at low pressure. Shear-induced diamondization of graphene was studied with MD in [613].

Very high ($10^{10} - 10^{12}/s$) strain rates suppress dislocation and twin nucleation and lead to overloading material with deviatoric stresses up to lattice instability. MD simulations in [412] demonstrated that under uniaxial compression of $< 111 >$Cu single crystal up to 204.9 GPa at $10^{12}/s$, melting occurs at 300 K, which is 5,134 K lower than the melting temperature at the same mean stress. This was in agreement with prediction of the thermodynamic theory in [412] which takes into account the effect of nonhydrostatic stresses and their energy. Similar results were obtained for Al. Since after melting nonhydrostatic stresses relax and melt is getting unstable, it recrystallizes at the ps time scale. That is why this was a virtual melting introduced in [410, 411, 614]. Shear-induced formation of amorphous band in boron carbide was investigated at atomic scale in [407, 408].

While highly 2D-disordered nanocrystalline hBN does not transform under hydrostatic pressure even at 52.8 GPa, it transforms under shearing in RDAC at 6.7 GPa [77]. One of the possible mechanisms for this reconstructive PT was related to the atomic rearrangements in large angle grain boundaries during plastic deformation. They occur in the localized shear zones, where $\tau$ reach the half of theoretical strength in shear. This atomic rearrangements play the role similar to thermal fluctuations at high temperature and may lead to PT to wBN. Since strength of wBN is essentially larger than that of hBN, shear zones occur in the weaker regions, i.e., they may shift into new hBN-wBN and grain boundaries.

Disordered materials, like amorphous inorganic solids or polymers, cannot posses dislocations, and consequently, must have different mechanisms for promotion of PTs/CRs. There is no theory for such cases, so we will make just few comments. As we mentioned in Section 3.1, some qualitative mechanisms were suggested, like "ROLLER" [278] and "CONMAH" [339] (see also [113]); that chemical bonds are chemically activated and react with other molecules under flow-induced collision [14]; bond breakage that produce nucleation sites [8]; that *elastic* shear strain can promote CR by reducing the highest occupied bonding molecular orbital – lowest unoccupied anti-bonding molecular orbital energy gap [335–338], etc.

Formation of amorphous shear band in Si-I single crystal via the virtual melting under shear was revealed with MD in [403] (Section 6.8.1). Further shear lead to cyclic PTs between a-Si, Si-I, and Si-IV. Shear-induced crystallization of a-Si was also related to the atomic jump-like rearrangements in the shear zones, which are similar to strong thermal activations and offer multiple chances for crystallization to occur during shearing. There are numerous examples of shear-induced crystallization in amorphous materials, which may follow the same mechanism but will not be considered here. Finding theoretical description for promoting of PTs/CRs in disordered materials is still an outstanding problem.

6.10. Summary and perspectives

The results discussed in this Section provide a nanoscale mechanisms and basis for understanding and description of strain-induced PTs under compression and shear in DAC and RDAC, during HPT, and other SPD technologies (e.g., ball milling), surface treatment, wear, friction, and projectile/meteorite penetration. Strain-induced PTs are nucleation controlled. It is shown that strong concentrator of all components of the stress tensor at the tip of the dislocation pileup (proportional to number of dislocations, which could exceed 10-20 for nanograined materials) can reduce the PT pressure by a factor of 10 to 100 (like in experiments [10, 64, 77, 79, 81, 82, 328, 375]), in many cases well below $p_e$ (like in [10, 64, 77, 79, 81, 82]), and even down to zero pressure, like in experiments where PT were obtained at sub-GPa [79, 81, 82].

We are not aware of any other defects and corresponding mechanisms which are able to reduce the PT pressure by such a magnitude. Strong but still weaker potency have compressive stresses due extra atomic planes in the pileups within the same grain, which also stabilise the HPP by producing semicoherent phase interfaces. While $\tau = \tau_y$ produce relatively minor contribution to the PT conditions, shear/deviatoric stresses at the tip of the dislocation pileup are limited by the theoretical strength, which could be 10-100 times $\tau_y$. Since stresses reach the lattice instability limit (determined, in particular, for Si I-Si II PT in [444, 513, 523] by PFA analytical derivations in combination with atomistic simulations, see Section 5.2), instability occurs, which may cause barrierless nucleation of HPP(s), dislocations or twins, or some combination of these processes. Such highly-deviatoric stresses cannot be obtained in bulk and may lead to new phenomena and phases that were not or cannot be produced (or inaccessible) without SPD. Plasticity plays a dual part: it promotes PT by creating stress concentrators but the stress concentrators may generate dislocations and twins, relaxing stresses and suppressing PT. Some combinations of normal stresses and shears were found for which dislocation activity in the transformed grain is minimized or do not hurt the PT, i.e., the PT wins a competition over dislocations. The most important experimental confirmation of the dislocation pileup-based mechanism is finding a correlation between grain-size dependence of $\sigma_y$ and $p_\varepsilon^d$ and revealing strong pressure growth (by 5-7 GPa) in nuclei of HPPs of Si.

Growth of the nuclei is governed by the same rules as for stress-induced PTs coupled with evolving dislocations. Equilibrium configuration of the HPP is controlled by the transformation work and phase equilibrium Eq. (30), in which stresses include contributions from all dislocations. In particular, dislocations relax internal stresses but produce an athermal interfacial friction $K$. Eq. (30) was confirmed by nanoscale and scale-free PFAs, MD simulations, and CAC approach and coincide with that for the sharp interface Eq. (16). Note that in the sharp-interface approach to PT and continuum plasticity [93, 95–97, 99, 461] and corresponding FEM solutions, plastic work also does not contribute to the driving force for PT, and plasticity affects the PT by changing the stress field and increasing the athermal threshold, similar to the discrete dislocations approach. This is in contrast to the continuum approach based on the Eshelby driving force that includes plastic work [452–455]. Of course, it is tempting to include plastic work in the driving force for a PT, which would explain promoting of PTs by plastic strains. However, this is the worst what could be done, to explain experiments with wrong theory which contradicts thermodynamic consideration in [93, 95–97, 99, 461] and results of the PFA in Eq. (30).

In addition, a similar thermodynamic equilibrium Eqs. (33), averaged over the entire polycrystal or HPP, or transforming grain of a bicrystal, is valid, which represent a basis for the future development of a nano- to microscale transition via coarse-graining (see Section 7). Important point is that in discrete dislocation approach, back stresses from the dislocation pileups and grain boundaries appear automatically, increasing the shear stresses much above the critical resolved shear stress for a single crystal. Also, shear stress in nanograined materials should be controlled by dislocation nucleation rather than propagation, because after nucleation at the grain boundary dislocations can easily propagate and disappear at the opposite grain boundary. Macroscopically, this is expressed in significant increase in $\sigma_y$ with reduction in the grain size according Hall-Petch effect, until reverse Hall-Petch relationship takes over for very small grain size. All this increases the contribution of shear stresses to the phase equilibrium Eqs. (30) and (33). The bottom line is that if there is essential growth away form the nucleating defect, it should be governed by rules of stress-induced PTs. This is still missing point in the current microscale kinetic equations (Section 7), which must be resolved. Both coherent interfaces and semicoherent interfaces are obtained in simulations, similar to experiments [21, 33]. However, problem of retaining metastable HPP at normal pressure within PFA remain unsolved and must be attacked in future. Note that currently more advanced large-strain PFA models and 3D simulations for PTs appeared at the nanoscale [520, 529, 530, 615–617] and in scale-free versions [531, 532]. Their combination with dislocation evolution along with 3D crystallography and solutions should lead to revealing new features of the interaction of PT and discrete plasticity and promoting PTs by plastic flow. However, it is not trivial and computationally expensive to expand contact problem formulation for dislocation pileups for 3D. Problem of intersection of different slip systems remains open as well. Different mechanisms of promotion of PTs/CRs should be found for noncrystalline or highly disordered materials.

To promote the dislocation pileup-based nucleation mechanism, one has to produce strong obstacles for dis-location motion, like grain and twin boundaries. For grain boundaries, this is achieved by producing nanograin initial material by SPD. Alternatively, starting material can be compacted nanopowder (like for Si [81]), leading to high-angle grain boundaries. It is important to suppress or eliminate the stress-induced or PT-induced [430] grain growth during the PT/CR [618, 619].

The above results may serve as a nanoscale basis for the developing defect-induced material synthesis of known and new HPPs under moderate and low pressures. Instead of increasing pressure, one can increase plastic straining and successively fill the material with strain-induced defects (mostly, dislocation pile-ups) with high concentrators for all $\sigma_{ij} \sim N$. Deviatoric stress is limited by the theoretical strength $= (10 - 100)\tau_y$. Such an unique, highly deviatoric, stress state at the limit of lattice stability may lead to new phases inaccessible under different conditions and to drastic reduction in PT pressure. This is similar, to some extend, to the successive local melting of materials by a moving laser beam in the additive manufacturing processes.

## 7. Microscale kinetics for strain-induced phase transformations and chemical reactions

### 7.1. Thermodynamically consistent strain-controlled kinetic equation

It became clear from the earlier studies that $c$ is a growing function of the plastic shear strain; When the straining stops, PT and reaction stop as well. This was reported, in particular, in [330] for semiconductor-metal PT in $InTe$, cubic-orthorhombic PT in $BaF_2$ [376], and for various materials in [23, 64]. A similar conclusion was made for various CRs, e.g., for the polymerization of maleic anhydride, trioxan, acrylamide, and methacrylamide [14, 16]. Later in situ XRD experiments with DAC and RDAC confirmed this for PTs in BN [74, 86, 87], iron [76], Zr [80, 108, 109], Si [81], and olivine [83]. Consequently, the kinetics of strain-induced PTs is independent of time, and plastic strain is a time-like parameter. The simplest formal strain-controlled kinetic equations were suggested in [14, 16] in terms of shear and can, in principle, be fitted to experiment. During HPP, $c$ during $\alpha - \omega$ PT was post-mortem measured versus $\gamma$ for one of the radii [29]. The pressure effect was not incorporated, and it was not clear whether the kinetics is applicable for other radii. Kinetics of trimerization reaction during ball milling was determined in terms of the number of balls impacts in [118]. More generally, strain-controlled kinetic equation could be expressed as $\frac{dc}{dq} = f(q, p, c)$, and the effects of $\boldsymbol{\varepsilon}_p$ mode and path, and $\boldsymbol{s}$ are neglected. As it is mentioned in Section 4.5, the microscale kinetics should be formally the same for PTs and CRs.

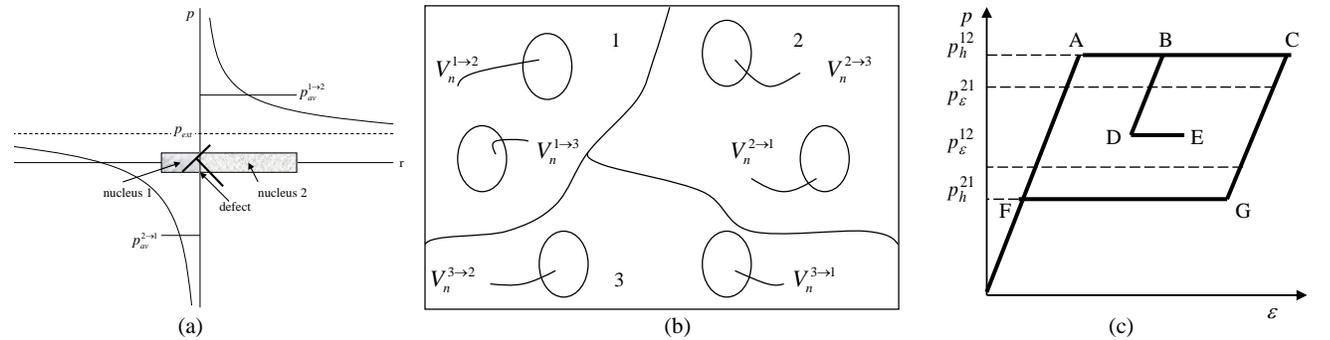

**Figure 44:** (a) Schematic of a typical strain-induced defect (superdislocation) producing strong concentration of stresses of opposite signs from different sides of the defect, decaying rapidly away from the defect. Stress concentrators produce a nucleus of phase 2 within phase 1 and phase 1 within phase 2. The transformation work is proportional to the pressure $p_{av}^{1\to 2}$ averaged over nucleus 2 and $p_{av}^{2\to 1}$ averaged over nucleus 1. (b) Schematic of all possible strain-induced PTs for three-phase system. (c) Pressure - compressive volumetric strain plot for hydrostatic loading, showing PT $1 \to 2$ along the line AC and PT $2 \to 1$ along the line GF. Strain-induced PT $1 \to 2$ proceeds along the line DE. Reproduced from [378].

The first mechanism-based kinetics for strain-induced PTs was derived in [100, 102] for two-phase material and generalized in [378] for $n$ phases. This was not a strict nano- to macroscale transition. Instead, all that was learned in [100, 102] for nucleation at dislocation pileup was conceptually introduced in the thermodynamic treatments at the microscale. Thus, a typical nucleating defect, like a superdislocation, was considered (Fig. 44a). It generates very large stresses, which rapidly reduce with distance from the defect. These stresses are compressive on one side of the defect and tensile on another, thus promoting simultaneously nucleation of the HPP within LPP and LPP within HPP. Thus, kinetic equation should have at least two terms, one for the direct and another for the reverse PT. For $n$-phase transformations, kinetic equation contains $n-1$ contributions from all other phases to the considered phase and $n-1$ contributions from the considered phase to all other (Fig. 44b). As the starting point, the PT criterion (15) was used for a small nucleus that appears during the small plastic strain increment $\Delta q$, and transformation work $w_t$ was decomposed into work of the external pressure $p_{ext}$ and of the remaining deviatoric stresses and internal pressure. The remaining part was approximated as a decreasing function of $\frac{dc}{dq}$. Indeed, this part of $w_t$ is proportional to the area enclosed between the overpressure $p - p_{ext}$ curve and the nucleus 2, since $fvep_t$ is constant within nucleus. Due to the reduction of stresses with distance from the defect, the larger nucleus (i.e., $\Delta c$) is, the smaller $p_{av}^{1\to 2}$, and transformation work is. Thus, the transformation work reduces with growing $\frac{\Delta c}{\Delta q} \simeq \frac{dc}{dq}$, which was included in thermodynamic criterion (15). Such dependence was also deduced in [100, 102] based on FEM results in [98] on the nucleation in TRIP steels at the shear-band crossing.

Resolving for direct and reverse PT, such a PT criterion for $\frac{dc}{dq}$ results in the kinetic equation

$$\frac{dc}{dq} = \frac{k_d (1-c)^\zeta}{c(\sigma_{y1}/\sigma_{y2})^w + (1-c)} \left(\frac{p - p_\varepsilon^d}{p_h^d - p_\varepsilon^d}\right)^\chi H\left(p - p_\varepsilon^d\right) - \frac{k_r c^\upsilon}{c + (1-c)(\sigma_{y2}/\sigma_{y1})^w} \left(\frac{p_\varepsilon^r - p}{p_\varepsilon^r - p_h^r}\right)^m H\left(p_\varepsilon^r - p\right). \quad (37)$$

Here, the Heaviside unit step function $H$ ($H(x) = 1$ for $x \geq 0$; $H(x) = 0$ for $x < 0$) ensures that the term for the direct PT contributes at $p > p_\varepsilon^d$ only, and the term for the reverse PT is nonzero for $p < p_\varepsilon^r$; $k_d$, $k_r$, $w$, $\chi$, $\zeta$, $\upsilon$, and $m$ are material constant. Terms proportional to $1-c$ and $c$ are included to stop the PT when the parent phase disappears. The terms with $\sigma_{y1}/\sigma_{y2}$ are based on a simple strain partitioning model that takes into account that plastic strain is larger in a weaker phase. Nanoscale determination of $p_\varepsilon^d$ was described in Section 6; similar treatment can be done for the $p_\varepsilon^r$. Since $p_h^d - p_\varepsilon^d$ and $p_\varepsilon^r - p_h^r$ are treated as constants, they can be included in $k_d$ and $k_r$, respectively. This is especially important for HPP that do not appear under hydrostatic loading, like Si III [81] or ringwoodite [83]. However, keeping these terms makes analysis below more informative.

Independence of the kinetic Eq. (37) of time has two sources. First, barrierless nucleation occurs at the ps-ns time scale and does not require waiting time for proper thermal fluctuations. Second, growth is arrested fast because of reducing stresses away from the defect. If time for the experimental measurements (e.g., collecting XRD patterns) is larger than the growth time, this resembles an instantaneous PT. When straining stops and there is no low-temperature dislocation creep during the observation time, new defects and HPP nuclei do not appear, and the interface propagation is thermodynamically prohibited. Fig. 44c shows pressure-volumetric strain curve for hydrostatic loading-unloading (FACGF) and strain-induced PT along line DE and correlates with Figure 18.

### 7.2. Stationary solution for two-phase system

1. For $p_\varepsilon^{12} > p_\varepsilon^{21}$, the effect of plastic strain on PT is relatively weak. PT does not occur for $p_\varepsilon^{21} < p < p_\varepsilon^{12}$, and $c$ remains steady. For $p > p_\varepsilon^{12}$ (or $p < p_\varepsilon^{21}$) the direct (or reverse) PT occurs only till completion.

2. For $p_\varepsilon^{12} < p_\varepsilon^{21}$, the effect of plastic strain on the PT pressure is stronger. For $p \geq p_\varepsilon^{21}$ (or $p \leq p_\varepsilon^{12}$), the direct (or reverse) PT takes place only till completion (or $c_s = 0$) For $p_\varepsilon^{12} < p < p_\varepsilon^{21}$, direct and reverse PTs occur simultaneously. For $\zeta = w$, an analytical stationary solution to Eq.(37) is

$$c_s = \frac{1}{1 + M\frac{(1-\tilde{p})^{m/\zeta}}{\tilde{p}^{\chi/\zeta}}}; \quad \tilde{p} = \frac{p - p_\varepsilon^d}{p_\varepsilon^r - p_\varepsilon^d}; \quad M = \left(\frac{\sigma_{y1}}{\sigma_{y2}}\right)^{w/\zeta} \left(\frac{k_r}{k_d}\right)^{1/\zeta} \frac{(p_h^d - p_\varepsilon^d)^{\chi/\zeta}}{(p_\varepsilon^r - p_h^r)^{m/\zeta}} \left(p_\varepsilon^r - p_\varepsilon^d\right)^{(m/\zeta - \chi/\zeta)}, \quad (38)$$

which is presented in Fig. 45a. Thus, for $M = 1$, corresponding, in particular, to equal material parameters for both phases and $m = \zeta = \chi = 1$, one has $c_s = \tilde{p}$. For small $M$ and limit $M \to 0$ (in particular, much stronger

HPP and/or suppressed kinetics of the reverse PT), one has $c_s \to 1$. For large $M$ and limit $M \to \infty$, one has $c_s \to 0$. For $c < c_s$ (or $c < c_s$), direct (or reverse) PT should take place for straining under constant pressure in the range $p_\varepsilon^d < p < p_\varepsilon^r$. Since stationary solution is unambiguous, it is independent of the loading path. Several other important implications follow from the analysis of Eq. (38).

(a) Since for some materials (e.g., for $\alpha \leftrightarrow \omega$ PT in Zr and Ti at pressure of 2-2.5 GPa [64, 322] and $B1 \to B2$ PT in $KCl$ at 1.8 GPa [62, 64]), pressures in RDAC for direct and reverse strain-induced PTs were the same, it was concluded that this is $p_e$. However, this is not true. Indeed, if the initial state corresponds to a steady state $\bar{c}_{2s}$ in Fig. 45a, then any infinitesimal pressure growth (or reduction) and plastic deformation will result in LPP→HPP (or HPP→ LPP) PT and zero pressure hysteresis. This can be achieved for any $p_\varepsilon^d < p < p_\varepsilon^r$, implying that zero-hysteresis has nothing to do with $p_e$. This is not surprising, because $p_e$ is not present in Eqs. (37)-(38) and, therefore, cannot be found from the macroscopic strain-induced experiment. However, as was found with PFAs, $\boldsymbol{\sigma}_e$ (including all internal stresses) can be found at the points of the interface and from the stresses averaged over the macroscopic sample with uniform or periodic boundary conditions (Sections 6.6.3 and 6.7).

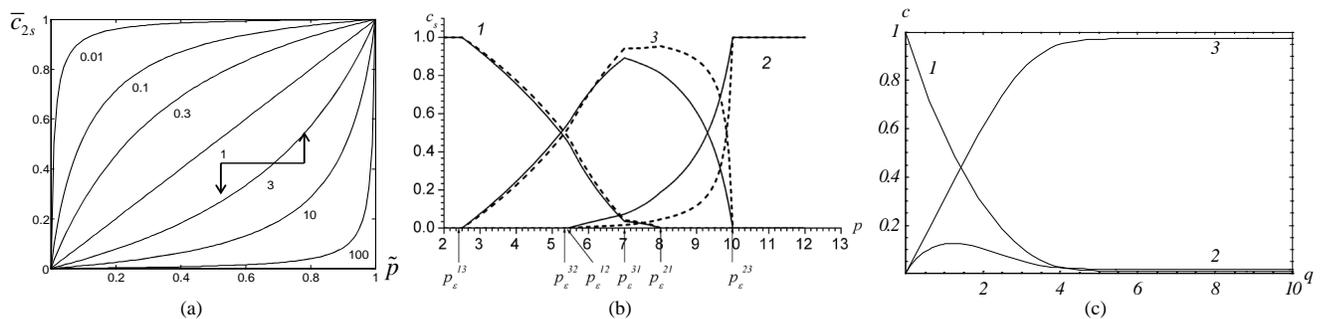

**Figure 45:** (a) Stationary volume fraction of the HPP 2 $\bar{c}_{2s}$, within the mixture of phases 1 and 2, and vs. the dimensionless pressure $\tilde{p}$ for three-phase system with non-transforming phase 3. Numbers near curves show $M$ values. (b) Stationary solution for the volume fractions of Si-I (1), Si-II (2), and Si-III (2) for strain-induced PTs for different pressure ranges. Solid lines are for $S_{31} = 1.33$, $S_{21} = 0.53$, and $S_{23} = 0.40$, and dotted lines are for $S_{31} = 1.20$, $S_{21} = 0.10$, and $S_{23} = 0.08$, where $S_{ij} = \frac{\sigma_{yi}}{\sigma_{yj}}$. (c) Kinetics of strain-induced PTs between three Si phases at $p = 7$ GPa, $S_{31} = 6.0$, and $S_{21} = 0.533$. Si-II appears at small strains but disappears at larger strains. Reproduced from [378].

(b) For comparable $\sigma_{yi}$ and kinetics of the direct and reverse PT, and for $M \leq 1$, relatively small $\bar{c}_{2s}$ can be obtained at a low pressure around $p_\varepsilon^d$. A significant $\bar{c}_{2s}$ can be produced close to $p_\varepsilon^r$ only. Large $\bar{c}_{2s}$ or complete PT can be achieved at low pressure only if the reverse PT kinetics is suppressed (i.e., for a small $M$). Another condition for a small $M$ is $(\sigma_{y2}/\sigma_{y1})^{w/\zeta} \gg 1$. Then, plastic strains localize in the weak LPP and promote the appearance of hard phases, which deform and, consequently, transform back much less. This is the case for strain-induced PT to diamond [64, 69, 79], cBN and wBN [74, 86, 87, 328] from the much weaker graphite or graphite-like BN. Even for single-phase strong HPP, it is difficult to start strain-induced PT to LPP because plastic strain is low or very localized, like for diamond to graphite PT obtained by indentation with a diamond indenter [542, 543].

(c) Realistically, complete PT can be obtained for $p > p_\varepsilon^d > p_\varepsilon^r$ only, when the reverse PT is impossible. For example, $\alpha - \omega$ PT in ultra-pure Zr, with $p_\varepsilon^d = 1.2$ GPa, completes at 2 GPa in RDAC for rotations by $20^o$ and $40^o$ (Fig. 5 in [80]). Even at rotation in RDAC by $180^o$ at 0.2 GPa reverse $\omega - \alpha$ PT was not observed in situ in [80].

(d) PT is promoted by plastic straining at $p > p_\varepsilon^d$ only. This explains why large plastic compression of materials does not cause the PT, but when followed by small shear at higher pressure, it essentially promotes PT. This question was raised in [92] based on available experiments at that time, experiments in which shear decreased the PT pressure in comparison with compression for PT in KCl, PbTe, InTe, Ge, and Si in [62, 330, 332, 383, 620]. It was mentioned that plastic strain for $p < p_\varepsilon^d$ suppresses PT because of strain hardening and growth of the athermal threshold $K_d$ and, consequently, $p_h^d$. However, as it was found later, $p_\varepsilon^d$ reduces with increasing plastic strain for Zr and olivine (which promotes the PT) until it reaches a steady state [80, 83]. Criticism of the statement that the shear decreases the PT pressure compared to plastic compression is presented in Section 7.5.2.

The existence of a steady two-phase solution (n-phase solution in [378], Section 7.3) and its independence of the initial state and straining path was confirmed in SPD experiments in [35, 36] by observed incompletion of various PTs and independence of $c_s$ of the processing path (principle of equifinality). Also, the steady solution with incomplete PT in [35, 36] implies that $p_\varepsilon^{12} < p_\varepsilon^{21}$ and these PTs are significantly affected by plastic straining. Another reason for the incomplete PTs may be nonuniform stress-strain fields in a sample deformed in DAC or RDAC.

*7.3. Analysis for three-phase system*

Eq. (38) is also valid for three-phase system with the nontransforming component 3, if volume fraction $c_{2s}$ is determined within 1+2 mixture. Thus, the steady solution is independent of the presence of nontransforming component. However, a nonstationary solution, especially at the initial stage, is essentially affected by the ratio $S_{13} = \frac{\sigma_{y1}}{\sigma_{y3}}$. Namely, adding stronger nontransforming component increases $\dot{c}$ by localizing plastic strain in the LPP, but adding a weaker component reduces $\dot{c}$ by localizing plastic strain in the nontransforming component and reducing it in phase 1. More intense plastic straining within stronger matrix also reduces $p_\varepsilon^d$ by reducing grain size and increasing defect densities, as it was observed for Zr [80, 82, 89], Si [81], and olivine [83].

This result explains known experiments that a matrix with $\sigma_y$ higher than that for reagents significantly reduces the reaction initiation pressure and allows us to run the reactions which otherwise did not start [15, 382] (Section 3.4). In particular, adamantane matrix reduces the initiation pressure for polymerization of naphtaline from 4 to 2 GPa and of succinic acid dinitrile from 10 to 3 GPa; for decomposition of $W(CO)_6$ from 5 to 2.5 GPa and of $Mo(CO)_6$ from 7 to 3.5 GPa [15]. Inversely, a matrix with lower $\sigma_y$ decelerates the reaction [15, 382]. The problem was qualitatively resolved in [100, 102, 378]. However, the quantitative rule found in [15, 382] that the rate constant for reactions in organic compounds within nonreactive matrix depends linearly on $\sigma_y$ of a matrix remains to be explained. Also, essential portion of the promoting effect of the nontransforming matrix with higher $\sigma_y$ or suppressing effect of the matrix with lower $\sigma_y$ may come from the pressure growth or decrease at the center of the sample during torsion at the fixed applied force Section 8, since only averaged pressure was reported in [15, 382].

Based on experimental data from [332, 480, 620] and assumed data for unavailable properties, modeling of strain-induced kinetics for PTs between phases I, II, and III in Si and Ge was performed in [378]. Fig. 45b shows stationary solutions for the volume fractions of phases vs. pressure for two sets of the selected ratios $S_{ij}$. In addition to single-phase regions, for Si I for $p \leq p_\varepsilon^{13}$ and Si II for $p \geq p_\varepsilon^{23}$, there are two two-phase regions. For $p_\varepsilon^{13} \leq p \leq p_\varepsilon^{32}$, $c_3$ increases with pressure at the expense of $c_1$ before appearance of phase II. For $p_\varepsilon^{21} \leq p \leq p_\varepsilon^{23}$ and a reducing pressure, $c_3$ increases at the expense of $c_2$. All three phases coexist for $p_\varepsilon^{32} \leq p \leq p_\varepsilon^{21}$. In the narrow range $p_\varepsilon^{32} \leq p \leq p_\varepsilon^{12}$, the PT $1 \rightarrow 3 \rightarrow 2$ takes place with growing $c_{3s}$ and very small increment of $c_{2s}$. For $p_\varepsilon^{12} \leq p \leq p_\varepsilon^{31}$, direct PT $1 \rightarrow 2$ also occurs. Both PT paths ($1 \rightarrow 3 \rightarrow 2$ for lower pressure range and $1 \rightarrow 2$ for higher pressures) were found in experiments in [332, 620], while under hydrostatic loading, Si III does not appear at all. Since above $p_\varepsilon^{31}$ and then $p_\varepsilon^{21}$, the reverse PT to phase 1 from phase 3 and then 2 is getting impossible, phase 1 disappears at $p \geq p_\varepsilon^{21}$. Comparing the dashed and solid lines in Fig. 45b for different $S_{ij}$, one can conclude that $c_{3s}$ reaches its peak exceeding 0.9 at $p_\varepsilon^{31}$ for the solid line and at $p_\varepsilon^{21}$ for the dashed line, respectively. Since $S_{31}$ is increased slightly, increase in $c_{1s}(p)$ for $p < p_\varepsilon^{31}$ is small. Large increase of $S_{21}$ by a factor of 5.3 significantly suppresses PT $I \rightarrow II$ and Si-I transforms mostly to Si-III. Similarly, large decrease in $S_{23}$ by a factor of 5 significantly suppresses PT $III \rightarrow II$.

As can be seen from Fig. 45c, $c_2$ reaches the maximum at small $q$ and then reduces to a value not detectable in the experiment. Should phase II be unknown, small strains could reveal it, but large strains do not. This could be used for the *searching unknown phases* for various multiphase systems. Thus, the general wisdom that the larger the shear, the stronger the promotion of PT, does not work in this case; the *straining must be optimal.* Also, the critical strain required to reach the stationary solution for Si depends on the $S_{ij}$. In particular, raising $S_{31}$ from 1.333 to 6.0 drops the critical strain from $\sim 10$ to $\sim 5$.

Note that in experiments in [332, 620], the optical and electric resistivity measurements, without in situ XRD, where utilized to detect the PTs; ruby was used to determine pressure, and $c_i$ were not measured. The PT sequence $Si - I \rightarrow III \rightarrow II$ was claimed. In situ experiments in [81] do not confirm such a sequence, and also show that PT strongly depends on the particle size (Section 9.2). Still, data in [81] is not sufficient to complete the kinetic model, especially since strain-induced Si-XI and V appear as well. Nevertheless, results in [378] give very useful generic information on the kinetics of strain-induced PTs for multiphase systems for a model material.

### 7.4. Strain-controlled kinetics of phase transformation and 2D disordering in BN

Some developments of the kinetics of strain-induced PT were suggested in [74] for PT hBN→wBN. The PT pressure in graphite-like systems significantly grows with the degree of 2D disordering described by the concentration of TSF $s$. These faults are produced by relative displacement or rotation of two parts of the lattice in (001) planes to arbitrary positions [621]. Also, large $\varepsilon_t$, in combination with applied $s$, cause large TRIP (see Section 3.7), which in turn promotes the strain-induced PT.

The kinetics for strain-induced PT for this system was derived in a similar way, as in [100, 102], but with several changes. Plastic strain was characterized by the angle of anvil rotation $\phi$ instead of $q$ to connect directly to some measurements. Reverse PT did not occur and was excluded. The $p_h^d$, was accepted to be a linear increasing function of $s$ and $c$, which was calibrated. An additional contribution to the thermodynamic driving force due to producing the nucleating defects by TRIP (similar to traditional plasticity) was mimicked by the linearly increasing function of $c$, which almost compensated the increase in $p_h^d$ with $c$. All these resulted in the following kinetic equations

$$\frac{dc}{d\phi} = J - bc - N\phi; \qquad s = s_0 + m(\phi - \phi_0) + n(c - c_0) \tag{39}$$

with material parameters $J$ (which linearly increases with pressure), $b > 0$, $N > 0$, $m > 0$, and $n > 0$; $b$ and $N$ include $n$ and $m$, respectively. TSF concentration $s$ linearly increases with applied plastic strain $\phi$ and TRIP, which is proportional to $c$. That is why the right side of Eq. $(39)_1$ also linearly reduces with $\phi$ and $c$ down to zero, leading to arrest of the PT. During the PT, the contribution of TRIP to $s$ is much more pronounced than due to applied plastic strain, like in experiments in [74], which actually led to revealing TRIP. Since $s$ is proportional to the plastic strain (traditional plus TRIP), it was suggested to use $s < 1$ as a measurable (in situ with XRD) characteristic of plastic strain.

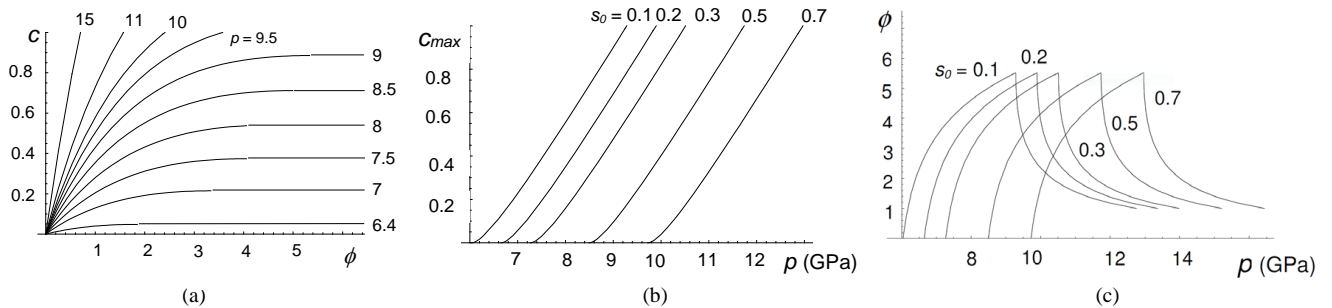

**Figure 46:** (a) Calculated volume fraction of wBN $c$ versus rotation angle of an anvil $\phi$ for $s_0 = 0.107$ and different pressures placed in GPa near the curves. (b) Maximum volume fraction of wBN versus pressure for different initial concentrations of the TSFs. (c) Rotation angle of an anvil necessary to reach $c_{max}$ versus pressure for various initial concentrations of TSF. Reproduced from [74].

Analysis of the analytical solution to Eq. (39) is presented in Fig. 46. For any pressure, kinetics tends to steady state; it reaches steady state before completion for $p \leq 9$ GPa, and completes before reaching steady state for $p \geq 9.5$ GPa (Fig. 46a). For all considered loadings, $s \leq 0.4$. Condition $J - bc - N\phi = 0$, along with solution for $c$, determines the maximum volume fraction of wBN $c_{max}$ and rotation of anvil $\phi$ necessary to reach this maximum versus $p$ and $s_0$, see Fig. 46b and c. It is clear that with increasing initial disordering pressure required to start and run, the PT increases. Points with $c_{max} = 0$ determine $p_\varepsilon^d(s_0)$.

The growing branch of the curves $\phi(p)$ is due to increase in $c_{max}(p)$. The peak points and the decreasing portions of the curves $\phi(p)$ correspond to $c_{max} = 1$. For any $s$, $\phi_{max}$ is smaller than 5.5 radians or $315^o$; again, the traditional wisdom, the larger plastic strain, the better, is not applicable because of increasing 2D disordering.

However, for completely disordered hBN, when disordering no longer increases, the larger $\phi$ is, the better the PT progress. Indeed, a highly disordered hBN was transformed to wBN at large shear at 6.7 GPa, while no PT was observed under hydrostatic loading up to 52 GPa in [77]. The above theoretical results are in qualitative agreement with the limited experiments in [74]. They also allowed to correctly interpret a paradoxical result that for PT hBN→wBN that $p_\varepsilon^d \simeq p_h^d \simeq 10$ GPa. While $s$ does not change under hydrostatic loading, it increases during plastic straining, thus suppressing the PT, which compensates the promoting effect of the strain-induced defects. However, in the modeling, for the same $s$, plastic shearing indeed lowers the PT pressure by 3 times for the beginning and 4 times for finishing the PT compared to the hydrostatic conditions.

### 7.5. Experimental confirmation of the strain-induced kinetics for Zr
### 7.5.1. Experiments with traditional diamond anvils

It took 16 years before the kinetic Eq. (37) was experimentally confirmed and quantified in [80] for $\alpha - \omega$ PT in strongly predeformed ultra-pure Zr during compression and then torsion in RDAC. The reverse PT did not occur at torsion, even at 0.2 GPa, and was not included in the simplified equation:

$$\frac{dc}{dq} = k\frac{(1-c)^a B}{c + (1-c)B} \frac{p_\alpha(q) - p_\varepsilon^d}{p_h^d - p_\varepsilon^d}; \quad p_\alpha > p_\varepsilon^d; \quad B = \left(\frac{\sigma_y^\omega}{\sigma_y^\alpha}\right)^w. \tag{40}$$

Here $p_\alpha(q)$ is the pressure in the $\alpha$ phase - accumulated plastic straining path for material particles. It was assumed $a = 1$, i.e., this is the first-order kinetics (neglecting in this classification a weak $c$-dependence due to $B \neq 1$).

Rietveld refinement [622, 623] using GSAS II [624] and MAUD [260] softwares of the in situ obtained XRD patterns was used for determining $c$, and lattice and texture parameters of $\alpha$ and $\omega$-Zr; pressure distributions in each phase were determined using volume of crystal cell and EOS under hydrostatic loading; sample thickness profile was determined in situ utilizing X-ray absorption. Pressure $p_\alpha$ was used in Eq. (40) because PT occurs within $\alpha$ phase and it is measurable directly; however, using pressure in mixture $p$ instead does not make essential difference. Since with axial diffraction all measured parameters are determined averaged over the sample thickness, $q = ln(h/h_0)$ has a similar sense and was determined at the sample center, where this equation is applicable. Eq. (40) was also checked at the sample center only, i.e., for approximately uniaxial compression without shears, even during the torsion. Evolution of experimental radial distributions of pressure in Zr phases and $c$ is presented in Fig. 47a. Based on this and two other experimental runs, the following conclusions have been derived.

Pressure $p_\varepsilon^d = 1.2 \pm 0.2$ GPa was the same for all three loading programs (without and with torsion) and all radial points where $\omega$-Zr was just observed; it is by a factor of 4.5 lower than $p_h^d = 5.4$ GPa and even well below than $p_e = 3.4$ GPa. Since the plastic strain tensor, its mode, and path vary significantly with radius (with shears growing with radius and zero shears at the center, see FEM simulations [110, 125, 126, 251, 360]), we arrive at the first rule: *for strongly plastically predeformed till steady state Zr, $p_\varepsilon^d$ for $\alpha - \omega$ PT is independent of the plastic strain tensor, its mode, and path.*

Because of the independence of the $p_\varepsilon^d$ of the radius and load in Fig. 19b and c, the same rule is valid for I→II, and II→III PT in PbTe in [358] and II→III PT in $C_{60}$ in [66]. The immediate consequence of this rule is a lack of a fundamental difference between the strain-induced PTs under *compression in DAC and shear in RDAC* in terms of reduction in PT pressure, physical mechanisms, and modeling, provided that steady $\sigma_y$ (and microstructure) are reached. Contrary to conventional beliefs [5–7, 23, 31, 32, 64], not only plastic shear, but all plastic straining paths produce the same effect. Note that the author's group used this result during the COVID-2019 time, when in situ XRD experiments could be performed remotely at APS with traditional DAC but not with RDAC. While this result drastically broadens the community that can study strain-induced PTs, RDAC has 3 advantages over DAC: it allows (a) (almost) independently controlling pressure and plastic strain, (b) much larger, unlimited shears, and (c) not only initiating, but also completing the PT at low pressure.

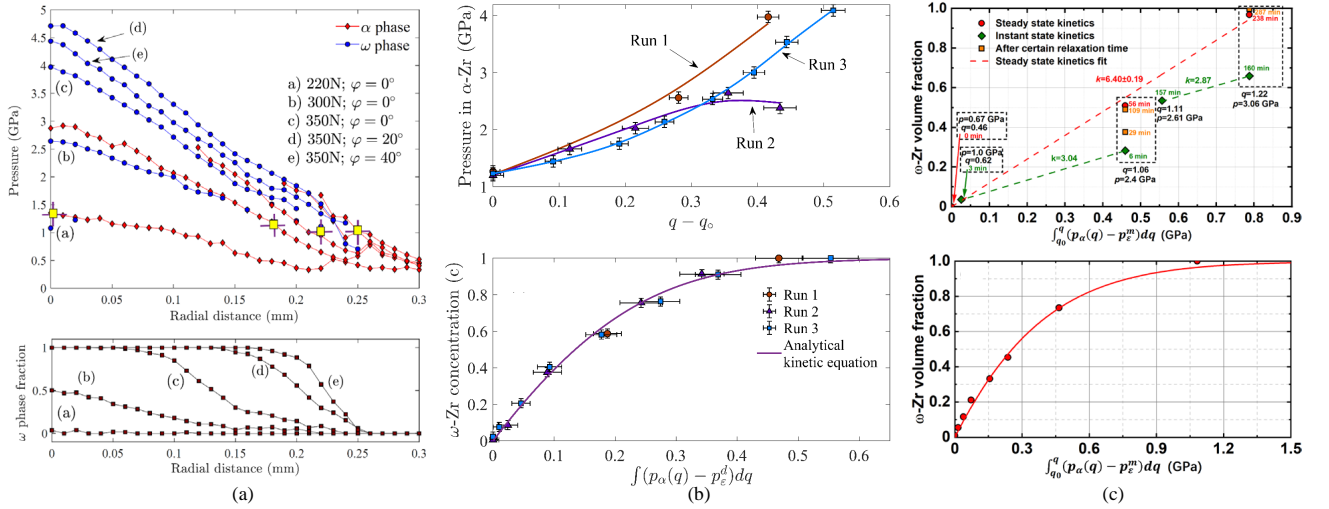

**Figure 47:** (a) Evolution of experimental radial distributions of pressure in ultra-pure Zr phases (top) and phase fraction of $\omega$-Zr (bottom). Inset presents applied force and angle of an anvil rotation $\phi$ [80]. Squares show the minimum pressure $p_\varepsilon^d$ for different radii, which is practically independent of the radius, and, consequently, $\boldsymbol{\varepsilon}_p$ and $\boldsymbol{\varepsilon}_p^{path}$. (b) Three experimental pressures - plastic strain $q - q_0$ loading paths (top); all experimental points from these paths are close to the analytical kinetics of strain-induced $\alpha - \omega$ PT versus weighted plastic strain-related parameter $I$ in ultra-pure Zr (bottom) [80]. (c) (Top) The phase fraction of $\omega$ phase versus $I$ and time during compression of commercially-pure Zr with rough diamond anvils. Red circles correspond to reaching the steady-state in time. Green diamonds are measured immediately after compression; they are shifted by time-dependent growth at fixed $q$ up to a steady state; time labels designate time from the beginning of the first measurement. Lines are theoretical curves for zero-order kinetics. (Bottom) Kinetics of $\alpha - \omega$ PT in commercially-pure Zr compressed with smooth diamond anvils. Line is the theoretical curve for the first-order kinetics. Reproduced from [82].

This rule, however, may not be universal for all material systems. For example, for strongly anisotropic crystals, like graphite and graphite-like BN, plastic shear is much different from other deformation modes and causes the largest reduction in PT pressure for PT from graphite to hexagonal (from 20 to 0.4 GPa) and cubic (from 70 to 0.7 GPa) diamonds or from highly disordered hBN to wBN (from more than 52 to 6.7 GPa).

Kinetic Eq. (40) can be integrated, resulting in function of $c$ on one side and weighted plastic strain-related parameter $I = \int_{q_0}^{q}(p_\alpha(q) - p_\varepsilon^d)dq$ on the right side, where $q_0$ is the value of $q$ at the beginning of the PT. Three experimental $p$ versus $q - q_0$ loading paths at the sample center are shown in Fig. 47b (top). All experimental points from these 3 paths are close to the analytical kinetics of strain-induced $\alpha - \omega$ PT in $c - I$ coordinates, Fig. 47b (bottom). This strongly confirms the theoretical kinetic equation and allows determination of all material parameters in it. Since preliminary plastic straining by multiple rolling was different for each of the 3 samples and led to different $q_0$ (but to the same steady hardness), results in Fig. 47b show that $p_\varepsilon^d$ and the entire *kinetic equation are independent of plastic strain $q_0$ below $p_\varepsilon^d$, i.e., before the PT. Also, strain-controlled kinetics is independent of pressure - plastic strain $q$ path.*

### 7.5.2. Explanation of misinterpretation of PT pressure for comparison and shear

For Zr annealed in inert Ar atmosphere at $650°C$ for 2 hours, a larger $p_\varepsilon^d = 2.3$ GPa was obtained at the center during compression [80]. However, away from the sample center, after much more intense compression-torsion plastic straining, the same $p_\varepsilon^d = 1.2$ GPa as at the sample center of strongly pre-deformed Zr was obtained.

This result corresponds to the traditional wisdom [5–7, 23, 31, 32, 64] that plastic shear lowers the PT pressure as compared to the plastic compression. However, because this contradicts the above results for strongly predeformed Zr, an alternative interpretation was suggested in [80] and advanced here based on later works [81–83, 89]. The minimum pressure $p_\varepsilon^d$ decreases with growing plastic strain and corresponding growing dislocation density, and reducing grain size, until steady states in dislocation density, grain size, and $p_\varepsilon^d$ is reached. Thus, it is not surprising that for annealed sample, during compression, plastic strain is not sufficient to reach the steady microstructure, and $p_\varepsilon^d = 2.3$ GPa is higher. At the periphery, where pressure is lower and plastic strains are much larger during torsion, the microstructure reached the steady state and so did the $p_\varepsilon^d$, resulting in the same $p_\varepsilon^d = 1.2$ GPa as everywhere in the strongly predeformed sample. Eq.(40) should be extended for the non-steady initial microstructures, e.g., by making all parameters dependent on $q$ evaluated from the annealed state, or, better, on the measurable current dislocation density and/or grain size.

*7.5.3. Experiments with rough diamond anvils*

Here, commercially-pure Zr was studied in [82, 89]. For traditional smooth anvils, $a = 1$ (like for ultra-pure Zr), $p_\varepsilon^d = 1.36$ GPa, and $p_h^d = 6.0$ GPa, i.e., transformation pressures are slightly higher than for ultra-pure Zr. Kinetic curve in Fig. 47c (bottom) is qualitatively similar to that for Fig. 47b (bottom), but kinetics is essentially slower.

*Promotion of grain refinement and PT.* It looks like it is typical for strain-induced PTs: as soon as a logical understanding and quantitative description is achieved, some small changes produce new phenomena. To increase contact friction between sample and anvil up to the maximum possible value equal to $\tau_y$, and reduce the relative sliding, diamond anvils with rough cullet were introduced in [81–83, 89]. Rough anvils somehow intensify plastic flow, microstructure evolution, and strain-induced PTs. Thus, in comparison to smooth anvils, crystallite size in $\alpha$-Zr before initiating the PT reduced from 65 to 48 nm, dislocation density increased from $1.26 \times 10^{15}$ to $1.83 \times 10^{15} m^{-2}$, which leads to the reduction in $p_\varepsilon^d$ from 1.36 to 0.67 GPa. This is 9 times lower than $p_h^d$ and 5 times lower than $p_e = 3.4$ GPa. The $p_\varepsilon^d$ was practically the same at the center and edge of the sample, i.e., it *is independent of $\varepsilon_p$ and $\varepsilon_p^{path}$*. However, different steady $p_\varepsilon^d$ for smooth and rough diamonds, i.e., for different steady crystallite size and dislocation density, led to the same problem as was discussed in [89, 90] and Section 2.8 for crystallite size and dislocation density: *for which classes of plastic strain $\varepsilon_p$, $\varepsilon_p^{path}$, $p$, and $p^{path}$ material exhibits one steady $\sigma_y$, crystallite size, dislocation density, and $p_\varepsilon^d$ and for which of $\varepsilon_p$, $\varepsilon_p^{path}$, $p$, and $p^{path}$ classes there are jumps from one steady microstructure, $\varphi^i(\boldsymbol{\sigma}) = 0$ surface, and $p_\varepsilon^d$ to another?*

However, the main surprise was in the discovery of *time dependence of strain-induced PT* at $q = const$. This challenged the main claim for strain-induced PTs that there is no PT progress without plastic straining. As can be seen in Fig. 47c (top), there are two limit kinetic curves versus $I$. The lower curve shows the instantaneous kinetics, measured immediately after compression. The upper curve is obtained by waiting without changing the load until steady state is reached in time. The sample thickness does not evolve during waiting time, i.e., creep-induced PT is excluded. The time-dependent growth of $c$ reaches 25-35% on the $\sim 1$ hour scale at pressure 2.4-3.1 GPa, i.e., even below $p_e$.

*Zero-order kinetics in strain.* In addition, the plastic strain-dependent part of the kinetics is a linear function of I, both for steady and instantaneous kinetics. This results in zero-order kinetics, $a = 0$, $B = 1$ in Eq.(40), with $k = 2.96$ for instantaneous and $k = 6.4$ for steady-state kinetics. The time-dependent portion is described by simplest exponential kinetics connecting instantaneous and steady-state branches for each $q$, with a characteristic time of 43.13 minutes.

PFA simulations in [380, 381], in particular, Figs. 34 and 36, have been utilized for interpretation of the time-dependent kinetics at fixed strain. Despite the very low pressure, the following processes occur after nucleation: growth in the of number of dislocations in the pileup, increasing the driving force for growth of an HPP, and in the transforming grain; growing HPP reaches the opposite grain boundary, and then thickens; the second nucleus emerges at the dislocation pileup within the transforming grain; nuclei coalesce, some dislocations disappear, reverse PT occurs, and only then are the steady dislocation and phase configurations reached. It was traditionally assumed that growth time until steady state is much shorter than the measuring time, and only steady states were detected. Due to low pressure in experiment with rough diamonds, the driving force for PT is small, growth is slower, and it may take about an hour after a strain increment instead of seconds, as was assumed. Also, it is clear that time-dependent kinetics could not be discovered in HPT experiments because postmortem studies require time for pressure release, sample preparation, and measurements. This underscores the necessity of in-situ measurements. Obtained results implied necessity of the developing advanced *combined strain- and time-dependent kinetics*, i.e., *theory for combined strain- and stress-induced PTs*.

First-order kinetics assumes nucleation of HPP nuclei and their growth within an LPP. In contrast, zero-order kinetics assumes propagation of the interfaces between fully formed HPP and LPP without a significant change in the interface area, like for broadening of the PT bands. Such a scenario was indeed observed in FEM simulation of the sample behavior in [625], but for the HPP essentially weaker than the LPP. This is not the case for $\alpha$ and $\omega$ phases of Zr, and *transition to zero-order kinetic for compression with rough diamonds remains a mystery*.

### 7.6. Coupled analytical-computational-experimental methods for Zr

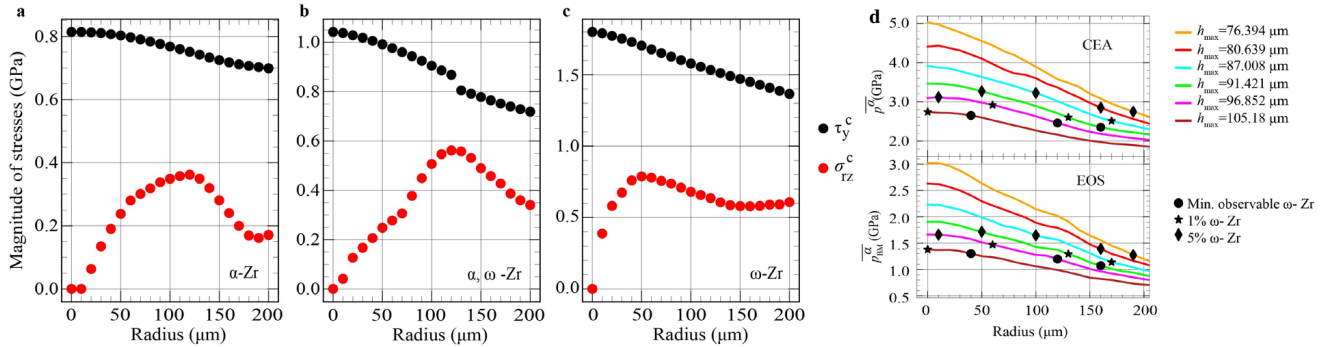

**Figure 48:** Distributions along the radius of the friction shear stress $\tau_f$ (red) and $\tau_y$ (black) for (a) for $\alpha$-Zr, (b) $\alpha - \omega$ mixture, and (c) $\omega$-Zr extracted with an analytical model and experimentally determined $\bar{E}^e_{rr} \simeq \bar{E}^e_{\theta\theta}$ distributions. (d) The radial distribution of pressure $p^\alpha$ in $\alpha$-Zr calculated with CEA (top) and hydrostatic EOS using XRD-measured volumetric strain. Reproduced from [108].

Qualitative higher progress was achieved in [108, 109] by introducing iterative CEA and CEA-FEM approaches to determine all heterogeneous fields in the sample and material parameters in the material models, i.e., material properties. The most advanced determination of the radial distribution of pressure $\bar{p}$ averaged over the sample thickness was done by utilizing the crystal cell volume measured by axial XRD and EOS defined at hydrostatic pressure [80, 82, 89, 131, 132, 257, 258, 626]. However, there is a significant difference between EOS under hydrostatic and non-hydrostatic conditions [222, 229, 477]. Also, for the axial XRD, only lattice planes almost parallel to the beam affects the measured XRD patterns. A significant error arises because elastic axial strain $\bar{E}^e_{zz}$ (and therefore stress $\sigma_{zz}$) do not contribute to the measured $\bar{p}$. To reduce error, distributions of the radial $\bar{E}^e_{rr}$ and circumferential $\bar{E}^e_{\theta\theta}$ elastic strains in $\alpha$ and $\omega$-Zr were determined in [108, 109] and used as an input for CEA. Pressure-dependence of $\sigma_y$, $\sigma_{y\alpha} = 0.82 + 0.190p$ (GPa) and $\sigma_{y\omega} = 1.66 + 0.083p$ (GPa) was determined using XRD peak broadening. The PT kinetics was not modeled in CEA-FEM, but the volume fraction was introduced homogeneously along the thickness from the experiments and evolution of all tensorial stress and plastic strain fields was determined for compression in DAC with smooth anvils.

The first step was to determine the friction stress $\tau_f$ distribution at the Zr-diamond contact with CEA. The stress field from the analytical plane-strain Prandtl's solution for compression with constant friction stress of an infinite rigid-plastic layer [476] was used as the starting point. The problem is statically determinate, i.e., determines all stress tensor component fields without involving kinematics and flow rule. It was generalized in several ways using a number of assumptions, each of which violated mechanical equilibrium conditions. To use obtained analytical solution, one needs to find $\tau_f(r)$. To utilize measured elastic strains along the radius, the nonlinear elastic rules were transformed to the modified pressure-dependent Hooke's law with consistent elastic moduli [478, 495, 496] for single crystal of each phase, then were averaged over the phase mixture. Based on experimental strong texture ($c$-axis is predominantly along the $z$ axis for $\alpha$-Zr and $r$ axis for $\omega$-Zr), these equations defined stress tensor field; equating this field to that from the analytical solution allowed determination of $\tau_f(r)$ shown in Fig. 49. It is evident that $\tau_f(r) \ll \tau_y(p)$, which shows the error in determination of $\tau_y(p)$ with smooth-DA using pressure gradient method [4, 47, 161–164] (Section 2.2).

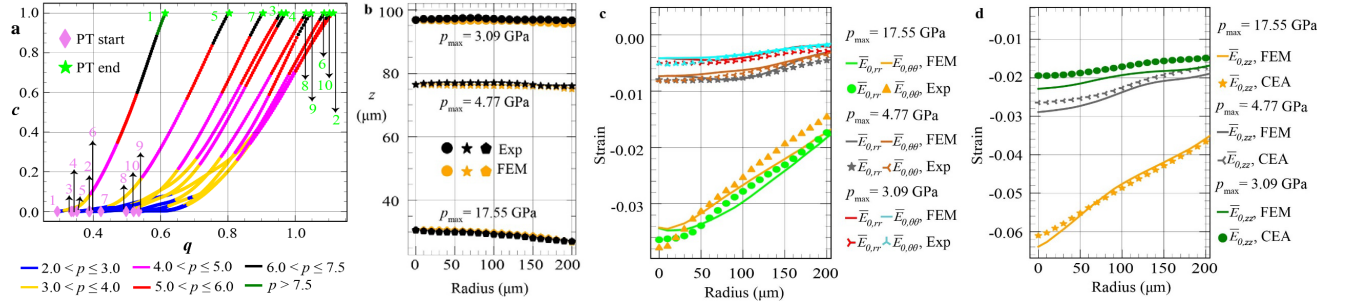

**Figure 49:** (a) CEA-FEM-based phase fraction $c$ of $\omega$-Zr versus accumulated plastic strain $q$ from the initiation until end of PT with superposed colors for pressure (in GPa) for different material points shown in Fig. 50b [109]. (b) Comparison of the sample thickness profiles from CEA-FEM and X-ray absorption experiments. (c) Comparison of FEM and experimental distributions of radial $\bar{E}^e_{rr}$ and circumferential $\bar{E}^e_{\theta\theta}$ elastic strains for 3 different maximum pressures, in $\alpha$-Zr, mixture, and $\omega$-Zr. (d) Comparison of CEA and FEM distributions of axial elastic strain $\bar{E}^e_{zz}$. Reproduced from [108].

The CEA changed $p^d_\varepsilon$ from 1.36 GPa, obtained using pressure field from the EOS without any simulations, to 2.70 GPa (Fig. 48d). Kinetic coefficient $k$ based on the fitting at the center of the sample was reduced from 11.45 to 6.14. Thus, with traditional approach, kinetic results contain significant errors. Like with pure experimental procedure in [80] (Section 7.5.1), $p^d_\varepsilon$ was found to be *independent of $\boldsymbol{\varepsilon}_p$ and its path and mode;* the entire *kinetic equation is independent of the plastic strain $q_0$ for $p < p^d_\varepsilon$ and $p - q$ path.*

In CEA-FEM, distributions of $\tau_f$ determined for experimental thicknesses are interpolated as a function of all intermediate thicknesses and implemented into FEM problem formulation as the friction stress for the contact problem, serving as the only experimental boundary conditions. Then FEM solution reproduces all tensorial stress, elastic, and plastic strain fields. Fig. 49b,c shows good correspondence between experimental and FEM-based sample thickness profiles and elastic radial $\bar{E}^e_{rr}$ and hoop $\bar{E}^e_{\theta\theta}$ strains for 3 different maximum pressures. Fig. 49d demonstrates good agreement between CEA and FEM distributions of elastic axial strain $\bar{E}^e_{zz}$. Surprisingly, despite the multiple quite strong assumptions involved in CEA, CEA reasonably matches all 2D tensorial stress component fields, especially for large $R/h$. In the following work [109], kinetic equation was upgraded to

$$\frac{dc}{dq} = k(1+\delta_1 q)(1+\delta_3 p)\frac{(1-\delta_2 c)B}{c+(1-c)B}\frac{p_\alpha(q) - p^d_\varepsilon}{p^d_h - p^d_\varepsilon}; \quad p > p^d_\varepsilon; \quad p^d_\varepsilon = \delta_4 - \delta_5 q_0; \quad B = \frac{\sigma^\omega_y}{\sigma^\alpha_y} \qquad (41)$$

with 6 fitting parameters $\delta_i$ and $k$. Here, $q_0 = q$ before PT starts; after $p$ reaches $p^d_\varepsilon = \delta_4 - \delta_5 q$, $q_0$ is determined from this equation and remains constant during the PT; thus, $p^d_\varepsilon$ is heterogeneously distributed (Fig. 50a). This heterogeneity depends on $p - q$ paths in each point: the lower $p$ and larger $q$ are, the smaller $p^d_\varepsilon$ is.

The key point was that since all fields have been calculated in the entire sample, phase fraction of $\omega$-Zr averaged over the deformed sample thickness, $\bar{c}$, was fitted to all experimental radial distributions of $\bar{c}$ by minimizing the selected error measures. These gave many more data points with much more complex and versatile $c - q$ paths (Fig. 49a) and straining paths (Fig. 50b,c). In the previous model (40), $\delta_1 = \delta_3 = \delta_5 = 0$, $\delta_2 = 1$, and $p^d_\varepsilon = \delta_4 = const$. Fig. 51a shows that this model catches the main trends (i.e., the physics) but with essential quantitative errors.

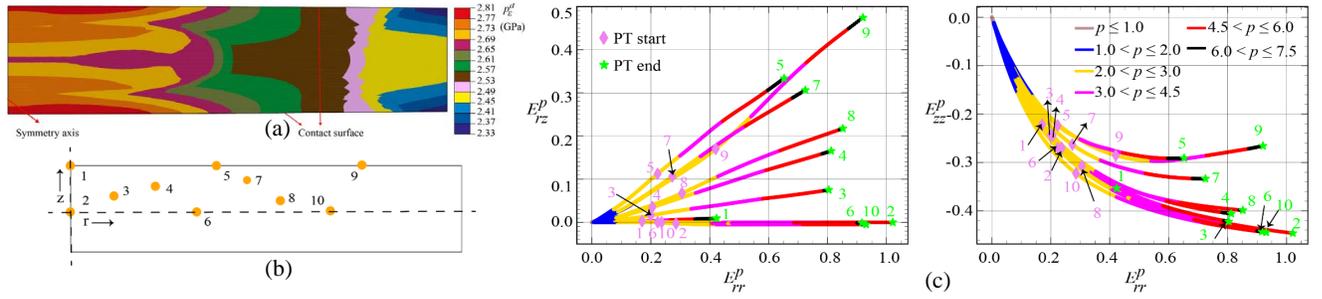

**Figure 50:** (a) Distribution of $p_\varepsilon^d$ within the sample due to heterogeneity of obtained $q_0$. (b) Cross-section of a sample in the undeformed state with the positions of the selected material points. (c) Two projections of the plastic straining paths in 3D space of shear $E_{rz}^p$ - radial $E_{rr}^p$ - axial $E_{zz}^p$ strains into planes $E_{rz}^p$-$E_{rr}^p$ and $E_{zz}^p$-$E_{rr}^p$ for material points shown in (b). Colors encode corresponding pressure evolution (in GPa). Reproduced from [109].

Condition $\delta_2 = 1$ in Eq. (40) provided $\dot{c} = 0$ for $c = 1$; however, it appeared that $\delta_2 = 0.925$ gives a better fit. Parameters $\delta_1$ and $\delta_3$ provide the simplest linear $q$- and $p$-dependence of the kinetic coefficient. However, Fig. 51b demonstrates that very good correspondence between experiments and FEM simulations is possible with $\delta_1 = \delta_3 = 0$, if $p_\varepsilon^d = 2.65 - (q_0 - 0.42)$ GPa is used. For this case, maximum absolute error in $\bar{c}$ for all points does not exceed 0.071, and integral relative error is very close to the minimal. Eq. (41) is much more reliable than Eq. (40), since it has much stronger verification for the entire sample.

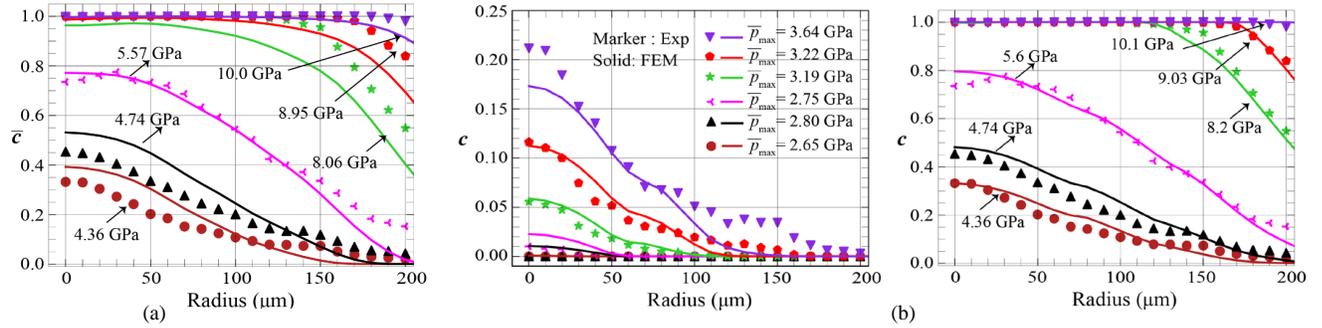

**Figure 51:** Comparison of calculated and experimental radial distributions of $\bar{c}$, designated with pressure at the symmetry axis, both averaged over the sample thickness [109]. (a) Simple model [100, 102] calibrated in [108] using data near the symmetry axis with parameters $\delta_1 = \delta_3 = \delta_5 = 0$, $\delta_2 = 1$, $k = 5.87$, and $p_\varepsilon^d = \delta_4 = 2.70$ GPa. (b) Advanced model (41) with parameters $\delta_1 = \delta_3 = 0$, $\delta_2 = 0.803$, $k = 5.20$, and $p_\varepsilon^d = 2.65 - (q_0 - 0.42)$ GPa for small and large volume fractions of $\omega$-Zr.

Number 0.42 in $p_\varepsilon^d = 2.65 - (q_0 - 0.42)$ is separated from 2.65 because at $q_0 = 0.42$ the PT initiates for the first time at the center of the sample. Constant $p_\varepsilon^d$ independent of plastic strain tensor and its path claimed in [80, 108] and linear dependence of $p_\varepsilon^d$ on $q_0$ in [109] are not conceptually contradictory. All independence statements are correct to within some scatter and range of data. Indeed, in [80] $p_\varepsilon^d = 1.2 \pm 0.2$ GPa was obtained based on parameters averaged over the sample thickness. In Fig. 50a, local values of $p_\varepsilon^d$ span from 2.33 to 2.81, and, averaged over the thickness, values span less, which gives the same absolute and smaller relative scatter than in [80]. Since data for a broad variety of straining paths in 4D space of plastic strain components and $p$ and corresponding varieties of the deviatoric stress paths is described by Eq. (41) in terms of $p$ and $q$ only, this confirms that the *kinetic equation (41) is independent of $\boldsymbol{\varepsilon}_p$, its mode and path and stress deviator $\boldsymbol{s}$, as well as $p - q$ paths.*

Presented in [109], complex and heterogeneous plastic strain tensors and their paths also confirm the validity of rules obtained in [89], i.e., that dislocation density, crystallite size, and $\sigma_y(p)$ are getting stationary and independent of $\boldsymbol{\varepsilon}_p$ and its paths. Also, it is determined in [82] that during $\alpha - \omega$ PT in Zr, crystalline size and dislocation density are unique functions of $c$, which are independent of $\boldsymbol{\varepsilon}_p$, its paths, and $p$. Eq. (41) allows one to obtain explicit kinetic equations for these parameters. Note that Eq. (41) should be generalized for the materials for which direct and reverse PTs occur simultaneously, annealed initial states, high strain rates, and 4D straining paths in RDAC.

*7.7. Summary and perspectives*

Thermodynamically consistent strain-controlled kinetic equations for PTs/CRs in two- and multiphase systems were derived that conceptually include the main consequences of the nanoscale mechanism of the nucleation at the tip of the defect with strong stress concentration (like dislocation pileup) [100, 102, 378]. Derivative of $c_i$ is evaluated with respect to $q$; in the first approximation, time does not participate in this equation, provided that for each strain increment, steady phase state is reached during the time smaller than the measurement time. Both direct and reverse PTs between phases are included. Ratio of the yield strengths participates in equations reflecting that plastic strain localizes in the weaker phases. Stationary solution to the kinetic equation revealed various important phenomena. When one PT can occur only (direct or reverse), PT can be completed. In the pressure range where direct and reverse PTs occur simultaneously, stationary volume fraction $c_s$ depends on the pressure in the range $p_\varepsilon^d < p < p_\varepsilon^r$. The larger $\sigma_{y2}/\sigma_{y1}$ and the faster kinetics of the direct PT relative to that for the reverse PT are, the larger the $c_s$ is. The following experimental phenomena are interpreted, and predictions are made:

(a) Stationary solution $c_s(p)$ explains that equal pressure for direct and reverse strain-induced PTs in Zr, Ti, and KCl [62, 64, 322] does not mean that this is $p_e$. Such an equality can be achieved for any $p_\varepsilon^d < p < p_\varepsilon^r$ if initial state corresponds to $c_s(p)$. Since $p_e$ is not present in the kinetic equation, it cannot be determined from the macroscopic strain-induced experiment.

(b) Nonstationary solution for a three-phase system with one non-transforming (inert) phase explains known experiments [15, 382] that an inert matrix stronger than reagents significantly accelerates the reaction, and a weaker matrix decelerates the reaction in organic substances. However, a stationary solution is independent of the inert matrix. Also, SPD within stronger matrix reduces $p_\varepsilon^d$, which is observed in [15, 382]. A macroscopic reason for intensification of the CR is that a stronger matrix produces a larger pressure gradient and pressure at the sample center for the same applied force (Section 8.3). Since $\sigma_{yi}/\sigma_{yj}$ is not involved in the kinetics of pressure-induced PTs but makes a nontrivial contribution to the kinetics of the strain-controlled PTs, it can be used for controlling the PT and phase selection. Generally, plastic strain-induced kinetics promotes appearance of stronger phases.

(c) Nonstationary solution for a three-phase system also shows that one of the HPPs can appear at small strains but then disappear. This could be used for searching for unknown phases for various multiphase systems. The general perception that the larger the shear is, the stronger the promotion of the PT is, fails; the straining should be optimal.

As it was mentioned in Section 2.11, discussing plasticity, the main experimental challenges in studying materials' behavior in DAC and RDAC are the strong heterogeneity of stress and plastic strain tensor fields, and only a very limited number of parameters that can be measured. This is even more important for studying strain-controlled kinetics, because plastic strain field is not measurable. The following CEA and CEA-FEM iterative approach for determination of all $\boldsymbol{\sigma}$ and $\boldsymbol{E}_p$ fields, PT kinetics, plastic properties, and contact friction rules were suggested in [108, 109]. All fields that can be measured have been measured. Physics-based models for coupled elastoplastic and transformational behavior and contact friction have been iteratively established and refined, and all material constants and functions were calibrated by fitting to some measured experimental fields and then confirmed by comparing to other measured fields. With such robust models, FEM simulations provided all fields, including components of $\boldsymbol{\sigma}$ and $\boldsymbol{E}_p$, $c$, and friction stress. In this process, distributions of the radial $\bar{E}_{rr}^e$ and circumferential $\bar{E}_{\theta\theta}^e$ elastic strains in $\alpha$ and $\omega$-Zr were measured and utilized as an input for CEA instead of pressure distribution, which essentially reduced error.

The kinetic equation was experimentally confirmed, quantified, and generalized for $\alpha - \omega$ PT in strongly predeformed ultra and commercially pure Zr, for compression with smooth and rough anvils in DAC and torsion in RDAC, and using various post-processing and analysis methods [80, 82, 89, 108, 109]. With CEA-FEM approach, for broad variety of straining paths in 4D space of plastic strain components and $p$, it was found that *kinetic equation (41) is independent of $\boldsymbol{\varepsilon}_p$, its mode and path, and stress deviator, $p - q$ paths, and also straining before initiation of the PT, provided that the stationary microstructure is reached.* The *minimum PT pressure $p_\varepsilon^d$ is also independent of $\boldsymbol{\varepsilon}_p$, its mode and path and stress deviator,* in addition to the similar independence of the steady yield strength, crystallite size, microstrain, and dislocation density (Section 2.7). Similar to the findings for microstructural parameters and $\sigma_y$, there are multiple steady values of $p_\varepsilon^d$. The main problem is to determine for which classes of $\boldsymbol{\varepsilon}_p$, $\boldsymbol{\varepsilon}_p^{path}$, $p$, and $p^{path}$ the PT starts with each specific $p_\varepsilon^d$, and for which classes of $\boldsymbol{\varepsilon}_p$, $\boldsymbol{\varepsilon}_p^{path}$, $p$, and $p^{path}$ there are changes from one steady $p_\varepsilon^d$ to another.

Before reaching steady microstructure, $p_\varepsilon^d$ reduces with increasing plastic strain and dislocation density, decreases with crystallite/grain/particle size in the region of the direct Petch-Hall effect, and increases with the grain size in the region of the inverse Petch-Hall effect [80–82, 89, 109]. After reaching steady dislocation density and crystallite size, $p_\varepsilon^d$ still slightly reduces with $q$. Since $p_\varepsilon^d$ is determined by the dislocation pileup with the largest number of dislocations $N$, i.e., by tail in distribution of $N$ in different dislocation pileups, the local $N_{max}$ may reach stationary state at greater $q$ than the averaged $N$. For different steady microstructures, $p_\varepsilon^d$ also reduces with growing dislocation density and reducing crystallite size [89].

This is the first general rule for the strain-controlled kinetics, which strongly simplifies theory and experimentation, and allows to apply this kinetics for various processes, like (a) defect-induced synthesis of nanostructured phases by SPD with HPT [23, 26, 27, 35] and ball milling [118], (b) surface processing (polishing, turning, etc.) [400, 627], (c) friction and wear [628], (d) geological processes (the deep earthquakes [83, 390, 391, 393], microdiamond appearance [79], and study of multiple PTs during plastic flow, which are currently described as pressure-induced), and (e) appearance of life in astrogeology [629–632]. CEA-FEM approach also led to the first non-trivial rules for friction shear stress between diamond and W [143] and Zr [108], laying down foundation of *high-pressure tribology*.

For hBN→wBN PT, the kinetics for PT was coupled to the kinetics for the evolution of TSF, $s$, which suppresses the PT; also, increasing $p_h^d$ with $c$ and $s$ and promoting effect of the TRIP were included [74]. Results qualitatively describe experiments in [74] and also lead to the conclusion that the plastic straining should be optimal instead of maximal. The effect of evolving microstructure, e.g., dislocation density and grain size, must be incorporated in the kinetic equation for any material when microstructure has not reached the steady state.

In situ experiments in DAC with rough diamonds on Zr revealed that due to low $p_\varepsilon^d$, significant time-dependent growth of $c$ occurs at constant $q$, violating the main claim for strain-induced PT. Thus, strain-controlled kinetics for the instantaneous and steady branches are connected by exponential time-dependent kinetics with the characteristic time of 43 min. Some rationale for the time-dependence of the strain-controlled kinetics are presented. Also, the order of the plastic strain-dependent part of the kinetics was changed from the first one to the zero order, which at the moment does not have a satisfactory explanation. Note that the results of the scale-free PFA simulations (Section 6.7) can be effectively utilized for further developing microscale kinetics.

## 8. Simulation of the Macroscopic Plastic Flow and Phase Transformations in a Sample

To analyze experiments in DAC and RDAC, extract material information from them, develop coupled experimental-FEM approaches, plan improved tests, and explain phenomena described in Section 3, one needs to have a solution for the plastic flow and strain-induced PTs in a specimen.

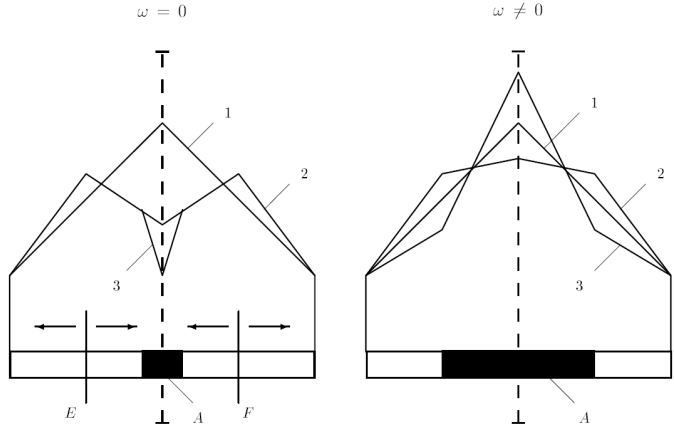

**Figure 52:** Sketch of an analytical solution for pressure distribution (a) at constant force without torsion (1 - before PT; 2 - after PT at $\sigma_{y1} = \sigma_{y2}$; 3 - after PT at $\sigma_{y1} < \sigma_{y2}$) and (b) during torsion at constant force (1 - $\sigma_{y1} = \sigma_{y2}$; 2 - $\sigma_{y1} > \sigma_{y2}$; 3 - $\sigma_{y1} < \sigma_{y2}$) [92].

*8.1. Simple analytical solution for stress-induced PTs*

The first simple analysis of PTs based on analytical solution for the unconstrained cylindrical sample twisted between rigid rough anvils (i.e., $|\boldsymbol{\tau}_f| = \tau_y$) was presented in [92]. At that time, the concept of strain-induced PTs under high pressure was unknown and criterion (15) for stress/pressure-induced PTs was utilized. Similar to Section 4.4, if transformation volumetric $\varepsilon_0$ and shear $\gamma^t$ strains are non-zero only and uniform in the HPP, and surface energy is neglected due to macroscopic size of the HPP, Eq. (15) can be simplified to

$$w_t = \bar{p}\varepsilon + \bar{\tau}\gamma^t = \Delta\psi^\theta + K, \tag{42}$$

where $\bar{p}$ and $\bar{\tau}$ are averaged over the PT region and process pressure and shear stresses. In [100, 102], the concept of strain-induced PTs under high pressure was suggested along with the kinetic Eq. (37) and an improved and more elaborated treatment of the surface sliding was implemented. Solution without the PT was discussed in Section 2.10 and will be utilized here. The main result used in [92] for explaining the promotion of PT due to torsion was that the torsion at the fixed force $P$ reduces the sample thickness. A sketch of the analytical solutions for $\gamma^t = 0$ is shown in Fig. 52. The starting point is the compressed sample with the linear pressure distribution with the maximum at the center. Let us consider PT in the central part of the sample with a phase fraction of HPP $c$. Without torsion, part of the material flows to the center of the sample. A neutral circle $EF$ with zero radial velocity can be found from the volume balance. Equilibrium Eq.(3) is applicable, but the friction stress in the region $EF$ reverses sign. It is evident that under $P = const$, pressure in the HPP and the $w_t$ in Eq.(42) decrease essentially. The larger $\sigma_{y2}$, the greater the pressure reduction in the HPP. Torsion, decreasing the sample *thickness* due to reduction in the radial friction $\tau_{rf}$, compensates the volume decrease due to PT, eliminates flow to the center, and *increases* pressure and $w_t$. This qualitatively explains why the torsion promotes the PT.

*"Reproducing" experimentally observed kinetics.* It was rationalized in [92] referring to the postulate of realizability [450] that the maximum $w_t$ and $c$ will be when infinitesimal radial flow from the center takes place. Then $\tau_{rf}$ does not reverse the sign, pressure increases monotonously toward the center and volume drop due to PT is fully compensated by the reduction in thickness. The last condition is described by

$$\varepsilon\,dc = -\frac{dh}{h} = d\varphi\Big(\Big(\frac{h_o}{h}\Big)^2 - 1\Big)^{-0.5}, \tag{43}$$

where relationship between $dh$ and $d\varphi$ from [92] (similar to Eq. (9), but less precise) was taken into account. This equation "explains" the numerous experiments relating $c$ and anvil's rotation angle [23, 64, 80, 83, 330, 376]. However, this does not explain later experiments in which the same $p_\varepsilon^d$ can be achieved during compression and torsion [80, 83].

*Pressure self-multiplication and demultiplication effects during PTs.* According to Eq.(3), for $\sigma_{y2} = \sigma_{y1}$, pressure distribution does not alter during the PT (Fig. 52). For weaker HPP, pressure decreases in the transforming region, which was observed for PT from stronger semiconducting Ge-III to weaker metallic Ge-II phase under shearing [332] and at during appearance of new $CuI-VIII'$ [359] (Section 3.5). For for stronger HPP, pressure increases at the center, despite the transformational volume reduction. This explains the pressure self-multiplication effect during PT in *KCl* [62, 64, 66], fullerene $C_{60}$ [64, 66, 71], and PbTe [383] (Section 3.5). Thus, there is no violation of the Le Shatelie principle; pressure growth is explained by a specific character of plastic flow, which is neglected in classical thermodynamics. Obtained solution also contributes to the phase selection rule. Let two different phases appear during the PT that have different yield strengths only. Then, under compression, the phase with the smaller $\sigma_y$ will be obtained (since pressure and $w_t$ are higher for $\sigma_{y1} > \sigma_{y2}$), and under torsion the phase with larger $\sigma_y$ should appear (since pressure and the $w_t$ are greater for $\sigma_{y2} > \sigma_{y1}$). Thus, torsion may lead to new phases which were hidden under compression test (Section 3.3).

In [62], the pressure self-multiplication was explained by an increase in elastic moduli after PT. This is not correct, because plasticity and equilibrium equations were completely neglected. Even at infinite moduli (like in our rigid-plastic model), the pressure is limited by the solution of the plastic equilibrium problem. For $\sigma_{y1} > \sigma_{y2}$, the pressure in the HPP reduces independent of the growth in elastic moduli. However, that group still relies in [65] on the interpretation from [62], ignoring more advanced theoretical approaches.

Let us estimate the effect of $\gamma^t \neq 0$. While $|\boldsymbol{\tau}_f| = \tau_y$ for both cases, for compression it varies to zero at the symmetry plane. During torsion, only radial shear stress $\tau_{rz}$ varies linearly to zero at the symmetry plane; azimuthal shear stress $\tau_{\theta z}$, producing torsion, is constant along the thickness. That is why averaged $\bar{\tau}$ is larger during torsion, increasing the $w_t$ and promoting the PT.

Also, this solution answers to some extent the following questions: Why large compression, which includes severe plastic shear at the contact surface due to friction, does not cause the PT, but small plastic strains during torsion cause the PT? Why a PT starts at the sample center where shear stresses and strains are absent or minor? These questions raised doubts that plastic strain governs the PTs under torsion, and theory in [92] supports them: no new physical assumptions were introduced; plastic strain does not contribute to the PT criterion (15) (and (42)), and explanations were based on simple mechanics of plasticity and contact friction.

Moreover, based on the gained understanding that the main reason for promotion of the PT is in additional axial displacement (rather than in plastic straining), several alternative methods to obtain additional displacement, even without torsion have been suggested in [92]. One can decrease $\sigma_y$ under $P = const$, e.g., due to heating of the external part or entire sample, causing reduction in thickness. Similar to the torsion, if an HPP is stronger, pressure should grow in it. This was observed in experiments in [384]: the pressure at the sample center increased by 30 % during $B1 \to B2$ PT in KCl caused by heating from 300 to 600 K.

Another opportunity was to utilize TRIP (Section 3.7). One can add to sample or gasket particles of materials transforming with sufficiently large $\varepsilon$ back and forth during temperature cycling. Under deviatoric stresses, this will lead to large TRIP proportional to number of cycles, which will cause axial displacement that compensates the volume reduction due to PT. Note that at megabar pressure, thickness of the sample cannot be reduced by increasing load due to capping of the anvils and elastic region at the center of the sample. Utilizing TRIP in embedded transforming particles or torsion should lead to additional axial displacements. Thus, simple solution in [92] was very promising since it qualitatively explained various nontrivial experimental phenomena and suggested some methods for promoting the PT and finding new phases via torsion compared to compression.

Another simple analytical solution in [328] based on the PT criterion (15) was obtained and successfully and quantitatively explained the reduction in the PT pressure rBN→cBN PT from 55 under hydrostatic loading to 5.6 under uniaxial compression (Section 3.6). It also suggested a way to produce an additional axial displacement under fixed force and without torsion by utilizing material softening due to rotational lattice instability.

All these created the impression that the theory is going in the correct direction. The author, together with S.B. Polotnyak, started development of a more detailed analytical model for PT during torsion [280] and FEM simulations for PT during compression in DAC [633, 634] based on criterion for stress-induced PT (15). However, after realizing that the concept of strain-induced PTs should be developed and applied, the author focused on this and suggested to Polotnyak to finish papers alone. Since it was revealed for Zr [82] that after strain-induced nucleation, the stress-induced growth occurs, results in [92, 280, 633, 634] may recover some interest.

In [635], pressure-induced PT in a plane layer was studied under compression. An oversimplified PT criterion ($p = p_e$ at the interface boundary), combined with another unrealistic assumption $\sigma_y = 0$ in two-phase region during PT (see discussion in [100, 102]), were used to reproduce a plateau at the pressure distribution.

Since this solution in [92] was one of the several different problems to illustrate the new theory for stress-induced PTs, no quantitative comparison with experiments, at least to order of the magnitude, were attempted. After simple estimates have been made, the author concluded that the approach in [92] could reduce the PT pressure due to torsion by $(1-2)\sigma_y$ only, i.e., a few GPa. This is not sufficient for explanation of experiments in Section 3. Thus, as often happens, small, ugly results destroy a major part of the simple, elegant theory. On the optimistic note, it led to introducing the concept of plastic strain-induced PTs [100, 102]. As Lev Landau (1962 Nobel Prize in Physics) said: "I need to read wrong paper to write the correct one."

*8.2. Simple analytical solution for strain-induced PTs*

In [100, 102], an analytical solution for torsion of a disk was significantly generalized, in particular, by more precise description of the kinematics of contact sliding. Without PTs, it was discussed in Section 2.10. What is more important than it was a part of a three-scale theory, and strain-controlled kinetic Eq. (37) was derived and utilized. Some conclusions, however, remained similar to those in [92], but interpreted in terms of additional or increasing plastic straining. Thus, reduction in thickness during torsion indeed promotes the PT. However, it happens not only due to compensating for the volume reduction during the PT, but also by adding additional plastic strains. This explains the seeming contradiction for strain-controlled kinetics: why does the torsion reduce the PT pressure and promote the PT at the center of the disk where there is no plastic shear? Pressure at the center is the highest and reaches $p_\varepsilon^d$ first; the thickness reduction during torsion produces $q = \ln(h/h_0)$, which transforms stress-induced PTs into strain-induced PTs and, according to Eq.(37), induces the PT. If one neglects the reduction in volume during the PT, Eq. (10) and Fig. 15a can be used, which qualitatively agree with experiments [23, 66, 280]. The thickness reduction is very large at small rotation angles, since for $\bar{\varphi} \to 0$, $\frac{dh}{d\bar{\varphi}} \to \infty$. However, it decelerates sharply with growing $\bar{\varphi}$, and promotion of the PT at the center decelerates as well. For the same $\varphi_a$, sliding and the thickness reduction grow with increasing $\tilde{m}$, which promotes the PT due to thickness reduction, but suppresses it due to reduced shear of material. For $\tilde{m} \to \infty$, $\bar{\varphi} \to \varphi_a$, only sliding without the torsion of the material occurs; still, the maximum thickness reduction occurs, which causes some and relatively uniform increase in $c$. For large $m$ and $\varphi_a$, both the compressive and shear plastic strains are small, i.e., torsion at $P = const$ does not lead to essential PT progress. An increase in $P$ is required to render the torsion effective again.

The above theory does not show the different effects of the torsion in one direction and forward and back torsion, like in some experiments [8, 9, 330]. It is mentioned and qualitatively discussed that the reduction in $\sigma_y$ and friction stress at the initial stage of the reversal due to the Bauschinger effect is possible, until they recover after increment $\Delta q \simeq 0.1$ (Fig. 4) [4, 90]. This should lead to larger reduction in the sample thickness and intensification of PT. However, smaller contribution $\bar{\tau}\gamma^t$ to $w_t$ was not discussed. It was also mentioned that the change in the sign of $\bar{\tau}$ may suppress the direct PT and promote the reverse PT. It is written in [5, 23] that the oscillating torsion leads to the same steady microstructure but requires a larger total rotation angle. One more important implication from the kinetic Eq. (37) is that the rotation angle of the material, rather than the anvil, contributes to $q$ and has to be measured. This was done in [245, 246, 250] using different methods. Qualitatively, [100, 102] presented the same explanation as in [92] for the pressure self-multiplication/demultiplication effects and the selection of stronger phases by torsion, though only as one aspect of the phenomena and linked to plastic strain rate and strain fields.

By approximate integrating kinetic Eq. (37) at the center of the sample, PT under compression and torsion were compared by reaching "experimentally detectable" volume fraction $c_d = 0.2$ (at that time, without XRD). While $p_\varepsilon^d$ and entire Eq. (37) do not distinguish different straining modes, the PT pressures for compression and torsion were essentially different. While during torsion, one can run the PT by increasing shear and $q$ at any constant pressure slightly higher than $p_\varepsilon^d$, to produce plastic strain during compression, one must increase $P$ and $p$, even if it is not necessary for PT. This explains why "shear" promotes the PTs compared to compression. It was concluded that the obtained PT pressures under compression and torsion are not material constants because they strongly depend on the straining path in the $p - q$ plane. This conclusion was later confirmed by FEM simulations [111, 386], especially in the presence of the gasket, since different elastoplastic gasket's properties allow a broad variation in the $p - q$ paths.

At the current level of knowledge, comparison of the PT pressure between compression and torsion depends on the initial grain size and value $c_d$. According to the solution in [100, 102], the difference between the PT pressure for compression and torsion reduces significantly with reduction of $c_d$. With modern XRD with synchrotron radiation, $c_d$ is within few percent, which makes the difference mentioned in [100, 102] small. On the other hand, $p_\varepsilon^d$ depends on the microstructure (grain size and dislocation density). If steady microstructure is reached, then $p_\varepsilon^d$ is also constant. That is why the same PT pressure was obtained in experiments under compression and torsion for $\alpha - \omega$ PT in strongly predeformed Zr [80, 82, 89] (Sections 7.5.1 and 7.5.3) and for Si-I→Si-II PT for 100 nm particles [81] (Section 9.2). However, for annealed material or when the steady microstructure is not reached, large shear strain at torsion reduces $p_\varepsilon^d$ compared to small plastic strain during compression. This explains why torsion led to a reduction in the PT pressure compared to compression, which was observed in most of the experiments [5–7, 23, 31, 32, 64] and explained in [80] (Section 7.5.2).

Two possible reasons for 'steps' (almost horizontal portions) on the pressure distribution, observed experimentally in [62, 66, 359, 620] (Section 3.2), were suggested in [100, 102] with some analytical support. Zero pressure gradient implies $\tau_r = 0$ according to the equilibrium Eq.(3). Since steps were found in a two-phase mixture, where a PT occurs, due to reduction in volume, material tends to flow toward this region from both sides. That means, that the neutral circle with zero radial velocity $V_r$ is located within this step, and the step represents the stagnation region. The pressure gradient in the stagnation region near the place with $V_r = 0$ zone (which is usually the symmetry axis) is close to zero, which partially explains the step. Another reason was the shear stress relaxation due to TRIP for PTs or RIP for CRs. We will revisit this problem based on FEM solution in Section 8.3. Some methods to promote and characterize strain-induced PTs and CRs have also been suggested in [100, 102].

*8.3. FEM simulations of strain-induced phase transformations in DAC and RDAC*

The first FEM simulations of a strain-induced PT in sample were published in a short letter [385] followed by comprehensive papers in [110] for compression in DAC and in [251] for torsion in RDAC. The pressure-independent $\sigma_y$, small $\boldsymbol{\varepsilon}_e$ and $\boldsymbol{\varepsilon}_t$ but finite $\boldsymbol{\varepsilon}_p$ and rotations, and linear elastic rules are utilized in them and in [252, 253, 360, 636–638] for a model material, and in [125, 126] for $\alpha - \omega$ PT in Zr for compression and torsion, respectively. Strict large strain constitutive equations with large all strain contributions and rotations, pressure-dependent $\sigma_y$, and the third-order Murnaghan potential have been derived and utilized for hBN→wBN PT in BN in [111, 386] and or $\alpha - \omega$ PT in Zr in [108, 109]. In all these papers, the kinetic Eq. (37) and the perfectly plastic model were utilized, assuming that plastic strains are large enough, and that a steady microstructure and the limit surface of perfect plasticity are reached.

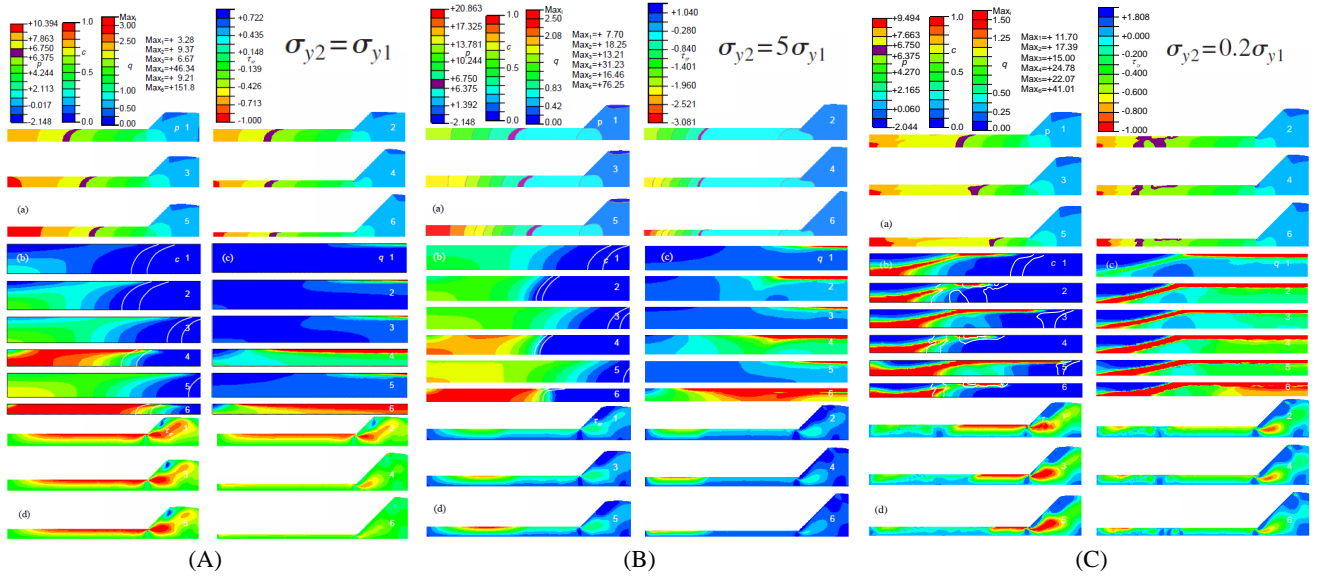

(A)          (B)          (C)

**Figure 53:** Comparison of fields of pressure $p$ (a), volume fraction of HPP $c$ (b), accumulated plastic strain $q$ (c), and shear stress $\tau_{zr}$ during compression (1, 3, and 5) and torsion at constant force (2, 4, 6) for $\sigma_{y2}=\sigma_{y1}$ (A), $\sigma_{y2}=5\sigma_{y1}$ (B), and $\sigma_{y2}=0.2\sigma_{y1}$ (C), respectively [110]. The pressures at the diamond-sample contact at $r=0$, $p_{max}$, after compression and torsion are the same and equal to 7.95 (1 and 2), 8.10 (3 and 4), and 8.80 (5 and 6). All legends are uniform except the inserted magenta color for $p_\varepsilon^r < p < p_\varepsilon^d$, where PTs do not occur. Magenta region in $p$ fields and white lines in $c$ fields separate this region in a sample. $Max_i$ is the maximum value of $q$. All normal and shear stresses are normalized by $\sigma_{y1}$ and $\tau_{y1}$, respectively.

Complete adhesion at the contact surface between a sample and diamond was implemented in [110, 251–253, 385]; contact sliding with a combined Coulomb and plastic friction has been implemented in the other above-mentioned papers, but [108, 109], where the distribution of $\tau_f/\tau_y$ was fitted to experimental data (Section 7.6). The contact equations are the same as for the case without PT (Section 2) but with $\tau_y$ for different phases or mixture. Let us first discuss generic results from [110, 251, 385], some of which are shown in Figs. 53 and 54 for 3 different $\sigma_{y2}/\sigma_{y1}$. Compressive normal stresses are assumed to be positive. Among important parameters, $\varepsilon_0 = -0.1$, $p_\varepsilon^d = 6.75$, $p_\varepsilon^r = 6.375$, which means that for $p_\varepsilon^r < p < p_\varepsilon^d$ the PTs do not occur.

For all cases, the PT initiates at the sample center, where pressure first exceeds $p_\varepsilon^d$ (like in most experiments [64, 66, 76, 80–83, 89]) and some plastic strain localizes along the slip line separating strongly and weakly plastically deformed regions. Second local maximum in $c$ is visible at the contact surface where plastic strains are large. For $\sigma_{y2}=5\sigma_{y1}$, pressure grows essentially at the sample center during torsion, reproducing pressure self-multiplication effect observed experimentally [62, 64, 66]. Even for $\sigma_{y2}=\sigma_{y1}$, there is some increase in pressure at the center during torsion without essential changes in the rest of the contact surface due to shrinking the region with reducing to zero radial friction $\tau_{zr}$.

Reduction in thickness during torsion is the same for $\sigma_{y2}=\sigma_{y1}$ and $\sigma_{y2}=0.2\sigma_{y1}$ and slightly smaller for $\sigma_{y2}=5\sigma_{y1}$. While analytical Eq. (10) describes FEM results without PT well, it underestimates thickness reduction for the case with PT. The difference is larger than would be expected due to $\varepsilon_0 = -0.1$. It was concluded in [251] that "volume change induces additional plastic flow called TRIP." Note that while TRIP is not included explicitly at the material point level via constitutive equation, large gradients of $c$ and $\varepsilon_0 c$ cause large internal stresses, which produce TRIP at the scale of the sample, leading to relaxation of stresses and larger plastic strain than without PT. Note that huge TRIP was observed experimentally for hBN→wBN PT due to large $\varepsilon_0 = -0.34$ [74, 86, 87] (Section 3.7).

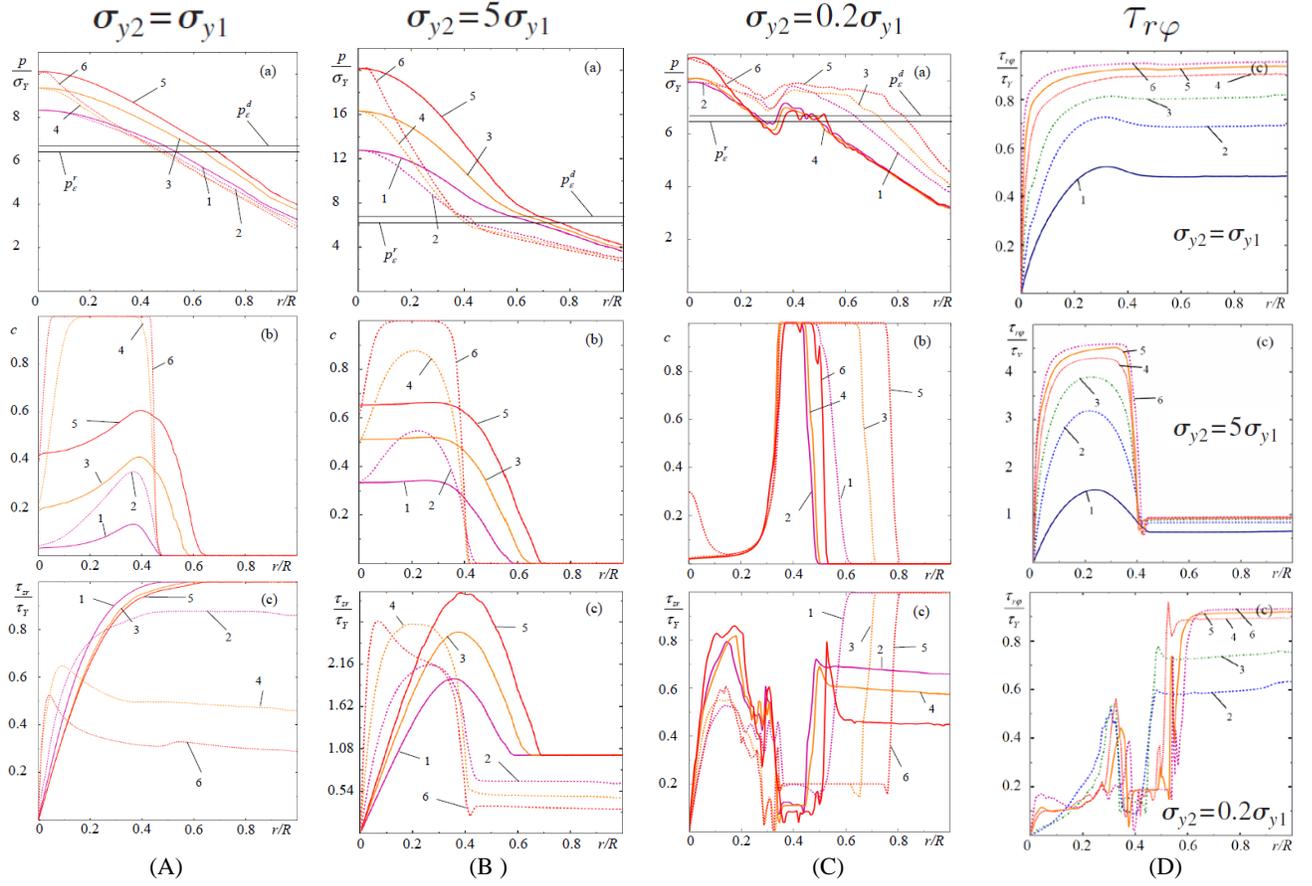

**Figure 54:** Radial distributions of pressure $p$ (a), volume fraction of HPP $c$ (b), and shear stress $\tau_{zr}$ (c) at the sample-diamond contact surface during compression (1, 3, and 5) and torsion at constant force (2, 4, 6) for $\sigma_{2y} = \sigma_{1y}$ (A) $\sigma_{y2} = 5\sigma_{y1}$ (B), and $\sigma_{y2} = 0.2\sigma_{y1}$ (C), respectively. The pressures $p_{max}$ are the same after compression and torsion. (D) Radial distribution of shear stress $\tau_{r\varphi}$ for different yield strength ratios of phases. Reproduced from [110].

Thickness reduction during torsion causes radial flow of the HPP to the LPP region, leading to reverse strain-induced PT. For slow kinetics of the reverse PT, the HPP may be found in the region with $p < p_\varepsilon^r$, leading to an incorrect identification of $p_\varepsilon^d$. Such a situation does not happen under compression.

Experimentally observed for KCl and fullerene $C_{60}$ plateaus (steps) in the pressure curves [62, 64, 66, 71, 358] (Fig. 19) are reproduced by FEM under torsion for $\sigma_{y2} = 5\sigma_{y1}$, but not under compression. The steps correspond to the localized two-phase region, like in experiments. For KCl and fullerene $C_{60}$, indeed $\sigma_{y2} \gg \sigma_{y1}$. Pressure at the step corresponds to $p_\varepsilon^d$ and then sharply reduces to $p_\varepsilon^r$. Thus, it indeed characterizes the PT but is not equal to $p_e$, as was suggested in [62, 64]. This finding allows us to identify these two key material parameters from the step. Torsion leads to a sharper two-phase region, like experiments on KCl [62, 64].

For $\sigma_{y2} = 0.2\sigma_{y1}$, the pressure (and all stress) distributions look messy with multiple oscillations, like numerical instabilities; however, similar oscillations are observed in experiments for ZnSe [63] (Fig. 19d) and CuI [359]. The simplified model [100, 102] suggests a pressure reduction in the PT region, not supported by FEM simulations. For this case, unstable flow with appearance of shear-PT bands is observed. Steps appear in the ring of fully transformed HPP near the contact surface, with pressure for compression higher than for torsion, i.e., like in experiments [63, 359]. The problem is that the pressure at the plateau for compression in FEM is above $p_\varepsilon^r$ and $p_\varepsilon^d$, does not have a specific meaning, and is determined by the coupled mechanics of plastic deformations and PT. For torsion, a step corresponds to $p_\varepsilon^d$ when PT is completed at the contact surface. With further torsion, pressure increases, but this is not relevant for PT.

In analytical treatment [100, 102], possible reasons for a step are: (a) zero $\tau_{zr}$ and, consequently, $\sigma_y$ in the PT region, (b) TRIP, and (c) zero radial velocity due to volume decrease during the PT. None of the above is confirmed by FEM simulations: $\tau_{rz} \neq 0$, material flows from the center, and TRIP was not introduced into constitutive equations. Thus, *Eq. (3) is not applicable in the region with a large gradient of c and q,* which led to wrong conclusions. Comparison of $p(r)$ and $\tau_{rz}$ in Fig. 54 for $\sigma_{y2} < \sigma_{y1}$ implies that *Eq.(3) is not applicable in the entire sample due to localized strain-PT bands.* This conclusion raises the question: how can equilibrium equation, which is fundamental and independent of the constitutive and all other equations, be violated? The answer is now clear. Strict equilibrium equation is Eq. $(11)_1$ in terms of normal stresses averaged over the thickness and shear stress at the contact; it is assumed in Eq.(3) that $\sigma_{\theta\theta} = \sigma_{rr}$ and $\sigma_{rr}$ averaged over the thickness is substituted with $p$ at the contact, which caused significant error. Thus, traditional wisdom and intuition should be changed. At the same time, Eq.(3) describes well FEM results for pressure distribution in a single phase regions for $\sigma_{y2} = 5\sigma_{y1}$ and independence of $p(r)$ of rotation angle for $\sigma_{y2} = \sigma_{y1}$. There is no signature in $p(r)$ due to the PT for $\sigma_{y2} = \sigma_{y1}$, even in the narrow two-phase region; thus, TRIP does not cause stress changes but leads to additional thickness reduction in this case.

FEM results give a more complete answer to the question of whether torsion promotes the PT compared to compression. For the same $p_{max}$, torsion leads to larger $c$ and faster completion of the PT for $\sigma_{y2} \geq \sigma_{y1}$ under smaller force. For completing the PT in some region, torsion essentially reduces $p_{max}$. However, for torsion, thickness is much smaller, and, for $\sigma_{y2} = 5\sigma_{y1}$, radius in which HPP is obtained is smaller; that means that the total HPP mass is larger under compression. Unexpectedly, for $\sigma_{y2} = 0.2\sigma_{y1}$, torsion suppresses PT: it reduces the HPP mass and slightly increases $p_{max}$. To complete the PT for $\sigma_{y2} = 5\sigma_{y1}$, $p_{max} > 50$. This number does not characterize the material properties (there is no such constant in the kinetic Eq. (37)), but rather sample behavior in DAC for the realized $p - q$ path. According to Eq. (37), the PT can be completed at $p$ slightly above $p_\varepsilon^d$, if it can be kept constant during straining, which can be realized in RDAC utilizing gasket (like for BN in [74]). For a weaker HPP, FEM shows the shear-PT band at the sample-diamond contact. Thin transformed layer without PT within the rest of the sample in DAC was observed for PT Si-I→Si-II in [441] and olivine to $\beta$-spinel in [392]. This observation also confirms the strain-induced (rather than pressure-induced) PT for under compression in DAC.

At the major part of sample-diamond contact, shear stress $\tau_{rz}$ reduces during torsion along with increase in $\tau_{r\varphi}$ to keep $\tau_f = \sqrt{\tau_{rz}^2 + \tau_{r\varphi}^2} = \tau_y$ for the corresponding phase. For partial *unloading* and torsion for $\sigma_{y2} = 5\sigma_{y1}$, a step corresponding to $p_\varepsilon^r$ forms, which may be used to determine $p_\varepsilon^r$ experimentally.

Most of the above surprising FEM results are due to nontrivial and strongly nonlinear interaction between PT, plastic deformations, and stress redistribution that is not describable within a simplified analytical model. Thus, the obtained FEM results altered the basic knowledge and nonlinear intuition in understanding and interpretation of experimentally observed PT and mechanical phenomena, measurements, and extracting data for material constitutive equations from the heterogeneous sample behavior. Several important future directions have been outlined in [110], including allowing for the defects evolution and combination of strain-induced and pressure-induced PTs in the kinetic equation, which are currently pursued [80–83, 88, 89].

*Effect of accelerated kinetics on compression in DAC.* Parametric study of processes in DAC with the multiplier $k$ in kinetic Eq. (37) increased by factors of 5, 10, and 30 was performed in [252]. Increase in $k$ significantly changes the evolution of the morphology of the PT regions, character of stress distribution, plastic flow, and the interpretation of experimental results. For all increased $k$, plastic flow to the sample center takes place at certain compression stage, in agreement with multiple experiments. While it changes, in some cases, direction of the friction stress, the pressure gradient does not reverse but is significantly reduced, i.e., *Eq.(3) is violated* and cannot be used for determination of $\tau_y$ for mixture. This warns that the reduction of $\sigma_y$ found in experiments [162] and similar works (Section 3.12) during the PT may just be a result of the application of the wrong method.

Alternatively, such a reduction may be a consequence of the very small grain size at the beginning of the PT, which falls in the region of the inverse Hall-Patch relationship (Section 3.9 and [81]).

Acceleration of PT essentially changes the morphology of the PT region for $\sigma_{y2} = 0.2\sigma_{y1}$. Thus, PT finishes away from the center at the symmetry plane, and an interface between a completely transformed HPP and LPP propagates and reaches boundary with diamond. Several plateaus were found in a highly oscillating pressure distribution at the contact surface, some corresponding to $p_\varepsilon^d$ (in the region where $c = 0$ at the contact), while others did not. Consequently, determination of $p_\varepsilon^d$ from the pressure at plateaus is not reliable. For $\sigma_{y2} = 5\sigma_{y1}$ and even for $k = 30$, pressure still increases in the entire transforming region despite the more intense volume decrease. A kink and then step in pressure distribution was reproduced with increasing $k$, like in experiments for KCl, PbTe, and fullerene $C_{60}$ [62, 64, 66, 71, 358] (Fig. 19), which was not present for $k = 1$. This narrow step is located in the quite broad two-phase region and is slightly above $p_\varepsilon^d$ (i.e., above pressure at the step for torsion, like in experiment [62, 64, 66, 71]). Reduction in pressure at step in experiments [62, 64, 66, 71] was interpreted as reduction in PT pressure due to shear. However, in FEM we have the same $p_\varepsilon^d$ for compression and torsion, i.e., such an interpretation is not justified. There are other plateaus with higher pressures at the center of the sample.

*Kinetics with $p_\varepsilon^d < p_\varepsilon^d$, loading, unloading and reloading in DAC [252] and RDAC [253].* In contrast to the above papers, here $p_\varepsilon^d = 6.375$ and $p_\varepsilon^r = 6.375$ (i.e., they are switched). This leads to simultaneous occurrence of direct and reverse PTs between these pressures and a steady $0 < c < 1$ at SPD, which was experimentally observed, e.g., under HPT of metals and ceramics [31, 33] and ball milling [118, 120]. For compression, the geometry of the transformed region visibly differs for $\sigma_{y2} = 0.2\sigma_{y1}$ compared with that in [252]. Steps and then kinks in $p(r)$ in the region with large gradient of $c$ appear even for $\sigma_{y2} = \sigma_{y1}$ but identification of $p_\varepsilon^d$ and $p_\varepsilon^r$ from the steps is less clear.

During unloading for fast kinetics, surprisingly, plastic flow and reverse PT occur for $\sigma_{y2} = \sigma_{y1}$ and more pronouncedly for $\sigma_{y2} = 0.2\sigma_{y1}$, but do not occur for $\sigma_{y2} = 5\sigma_{y1}$, which must be considered in experiments. At the very beginning of unloading for $\sigma_{y2} \leq \sigma_{y1}$, a small increase in $c$ is observed.

Two reasons for plastic flow during load decrease are discussed: (a) heterogeneous plastic strain, stress, and $c$ fields and (b) TRIP due to volume change during the PTs, leading in turn to PTs, i.e., positive mechanochemical feedback (Section 7.4 and [74, 86, 87]). Thus, PT should be monitored in situ during unloading rather than assuming it will not occur. To avoid reverse PT during unloading, problem on minimization of plastic straining should be solved, probably, with utilization of different gaskets. For reloading to the same force, PT practically does not take place for $\sigma_{y2} = 5\sigma_{y1}$ due to very small plastic strains. For $\sigma_{y2} \leq \sigma_{y1}$, PT occurs in slightly different regions but does not reach the same $c$ as for initial loading. Similar problems for torsion in RDAC for the same $p_\varepsilon^d$, $p_\varepsilon^r$, 3 ratios of $\sigma_{y2}/\sigma_{y1}$, and for $k =$1, 5, and 10 were solved in [253]. Two unloading pathways were treated: (a) applied force is reduced at a fixed rotation angle, which is accompanied by reduction in torque and (b) torque is reduced down to zero first and then force is reduced. For (a) and $k =$5, reverse PT is less pronounced than after compression in DAC and practically absent for $\sigma_{y2} = 5\sigma_{y1}$. For $\sigma_{y2} = 5\sigma_{y1}$ and $k =$1, the reverse PT is practically absent, and is getting more profound with increasing $k$, both due to accelerated kinetics and TRIP. For (b), reverse PT practically does not occur during reduction of torque due to high pressure and is less pronounced than in (a) during the force reduction. These results can be used in designing unloading pathways to control the reverse PTs and retain HPPs.

Two reloading paths were applied. (i) After unloading in (a), $P$ increases to the same value before unloading, and sample is then twisted at the $P = const$. Increase in $P$ alone does not recover completely the reverse PT during unloading. However, a small torsion practically recovers the HPP geometry and the PT completeness in it, which only slightly increases with further torsion. (ii) After loading, $P$ is slightly reduced at a constant angle, followed by torsion at $P = const$. Reverse PT occurs, and the rate of reverse PT reduces with increasing $\sigma_{y2}/\sigma_{y1}$.

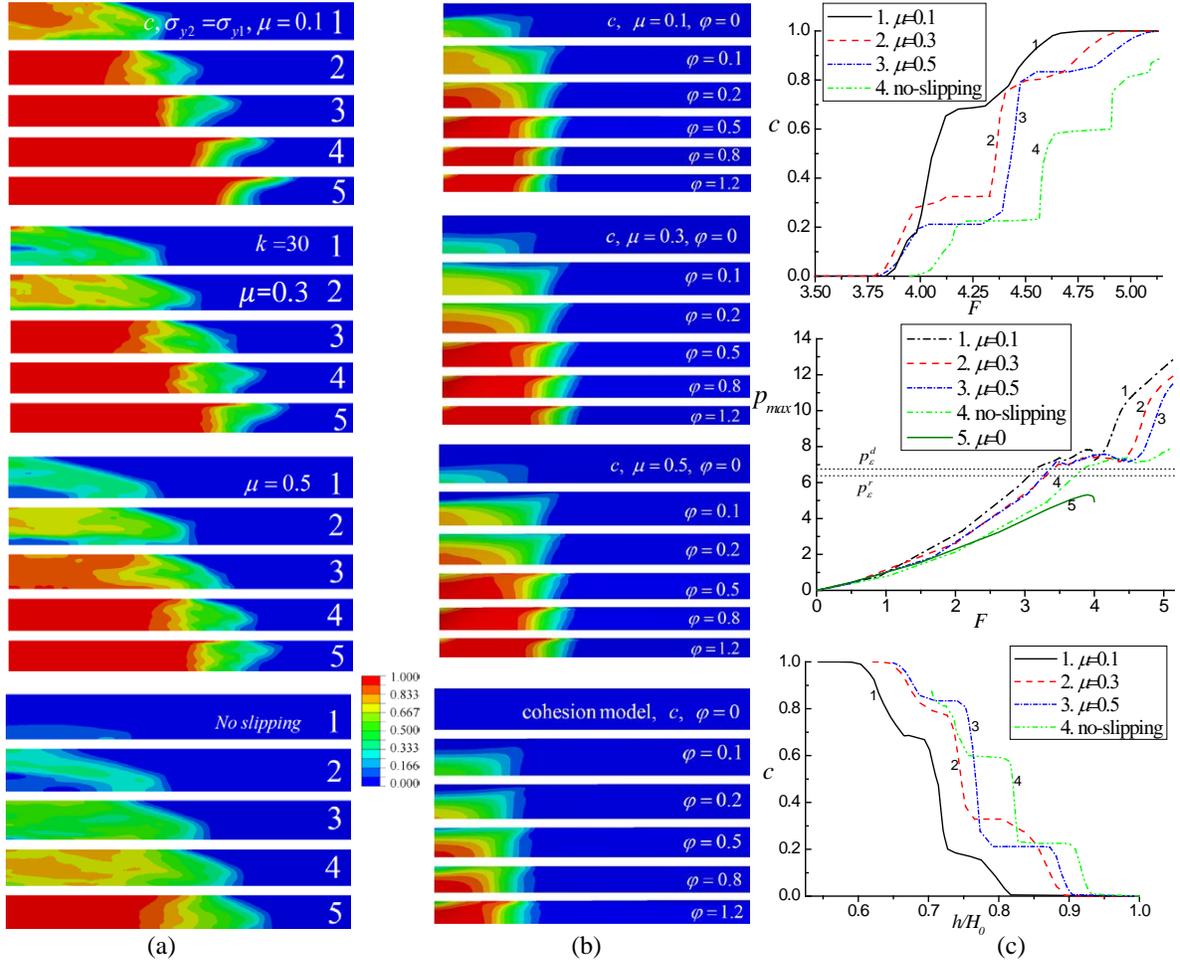

**Figure 55:** (a) Evolution of volume fraction of HPP during compression in DAC for $k = 30$ and $\sigma_{y2} = \sigma_{y1}$ to the same forces but with different friction coefficients $\mu$ [636]. (b) Evolution of volume fraction of HPP during torsion in RDAC for $k = 5$ and $\sigma_{y2} = \sigma_{y1}$ at the force slightly lower (3.75 vs. 4) than in #1 in (a) with different friction coefficients $\mu$ [637]. (c) Volume fraction of the HPP phase $c$ (at the point with $r/R = 0.3$ and $z = 0$) and $p_{max}$ versus applied dimensionless force $F$ and volume fraction of the HPP phase $c$ versus ratio $h/H_0$ for the compression in DAC in (a) [636, 637].

For $\sigma_{y2} = 0.2\sigma_{y1}$ (or $\sigma_{y2} = 5\sigma_{y1}$), torsion increases (or reduces) the pressure at the contact surface in the central part of sample and reduces at the periphery (or inclined surface), like in the experiments on ZnSe [63] (or KCl [62, 64]).

*The effect of contact friction.* While complete adhesion between sample and rigid anvils was accepted in the previous papers, contact sliding combining Coulomb friction at lower normal stresses and plastic friction with $\tau_f = \tau_y(c)$ in DAC and RDAC was introduced in [636, 637]. The above simulations with complete adhesion at the contact surface have two drawbacks: (a) There is an unrealistic shear band at the external part of the sample/culet contact surface and, especially, at the conical surface. Thickness of a band is equal to the thickness of one finite element (i.e., it is mesh-dependent). If in this band $p > p_\varepsilon^d$, this artificially enhances the kinetics of strain-induced PT. (b) Adhesion artificially suppresses the radial plastic flow and reduction in thickness. Some of the typical results for comparison of the PT kinetics in DAC and RDAC for different friction conditions, including adhesion, are presented in Fig. 55. For $\mu \leq 0.01$, pressure does not reach $p_\varepsilon^d$ and PT does not occur. This result and the fact that friction increases pressure gradient and pressure intuitively suggest that large friction promotes PT; however, simulations suggest opposite. For each force in Fig. 55a for DAC, the smaller the friction, the more pronounced the PT is. The same is true for a specific point in Fig. 55c, but with some minor crossovers. The smaller the friction is, the larger the reduction in thickness (and plastic strain) is, which produces larger $p_{max}$, both promoting the PT. "Old" intuition works only while comparing for the same $h/H_0$, which shows that the smaller the friction is, the smaller $c$ is (with some minor crossovers). Actually, comparison for the same $h$ makes more sense because then the larger radius of the HPP region, the larger its mass; for the same force, this is not necessarily the case because of different thicknesses. However, the DAC experiments are routinely controlled by force, and only in some recent works [80, 82, 89, 425] the thickness during the loading was measured in situ systematically. Among other interesting features in Fig. 55c, the regions with almost constant $c$ and $p_{max}$ in the course of the loading are observed.

The intriguing *self-locking of sliding effect* is revealed in DAC in [636] for $\sigma_{y2} \geq \sigma_{y1}$. While the sliding condition is satisfied, still, sliding does not occur because this region is surrounded from both sides by regions in which the sliding criterion is not met. Surprisingly, large steps with $p \simeq p_\varepsilon^d$ appear for $k = 30$ and $\sigma_{y2} = \sigma_{y1}$ (while they were absent for adhesion condition) but they are much less clear for $\sigma_{y2} = 5\sigma_{y1}$ (contrary to the case with adhesion).

For torsion, sliding accelerates the PT compared to adhesion for all $\mu$ and $\varphi$, but the effect of friction is not monotonous (Fig. 55c). PT is accelerated in one region but suppressed in another for different $\mu$ and $\varphi$. Based on the radius of the HPP, for $\varphi > 0.5$, the PT region for $\mu = 0.3$ and $0.5$ are almost the same and larger than for $\mu = 0.1$. Reduction in thickness during rotation increases with reduction in friction, but results for $\mu = 0.3$ and $0.5$ are practically the same. For $\mu = 0.1$ and $\varphi > 0.5$, slight reverse PT occurs due to local pressure reduction. Since thickness also reduces with increasing $\varphi$, this implies that $\varphi \simeq 0.5$ *is an optimal rotation angle maximizing the HPP mass*, and traditional intuition, the larger torsion, the better, does not work. Back and forth torsion slightly suppresses PT progress as compared to monotonous rotation, in agreement with experiments [5, 23].

The main problem is that the actual friction coefficient is unknown. For smooth anvils, it was found in [108] that for compression of Zr up to 14 GPa, $\tau_f$ in the culet does not exceed $0.5\tau_y$, but friction does not follow the Coulomb rule and has essential scatter. To reach $\tau_y$ and eliminate the indeterminacy in friction condition, rough anvils were introduced in [81–83, 89, 425] (Section 2.7). For rough anvils, sliding is allowed when $\tau_f = \tau_y$ only, but FEM simulations in comparison with experiments are yet to be published.

Note that the complete adhesion condition was initially used to exclude the iterative process at the contact and achieve better convergence. However, it appeared that with including contact sliding, convergence surprisingly improved because of reducing large plastic strain and especially their gradients near the contact surface.

Similar models, but with elastic deformations of anvils and jump in elastic properties of phases, were utilized in [125, 126] for $\alpha - \omega$ PT in Zr for compression and torsion, respectively. Since data obtained in later in situ studies [80, 82, 89, 108, 109, 425] were unavailable, the best guess based on available data in literature was used. The specific features for this PT were very small $\varepsilon_0 = -0.014$ and large $\sigma_{y2}/\sigma_{y1} = 6.56$. Very low $\sigma_{y1} = 180$ MPa from [639] was for annealed material; that is why strain hardening neglected in [125, 126] should be included. FEM results are qualitatively similar to that obtained for a model material in [636, 637] but were used for interpretation of existing experiments, explaining a scatter in results from different groups, correcting experimental PT pressures based on force divided by area, discussing $\alpha - \beta$ and $\omega - \beta$ PTs, effect of texture, and effect of PTM on $\alpha - \omega$ PT in Ti, including combination of stress- and strain-induced PTs and their effect on the irreversibility of PT.

Based on fields similar to those in Figs. 53B and 54B, it was stressed that the PT process in RDAC for material with large $\sigma_{y2}/\sigma_{y1}$ is far from being satisfactory for producing large mass of HPP. Due to pressure self-multiplication, $p \gg p_\varepsilon^d$ in the HPP, which is not needed for PT but may damage anvil. During the torsion, the radius of the HPP region does not increase, but thickness reduces, decreasing the total mass of the HPP. To overcome these drawbacks, a strong gasket with optimized geometric parameters could be used. Indeed, close to uniform $p$ and $c$ and small pressure increase during torsion for hBN→wBN PT was obtained experimentally in [74] based on a simplified analytical design optimization. This was also confirmed by FEM simulations in [360, 386].

*The effect of gasket on the processes in DAC.* Utilizing a gasket in DAC and RDAC allows: (a) experimentation with gases, liquids, powders; (b) achieving a much higher pressure in the sample and larger thickness by increasing the strength of the gasket; (c) reducing the radial plastic flow of the sample toward periphery and changing it toward the center, when necessary; (d) reducing radial heterogeneity of all fields; (e) controlling deformation-PT processes and microstructure by changing gasket strength and geometry.

A detailed FEM study of the effects of the kinetic parameter $k$, $\sigma_{y2}/\sigma_{y1}$, $\sigma_{yg}/\sigma_{y1}$ ($\sigma_{yg}$ is for gasket), relative sample/diamond culet radius $R_s/R$, and initial thickness of a sample $H_0$ and the gasket at the periphery $H_1$ on the coupled PTs and plastic flow in DAC is presented in [638]. Combined Coulomb and plastic friction is applied at all contacts. Results for selected parameters are shown in Fig. 56. Comparison with the similar fields in Figs. 53 and 55 show the drastic effect of the gasket.

Plastic strain for the same pressure is strongly reduced compared to loading without gasket. Gasket changes radial plastic flow completely (for $\sigma_{y2} \leq \sigma_{y1}$) or partially (for $\sigma_{y2} = 3\sigma_{y1}$) toward the center, which leads to high quasi-homogeneous pressure for $\sigma_{y2} \leq \sigma_{y1}$. During the PT to a stronger HPP, plastic strain and $c$ are also quasi-homogeneous, which suggests a simple method of obtaining the strain-controlled kinetic equation from experiment. This is not the case for $\sigma_{y2} \leq \sigma_{y1}$, for which a completely transformed HPP region with a sharp interface appears at the sample-anvil contact and propagates to the symmetry plane through weakly transformed LPP. Geometry of the HPP for $\sigma_{y2} = \sigma_{y1}$ and $\sigma_{y2} = 0.3\sigma_{y1}$ is very close, i.e., multiple shear/PT bands and corresponding stress oscillations that appear without gasket for $\sigma_{y2} \ll \sigma_{y1}$ are suppressed by gasket. A novel mechanism of sliding at the contact line between the anvil, sample, and gasket, coined *extrusion-based pseudoslip*, is revealed (Fig. 57,a). Increase in $k$ from 6 to 30 significantly accelerates PT kinetics at the beginning of PT but much less at completing, because pressure reduces faster for faster kinetics. For a longer sample ($R_s/R = 0.7$) with $\sigma_{y2} = \sigma_{y1} = \sigma_{yg}/3$, radial heterogeneity of all fields is much higher than for a shorter sample ($R_s/R = 0.4$). Due to the smaller $\sigma_y$ under the culet, reduction in thickness, plastic strain, and $p$ in the central region is much higher for long sample for the same force, which strongly accelerates the PT. The same is true but to a lesser extent for volume fraction $c_0$ and plastic strain $q_0$ averaged over the sample thickness around $r = 0$ versus pressure at the center (Fig. 57,b).

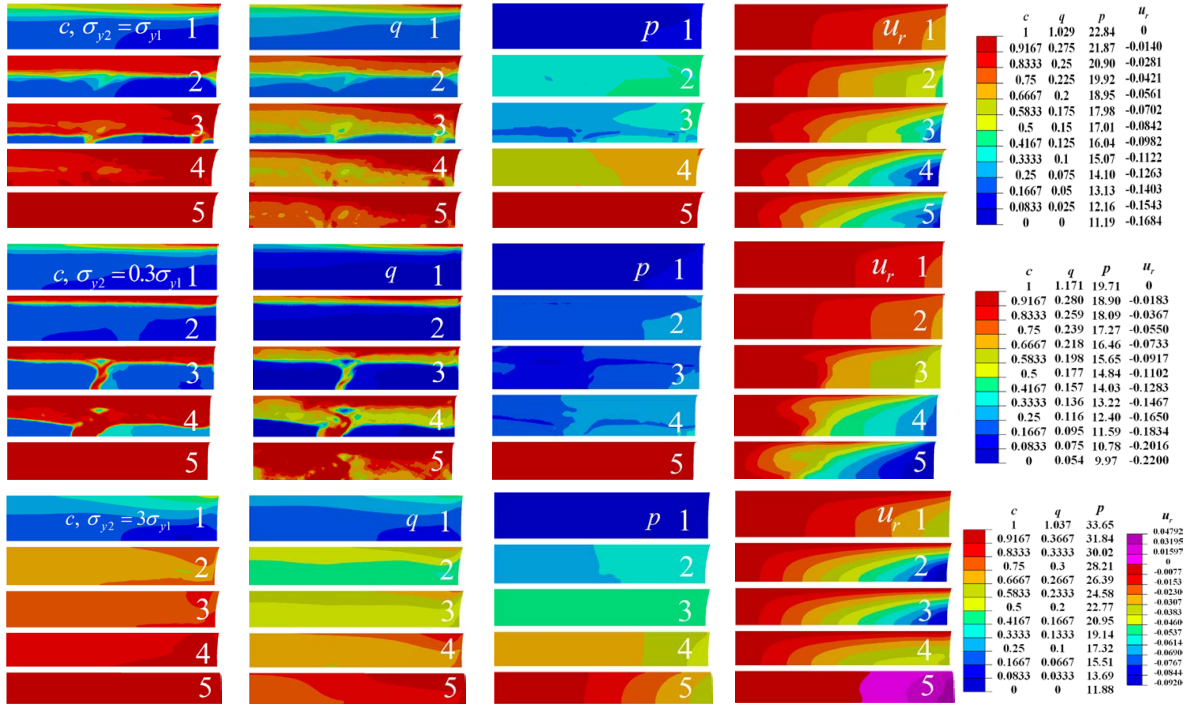

**Figure 56:** (a) Evolution of volume fraction of HPP $c$, accumulative plastic strain $q$, pressure $p$, and radial displacement $u_r$ in a sample compressed within gasket in DAC for $\sigma_{y2} = \sigma_{y1}$ (upper row), $\sigma_{y2} = 0.3\sigma_{y1}$ (middle row), and $\sigma_{y2} = 3\sigma_{y1}$ (lower row) for $\sigma_{yg} = 3\sigma_{y1}$, $k = 6$, $R_s = 2H_0$, $R = 5H_0$, and $H_1 = 2H_0$ [638].

This clearly shows that $p-c$ curve, including pressures for initiating (i.e., obtaining detectable $c$) and completing the PT, does not characterize the PT kinetics but rather the behavior of a sample-gasket system, which determines the $p - q$ loading curve. This explains why and how gaskets with different $\sigma_y$ and geometric parameters lead to different pressures for initiating and completing the PT, in particular, to significant scatter in results from different groups. There is no need for the pressure to grow to run and complete PT; this is just a consequence that in DAC, an increase in $q$ requires an increase in $p$. This could be overcome by using RDAC. While comparing cases with $H_0/R = 0.2$ and $H_0/R = 0.3$, despite the visible local differences with smaller pressure heterogeneity for larger $H_0/R$, for the same $P$ and $p_{max}$, the PT progress is practically independent of the $H_0/R$. Therefore, the mass of the HPP is larger for larger initial and, consequently, final thicknesses. Weaker gasket ($\sigma_{yg}/\sigma_{y1} = 0.2$ instead of 0.3) promotes plastic flow and heterogeneity of all fields. Therefore, it slightly promotes the PT for the same $P$ and $p_{max}$ (Fig. 57,c). However, due to larger thickness, the mass of the HPP is larger for a stronger gasket.

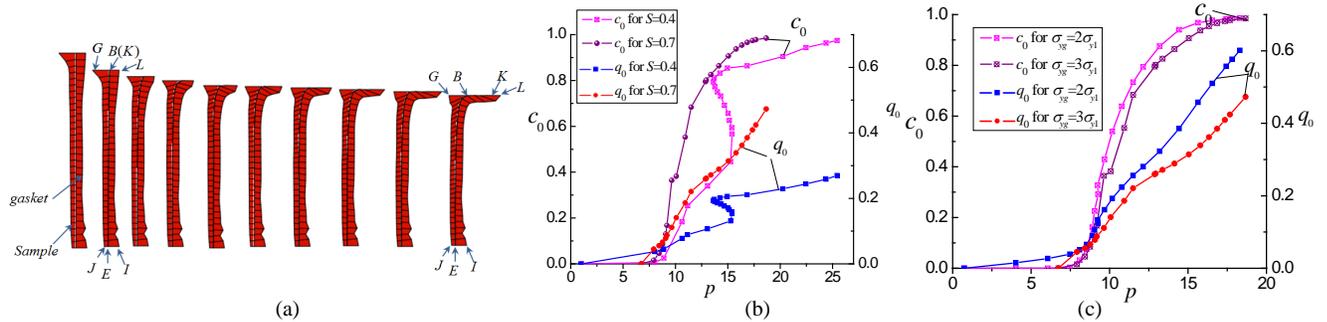

**Figure 57:** (a) Extrusion-based pseudoslip sliding mechanism in DAC. When sliding condition along the diamond is not met for gasket but met for sample, the gasket material slides up with respect to sample and extrudes, making a new contact with a diamond, thus allowing the sample material to slide toward the center. (b) Evolutions of volume fraction $c_0$ of the HPP and plastic strain $q_0$ averaged over the sample thickness around $r = 0$ versus pressure at the center contact surface, for $H_1 = 2H_0$, $\sigma_{y2} = 3\sigma_{y1}$, and different relative sample radii $S = R_s/R$. (c) The same for $H_1 = 2H_0$, $R_s = 0.7R$, and different gasket strengths. Reproduced from [638].

*The effect of gasket on the processes in RDAC.* A similar study for torsion in RDAC was performed in [360]. Based on results in Fig. 58, $p_{max}$ to produce the same $c_0$ (for $c_0 > 0.03$) is essentially smaller in RDAC, especially for large $c_0$. As we discussed multiple times, this does not mean that shear promotes the PT stronger than compression, because the kinetic Eq. (37) used in simulations is independent of the mode of plastic straining. Torsion just allows one to produce $p - q$ path with lower pressure and larger $q$. Still, with $p_\varepsilon^d = 6.75$, due to pressure growth during torsion, a significant amount of HPP in Fig. 58 can be obtained slightly below $p = 10$ only, i.e., there is significant room for optimization of the sample-gasket system. In fact, a simplified estimate used in the design of the gasket/sample system allowed for obtaining a uniform $c_0$ and light pressure increase during torsion for hBN→wBN PT in [74]. This was called the homogeneous pressure self-multiplication effect. For $\sigma_{y2} = \sigma_{y1}$, pressure required to reach the same $c_0$ is lower than for the stronger and weaker HPPs. For $\sigma_{y2} \geq \sigma_{y1}$, while pressure is slightly higher at the center, plastic strain is much larger at the periphery. Thus, $c_0$ is governed mostly by $q$ field and reduces from the periphery to the center. Sliding of sample relative to anvils reduces with increasing $\sigma_y$ at the center, and is almost independent of $\sigma_y$ for $r/R_s > 0.5$. For $\sigma_{y2} \geq \sigma_{y1}$, sliding in the central region is absent. At the boundary with the gasket, torsion angle of a sample is 0.4 of that of an anvil.

For the weaker HPP, a mesh of shear-PT bands is developed due to transformation softening. These results were elaborated and used in [83] for interpretation of the experiments on the olivine-ringwoodite PT as the mechanism of the deep-focus earthquake via shear-PT banding (Section 10.1). For a longer sample and weaker gasket, intensified radial plastic flow and thickness reduction accelerate the PT kinetics. This was used in [81] to intensify the PTs in Si by changing a stronger steel gasket with the weaker Cu gasket. The effect of gasket thickness on plastic flow and PTs is not significant, except at initial stages of torsion. An increase in kinetic coefficient $k$ from 6 to 12 accelerates the PT progress in the sample, despite a slight decrease in $p$ and $q$ due to accelerated volume reduction.

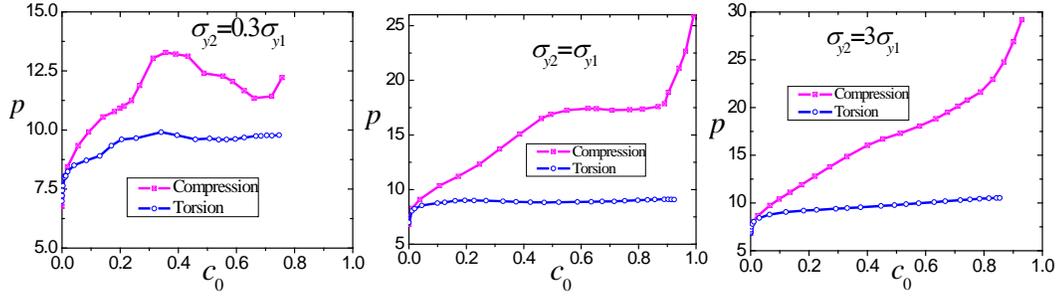

**Figure 58:** Maximum pressure $p$ in the sample required to obtain the volume fraction $c_0$ of the HPP averaged over the sample for compression in DAC and torsion in RDAC for different ratios of the yield strengths of phases [360].

*Fully nonlinear solutions for hBN→wBN in DAC [111] and RDAC [386].* A strict system of equations for large elastic, transformational, and plastic strains was derived and implemented for this study. The isotropic Murnaghan potential was used for sample and Re gasket, and elastic potential for cubic crystals was utilized for diamonds, both third-order in terms of Lagrangian strain. Detailed computational algorithm with consistent tangent moduli was presented in [111] and implemented in the ABAQUS code [211]. Specific features for these problems in comparison with those in [360, 638] are: much higher pressure up to 55 GPa causes finite elastic deformations in BN, Re, and diamond, leading, in particular, to capping of diamonds; $\sigma_y$ is pressure-dependent; very large $\sigma_{y2} = 30\sigma_{y1}$ =9 GPa and $\sigma_{yg}$ =8 GPa at zero pressure; large ratio of shear $G_2/G_1 = 19.2$ and bulk $B_2/B_1 = 10.7$ moduli, and large $\varepsilon_0 = -0.39$. Selected results for DAC are presented in Fig. 59. Due to the strong gasket, pressure in a sample grows much faster than plastic strain. That is why with $p_\varepsilon^d = 6.7$ GPa, detectable $c_0$ starts at ∼12.5 GPa only, and PT does not complete even at 50 GPa. These numbers do not characterize the hBN→wBN, because they can be easily changed by reducing strength and geometric parameters of the gasket or twisting in RDAC. They characterize behavior of the sample-gasket system and its not very smart design. The only way to characterize the strain-induced PT is to determine kinetic Eq. (37) or at least some parameters in this equation, like $p_\varepsilon^d$. Selected results for processes in RDAC are presented in Fig. 60. Due to complex redistribution of the applied force between sample and gasket during torsion, and change in strength of the sample and its diameter, pressure distribution evolves non-monotonously and heterogeneously. In particular, pressure at the center increases (pressure self-multiplication effect despite the large volume reduction) and then decreases during the torsion during the PT, and is much lower than under compression in DAC for same $c$. The adhesion zone between sample and diamond exists at the beginning of the torsion and reduces toward the center during torsion. For rotation angle of an anvil $\varphi = 1.2$, rotation angle of material near the boundary with gasket is more than 2 times smaller; it further reduces to 0.1 in the gasket at $r = R$. Increased sliding in a sample decelerates plastic strain and PT kinetics with increasing $\varphi$. Such a sliding in RDAC was observed experimentally in [245, 250]. An increase in $P$ intensifies the plastic flow and PT. Increase in sample radius from 70 to 100 $\mu m$ for $R = 150\mu m$ significantly increases pressure in a sample and $c$ for compression. During torsion, the difference reduces and disappears at $\varphi = 1$. Thus, with a strong gasket, a larger sample leads to larger mass of the HPP. Experiments on hBN→wBN in [74] within stainless steel gasket were also modeled. Uniform pressure distribution during compression and torsion without PT, and slight increase in pressure (higher at the center) during the PT were obtained, like in experiments [74]. However, plastic strain and $c_0$ at the beginning of PT are very heterogeneous in FEM simulations, while in experiments, a uniform completion of the PT was observed. Note that CEA-FEM for $\alpha - \omega$ PT in Zr with strong experimental verification was presented in Section 7.6.

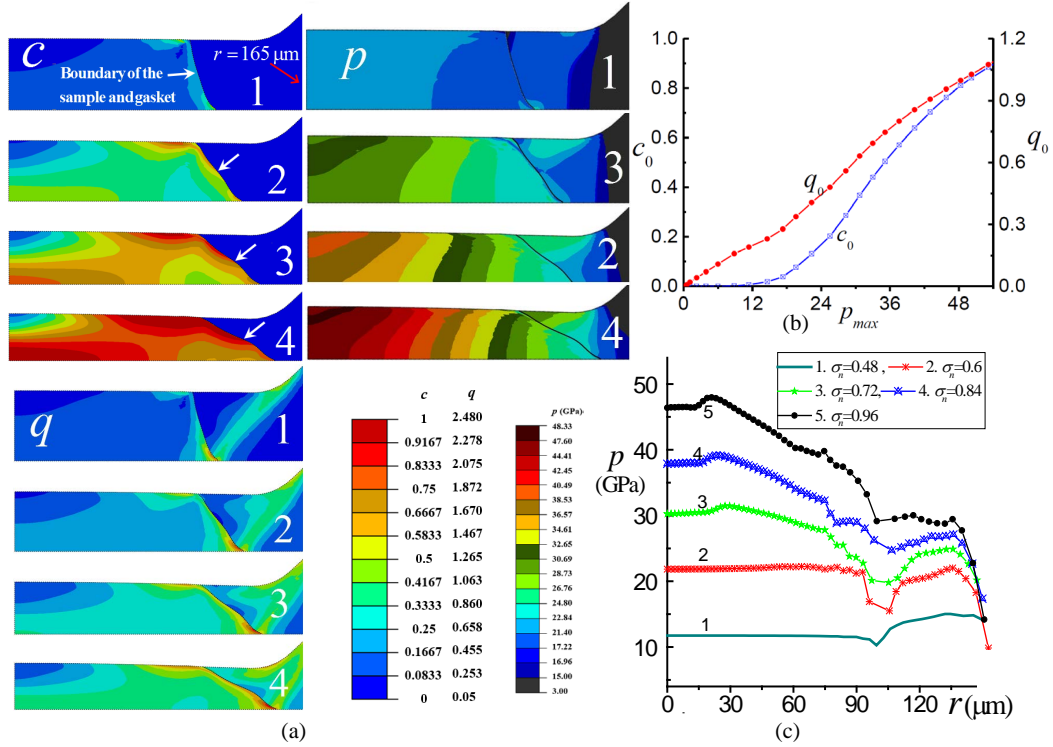

**Figure 59:** (a) Evolution of the main fields during compression of BN sample within Re gasket in DAC (sequence of the panels 2 and 3 is exchanged to the correct one compared to that in [111]). (b) Evolution of $c_0$ and $q_0$ as function of $p_{max}$ in the sample. (c) Variation of the pressure distribution at the contact between the sample and gasket with diamond; initial $R_s = 100 \mu m$. Reproduced from [111].

### 8.4. Summary and perspectives

While simple analytical model describes qualitatively well some of the features of deformation-PT processes, advanced FEM simulations give a much more precise picture. Comprehensive FEM studies of the processes in DAC and RDAC and during HPT have been reviewed here. The nontrivial effects of the ratio of the $\sigma_y$ of phases, contact friction conditions, difference $p_\varepsilon^d - p_\varepsilon^r$, proportionality factor $k$ in the kinetic equation, loading, unloading and reloading, and gaskets with different geometry and strength on stress-plastic strain and $c$ fields were parametrically studied. This led to the development of new nonlinear intuition in understanding of occurring complex and highly nonlinear and coupled processes. Experimentally observed pressure self-multiplication/demultiplication effects, the homogeneous pressure self-multiplication effect during PT, and steps in pressure distributions are reproduced and interpreted. New effects, the self-locking of sliding and extrusion-based pseudoslip mechanism of sliding in DAC when gasket is used, are predicted computationally. It is found that oscillating torsion slightly suppresses PT progress in comparison with monotonous rotation, in agreement with experiments [5, 23].

Simulations suggested a meaningful comparison of the compression and torsion for producing the PT without gasket. Unexpectedly, for $\sigma_{y2} = 0.2\sigma_{y1}$, torsion suppresses PT, reducing the HPP mass and slightly increasing $p_{max}$. For the same $p_{max}$ at the center, torsion leads to larger $c$ and faster completion of the PT for $\sigma_{y2} \geq \sigma_{y1}$ under smaller force. For completing the PT in some region, torsion essentially reduces $p_{max}$. However, for torsion, thickness is much smaller and, for $\sigma_{y2} = 5\sigma_{y1}$, radius of the HPP is smaller; i.e., the total HPP mass is larger under compression. Generally, the PT process in RDAC for material with large $\sigma_{y2}/\sigma_{y1}$ is far from being satisfactory for producing large mass of HPP. Due to the pressure self-multiplication effect, the pressure in the HPP is much higher than $p_\varepsilon^d$, which is not needed for PT but may damage anvil. During the torsion, the radius of the HPP region does not increase but thickness reduces, reducing the total mass of the HPP. Such high pressure does not characterize the transformational properties but rather sample behavior in DAC for the realized $p - q$ path. According to Eq. (37), the PT can be completed at $p$ slightly above $p_\varepsilon^d$, if it can be kept constant during straining. Traditional wisdom that plastic shear promotes PTs/CRs in comparison with compression was developed for large-grain materials. Since plastic straining reduces grain size, increases strength, and reduces $p_\varepsilon^d$ until steady state is reached, and plastic straining in RDAC can be much larger and less related to pressure growth than under compression in DAC, $p_\varepsilon^d$ in RDAC is lower than in DAC. For nanograined materials, steady $p_\varepsilon^d$ is reached and is independent of $\varepsilon_p$, its path and mode, which is used in the current models. That is why other factors dominate the results of simulations.

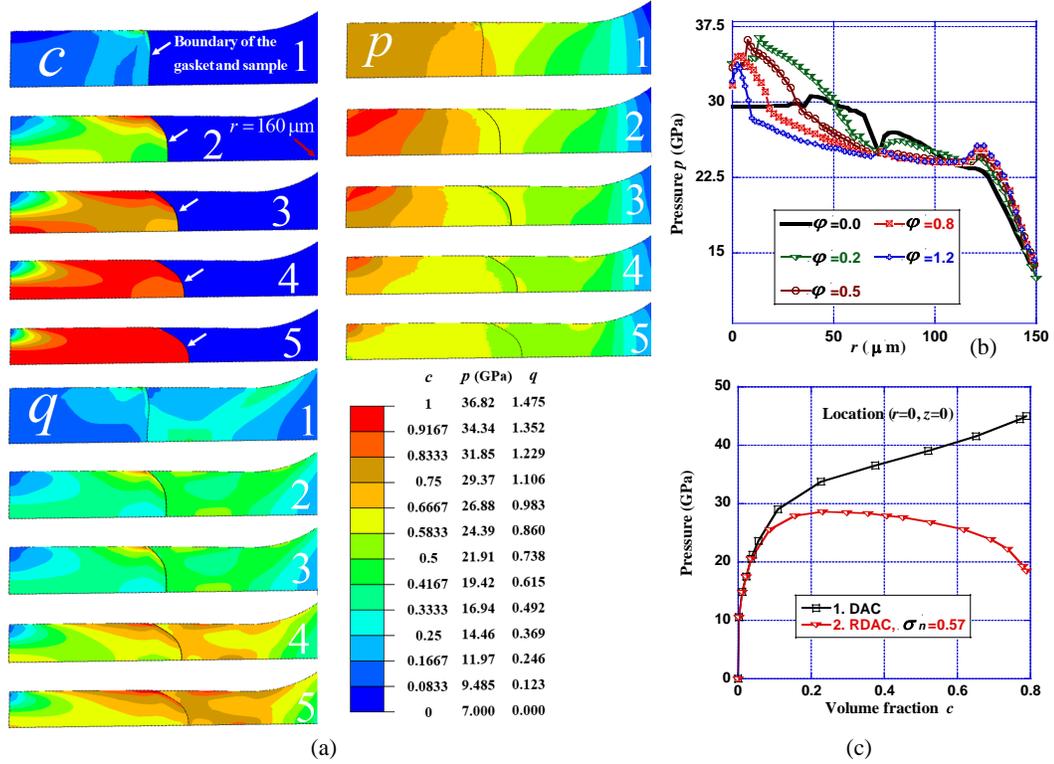

**Figure 60:** (a) Evolution of the main fields during torsion of BN sample within Re gasket in RDAC. (b) Variation of the pressure distribution at the contact of the sample and gasket with diamond; initial $R_s = 70\mu m$. (c) Pressure required to reach the same volume fraction of the wBN at the center of the sample for compression in DAC and torsion in RDAC [386].

To overcome the above drawbacks for $\sigma_{y2} \gg 5\sigma_{y1}$, a strong gasket with optimized geometric parameters could be used. Indeed, close to uniform $p$ and $c$ and small pressure increase during torsion for hBN→wBN PT was obtained experimentally in [74] and confirmed by FEM simulations in [360, 386]. With gasket, torsion significantly outperforms compression in terms of much lower pressure for obtaining the same mass of HPP for all $\sigma_{y2}/\sigma_{y1}$. The effect of gasket strength and geometric parameters on PT process was analyzed. However, for studying deformational-transformational behavior, experimental results without or with weak gaskets are much more informative because they involve much more heterogeneous, i.e., versatile, loading paths. Alternatively (or in addition) to a gasket, diamonds with grooves can be produced and utilized in DAC and RDAC, similar to those in dies for constrained HPT (Fig. 2b). This will limit radial plastic flow, produce more uniform fields in radial direction, and increase thickness of a sample. Diamonds with grooves will also allow for directly reproducing conditions for constrained HPT and study this process in situ.

The effect of contact friction in DAC is also not straightforward. While compare (for $k = 30$ and $\sigma_{y2} = \sigma_{y1}$, Fig. 55) for the same $h/H_0$, the smaller friction is, the smaller $c$ is, as expected. However, for the same force (which is usually controlled in an experiment), the smaller friction is, the larger $c$ is. Comparison for the same $h$ is more informative, because the larger the radius of the HPP region is, the larger its mass is; for the same force, this is not necessarily true because of different $h$. For torsion, sliding accelerates the PT compared to adhesion for all $\mu$ and $\varphi$, but the effect of friction is not monotonous. The best and practically the same results for the radius and thickness of the HPP for $\varphi \geq 0.5$ are for $\mu = 0.3$ and $0.5$. Reduction in thickness with growing $\varphi$ implies that $\varphi \simeq 0.5$ *is an optimal rotation angle maximizing the HPP mass*, and traditional claim, the larger torsion the better, is not supported. Of course, for other combinations of $k$ and $\sigma_{y2}/\sigma_{y1}$, results will be different, but any particular case is instructive.

The most valuable achievement with FEM simulations is the development of CEA-FEM approaches to determine all heterogeneous fields in the sample and material parameters in the material models (Sections 2.4 and 7.6). We believe that this is the most important future direction for quantitative study of strain-induced PTs/CRs. The key point is the development of more advanced physics-based kinetic equations, involving the effect of time and strain rate and combining strain-stress-induced PTs/CRs. With robust models, FEM simulations will be a vital tool for planning and analysing experiments.

## 9. Strain-induced PTs in selected material systems

Detailed in situ studies of strain-induced PTs and microstructure evolution in DAC and RDAC were performed for various material systems. Information for each specific material in this review is spread over the entire text and can be easily found by searching for the material symbol (Zr, BN, etc.).

*Fe.* Thus, for $\alpha - \epsilon$ PT in iron in RDAC [76], the PT start pressure was reduced from 15.4 GPa (hydrostatic experiment) to 10.8 GPa, and the PT was completed below 15.4 GPa. Also, strong pressure self-multiplication effect was observed due to stronger HPP. It is shown that shear stress could reduce the PT pressure by 1 GPa only, i.e., strain-induced contribution is significant. An earlier study in [324] of the effect of the PTM on this PT revealed that the increasing $\sigma_y$ of the PTM decreased the transition start pressure and the interval of the $\alpha - \epsilon$ and $\epsilon - \alpha$ PT. Actually, the lowest PT pressure was 10.6±0.4 GPa with $Al_2O_3$ as PTM, which is much harder than iron. This means that (a) for many of PTM media, the PT in iron was plastic strain-induced rather than stress-induced, (b) plastic straining reduces the PT hysteresis (like in Section 3.2), and (c) strong particles reduce the PT pressure by intensifying the plastic flow (like in Section 3.4).

*SiC.* High-density amorphous phase of SiC was reveled in RDAC at ~30 GPa and rotation angle of $2{,}160^o$ starting with hexagonal 6H SiC [78]. This PT is reversible, i.e., can be detected in situ only. Since amorphization is considered a damage mechanism in brittle ceramics, this finding is of importance, e.g., for armor ceramics.

*BN.* Highly ordered rBN→cBN PT at 5.6 GPa utilizing rotational plastic instability was described in [328] and Section 3.6. Strain-induced hBN→wBN PT in RDAC, its connection to the evolution of the TSF and kinetic modeling, revealing TRIP, and cascading structural changes were presented in [74, 86, 87] and Sections 3.7 and 7.4. While highly disordered nanocrystalline hBN does not transform under hydrostatic pressure even at 52.8 GPa, it transforms under deformation in RDAC at 6.7 GPa [77]. One of the possible mechanisms for this reconstructive PT was discussed in [77] and Section 6.9. Interesting low-pressure hBN→wBN and hBN→cBN PTs in RDAC were reported in [482]. The pressure was evaluated from the shifts of the Raman spectra of the diamond anvil [135], which may have essential error at low pressures due to various assumptions. Also, just single pressure is given while pressure distribution is quite heterogeneous, especially during PT to much harder HPPs. Thats is why specific PT initiation pressures must be double checked.

*Zr.* The most detailed in situ advanced studies are performed on $\alpha - \omega$ PT in Zr and they are reviewed here in detail. This covers studies of the steady $\sigma_y(p)$, microstructural parameters, and $p_\varepsilon^d$ [80, 89] (Sections 2.7 and 3.10), including problem with multiple steady values of each (Section 2.8); developing CAE-FEM approaches to determine tensorial stress and plastic strain fields and $c$ [108, 109] (Section 7.6), determination of quantitative kinetics of strain-induced PT [80, 108, 109] (Sections 7.5 and 7.6), including time dependence of the strain controlled kinetics [82] (Section 7.5.3); new rules for the microstructure evolution during the strain-induced PT [82] (Section 3.11) and pressure-induced PT [88] (Section 4.3).

*9.1. Strain-induced PTs from graphite to hexagonal and cubic diamonds and other superhard phases*

Numerical atomistic simulations predict that graphite can be directly (without catalyst) transformed into a variety of superhard C phases, including hexagonal and cubic diamond, M-, J-, Z-, bct-carbon, fullerenes, onions, and many others [640–645] under different conditions. Many of them (including nanotwinned diamond [646]) were obtained in experiments, but at pressures and temperatures that prevent their industrial applications [647–649]. Under quasi-hydrostatic compression, graphite supposedly transforms to diamond at 70 GPa [318], while this was not strictly proved. Reversible PT to hexagonal diamond (lonsdaleite) under quasi-hydrostatic compression was claimed in [650] in the range of 14-18 GPa for different graphite types and high-pressure apparatuses. Note that we will not discuss here conclusions from [651] that lonsdaleite is a faulted and twinned cubic diamond, which does not exist as a discrete phase. Irreversible PTs from graphite to superhard hexagonal and cubic diamonds were obtained in RDAC at 17-20 GPa [64, 69]. In atomistic simulations in [79], the perfect graphite under hydrostatic pressure losses lattice stability and transforms to a cubic diamond at 250 GPa; uniaxial compression along the graphite [0001] direction at zero lateral strains leads to the PT at 52 GPa. Adding shear stress of 6-8 GPa further decreases the PT pressure to 17-26 GPa; temperature increase from 0 to 300 K further reduces the PT pressure to 15 GPa. These pressures are consistent with experimental values obtained in RDAC in [64], and attempts in [64] to obtain diamonds at lower pressures were not successful. Still, our analytical estimates [100, 102] and PFA simulations [380, 381] of the PT at the tip of dislocation pileup show that the applied pressure can be lower than $p_e$ down to zero.

These simulations encouraged very-low-pressure RDAC experiments in [79]. Surprisingly, they resulted in reversible PT to hexagonal diamond at 0.4 GPa and irreversible PT to cubic diamond at 0.7 GPa (Table 1), both below $p_e = 2.45$ GPa under hydrostatic loading, 2.24 GPa under uniaxial compression, and 1.94 GPa under compression and shear. These are the records of plastic-straining-induced decreasing in PT pressure by a factor of 50 and 100, respectively. The formation and retrieving the cubic diamond was also confirmed by HRTEM and Electron Energy Loss Spectrum. Further torsion-induced pressure growth to 3 GPa led to the finding with HRTEM a *new orthorhombic diamond phase*. The retrieved samples, in addition to cubic and orthorhombic diamonds, contain amorphous phase, fullerenes, and fragmented graphite. These results, if properly scaled up and achieving reproducibility, may be a precursor of novel low-pressure technology of defect-induced diamond synthesis without a catalyst and at ambient temperature (Section 11.2). Strain-induced diamond synthesis also suggests a new mechanism of microdiamond appearance in the Earth's crust at low $p$ and temperature [79] (Section 10.2). PT from amorphous carbon to graphite and diamonds nanoclusters under dynamic shearing of Karelian shungites was revealed at 3 GPa and 1,450-1,650$^o$C [652], i.e., below $p_e$. This finding was interpreted utilizing results in [79].

Hexagonal diamond was obtained under nonhydrostatic compression in DAC at 100 GPa and 400$^o$C in [653, 654] in the external ring of the sample and recovered at ambient conditions. With reference to FEM simulations in [110], it was classified as strain-induced. Glassy carbon in [320] was transformed into a 200 nm band of hexagonal diamond, within which a band of nanocrystalline cubic diamond was formed under nonhydrostatic compression in DAC at room temperature to 80 GPa. These PTs were not observed under hydrostatic compression. Shear band formation and temperature rise above several thousands K was assumed for formation of the diamonds; MD simulations showed that heating to 2,000-3,000 K of hydrostatically and non-hydrostatically compressed carbon at 80 GPa leads to PTs. It seems to us doubtful that significant plastic deformation can be localized and produce significant heating because diamonds are much stronger than glassy carbon. The diamond band looks like a martensitic plate rather than a shear band, in contrast to PT-shear bands in fullerene [72] and olivine-ringwoodite [83, 393].

Note that structural transformations in a few-layer graphene in RDAC were obtained in [655] and shear-induced diamondization of graphene was studied with MD in [613]. A diamond-like structure (but not a diamond) with $sp^3$ bonds was obtained by HPT at 6 and 20 GPa after 50 turns [656]. The potential reason why (in addition to different initial structure, 2D disordering, and impurities in graphite) why cubic diamond was synthesised in [79] at sub-GPa and small shear and was not obtained at 20 GPa and 50 turns [656] or was obtained at 20 GPa and large shear only [64, 69] is the following. It is most probable that in [79], almost spherical nanoscale particles (instead of flakes) with a low 2D disordering were used. Their stochastic orientation suppressed easy sliding along the weak (0001) plane, resulting in large shear stresses. Small particle size reduces $p_\varepsilon^d$ (like for $\alpha - \omega$ PT in Zr [80, 82], Si-I→Si-II and Si-I→Si-III [81], and olivine-ringwoodite [83]), and high 2D ordering does the same (like for rBN-cBN PT [328]). Small plastic shear induces dislocation pileups with strong stress concentrators but not essential 2D disordering, which causes the PT at low pressure. These conditions evidently were not fulfilled in other papers. Detailed experiments relying on the modern understanding are needed to formulate optimal conditions for defect (strain)-induced diamond synthesis with large volume fractions. Recent MD simulations in [612] demonstrated that shearing reduces the PT pressure for nanocrystalline graphite to diamond down to 2.3 GPa at 950 K.

Some metastable phases in graphite, intermediate between the $sp^2$ graphites and $sp^3$ diamonds, were found in DAC with He and Ne PTM in the pressure range of 25-40 GPa in [657] and considered precursors for the diamond formation. It is possible that plastic straining promotes formation of the required intermediate phase at low pressures. Paper [657] also reviews PT in graphites under quasi-hydrostatic loading. In [658], diamond anvil breaks during small torsion of cBN gasket at 100 GPa by producing a conical band of the intermediate 3D covalently bonded carbon phase, which partially transforms to graphite. In [659], strong nanocrystalline graphite sphere compressed in DAC up to local pressure of 100 GPa with large shear stresses, locally transforms to ultrastrong metastable $sp^3$ phases, which transforms to diamond or graphite during decompression. PTs in the plastically predeformed materials during decompression is a special topic that must be studied much deeper. FEM simulation in [660] showed that plastic straining may continue during decompression, which may initiate strain-induced reverse PT or represent a new pathway for reaching desired HPP. Even under hydrostatic loading or nanoindentation, some HPP that do not appear at loading, emerge under unloading, e.g., Si III and XII [81, 441, 442].

*PT graphite-onions.* It was obtained in [65, 73] that in RDAC in the pressure range 55-115 GPa, graphite transforms not into diamond but to more dense onion-like $sp^2$ and $sp^3$ structures. Nano-diamonds within NaCl PTM in DAC transform to similar structures under laser radiation, leading to an estimated 2,000-3,000 K. Above 115 GPa, diamonds again form in RDAC.

*PTs in fullerene $C_{60}$.* Two new phases, of the fullerene, IV and V (the latter was claimed to be harder than diamond), were produced by shear in RDAC [64, 71]. The results were confirmed in [66], but, because of lower pressure, only phase IV was obtained, and it scratched the diamond anvil surface. After RDAC experiment on the fullerene in [72], localized shear-PT bands were observed containing 5 HPPs: nanocrystalline monoclinic, triclinic, and hcp $C_{60}$, linearly polymerized fullerene and polytypes, and fragments of amorphous structures.

Diamond-graphite PT was claimed in [542, 543] under the indentation of diamond by a diamond indenter. While high pressure reduces $w_t$ for this PT, it permits plastic straining without damage. Therefore, strain-induced PT could proceed under high applied pressure, because local pressure, due to new defects, could be much smaller or even tensile for $\phi < 0$. Alternatively, defects produced at high pressure cause PT during pressure release.

**Hydrostatic Compression** (PTM: He)

Micron Si: I $\xrightarrow{13.5\text{ GPa}}$ I+II+XI $\xrightarrow{14.6\text{ GPa}}$ I+XI $\xrightarrow{15.3\text{ GPa}}$ XI $\xrightarrow{16.5\text{ GPa}}$ V $\xrightarrow{0\text{ GPa}}$ XII+III     Retrieved at unloading phases do not appear under pressure

100 nm Si: I $\xrightarrow{16.2\text{ GPa}}$ I+II+XI $\xrightarrow{17.8\text{ GPa}}$ II+XI $\xrightarrow{18.1\text{ GPa}}$ XI $\xrightarrow{19.3\text{ GPa}}$ V $\xrightarrow{0\text{ GPa}}$ XII+III, I+III    Reverse PT to Si-I at ambient pressure

30 nm Si: I $\xrightarrow{14.6\text{ GPa}}$ I+XI $\xrightarrow{19.8\text{ GPa}}$ I+V $\xrightarrow{23.2\text{ GPa}}$ V $\xrightarrow{0\text{ GPa}}$ a    Avoiding Si-II    Pure a-Si at ambient pressure

**Non-hydrostatic Compression**

Micron Si: I $\xrightarrow[96\ \mu m]{2.6\text{ GPa}}$ I+II $\xrightarrow[56\ \mu m]{10.4\text{ GPa}}$ I+II+XI $\xrightarrow[24\ \mu m]{14.5\text{ GPa}}$ I+XI $\xrightarrow[56\ \mu m]{15.2\text{ GPa}}$ XI $\xrightarrow[32\ \mu m]{15.7\text{ GPa}}$ XI+V   (S.S.)
XII+III $\xleftarrow{0\text{ GPa}}$ V $\xleftarrow[16\ \mu m]{16.3\text{ GPa}}$

Obtaining Si-III, ×∞ !

100 nm Si: I $\xrightarrow[40\ \mu m]{0.6\text{ GPa}}$ I+II $\xrightarrow[40\ \mu m]{6.7\text{ GPa}}$ I+II+III $\xrightarrow[0\ \mu m]{10.6\text{ GPa}}$ I+II+XI $\xrightarrow[20\ \mu m]{15.4\text{ GPa}}$ II+XI $\xrightarrow[0\ \mu m]{17.5\text{ GPa}}$ V $\xrightarrow{0\text{ GPa}}$ III   (Cu)   Retrieving pure Si III

30 nm Si: I $\xrightarrow[30\ \mu m]{4.9\text{ GPa}}$ I+II $\xrightarrow[0\ \mu m]{10.4\text{ GPa}}$ I+II+XI $\xrightarrow[40\ \mu m]{13.2\text{ GPa}}$ XI+V $\xrightarrow[0\ \mu m]{13.7\text{ GPa}}$ V $\xrightarrow{0\text{ GPa}}$ I+II&II   (Cu)   Retrieving Si II

Obtaining Si-II, ×∞

**Torsion**

Obtaining Si-III, ×∞    Four-phase mixture

100 nm Si: I $\xrightarrow[70\ \mu m]{0.3\text{ GPa}}$ I+II $\xrightarrow[60\ \mu m]{10°,\ 0.6\text{ GPa}}$ I+II+III $\xrightarrow[0\ \mu m]{9.9\text{ GPa}}$ I+II+III+XI $\xrightarrow[50\ \mu m]{12.2\text{ GPa}}$ I+II+XI   (Cu)   PT pressure × 54 ↓
XII+III $\xleftarrow{0\text{ GPa}}$ II+XI $\xleftarrow[40\ \mu m]{12.6\text{ GPa}}$ XI+V $\xleftarrow[20\ \mu m]{13.9\text{ GPa}}$

100 nm Si: I $\xrightarrow[30\ \mu m]{1.6\text{ GPa}}$ I+II $\xrightarrow[40\ \mu m]{24.6°,\ 4.4\text{ GPa}}$ I+II+III $\xrightarrow{0\text{ GPa}}$ I+II, I+III, I+II+III+XII   Rough anvils   (S.S.)   Retrieving Si-II & Si I+III

100 nm Si: I $\xrightarrow[20\ \mu m]{10.2°,\ 3.5\text{ GPa}}$ I+II $\xrightarrow{0\text{ GPa}}$ I    Rough anvils   (S.S.)   Reverse Si-II → I PT

**Figure 61:** (a) PT sequence in Si between phases I (diamond cubic (S.G: Fd$\bar{3}$m)), II ($\beta$-tin tetragonal (I4$_1$/amd)), III (bcc (Ia$\bar{3}$)), XI (orthorhombic (Imma)), V (simple hexagonal (P6/mmm)), XII (rhombohedral (R$\bar{3}$)), and a (amorphous) for 3 particle sizes under hydrostatic and plastic compressions and torsion in RDAC [81]. The 0 GPa shows phases after pressure release. Different phases separated by & are for different regions in the sample. PT pressures and angles of an anvil rotation are presented above the arrows; distances from the sample center are given below the arrow. The PTM and gasket materials are shown at the right end. The most interesting findings are highlighted and commented on.

## 9.2. Multiple pressure- and strain-induced phase transformations in Si

Semiconducting Si is widely utilised in photovoltaics, microelectronics, integrated circuits, and MEMS/ NEMS. Single-crystal Si is important in high-power lasers. Polycrystalline Si is broadly implemented in thin transistors, solar panels, and very large-scale integration manufacturing. Many important points on PTs in Si have already been discussed above. Thus, drastic reduction in PT pressure due plastic straining for Si-I→Si-II and Si-I→Si-III PTs for particles of 3 different sizes [81] was summarized in Table 1. Section 3.3 provides multiple examples of changing in PT sequence (paths) in Si due to plastic straining and particle size. In particular, this includes separation of Si-II and Si-XI, retaining of Si-II and single-phase Si-III under normal pressure, and achieving reverse Si-II→Si-I PT. In dynamic loading, both retaining Si-II and reverse Si-II→Si-I PT was observed in [521], but quenching started at some elevated temperature and occurred much faster than in static experiments. The criterion for Si-I→Si-II PT in perfect crystal based on lattice instability condition under general $\boldsymbol{\sigma}$ determined by atomic simulations [444, 513, 545] is described in Section 5. Grain/particle size dependence of the pressure- and strain-induced Si-I→Si-II PT was experimentally determined and correlated with the direct and reverse Hall-Petch relationship for the $\sigma_y$ in [81] (Section 6.3). In situ experimental proof of the strong pressure concentration in strain-induced nuclei of Si-II and Si-III, supporting dislocation pileup mechanism was reported in [81] (Section 6.4). MD simulations of amorphization in Si-I under shear at the tip of dislocation pileup and passage/non-passage of different dislocations through different grain boundaries were analyzed in [566, 601] (Section 6.8.1). Formation of a shear band in Si-I via the virtual melting under shear and cyclic PTs between Si-I, Si-IV, and a-Si was revealed with MD in [403] (Section 6.8.1). Theoretical kinetics of strain-induced PTs between Si-I, Si-II, and Si-III was analysed in [378] (Section 7.3). Fig. 61 summarises results from [81]. Nontrivial evolution of the crystallite phase during nonhydrostatic compression was determined as well. EOS for different Si phases of different particle size were presented in [661]. Due to large and anisotropic $\varepsilon_t$, very nontrivial nanostructure, crystallography, and stress relaxation mechanisms even under hydrostatic loading were revealed in [662] by combining Laue diffractions, MD, and extended crystallographic theory.

Note that for nanoparticles and nanograined materials, the effect of a change in surface or grain boundary energy during the PT may play a crucial role in reducing PT pressure and phase selection, e.g., for multiple Si, carbon, and fullerene phases. PFA to surface-induced PTs with mechanics for melting [446, 548] and martensitic PTs [546, 547, 549, 550] revealed many interesting effects and phenomena. Similar works were performed for the void and grain boundary [663, 664]. An interesting effect of the ratio of two scale parameters on various surface- and interface-induced phenomena is reviewed in [551]. However, none of these works includes plasticity, which is required for application to strain-induced PTs.

Metastable semiconducting Si/Ge phases III, IV, and XII, due to their variety of properties, may become promising candidates for electronic and optical materials. Nanostructured phases Si-III and XII were obtained in [665–668] after 10 turns at 24 GPa, while in in-situ experiments in [81] they were obtained at few GPa and less than 1 turn. Low pressure makes it easier to scale up this process for industrial applications.

Similar HPT studies were performed on PTs in Ge [669–671]. Ge possesses some advantages over Si [672]: narrow band gap and high absorption coefficient; larger intrinsic electron mobilities; better quantum-confinement effects for photoluminescence and band-gap controlling the nanostructures; and compatibility with high-dielectric constant materials, allowing integration with modern Si-based semiconductor processing technology. This makes Ge promising for high-speed and optical devices. Nanograined Ge due to unique photoluminescence has a potential application in optoelectronics [669]. Some potential engineering application of the results obtained for Si are presented in Section 11.2.

## 10. Geological and astro-geological problems

*10.1. Resolving riddles of the PT-based mechanism of the deep earthquakes*

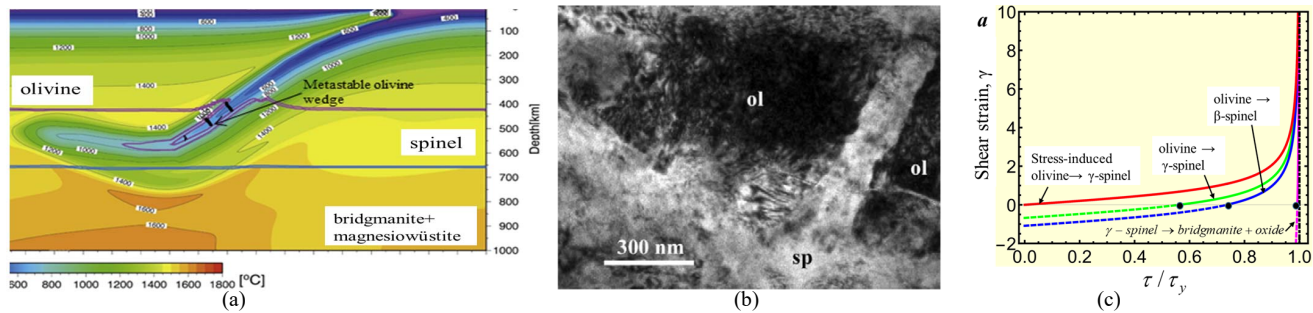

**Figure 62:** (a) Simulation results for subduction of the Pacific plate with metastable olivine wedge underneath Japan with the temperature contour lines [680]. Magenta lines correspond to 1% (upper line) and 99% (lower line) of volume fraction of $\beta$-spinel obtained from the olivine; blue lines correspond to the same for bridgmanite+magnesiowüstite obtained from $\gamma$-spinel. Black lines are for PT-deformation-heating bands. (b) Single PT-shear band in $Mg_2GeO_4$ olivine transforming to spinel [389]. (c) TRIP shear strain vs. $\tau/\tau_y$ for different PTs [393]. Dots designate $\tau_{min}$ for starting strain-induced PT. Curve for the stress-induced PT is similar to that in Fig. 21d.

Deep-focus earthquakes happen at 350-660 km, at 12-23 GPa, and 900-2,000 K, and are more than century-old puzzles in geophysics. The main hypothesis is that they are triggered by instability due to PT from the cold metastable subducted olivine (forsterite) to $\beta$-spinel (wadsleyite) or $\gamma$-spinel (ringwoodite) [389–391, 673–680] (Fig. 62a). However, there are many puzzles: (a) What makes strain rates to increase abruptly from geological ($10^{-17} - 10^{-15}/s$) to fast seismic ($10 - 10^3/s$) rates? (b) Can this occur without PT, i.e., with the help of alternative mechanisms analyzed in [393, 679, 681–685])? (c) What causes the metastable olivine that avoids PTs for over a million years, to abruptly transform to spinel on the second time scale and produce strong seismic waves? (d) What is the relationship between shear-dominated seismic signals with dilatation-dominated $\varepsilon_t$? (e) How to rationalize the distribution of the earthquake frequency with depth, and why do earthquakes cease to occur below 660 km? There are numerous publications devoted to each of these problems and their combinations. However, the only relevant paper for the current review is paper [393], in which the key point is that olivine→spinel PT is considered like *plastic strain-induced* (rather than pressure/stress-induced), which was not previously done for any geological material. The multi-step mechanochemical mechanism of coupled thermoplastic flow and PT in a shear-PT band is presented in Fig. 63.

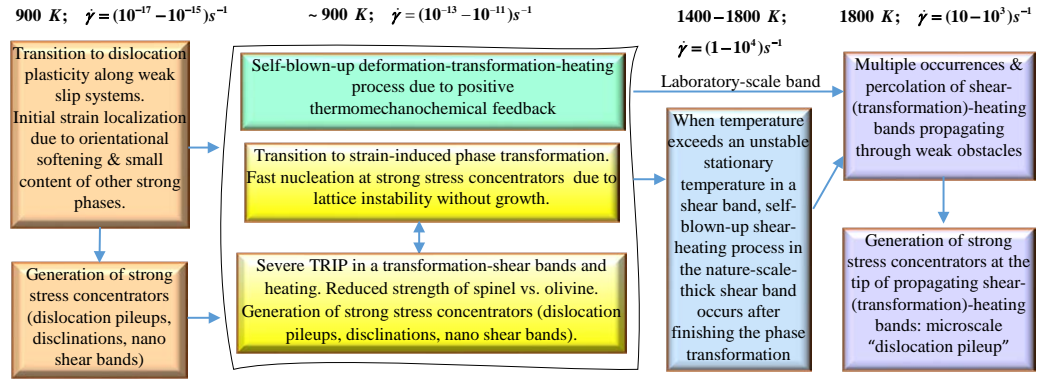

**Figure 63:** The multi-step mechanochemical mechanism of coupled thermoplastic flow and PT in a shear-PT band causing high strain and PT rates and temperatures [393]. Initial temperature and shear rate for each stage are placed above the scheme.

The first stage of strain localization is caused by the transition from diffusional creep to dislocation plasticity along the favorably oriented [001](010) slip systems with the critical shear stress much lower than for all other slip systems. This leads to orientational softening. Also, localization occurs along the path with a relatively small portion of other strong phases, like pyroxene, garnet, and others. In a shear band without PT, temperature evolution equation has two steady solutions. One of them with the temperature close to the initial one is stable. The The second solution with a much higher temperature is unstable, and if exceeded, unlimited accelerated plastic flow can occur. Thus, (a) localized growth in temperature and strain rate in a thin band is impossible without PT; high temperature leading to melting based on adiabatic approximation is wrong; (b) if due to additional heating source (e.g., PT and TRIP) temperature exceeding the unstable stationary temperature is reached, unlimited heating up to melting, and a drastic increase in the strain rate is possible.

Localized plastic flow generates dislocation pileups, which produce strong stress concentrators and strain-induced olivine-spinel PT. Fast nucleation at tips of dislocation pileups occur without essential growth, producing nanograined spinel (like in experiments), which experimentally determined $\sigma_y$ is several times smaller than that of the olivine. For strain-controlled kinetics of the type of Eq. (40), transformation rate is proportional to the strain rate. Indeed, multiplying Eq. (40) by $dq/dt$, we obtain that $dc/dt \sim dq/dt$. Thus, with very high strain rate, PT rate can be also very high, and PT can complete within seconds, see below.

Next, as it was mentioned in Section 3.7, based on an advanced 3D analytical solution for coupled straining-PT-heating problem in a shear-PT band with application to PT olivine-spinel, $\varepsilon_0$ causes severe TRIP and self-blown-up deformation-PT-heating caused by positive thermomechanochemical feedback between plasticity, strain-induced PT, and TRIP. This large TRIP shear rationalizes shear-dominated seismic signals. despite dilatation-dominated $\varepsilon_t$. In nature, after completing PT, the temperature in a shear band exceeds the unstable steady value, above which further progressive straining, with increased strain rate and heating, takes place due to shear flow without PT. In the laboratory, due to much smaller scale, heating does not practically occur. Similar processes should occur in multiple PT-shear bands and, after completing the PT, shear bands that find pathways through weak phases and may percolate. They produce strong stress concentrators at their tips, reproducing a larger-scale counterpart of dislocation pileups and repeating the above processes in larger volumes and magnifying the generated seismic waves.

It was obtained in [393] that the final strain rate $\dot{q}$ in a shear band at temperature $T$ is proportional to that before localization at the initial temperature $T_0$. Consequently, the final $\dot{q}$ is distributed with depth proportionally to the initial $\dot{q}$ before localization. This explains the correlation between distribution of the seismicity with the depth in the transition zone and $\dot{q}$ before localization found in [686, 687]. The most probable causes for the lack of seismicity below 660 km were explained in [393] by not meeting some of the conditions for the processes shown in Fig. 63. They include too high $T_0$, which promotes diffusional instead of dislocation deformation mechanism, reduction in the strain rate jump due to thermoplastic strain localization, and possibility of the PT without the necessity of strain-induced contribution.

Conceptual experimental proof of the strain-induced olivine-spinel PT was obtained in [83]. While this PT was never observed at room temperature under hydrostatic or quasi-hydrostatic compression, the olivine - $\gamma$-spinel PT was obtained in RDAC at 15-28 GPa and plastic strain of 2.3-9 within tens of seconds. This PT was reversible under pressure release, which demonstrates the possibility of misinterpretation of postmortem study of the deformation-PT processes in olivine. The $p_\varepsilon^d$ linearly decreases with growing $q$, dislocation density, and microstrain, and reducing crystallite size. FEM simulations of the sample twisting in the dRDAC reproduce some important experimental features and imply the development of the shear-PT bands, as required in the theory in [83]. Also, rules of the evolution of the crystallite size, microstrain, and dislocation density during SPD and PT are obtained.

Thus, plastic strain-induced olivine-spinel PT can occur within tens of seconds at strain rates of $3.5 \times 10^{-3} - 3.3 \times 10^{-2}/s$ instead of millions of years even at room temperature, which resolves the main puzzles. In the Nature, $\gamma$ reaches $10^6$ in a shear band [389], which can be provided with the help of TRIP and additional plastic flow after completing PT. If we assume the PT time of 10 s, even shear strain in the range $10 - 10^4$ provides the desired strain rate of $1 - 10^3/s$, observed in Nature. But for such strain rates, PT rates will be reduced proportionally and proportionally smaller shear strain is required. Results in [83, 393] demonstrate a potential for novel strain-induced PT mechanism versus a traditional $p - T$-induced mechanism for a quantitative (up to an order of magnitude) explanation of the deep earthquakes. They also suggest reconsidering the experiments required to study the PT-based mechanism of the deep-focus earthquakes. The temperature must be low enough to avoid pressure/temperature-induced PT, and SPD should be applied instead. Also, prior plastic straining at normal or low pressure, leading to significant grain refinement and an increase in dislocation density, can be utilized to essentially reduce PT pressure and is required for PT plastic strain.

10.2. Appearance of microdiamonds in the low-pressure cold Earth crust

In geology, diamonds are considered to be crystallized in the Earth's mantle from various melts according to graphite-diamond equilibrium line, namely, at $T > 900 - 1200^0$C and $p > 4$ GPa, i.e., below 100-150 km [688, 689] (Fig. 64a). They are delivered to the surface by deep-source volcanic eruptions. Independently, microdiamonds were discovered in the low-$p\&T$ continental crustal and non-kimberlite bearing rocks regions, e.g., in the Kokchetav Massif of Russia [690–692], the Dabie Shan mountains of China [693], and the Rhodope Mountains of Greece [694], and the Variscan French Massif Central [695]. These microdiamonds were interpreted as having a metamorphic origin, i.e., rocks from the continental crust (a thin shell of the outside of Earth) were subducted below 100 km into a high-pressure, high-temperature region and, after appearance of diamonds, uplifted to the surface by rapid exhumation to preserve them [690–696]. However, as mentioned in [688] "the presence of diamonds with a much more restricted carbon isotope signature than eclogitic diamonds does not support this origin." There are other significant inconsistencies in subduction-uplifting mechanism [697, 698].

Metastable growth of diamond, along with graphite in low-pressure and temperature crust from a carbonaceous fluid is advocated in [698] and references. It was stated in [697] that "Kumdikol microdiamond deposit has close association with ductile to semi-ductile strike-slip shear zones, rich in graphite," see Fig. 64b. With reference to metastable growth of coesite within quartz (and private communication with H.W. Green) and aragonite in limestone under plastic deformation, it was suggested in [697] that diamond could be formed in Earth directly from graphite (without fluids) due to high shear stresses and plastic deformations. Later, Dobrzhinetskaya rejected her own hypothesis [696] because of a lack of supporting data that graphite can transform to diamond at low pressure and shear stresses and strains. Subduction and uplifting hypothesis and crystallization of diamond from carbon-bearing fluids (COH, Si, and carbonate melt) at 6-9 GPa and 900-1200$^0C$ is currently the major mechanism [696]. Direct graphite-diamond PT is also assumed in Earth, but at $p =$12-25 GPa and $T = 2,000^0C$.

As it is discussed in [79], finding of graphite-diamond PT under shear at 0.7 GPa and 300 K opens possibility to explore an alternative pathway for diamond formation in the interior of the Earth, which we will elaborate here. Carbon sources (e.g., graphite, limestone ($CaCO_3$), marble, dolomite ($CaMg(CO_3)_2$), organic carbon, or particles of plant debris) in the low-$p\&T$ Earth crust may directly transform to diamond by PTs/CRs under large $\gamma$ during, e.g., tectonic plate collision, fracture and shearing events in shallow earthquakes. This could be, in particular, microdiamonds without any inclusions but graphite in Kokchetav massif [696] and from Jagersfontein (South Africa) [648] that cannot be explained by formation at high $p\&T$ in the mantle For example, the microdiamonds (10 to 250 $\mu m$) in the Kokchetav Massif could be produced within shear-PT bands during tectonic rifting in the Mesozoic Era. The microdiamonds in the Dabie mountain could be produced during the collision of the Yangtze and the North China plates. Such strain-induced PTs may resolve the puzzle that microdiamond sites lack a sedimentary lithosphere. Even assuming subduction, microdiamonds may appear at much lower depth due to generated shear strains. Shear-induced diamonds may also appear away from the craters made by asteroid or meteorite impact, where pressure and temperature are low but shear banding occurs. Focused in situ RDAC experiments and corresponding simulations should not only address different types of graphite, but also some other geological carbon-containing minerals (limestone, marble, dolomite or organic carbon) with addition of carbon-containing fluids, are required for more justified conclusions. The results in [79] have already been used in [652] for interpretation of PT from amorphous carbon to graphite and diamond nanoclusters under dynamic shearing of Karelian shungites at estimated 3 GPa and 1,450-1,650$^oC$, i.e., below $p_e$.

The fluxes and formation of carbon phases under geological environments are also crucial in geophysics [699, 700] to track path and history of the carbon from which diamond is formed and study geodynamics. For instance, diamond formed in nature through various geological processes provides information about processes occurring deep into the Earth interior [699, 700]. In particular, coesite in Dabie Shan mountains was deduced to be produced at 2.8 GPa and 740$^0$C; presence of diamonds in this coesite changed these numbers to 4 GPa and 900$^0$C [693]. The formation of diamond at tectonic boundaries and in the Earths interior provides hints to trace geological historical activities, but the understanding is far from satisfactory. The exploration of new conditions to promote carbon transformations is thus both of practical and fundamental importance in geophysics. Also, knowledge that diamonds may be produced by large shears at low $p\&T$ offers novel strategies for their seeking in the crust.

Generally, the most probable mechanisms of diamond formation are determined by analysis of C isotopes, surrounding materials, inclusions, and atomic elements in diamonds. However, in most of mechanisms, graphite is formed along with diamond. These graphite particles could be transformed to diamonds by shear in the low $p\&T$ crust. Thus, geological evidence of the formation of diamonds under high $p\&T$ conditions does not contradict that neighboring diamonds could be shear-induced directly in the crust.

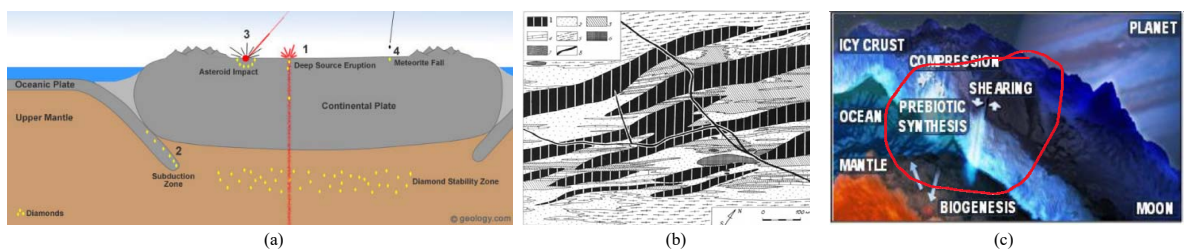

**Figure 64:** (a) 1 - diamonds crystallized in the Earth's mantle and delivered to the surface by deep-source volcanic eruptions. 2 - microdiamonds in subduction zones that will be uplifted to the surface. 3 and 4 - diamonds due to asteroid and meteorite impact [701]. (b) Tectonic map of the region of the Kumdikol microdiamond deposits with the strike-slip shear zones. 1 - severely deformed lenses with a high microdiamonds content [697]. (c) Potential mechanochemical synthesis of prebiotic species in the shear zone of icy crust of solar system planets and moons [630].

*10.3. Mechanochemical origin of life beyond the Earth*

There is a hypothesis [630, 631, 702, 703] that tidal and tectonic deformations result in the athermal mechanochemical (i.e., pressure-shear-induced) synthesis of prebiotic species (required to start life) in the shear bands in icy crust of solar system planets and moons (such as Europe, Enceladus, Triton, and Pluto). This hypothesis recently appeared in a NASA public announcement [630]. Similar life-building mechanochemical pathways might have occurred on early Earth. Life-building proteins are polypeptides that contain more than approximately fifty amino acids, in particular, glycines ($NH_2CH_2COOH$). While there is some DFT support [703] of forming polypeptides under $p > 10 GPa$ and shearing of glycine in "virtual RDAC", none confirming experimental results in RDAC were obtained. Some promising results have been obtained using ball milling [631]. Generally, organic reactions are strongly promoted by shear under pressure but were not studied in situ in RDAC [14, 16, 377] (see Section 3.4). Also, finding the glycine amino acid in comets and meteorites resulted in a hypothesis that their impacts may make a contribution to delivery and polymerization of amino acids in the early Earth for protein generation [629]. This hypothesis was checked in [632] utilizing HPT. While glycine was not polymerized, it was partially decomposed to ethanol at 1 and 6 GPa and $q = 120$. This demonstrated a possible origin of ethanol found in comets.

## 11. Summary and future directions

*11.1. Mechanism, general rules, and maximization of the promoting mechanochemical effect of SPD*

The main experimental results and mechanochemical phenomena and summary of their interpretations are presented in Section 3. Each Section ends with a summary and perspectives, so there is no need to repeat them here. Rephrasing a known quote about education popularized by Einstein, *knowledge is what remains when everything learned is forgotten,* we ask the question: What are the main messages from the current state of the art in the field that we must remember? We suggest three of them.

1. What is the main mechanism that promotes the PTs/CRs? Strain-induced PTs are controlled by nucleation at some defects, producing a strong stress concentration. The only strong stress concentrator that can reduce the PT pressure by one to two orders of magnitude (like in experiments) is the tip of the dislocation pileup. All components of the stress tensor at the tip are proportional to number of dislocations, which could exceed 10-20 for nanograined materials. While shear stresses equal to $\tau_y$ produce relatively minor contribution to the PT conditions, shear/deviatoric stresses at the tip of the dislocation pileup are limited by the theoretical strength, which is $(10-100)\tau_y$. Since stresses reach the lattice instability limit (determined by atomic simulations under stress tensor), instability occurs and causes barrierless nucleation of HPP(s), dislocations or twins, or some combination of these processes. Such highly-deviatoric and versatile stress states cannot be obtained in bulk and may lead to new phenomena and phases that are inaccessible without plastic deformation and to drastic reduction of the applied pressure required for PT. Plasticity produces a dual action: it promotes PT by creating stress concentrators but these stress concentrators may generate dislocations and twins, relaxing these stresses and suppressing PT. Some combinations of normal stresses and shears were found for which dislocation activity near the tip of the dislocation pileup, even if exists, does not hurt the PT, i.e., the PT wins a competition over dislocations. For realization of this mechanism, grain boundary must be non-penetrable by dislocations. Since dislocation pile-up produces both compressive and tensile stresses of the same magnitude in different regions, i.e., it simultaneously promotes both direct and reverse PTs, which is taken into account in the microscale kinetic equation. Also, some intermediate metastable phases may appear, which may promote appearance of the desired HPP during further loading or even unloading. Promotion of the PT by dislocation pileup should be much stronger if the PT is caused by phonon instability rather elastic instability since reduction in elastic moduli lead to the proportional reduction in stress concentration.

There are various experimental results consistent with this mechanism, e.g., lack of the effect of plastic straining on $p_\varepsilon^d$ for fcc-hcp PT in Pb above the recrystallization temperature [573] and reduction in direct PT pressure and increase in reverse PT pressure for strain-induced fcc-fcc PTs in Ce in [64] despite zero transformation shear. The most convincing are two strongest confirmations from [81] (Sections 6.3 and 6.4):

(a) For strain-induced Si-I→Si-II PT, $p_\varepsilon^d$ decreases with decreasing grain/particle size in the region of direct Hall-Petch relationship for $\sigma_y$, and then increases with further reducing grain/particle size in the region of the inverse Hall-Petch relation. For pressure-induced PT, $p_\varepsilon^d$ increases with decreasing particle size.

(b) At the very beginning of the strain-induced Si-I→Si-II and Si-I→Si-III PTs in 100 nm particles and Si-I→Si-II PT in micron particles, pressure in small HPP regions was 5-7 GPa higher than in Si-I, despite reduction by 2.7 GPa for pressure-induced PT due to 22% volume decrease. For 30 nm Si particles, because of suppressed dislocation activity in the region of the inverse Hall-Petch relationship, the pressure increase was 1.4 GPa only.

Growth of the nuclei is governed by the same rules as for *stress-induced PTs* coupled with evolving dislocations. Equilibrium configuration of the HPP is controlled by the transformation work and phase equilibrium both at interfaces, Eq. (30), and for the entire polycrystal.

2. What are the main rules of SPD, microstructure evolution, and strain-induced PTs? As it was described in Sections 3.10, after some critical level of plastic deformation, the yield strength, dislocation density, crystallite size, and microstrain, as well as the minimum pressure for strain-induced PT reach steady values (not necessarily simultaneously), which are independent of the plastic strain tensor $\boldsymbol{\varepsilon}_p$, including its mode and entire path $\boldsymbol{\varepsilon}_p^{path}$; dislocation density, crystallite size, and microstrain are also independent of pressure and pressure path; at monotonous and quasi-monotonous loading, materials also deform like isotropic and perfectly plastic with fixed surface of perfect plasticity $\varphi(\boldsymbol{\sigma}) = 0$. Also, for the strain rate from 0.004 to $20/s$ for some metals, steady hardness, shear stress, dislocation density, grain, and crystallite size are practically independent of the strain rate. However, there are multiple steady states caused by different SPD techniques or loading processes, which require further comprehension (Section 11.2).

As discussed in Section 3.11 that during $\alpha - \omega$ PT in Zr, the crystallite size and dislocation density (and, consequently, microstrain) in $\omega$-Zr depend on volume fraction of $\omega$-Zr $c$ only, and are independent of pressure, plastic strain tensor $\boldsymbol{\varepsilon}_p$ and its path $\boldsymbol{\varepsilon}_p^{path}$, rough- and smooth-DA, and different initial $\alpha$-Zr microstructures. The same rule is valid for $\alpha$-Zr, but this function of $c$ differs for rough-DA and smooth-DA ,and possesses essential scatter for $0.4 < c < 0.5$. During olivine-spinel PT, the crystallite size of spinel is steady and independent of $\boldsymbol{\varepsilon}_p$, its path $\boldsymbol{\varepsilon}_p^{path}$, and phase fraction of spinel.

As written in Section 7.6, a strain-controlled kinetic equation for PTs was derived, motivated by dislocation pileup-based mechanism and nanoscale theory. Using developed CEA-FEM approach, for broad variety of loading paths in 4D space of plastic strain components and $p$, it was found that kinetic equation (41) is independent of $\boldsymbol{\varepsilon}_p$, its mode and path and deviatoric stress tensor, pressure $p$ - plastic strain $q$ paths, and also straining before initiation of the PT provided that the stationary microstructure is reached. Similar to the finding for microstructural parameters and $\sigma_y$, there are multiple steady values of $p_\varepsilon^d$, and the main problem is to determine for which classes of $\varepsilon_p$, $\varepsilon_p^{path}$, $p$, and $p^{path}$ the PT starts with each specific $p_\varepsilon^d$, and for which of $\varepsilon_p$, $\varepsilon_p^{path}$, $p$, and $p^{path}$ classes there are jumps from one steady $p_\varepsilon^d$ to another.

The above rules drastically simplify the general theory, developing and experimental calibration of the specific models for plasticity, microstructure evolution, and strain-induced PTs, and can govern optimization of strain/defect-induced synthesis of nanostructured HPPs. To some extent, a similar rule for microstructure evolution was found in situ in [88] for pressure-induced $\alpha - \omega$ PT in Zr (Section 4.3 and Fig. 23): the microstrain, average crystallite size, and dislocation density in $\omega$-Zr for $c < 0.8$ are functions of the phase fraction of $\omega$-Zr only, which are independent of the pressure and plastic strain before the PT.

3. How to maximize promoting effect of the plastic straining on PT? Since promoting effects of SPD are more pronounced for nanograined materials with the grain size in the region of Hall-Petch relationship for $\sigma_y$, one has to prepare nanograined material with steady microstructure by SPD under normal pressure and apply plastic straining (even not necessary severe) under high pressure to promote PTs/CRs [80, 82, 83, 85]. Alternatively, one can use nanoparticles, producing high-angle grain boundaries under compression within a weak gasket, to maximize plastic strain [81]. Important point is to eliminate the stress-induced or PT-induced [430] grain growth during the PT/CR [618, 619]. For steady nanostructures, the same minimum pressure for strain-induced PT/CR can be reached during compression in DAC as under torsion in RDAC, significantly increasing community that works on strain-induced PTs/CRs. However, to complete the PT/CR under low pressure, torsion in RDAC is much more effective because pressure growth during torsion can be avoided or minimized. For RDAC, strong gasket can be used to minimize reduction in thickness during torsion. Rough diamonds intensify plastic flow, grain refinement, and PTs. Alternatively (or in addition) to gasket, diamonds with grooves can be produced and utilized in DAC and RDAC, similar to those in dies for constrained HPT (Fig. 2b). This will limit radial plastic flow, produce more uniform fields in radial direction, and increase thickness and volume of a sample. Diamonds with grooves will also allow directly reproducing conditions for constrained HPT and study this process in situ. It was also found in situ in [82] that PT during plastic straining is much more effective way to reduce the grain size and increase the dislocation density in HPP than plastic straining alone.

To increase chances to retain metastable HPP at normal pressure with the potential using them in engineering applications, one has to apply extra plastic deformation at high pressure to reach steady microstructure, $\sigma_y$, and, consequently, athermal threshold $K$, which according to Eq. (17) increases pressure hysteresis for pressure-induced PTs. Then pressure release path that minimizes or avoids plastic deformations must be found to avoid/minimize strain-induced reverse PT. Very fast unloading is one of the ways to minimize plastic strain and quench HPP.

*11.2. Future directions*

Next question is: What now? Or, in German, *Was nun*? This was the name of a pub (*eine Kneipe*) not far from the University of Hannover, Germany, which stayed open until morning. Here, students and assistants summarized all global problems that they had almost solved in the previous, already closed, pubs. We will do a similar job here.

1. Despite the clear success in developing, calibration, and quantitative verification of the time-independent strain-controlled kinetic Eq. (41), kinetic description of strain-induced PTs/CRs requires complete reconsideration. As optimistically mentioned by Sholem Aleichem, "Do not be upset. It was never so bad that later did not get worse." It was well accepted, starting from works on strain-induced PTs in steels [98, 105, 106] that strain-induced PTs occur by nucleation at some strain-induced defects with very limited growth. That is why when plastic strain stops, PT does not occur, and time does not matter. However, it became clear from PFA, MD, and CAC simulations (Sections 6.6, 6.7, and 6.8) that strain-induced PTs occur via two traditional processes: nucleation at strain-induced defects and growth away from the defect till steady state. The latter is clearly stress-induced PT, and steady state is governed by the same local and global thermodynamic equilibrium conditions (Sections 6.6.3 and 6.7) like for stress-induced PT. Thus, nucleation and growth must be treated separately and then integrated, and time must be included in description of the stress-controlled growth. This was confirmed in the first experiments for $\alpha - \omega$ PT in Zr [82]. Studying the effect of strain rate could shed further light on the kinetics. Speculating, for a very slow strain rate, plastic straining can be accommodated by diffusional creep instead of dislocation/twinning plasticity (Section 10.1), and no stress concentrators are available for strain-induced nucleation, i.e., PT is fully stress-controlled. For very high plastic strain rates, there is time for nucleation but no time for growth during the straining. One may expect very small volume fraction of the strain-induced HPP during straining followed by essential stress- and time-controlled growth on much longer time scale, even at fixed plastic strain. Thus, *generalizing current kinetic equation by allowing time-dependent growth governed by stress-induced PT kinetics, and broad range of strain rates is an outstanding fundamental problem with various applications from very slow geological processes to traditional HPT rotation rates to dynamic ball-milling processes and processes in shear band and shock loading.* This can be done by further development of multiscale modeling, from atomistic to macroscale, especially CAE-FEM approaches and developing new experimental methods to measure some parameters (stresses, elastic strains, displacements, etc.) to be used for calibration and verification of the models. Utilization of physics-informed machine learning methods should lead to further progress.

Next step is to include microstructure evolution in PT/CR kinetics (see examples in [74, 83] and Sections 7.4 and 10.1) and PT/CR kinetics in microstructure evolution (see examples in [82] and Section 3.11). Even the entire kinetics for strain-induced PTs/CRs could be more physical if some relevant microstructural parameter ($d$, $\rho_d$, or concentration of TSF for layered materials) is used as a time-like parameter instead of $q$, until these parameters reach a steady state. Allowing for defect distribution instead of just density may lead to qualitative leap in the kinetic description. The most important part plays the strongest stress concentrators, i.e., the tail of defect distribution, similar to that for fracture. The developed kinetic and macroscale models can be used for FEM-based planning, interpreting, and post-processing laboratory and technological experiments, and finding methods of controlling PTs and microstructure by SPD. The development of more advanced CAE-FEM approaches will continue transforming the good out of the bad, i.e., the challenges related to strongly heterogeneous fields in DAC/RDAC into opportunity to generate and process big data from a single experiment.

Note that all these future developments do not diminish the role of Eq. (41) because it describes very nontrivial experiments well. *All complex theories have a metastable nature: they look satisfactory for a while until new experimental results/phenomena force to transform the theory to more comprehensive ones.*

2. Traditional atomistic studies of phase equilibrium and instabilities of phases and search for new phases under pressure have to be extended for general stress tensor (see examples in [444, 513, 523, 535, 536, 539–541]).

3. Exploratory study in Section 6 on interaction between discrete dislocations, grain boundary, and PTs for mostly model systems with PFA, has to be brought to a qualitatively higher level. They should include twinning, 3D formulation, and performed for real materials. Material parameters for PFA can be determined from atomistic simulations and fitted to a macroscopic experiment, realizing the scale bridging. Finding methods for in situ study of the nucleation mechanisms is crucial. As an example, HPPs of strong materials like Si can be found under uniaxial compression of small-scale samples (e.g., nanopillars) using in situ TEM. Similar experiments for a polycrystalline nanopillar can be used for in situ detection of the dislocation pileups and corresponding nucleation mechanism or revealing new mechanisms for strain-induced PTs. Also, the concept of the defect phases, including dislocations, is currently under extensive development [715]. Developing multiscale (from atomistic to macroscale) dislocation pileup phase and transformation diagrams and dynamics, and their interaction with different phases, dislocations, grain boundaries, and twins' "phases" under high stress tensor represents a challenging and rewarding task.

4. As it was discussed in Section 2.8 and [82, 89, 90], there is a key outstanding problem in the theories of plasticity, microstructure evolution, and strain-induced PTs under SPD: for which classes of plastic strain $\varepsilon_p$, its path $\varepsilon_p^{path}$, pressure $p$, and pressure path $p^{path}$ does a material possess one steady value of dislocation density, crystallite size, microstrain, yield strength (yield surface $\varphi(\boldsymbol{\sigma}) = 0$), and minimum pressure for strain-induced PT, and for which of $\varepsilon_p$, $\varepsilon_p^{path}$, $p$, and $p^{path}$ classes are there jumps from one steady microstructure, yield strength, and minimum pressure for strain-induced PT to another? Some steps toward its solution were suggested in [90]. Solution to this problem will allow one to develop general quantitative theories for SPD, microstructure evolution, and strain-induced PTs/CRs, and explain why different SPD techniques lead to different microstructures, and how to control them.

5. While we did our best to simplify experiments, FEM simulations, and their coupling by preparing sample with steady microstructure, $\sigma_y$, and $p_\varepsilon^d$ by preliminary SPD at normal pressure, during the PT, microstructure and, consequently, $\sigma_y$ in each phase significantly evolved (Section 3.11). This is neglected in the current macroscale models. The distribution of the yield strengths of phases could be measured using XRD peak broadening or shifting (Section 2.5 and Fig. 9b,c [108, 208, 226–228, 233]). Obtained data can be used to develop a model for $\sigma_y(p, \rho_d, d)$ for each phase. This could bring the modeling techniques for multiphase materials with PTs to a qualitatively new level.

6. *Potential technological applications, controlling PT and microstructure.*

Obtained fundamental results may serve as a nanoscale basis for developing various technologies. Thus, defect-induced material synthesis can be used for obtaining known and new nanostructured HPPs under moderate and low pressures. Instead of increasing pressure, one can increase plastic strain and successively fill the material with strain-induced defects (mostly, dislocation pile-ups) with high concentrators of all components of stress tensor. Such a unique, highly deviatoric (limited by the theoretical strength only) stress state at the limit of lattice stability may lead to new phases inaccessible under different conditions and to drastic reduction in PT pressure. This is similar, to some extent, to the successive local melting of materials by a moving laser beam in the additive manufacturing processes.

For example, results for Si (Section 9.2) [81] can be used for advanced modeling and optimization of surface treatment (cutting, turning, polishing, scratching, etc.) of strong brittle semiconductors and ceramics and designing regimes of ductile machining by involving PT to ductile (Si/Ge-II and a-Si/Ge) phases [400, 401, 627, 704–707]. This allows to mitigate/avoid surface damage that is usually eliminated using chemical polishing. Current modeling does not treat PTs in Si as strain-induced and uses PT pressures more than 10 times higher than in experiments in [81]. This is because the atomistic simulations are constrained to very small grain sizes, where the inverse Hall-Petch relationship is valid (Fig. 31a). One has to take into account the effect of the grain size, especially in view that for 100 nm particles, the PT pressure reduced to sub-GPa, including for Si-III, which may also appear.

Since traditional PT pressures for Si and Ge exceeded 10 GPa, they were neglected in most of applications. Due to very low transformation pressures for strain-induced PT in [81], one must be warned that these PTs may occur during friction and wear, e.g., in NEMS and MEMS, and must be considered or avoided.

Due to the above potential applications, finding an economic way of controlling PTs, grain size, and dislocation density evolution to produce and retain the desired phases and nanostructures during straining and after pressure release are of great importance, and results for Si may be instructive for other material systems. Si and Ge can be considered as model materials for other strong and brittle semiconductors ($Si_xGe_{1-x}$, GaAs, InSb, GaSb, etc.), graphite-diamonds (cubic, hexagonal, orthorhombic, etc.), and similar BN systems with large anisotropic transformation strains.

Results for graphite-to-diamond PT at sub-GPa in Section 9.1 [79] may be a precursor of novel low-pressure technology of defect-induced diamond synthesis without heating and a catalyst. Traditional technologies of diamond synthesis utilize pressure of 4.5-5 GPa at 1,500-1,600 K and molten metallic catalyst [708–710] (similar for cBN). The initial mixture consists of graphite and metal catalyst in weight proportion from 1.5 to 1. After diamond synthesis, there is residual graphite (20%-50%), and diamond is extracted from the mixture with metal and graphite using special technology. Due to the high temperature, additional heating and thermal isolation elements of the high-pressure chamber have a volume ∼2 times larger than the mixture of graphite and metal. Thus, the effective volume fraction of diamonds with respect to chamber volume is relatively small. Chemical cleaning of diamonds from catalyst and other involved materials is the most expensive and complex stage in the technology [711, 712]. From the experience at the ISM, the author remembers that reduction in pressure from 4.4 to 4.3 GPa by utilizing nanoscale catalyst particles leads to essential economic effect due to increase in number of cycles that WC-6%Co dies/matrices can withstand before fracture.

Due to very large transformation strains, PTs in Si, Ge, BN, and C may absorb energy an order of magnitude larger than shape memory alloys; in addition, TRIP absorbs a comparable energy level. If one inserts inclusions of these LPPs in strong brittle materials, like armor ceramics, PT in these inclusions will absorb a large amount of energy and provide some plasticity. Appearance of stronger HPPs (like cBN, wBN, and diamonds) will lead to further strengthening. This is a high-pressure counterpart of transformation toughening phenomenon in ceramics.

PT during plastic straining is much more effective (i.e., at much lower pressure and smaller plastic strain) way to reduce the grain size and increase the dislocation density in HPP than plastic straining alone (Section 3.11 [82]). These results may lead to nanocomposites with optimal functioned properties.

Independence of steady microstructure, $\sigma_y$, and $p_\varepsilon^d$ of the loading process implies that the obtained equations are valid not only for HPT, but for any SPD processes, like extrusion, rolling, ball milling, shot pinning, friction, surface processing, etc.

7. Various phenomena, like PT induced by rotational plastic instability, shear banding, TRIP/RIP, and turbulent-like plastic flow, intensifying plastic flow also intensify strain-induced PTs/CRs and grain refinement. Some of them produce positive mechanochemical feedback, like TRIP/RIP, defect production, and PTs/CRs producing cascading structural changes [74, 388] and self-blown-up deformation-PT-TRIP-heating process in a shear band [393]. After better understanding, they can be utilized for strain- or defect-induced material synthesis.

8. Reconsideration of the many geological and astro-geological problems (like in Section 10) utilizing the concept of strain-induced rather than pressure/stress-induced PTs and CRs promises to revolutionize these fields. Important progress could also be obtained by including TRIP/RIP and self-blown-up deformation-PT/CR-heating phenomena for consideration of the processes in shear-transformation bands.

9. The first non-trivial results for distribution of friction shear stress between diamond and W [143] and Zr [108], obtained with combined experimental-FEM and CEA-FEM approaches, and an emerging method to measure contact normal and shear stresses [139] offers the tool for the development of the foundation of high-pressure tribology.

**Acknowledgments**

The author acknowledges the support of the U.S. Army Research Office (W911NF2420145), the U.S. National Science Foundation (DMR-2246991 and CMMI-2519764), and Iowa State University (Murray Harpole Chair in Engineering). The simulations were performed at Extreme Science and Engineering Discovery Environment (XSEDE), allocation TG-MSS170015. Experimental work was performed at HPCAT (Sector 16) at APS. HPCAT operations are supported by DOE NNSAs Office of Experimental Science. The APS is a U.S. Department of Energy (DOE) Office of Science User Facility operated for the DOE Office of Science by Argonne National Laboratory under Contract No. DEAC0206CH11357. The author greatly appreciates and enjoyed collaboration with all his coauthors of papers cited here, including graduate students, postdocs, and faculty and researchers from different organizations.